\documentclass[12pt]{article}
\usepackage{amsfonts,amssymb,epsfig,amsmath,mathtools,xcolor}
\usepackage{color}

\usepackage{mathrsfs}
\usepackage{graphicx}
\usepackage{subcaption}
\usepackage{hyperref}
\usepackage{multirow}
\usepackage{array}

\renewcommand{\baselinestretch}{1.2}

\newcommand{\tr}{\textrm{tr}}
\newcommand{\nn}{\nonumber}
\newcommand{\calC}{\mathcal{C}}
\newcommand{\eps}{\epsilon}

\newcommand{\q}{\tilde{q}}
\newcommand{\psit}{\tilde{\psi}}

\begin{document}

\makeatletter \@addtoreset{equation}{section} \makeatother
\renewcommand{\theequation}{\thesection.\arabic{equation}}
\renewcommand{\thefootnote}{\alph{footnote}}

\begin{titlepage}

\begin{center}
\hfill 
\\

\vspace{2cm}

{\Large\bf BPS phases and fortuity in higher spin holography}

\vspace{2cm}

\renewcommand{\thefootnote}{\alph{footnote}}

{\large Seok Kim$^1$, Jehyun Lee$^1$, Siyul Lee$^2$ and Hyunwoo Oh$^1$}

\vspace{0.7cm}

\textit{$^1$Department of Physics and Astronomy \& Center for
Theoretical Physics,\\
Seoul National University, Seoul 08826, Korea.}

\vspace{0.2cm}

\textit{$^2$Instituut voor Theoretische Fysica, KU Leuven,
Celestijnenlaan 200D, 3001 Leuven, Belgium.}

\vspace{0.7cm}

E-mails: {\tt seokkimseok@gmail.com,
ljs9125@snu.ac.kr, siyul.lee@kuleuven.be, 
hyunwoo1535@snu.ac.kr}

\end{center}

\vspace{1cm}

\begin{abstract}

We study the BPS states of $U(N)_k\times U(1)_{-k}$ vector Chern-Simons theory on a sphere
at weak coupling $\lambda=\frac{N}{k}\ll 1$, dual to an AdS$_4$ higher spin gravity. 
Higher spin currents are well known to be anomalous at $\lambda\neq 0$.
We show that these non-BPS higher spin particles form multi-particle `BPS bounds' at low energy, 
and interpret them as a primordial form of small black hole states. We also construct a new 
heavy BPS operator at $N=2$. We study the BPS phases of this system from the large $N$ index 
at Planckian `temperatures'.
The deconfined saddles at high temperature exist only above a threshold, similar to 
the BTZ black holes. The low temperature saddles are given by novel 2-cut eigenvalue
distributions. Their phase transition involves subtle issues like the holomorphic anomaly 
and the background independence, whose studies we initiate. In particular, we obtain a lower 
bound on the critical temperature by studying the eigenvalue instantons.

\end{abstract}

\end{titlepage}

\renewcommand{\thefootnote}{\arabic{footnote}}

\setcounter{footnote}{0}

\renewcommand{\baselinestretch}{1}

\tableofcontents

\renewcommand{\baselinestretch}{1.2}

\section{Introduction}

Studies of string theory in extreme conditions often provide insights into its 
fundamental aspects. Among others, higher spin gravity theories have been explored as 
the tensionless limit of string theories. In particular, 
some simple higher spin gravity theories in AdS$_4$ \cite{Fradkin:1987ks,Vasiliev:1990en} 
are known to be holographically dual to large $N$ vector models
\cite{Klebanov:2002ja,Sezgin:2002rt}.

In this paper, we study a 3d supersymmetric vector model gauged by the Chern-Simons fields, known as 
the ABJ vector Chern-Simons theory \cite{Aharony:2008gk}. This theory has $U(N)_k\times U(1)_{-k}$ gauge group and Chern-Simons levels, preserving $\mathcal{N}=6$ superconformal symmetry. This theory at large
$N$ 't Hooft limit is suggested to be dual to a supersymmetric higher spin gravity containing a coupling
constant  $\lambda=\frac{N}{k}$ \cite{Chang:2012kt} (see also \cite{Aharony:2011jz,Giombi:2011kc}). 
We study the BPS states of the field theory on $S^2\times \mathbb{R}$ at small nonzero $\lambda$ 
which might be regarded as
the `BPS black hole microstates' of this rather exotic gravitational system. 
Although there are known solutions of the AdS$_4$ higher spin gravity 
\cite{Didenko:2009td,Iazeolla:2011cb,Bourdier:2014lya}, it is highly 
unclear to which extent they physically behave like black holes. Rather, following the strategy of
\cite{Shenker:2011zf} (see also \cite{Gutperle:2011kf,Castro:2011fm} for similar studies 
in AdS$_3$), we rely on thermodynamic criteria to study the black hole like physics from 
field theory. We consider interacting theories because turning on and increasing $\lambda$ 
moves the traditional higher spin theory towards string theory and exhibits 
interesting physics.

In AdS string theories, black holes appear in two branches: large and small black holes. They have
positive/negative specific heats, respectively, and play different roles in characterizing 
the thermodynamics of quantum gravity in various ensembles. (Large/small black holes have BPS
analogues, characterized by positivity/negativity of certain susceptibility.) 
Large black holes are dual to the deconfined phase of the field theory \cite{Witten:1998zw}. 
Since deconfinement 
is rather universally expected in gauge theories at high temperature, one may  
identify the `large black holes' from QFT as the deconfined phase. 
On the other hand, small black holes seem to be less universal in large $N$ gauge 
theories.\footnote{It is suggested that small black holes are characterized by partial 
deconfinement \cite{Hanada:2018zxn,Berenstein:2023srv} in matrix QFT.} 
In fact, we will find large $N$ 
thermodynamic saddles which qualitatively behave like large black holes, 
but none which look like small black holes.

We have two major motivations to study this model. The first one is technical. 
In supersymmetric matrix field theories with AdS string duals,
the BPS states are roughly classified into graviton and black hole states. 
The former is well understood, while finding the black hole 
states with large matrices is hard: see
\cite{Chang:2022mjp,Choi:2022caq,Choi:2023znd,Chang:2023zqk,Budzik:2023vtr,
Budzik:2023xbr,Chang:2023ywj,Choi:2023vdm,Chang:2024zqi,deMelloKoch:2024pcs,Gadde:2025yoa,Chang:2025mqp} 
for recent progress. Similar studies with vector-valued matters are relatively simpler.
Also, the large $N$ thermal partition function/index are 
easier to study with vector matters than with matrices. We will take advantage of 
these technical simplifications to study the novel BPS operators and their large $N$ 
thermodynamics. Second, the physics of BPS states in the vector model 
is in a sense richer in that they have more subtle quantum structures.

In string theory, the entropy of large charge BPS states exhibits nontrivial (black hole like) 
behaviors when the charge $E$ scales like the inverse Newton constant $G^{-1}$ ($\sim N^2$ 
for 4d $\mathcal{N}=4$ Yang-Mills, and $\sim N^{\frac{3}{2}}$ for ABJM \cite{Aharony:2008ug}). 
The entropy $S(E)$ is a nontrivial function at the same order, 
$S(E)=\frac{f(EG)}{G}$ where the function $f$ does not have explicit $G$ dependence.
The transition between the large/small black holes (in the microcanonical ensemble) 
also happens in this region. At $E\sim \mathcal{O}(1)\ll \frac{1}{G}$, the entropy is independent 
of $G$, coming from the ideal gas of low energy gravitons.

In the ABJ vector model at $\lambda\ll 1$, $S(E)$ of BPS states exhibits new features beyond 
the graviton gas over a wider range of charges. 
To explain this, first note that at the single trace level, the only BPS operators are those 
in the graviton multiplet. Other single-trace operators belong to multiplets that contain 
higher spin currents and become anomalous at $\lambda\neq 0$
\cite{Maldacena:2011jn,Chang:2012kt}.
However, at the multi trace level, we find multi-particle BPS bound states 
beyond gravitons, even at low energies $E\sim O(1)$ when $S(E)$ is still microscopic. 
That is, some multi-particle states of non-BPS higher spin particles 
acquire binding energies and saturate the BPS bound. We use the term bound states to denote negative interaction energies, although there is no 
sharp notion of spatially bound wavefunctions.
The underlying algebraic structure is the trace relations of large $N$ vectors. 
(`Trace' and `trace relations' respectively mean an inner product of two vectors and 
the relations among multi-trace operators.) 
At larger charge $E$ scaling in $N$, $S(E)$ will see the $N$ individual `quarks' 
of the vector model, exhibiting the deconfined behavior. In the grand canonical ensemble, with 
the inverse `temperature' $\beta$ conjugate to $E$ fixed, the phase transition happens 
at $\beta\sim N^{-1}$ (at which $E\sim N^3$). We expect the high temperature BPS phase 
to be dominated by the BPS states constructed using the trace relations of 
finite $N$ vectors. We find one such cohomology in the $N=2$ theory, illustrating 
their existence.

To summarize, while nontrivial physics beyond the graviton gas happens in a rather 
definite region $E\sim \frac{1}{G}$ in string theory (matrix QFT), it happens 
in a wider range of charges $1\lesssim E\lesssim N^3$ in the higher spin gravity 
(vector CS theory). In string theory, we find three regions of $E$, 
each dominated by the graviton gas, small black holes and large black holes. 
In the ABJ vector model, we find two distinct regions, the low energy region 
dominated by gravitons and the higher spin BPS bounds, and the high energy region
accounted for by the new heavy operators.

To better understand the possible meanings of this spectrum, it is helpful to know 
the connection between the ABJ vector model and the SCFT with a string theory dual. 
The ABJ vector Chern-Simons model can be generalized to the $U(N)_k\times U(N^\prime)_{-k}$ 
quiver gauge theory. This theory holographically interpolates the higher spin theory and 
string theory as follows \cite{Chang:2012kt}. First, taking $N,k\rightarrow\infty$ 
(with $0\leq \lambda\leq 1$) and keeping $N^\prime$ fixed, one obtains a higher spin 
theory with the fields charged in the bulk $U(N^\prime)$ gauge field. 
The 't Hooft coupling of this bulk gauge interaction is $\lambda_B\equiv\frac{N^\prime}{N}$, 
and $\lambda$ is an extra bulk interaction parameter. 
As $\lambda_B$ grows, the $U(N^\prime)$ interactions 
are suggested to bind the higher spin particles into strings. 
Then in the limit $N,N^\prime,k\rightarrow\infty$ with 
$\lambda\gg 1$ fixed and $N-N^\prime<k$,
one finds the weakly coupled type IIA string theory on $AdS_4\times \mathbb{CP}^3$ 
as the holographic dual. Changing the couplings 
$(\lambda=\frac{N}{k},\lambda_B=\frac{N^\prime}{N})$ from $\lambda=0$, $\lambda_B=0$ 
to $\lambda\gg 1$, $\lambda_B=1$, the holographic gravity dual 
interpolates the weakly-coupled higher spin theory and the weakly-coupled string theory.

Deforming the higher spin theory into string theory by increasing $\lambda=\frac{N}{k}$ and
$\lambda_B=\frac{N^\prime}{N}$, we expect that the multi-particle BPS bounds of 
the non-BPS higher spin particles appear at higher delayed energies which scale in $N^\prime$. 
This is because the trace relations of large $N$ vectors which enabled these BPS bounds 
are replaced by the trace relations of $N\times N^\prime$ matrices. 
We conjecture that these delays will split the low energy branch of 
the vector model into the graviton region and the small black hole region at large enough
$N^\prime$.

We shall also study the large $N$ saddle points of the index and attempt to 
determine the BPS phase structures of the vector model. In matrix-valued QFTs, 
one had to make various guesses for the saddle points: see 
\cite{Cabo-Bizet:2018ehj,Choi:2018hmj,Benini:2018ywd} and references thereof.
In the vector model, one can derive the large $N$ saddles rather systematically.
As mentioned above, nontrivial large $N$ saddles and their phase transitions happen 
at $\beta\sim N^{-1}$ in the index. At fixed $N\beta$ of order $1$, we find two  
distinct phases at lower and higher temperatures. 
We construct the saddles for these two phases and discuss
aspects of the phase transition. We only partly clarify the nature of the 
transition, due to various technical/conceptual subtleties of the multi-cut eigenvalue 
distributions with filling fractions. We find that various fundamental issues 
of quantum gravity,  such as the background independence, holomorphic anomaly, etc., 
arise in this simple model.

The rest of this paper is organized as follows. 
In Section \ref{sec:2}, we study the local BPS operators at weak coupling $\lambda\ll 1$ in
the cohomology formulation. In particular, we consider the cohomologies of a classical
interacting supercharge $Q$ whose spectrum is in 1-to-1 map to the BPS states at the 2-loop level 
$\mathcal{O}(\lambda^2)$. We study three classes of cohomologies throughout three subsections,
namely BPS gravitons, higher spin BPS bounds, and a `heavy' cohomology at $N=2$. 
In Section \ref{sec:3}, we study the large $N$ approximation of the index and discuss its physics 
including the phase transition. 
We discuss the relation between the nature of the phase transition and the microstates which trigger it, for the index as well as the partition function of the vector model. Section \ref{sec:4} concludes with remarks and future directions. 
Appendix \ref{sec:appA} explains the counting and the constructions of BPS operators. 
Appendix \ref{sec:appB} explains the large $N$ saddle point calculations for the index.
Appendix \ref{sec:appC} explains the similar calculations for the free partition function.

\section{Cohomologies of ABJ vector model}\label{sec:2}

We consider the $U(N)_k\times U(1)_{-k}$ ABJ Chern-Simons-matter theory at 
$k\gg 1$. This theory has $\mathcal{N}=6$ superconformal symmetry. 
Among the symmetry generators in $OSp(6|4)$, 
the Poincare supercharges $Q_{IJ\alpha}$ (with $IJ$ antisymmetric and $I,J=1,\cdots,4$) 
and the conformal superchares $S^{IJ}_\alpha$ are Hermitian conjugate to 
each other in the radial quantization: we shall often write $S=Q^\dag$.
We also define $\overline{Q}^{IJ}_\alpha\sim\frac{1}{2}\epsilon^{IJKL}Q_{KL\alpha}$.
Some important algebra is schematically given by 
\begin{equation}
  \{Q_{IJ\alpha},\overline{Q}^{KL}_\beta\}\sim \delta_{[I}^{[K}\delta_{J]}^{L]}P_{\alpha\beta}\ , \quad 
  \{Q_{IJ\alpha},S^{KL\beta}\}\sim \delta_\alpha^\beta\delta_{[I}^{[K}\delta_{J]}^{L]}H
  -2\delta_\alpha^\beta \delta^{[K}_{[I}{R^{L]}}_{J]}
  -\delta_{[I}^{[K}\delta_{J]}^{L]}J_\alpha^{\beta}\ ,
\end{equation}
where $P_{\alpha\beta}$ and $J_{\alpha\beta}$ are the translation and rotation generators on $\mathbb{R}^3$, respectively, and ${R^I}_J$ (satisfying ${R^I}_I=0$) are the $SU(4)_R\sim SO(6)_R$ R-symmetry generators. 
The BPS states that we study in this paper are annihilated by $Q\equiv Q_{34-}$ and 
$S=Q^\dag\equiv S^{34-}$, making them $\frac{1}{12}$-BPS. From the algebra 
\begin{equation}
  \{Q,Q^\dag\}=E-({R^3}_3+{R^4}_4+J)\equiv E-\frac{R}{2}-J\ , 
\end{equation}  
the energies (scaling dimensions) of BPS operators are 
given by $E={R^3}_3+{R^4}_4+J$. Note that ${R^3}_3+{R^4}_4 = -{R^1}_1-{R^2}_2$ 
from the traceless condition of $SU(4)$. 
In the matrix form, ${R^3}_3$ and ${R^4}_4$ in the fundamental representation 
are given respectively by
\begin{equation}
  {R^3}_3={\rm diag}({\textstyle -\frac{1}{4},-\frac{1}{4},\frac{3}{4},-\frac{1}{4}})\ ,\quad 
  {R^4}_4={\rm diag}({\textstyle -\frac{1}{4},-\frac{1}{4},-\frac{1}{4},\frac{3}{4}})\ .
\end{equation}
Therefore, ${R^3}_3+{R^4}_4={\rm diag}(-\frac{1}{2},-\frac{1}{2},\frac{1}{2},\frac{1}{2})$, 
or $R={\rm diag}(-1,-1,1,1)$. The supercharges that commute with $Q$ and $Q^\dag$ are 
$Q_{13+}$, $Q_{14+}$, $Q_{23+}$, $Q_{24+}$ and their Hermitian conjugates. 
The bosonic generators that commute with $Q$ and $Q^\dag$ are 
$SU(2)\times SU(2)\sim SO(4)\subset SU(4)_R$ and $Sp(2)\subset Sp(4)$. 
In the former, the two $SU(2)$ act on 
$I=1,2$ (call $i=1,2$) and $I=3,4$ (call $a=1,2$) respectively. 
The full subalgebra of $OSp(6|4)$ which commutes with $Q,Q^\dag$ 
is $OSp(4|2)$. The BPS operators preserving a definite pair $Q,Q^\dag$ 
of supercharges form $OSp(4|2)$ multiplets.

The ABJ theory has the following $U(N)_k\times U(1)_{-k}$ bifundamental scalars and fermions, 
\begin{equation}
  \Phi_I=(\phi_i,\tilde{\phi}^{\dag}_a)\ ,\quad 
  \Psi^I_{\alpha}=(\psi^i_{\alpha},\tilde\psi^{\dag a}_{\alpha})\ .
\end{equation}
Also, there are $U(N)_k\times U(1)_{-k}$ Chern-Simons gauge fields $A_\mu$ and $A_\mu^\prime$. 
We first consider those fields which are BPS (with respect to $Q=Q_{34-}$)
in the free limit $k\rightarrow \infty$. Forming all possible gauge invariants of 
these free BPS fields, we will have a complete list of BPS gauge-invariant operators 
in the free theory. Then we turn to the theory with large but finite $k$, and consider 
the subset of free BPS operators which remain BPS at the leading nontrivial order in 
$\frac{1}{k}$. It turns out that nonzero anomalous dimensions can appear from the 
$2$-loop level, $\sim \frac{1}{k^2}$. The spectrum of the 2-loop BPS states
is the main interest of this section.

The free BPS letters are given by (see Tables 1 and 2 of \cite{Bhattacharya:2008bja}
for their quantum numbers) 
\begin{equation}\label{free-BPS-letter}
  (D_{++})^j \, \phi^{\dag}_i\ , \quad (D_{++})^j \, \tilde\phi^{\dag}_a\ , \quad 
  (D_{++})^j \, \psi_{i+}\ , \quad (D_{++})^j \, \tilde\psi_{a+}\ .
\end{equation}
$D_{\alpha\beta}$ with $\alpha,\beta=\pm$ are the three derivatives, 
which will be promoted to covariant derivatives in the interacting theory.
In the classical interacting theory, the $Q$ transformations of (\ref{free-BPS-letter}) 
no longer vanishes. Note that the full supersymmetry transformation rules for $Q_{IJ\alpha}$ 
can be found in, e.g. \cite{Gaiotto:2008cg,Hosomichi:2008jb,Bandres:2008ry}. 
Below, we will only use a subset of these rules, with suitably rescaled fields:
\begin{equation}
(q_i,\tilde{q}_a)\sim (\phi^\dag_i\ ,\tilde\phi^\dag_a)\ , \quad
(\psi_i, \tilde\psi_a)\sim (\psi_{i+}\,\tilde{\psi}_{a+})\ , \quad
D\sim D_{++}\ .
\end{equation}
The $Q$ transformations of these free BPS letters
in the interacting theory are can be written
(after rescaling the letters to absorb the $\frac{1}{k}$ factors) as
\begin{equation}\label{Q-free-BPS}
  Qq_i=0\ , \quad
  Q\tilde{q}_a=0\ , \quad
  Q\psi_i= (\tilde{q}_a\cdot q_i) \tilde{q}^a\ , \quad 
  Q\tilde\psi_a= q^i(\tilde{q}_a \cdot q_i)\ .
\end{equation}
$q_i$, $\tilde\psi_a$ are in the fundamental representation of $U(N)$ (column vectors),
while $\tilde{q}_a$, $\psi_i$ are in the antifundanemtal representation (row vectors).
Pairs of fundamental/antifundamental fields are contracted by inner products
that we denote with a dot.
$Q$ acting on the covariant derivative $D$ is schematically given by
\begin{equation}
  [Q,D] \sim \lambda_+ \ ,
  \qquad \text{where}~~~\lambda_+ = q^i\otimes\psi_i-\tilde{\psi}_a\otimes\tilde{q}^a\ .
\end{equation}
More precisely, 
\begin{eqnarray}
Q(Dq_i) &=& \lambda_+ \cdot q_i - q_i v ~, \nn\\
Q(D\tilde{q}_a) &=& v \tilde{q}_a - \tilde{q}_a \cdot\lambda_+ ~, \nn\\
Q(D\tilde\psi_a) &=& \lambda_+ \cdot\tilde\psi_a + \tilde\psi_a v 
+ D(q^i(\tilde{q}_a \cdot q_i))~, \nn\\
Q(D\psi_i) &=& v \psi_i + \psi_i \cdot \lambda_+ + D((\tilde{q}_a\cdot q_i) \tilde{q}^a)~,
\end{eqnarray}
where $v\equiv{\rm tr}(\lambda_+)$.

Now we consider the $OSp(4|2)$ commuting subalgebra.
The Poincare supercharges in this subalgebra are $Q_{ia+}\equiv Q_{ia}$. 
They act on the BPS fields as 
\begin{equation}
  Q_{ia}q_j=\epsilon_{ij}\tilde\psi_a\ , \quad
  Q_{ia}\tilde{q}_b=-\epsilon_{ab}\psi_i\ , \quad
  Q_{ia}\psi_j=-\epsilon_{ij}D\tilde{q}_a\ , \quad
  Q_{ia}\tilde\psi_b=\epsilon_{ab}Dq_i\ ,
\end{equation}
up to an overall constant which does not matter to us. Furthermore, one finds 
\begin{equation}
    Q_{ia}(Dq_j)=D(Q_{ia}q_j)~,
\end{equation}
and so on. This is because 
\begin{equation}
  Q_{ia\alpha}A_{\beta\gamma}\sim\epsilon_{\alpha(\beta|}
  \left[q_i\otimes \tilde\psi^\dag_{a|\gamma)}-\psi^\dag_{i|\gamma)}\otimes \tilde{q}_a\right]\ ,
\end{equation}
which involves non-BPS fields. 
Restricting to the BPS spin component $\alpha,\beta,\gamma=+$, 
the right hand side vanishes. This means that $Q_{ia}$ and 
$D\sim D_{++}=\partial_{++}-iA_{++}$ commute.

On these BPS fields, the R-charges $R=2({R^3}_3+{R^4}_4)$ of the elementary fields 
$q_i$, $\tilde{q}_a$, $\psi_i$, $\tilde\psi_a$ are all equal to $1$.
So $R$ may be regarded as the number of `letters' in the operator.

In the strictly free theory, all gauge invariant combinations of the letters (\ref{free-BPS-letter}) 
are composite BPS operators because all the cubic terms appearing on the right hand sides 
of the $Q$ transformations are zero in the $k\rightarrow \infty$ limit. (The covariant 
derivatives are also replaced by ordinary derivatives in the limit.) These free BPS 
operators are arranged into a tower of (mostly higher spin) supermultiplets.
Let us review this tower before we discuss the interacting theory.

Consider the bosonic single-trace operators of this theory: 
\begin{eqnarray}\label{single-trace-boson}
  (\tilde{\mathcal{J}^I}_{J})_{\mu_1\cdots \mu_s}&=&
  \Phi^{\dag I}\cdot{\partial}_{\mu_1}
  \cdots{\partial}_{\mu_s}\Phi_J+\cdots\qquad (s\geq 0)
  \nonumber\\
  (\tilde{\mathcal{K}^I}_J)_{\mu_1\cdots\mu_s}&=&
  \overline\Psi_J\cdot \gamma_{(\mu_1}\partial_{\mu_2}\cdots \partial_{\mu_s)}\Psi^I+\cdots
  \qquad (s\geq 1)
  \nonumber\\
  \tilde{\mathcal{K}^I}_J&=&\overline{\Psi}_J\cdot \Psi^I\qquad (s=0)\ .
\end{eqnarray}
We will not discuss the fermionic single-trace operators here. 
(Some features of these fermionic operators will be discussed below.) 
Hidden behind the dots are extra terms with some derivatives acting to their left and/or 
with subtractions of the trace parts of the Lorentz indices: they ensure that the 
operators with $s\geq 1$ are conserved and that they are traceless with regards to
the Lorentz indices, making them the proper spin $s$ representations.
For instance, see \cite{Giombi:2011kc,Aharony:2012nh} for some
examples with low $s$. Other single-trace operators can be written as linear combinations of 
these operators and their conformal descendants.
In the free theory, the operators for $s\geq 1$ are all conserved currents, e.g. 
$\partial^{\mu_1}(\tilde{\mathcal{J}^I}_J)_{\mu_1\mu_2\cdots \mu_s}=0$.
In the ABJ vector theory, (\ref{single-trace-boson})
are all parts of suitable $OSp(6|4)$ multiplets.
We separate the $SU(4)_R$ singlet parts 
$\mathcal{J}_{\mu_1\cdots \mu_s}\equiv (\tilde{\mathcal{J}^I}_I)_{\mu_1\cdots \mu_s}$, 
$\mathcal{K}_{\mu_1\cdots \mu_s}$ from
the traceless adjoint parts $({\mathcal{J}^I}_J)_{\mu_1\cdots \mu_s}$, 
$({\mathcal{K}^I}_J)_{\mu_1\cdots\mu_s}$ for the discussions below. 
We also schematically write these spin $s$ operators as $\mathcal{J}_{(s)}$,
$({\mathcal{J}^I}_J)_{(s)}$, $\mathcal{K}_{(s)}$, $({\mathcal{K}^I}_J)_{(s)}$.
Among these operators, those that fall in our BPS sector 
(annihilated by $Q=Q_{34-}$ and $Q^\dag$) are
\begin{eqnarray}\label{single-trace-boson-BPS}
  ({\mathcal{J}^i}_{a+2})_{+_1\cdots +_s} &\sim& 
  q_i\cdot{\partial}^s \tilde{q}_a+\cdots~, \qquad (s\geq 0) \nonumber\\ 
  ({\mathcal{K}^{i}}_{a+2})_{+_1\cdots +_s} &\sim& 
  \tilde\psi_{a}\cdot \partial^{s-1}\psi_{i}+\cdots~, \qquad (s\geq 1)
\end{eqnarray}
with $i=1,2$, $a=1,2$ and $\partial\equiv\partial_{1+i2}$,
all belonging to the $SU(4)_R$ adjoint part ${\mathcal{J}^I}_J$, ${\mathcal{K}^I}_J$.

We first discuss the scalar operators at $s=0$. 
In the notion of \cite{Cordova:2016emh}, ${\mathcal{J}^I}_J$ and $\mathcal{J}$ are the
superconformal primaries of the multiplets $B_1[0]_1^{(0,1,1)}$ and $A_2[0]_1^{(0,0,0)}$,
respectively.\footnote{See Table 8 there. In $N_n[2J]_E^{(R_1,R_2,R_3)}$, 
$N=B,A,L$ is the type of the multiplet, $n$ labels the sub-types, 
and $J$, $E$, $(R_1,R_2,R_3)$ are the angular momentum, scaling dimension,  
$SO(6)_R$ Dynkin labels of the primary.} 
$\mathcal{K}\sim QQ\mathcal{J}$ is a descendant in the multiplet $A_2[0]_1^{(0,0,0)}$, and  
${\mathcal{K}^I}_J\sim QQ{\mathcal{J}^I}_J$ is a descendant in $B_1[0]_1^{(0,1,1)}$.
The multiplet $B_1[0]_1^{(0,1,1)}$ contains the stress tensor, 
which is absolutely protected. So the BPS operators within 
this multiplet will remain so even after turning on interactions. 
(However, their multi-traces may be lifted by interactions: see below.)
We call it the graviton multiplet. This multiplet also contains 
the $s=1$ conserved current for the $SU(4)_R$ symmetry, which is a linear 
combination of $({\mathcal{J}^I}_J)_\mu$ and $({\mathcal{K}^I}_J)_\mu$: see next paragraph. 
$A_2[0]_1^{(0,0,0)}$ that hosts $\mathcal{J}$ and $\mathcal{K}$ also 
contains higher spin currents ((5.68) of \cite{Cordova:2016emh}) 
and will be anomalous \cite{Maldacena:2011jn} in the interacting theory  
by combining with another short multiplet of multi-trace operators. 
The $\mathcal{N}=6$ higher spin gravity dual has $2^{\mathcal{N}-1}=32$ scalars with 
mass $m^2=-2$ \cite{Chang:2012kt}. $16$ of them are given the regular boundary condition 
with scaling dimension $E=2$, which are dual to ${\mathcal{K}^I}_J$ and $\mathcal{K}$. 
The other $16$ are given the alternate boundary condition 
with $E=1$, which are dual to ${\mathcal{J}^I}_J$ and $\mathcal{J}$.

Now we consider the supermultiplets that contains the operators
(\ref{single-trace-boson}) at $s\geq 1$.
It is more convenient to include the multiplets for the $s=0$ operators that
we already explained in the previous paragraph and discuss altogether. 
The superconformal multiplets of single-trace operators and their bosonic contents 
are given by (see \cite{Cordova:2016emh}, Section 5.4.6):
\begin{eqnarray}
  B_1[0]_1^{(0,1,1)}&:&{\mathcal{J}^I}_J\in [0]_1^{(0,1,1)} \nonumber\\
  &&Q^2{\mathcal{J}^I}_J\sim ({\mathcal{K}^I}_J, 
  {(\mathcal{J}+\mathcal{K})_\mu},{({\mathcal{J}^I}_J+{\mathcal{K}^I}_J)_\mu})
  \in [0]_2^{(0,1,1)}\oplus [2]_2^{(0,0,0)\oplus(0,1,1)}\nonumber
  \\
  &&Q^4{\mathcal{J}^I}_J\sim (\mathcal{J}+\mathcal{K})_{\mu\nu}\in [4]_3^{(0,0,0)}~,
  \nonumber\\
  A_2[0]_1^{(0,0,0)}&:&\mathcal{J}\in [0]_1^{(0,0,0)}
  \nonumber\\
  &&Q^2\mathcal{J}\sim (\mathcal{K},
  ({\mathcal{J}^I}_J-{\mathcal{K}^I}_J)_\mu)\in [0]_2^{(0,0,0)}\oplus [2]_2^{(0,1,1)}
  \nonumber\\
  &&Q^4\mathcal{J}\sim ({\mathcal{J}^I}_J+{\mathcal{K}^I}_J)_{\mu\nu}\in [4]_3^{(0,1,1)}
  \nonumber\\
  &&Q^6\mathcal{J}\sim (\mathcal{J}+\mathcal{K})_{\mu\nu\rho} \in [6]_4^{(0,0,0)}~,
  \nonumber\\
  A_1[2s]_{s+1}^{(0,0,0)}\ \ (s\geq 1)&:&(\mathcal{J}-\mathcal{K})_{\mu_1\cdots\mu_s}
  \in [2s]_{s+1}^{(0,0,0)}\nonumber\\
  &&Q^2(\mathcal{J}-\mathcal{K})_{\mu_1\cdots\mu_s}\sim 
  ({\mathcal{J}^I}_J-{\mathcal{K}^I}_J)_{\mu_1\cdots \mu_{s+1}} 
  \in [2(s+1)]_{s+2}^{(0,1,1)}
  \nonumber\\
  &&Q^4(\mathcal{J}-\mathcal{K})_{\mu_1\cdots\mu_s}\sim 
  ({\mathcal{J}^I}_J+{\mathcal{K}^I}_J)_{\mu_1\cdots \mu_{s+2}} 
  \in [2(s+2)]_{s+3}^{(0,1,1)}
  \nonumber\\
  &&Q^6(\mathcal{J}-\mathcal{K})_{\mu_1\cdots\mu_s}\sim 
  (\mathcal{J}+\mathcal{K})_{\mu_1\cdots \mu_{s+3}} \in [2(s+3)]_{s+4}^{(0,0,0)}\ .
\end{eqnarray}
By $\mathcal{J} \pm \mathcal{K}$ (or ${\mathcal{J}^I}_J \pm {\mathcal{K}^I}_J$),
we schematically denote two different linear combinations of the pair of operators: 
the actual coefficients of these combinations may differ from 1, such as those in (\ref{BPS-B1}).
The fermionic single-trace operators that we did not list  
take the form of $Q^{n}(\textrm{primary})$ in these multiplets with odd $n$.

We also explain how the BPS single-trace operators (that preserve $Q$ and $Q^\dag$)
of the free theory are located in the supermultiplets of the previous paragraph.
First, the BPS states within 
the graviton multiplet $B_1[0]_1^{(0,1,1)}$ are given (up to conformal descendants) by
\begin{eqnarray}\label{BPS-B1}
  u_{ia}&\equiv &{\mathcal{J}^i}_{a+2}\sim q_i\cdot \tilde{q}_a\in [0]_1^{(0,1,1)}~,
  \nonumber\\
  Qu_{ia}&\sim& (q_{(i}\cdot\psi_{j)},\tilde{q}_{(a}\cdot\tilde\psi_{b)},
  q^i\cdot\psi_{i}-\tilde{q}^a\cdot\tilde\psi_{a})\equiv (v_{ij},\tilde{v}_{ab},v)
  \in [1]_{\frac{3}{2}}^{(0,2,0)\oplus (0,0,2)\oplus (1,0,0)}~,
  \nonumber\\
  Q^2u_{ia}&\sim& \tilde{q}_a\cdot \partial q_i-\psi_{i}\cdot\tilde\psi_{a}\equiv w_{ia}
  \in [2]_2^{(0,1,1)}~,
  \nonumber\\
  Q^3u_{ia}&\sim& 3\partial q^i\cdot\psi_{i}-q^i\cdot\partial \psi_{i}
  +3\partial \tilde{q}^a\cdot \tilde\psi_{a}-\tilde{q}^a\cdot \partial\tilde\psi_{a}
  \equiv x\in [3]_{\frac{5}{2}}^{(1,0,0)}~.
\end{eqnarray}
Here $Q$ schematically denotes all possible $Q_{ia}$'s in $OSp(4|2)$.
Then within $A_2[0]_1^{(0,0,0)}$, one finds an $OSp(4|2)$ multiplet with the primary 
$Q^\prime\mathcal{J}\sim q^i\cdot\psi_{i}+\tilde{q}^a\cdot\tilde\psi_{a}\in [1]_{\frac{3}{2}}^{(1,0,0)}$:
\begin{eqnarray}\label{BPS-A2}
  &&\{Q^\prime\mathcal{J}\in [1]_{\frac{3}{2}}^{(1,0,0)}\} \stackrel{Q}{\longrightarrow}
  \{Q^\prime Q\mathcal{J}\in [2]_2^{(0,1,1)}\} \stackrel{Q}{\longrightarrow} 
  \{Q^\prime Q^2\mathcal{J}\in [3]_{\frac{5}{2}}^{(0,2,0)\oplus (0,0,2)}\}\nonumber\\
  &&\stackrel{Q}{\longrightarrow} \{Q^\prime Q^3\mathcal{J}\in [4]_3^{(0,1,1)}\}
  \stackrel{Q}{\longrightarrow}
  \{Q^\prime Q^4\mathcal{J} \in [5]_{\frac{7}{2}}^{(1,0,0)}\}\ .
\end{eqnarray}
Here $Q^\prime\equiv Q_{34+}$ \cite{Bhattacharya:2008zy},
and other $Q$'s again denote $Q_{ia}$'s in $OSp(4|2)$.
Finally, in $A_1[2s]_{s+1}^{(0,0,0)}$, one finds the $OSp(4|2)$ multiplets with 
the primary $Q^\prime (\mathcal{J}-\mathcal{K})_{\mu_1\cdots \mu_s}
\in [2s+1]_{s+\frac{3}{2}}^{(1,0,0)}$:
\begin{eqnarray}\label{BPS-A1}
    &&\{Q^\prime(\mathcal{J}-\mathcal{K})_{(s)}\in [2s+1]_{s+\frac{3}{2}}^{(1,0,0)}\} \stackrel{Q}{\longrightarrow}
  \{Q^\prime Q(\mathcal{J}-\mathcal{K})_{(s)}\in [2s+2]_{s+2}^{(0,1,1)}\} \nonumber\\
  &&\stackrel{Q}{\longrightarrow} 
  \{Q^\prime Q^2(\mathcal{J}-\mathcal{K})_{(s)}\in [2s+3]_{s+\frac{5}{2}}^{(0,2,0)\oplus (0,0,2)}\}
  \stackrel{Q}{\longrightarrow} \{Q^\prime Q^3(\mathcal{J}-\mathcal{K})_{(s)}\in [2s+4]_{s+3}^{(0,1,1)}\}
  \nonumber\\
  &&\stackrel{Q}{\longrightarrow}
  \{Q^\prime Q^4(\mathcal{J}-\mathcal{K})_{(s)} \in [s+5]_{s+\frac{7}{2}}^{(1,0,0)}\}\ .
\end{eqnarray}
All the free BPS operators of (\ref{BPS-A2}) and (\ref{BPS-A1}) will be lifted in the interacting theory.

We have explained the single trace operators in the free limit. Morally, 
they are single particle states in the AdS$_4$ dual. 
Multiplying them, the multi-trace operators are multi-particle states in AdS.
In particular, multiplying the single-trace BPS operators that we explained above, 
one obtains the general set of BPS operators in the free limit.

Turning on the interactions, $\lambda\neq 0$, one has to promote all the derivatives 
in these operators to covariant derivatives. 
Most of these single trace operators fail to be BPS in the interacting theory, except 
those in the graviton multiplet $B_1[0]_1^{(0,1,1)}$. This is expected because all other 
multiplets contain higher spin currents which are not conserved in the interacting theory
\cite{Maldacena:2011jn,Chang:2012kt}. That is, due to the lack of their conservation, 
the divergences of these currents are nonzero and given by certain multi-trace operators. 
As a result, the single-trace higher spin currents mix with certain multi-trace operators 
and form long multiplets, whose scaling dimensions are no longer protected. At the leading 
order in the small coupling $\lambda$, $Q$ and $Q^\dag$ acting on the free BPS fields starts from the 
$\frac{1}{k}$ order, i.e. at $1$-loop. In particular, the supercharge operators at this 
1-loop is completely given by the supercharges of the classical interacting theory.
The leading anomalous dimension is given by
$\{Q,Q^\dag\}=E-\frac{R}{2}-J$, which starts from $\frac{1}{k^2}$ and is thus 2-loop.  
In this paper, we are interested in 
the subset of the free BPS operators, at both single- and multi-trace levels, which remain 
BPS at the 2-loop level in $\lambda$ (and exactly in $\frac{1}{N}$).

To study the spectrum of these 2-loop BPS operators, we employ a cohomological 
formulation \cite{Grant:2008sk,Chang:2013fba}. The local BPS operator
$\mathcal{O}$ with vanishing 2-loop anomalous dimension satisfies
\begin{equation}
  (QQ^\dag+Q^\dag Q)\mathcal{O}=0\ .
\end{equation}
(The action of $QQ^\dag+Q^\dag Q$ on $\mathcal{O}$ is implemented by commutators, 
which we write as above for the simplicity of notation.) Here note that the supercharges  
are nilpotent, $Q^2=0$, from the algebra. So $\mathcal{O}$ can be 
formally regarded as a harmonic differential form, regarding $Q$ formally as a nilpotent 
exterior derivative. The Hodge theory states that these harmonic forms are in 1-to-1 map 
to the cohomology classes of $Q$, i.e. the set of $Q$-closed operators satisfying 
$Q\mathcal{O}=0$ with the identifications $\mathcal{O}\sim \mathcal{O}+Q\Lambda$ 
of $Q$-exact shifts. So to understand the spectrum of BPS operators, one can study the cohomology 
classes of $Q$.

We will study the theory, especially the cohomologies, in the `weakly coupled' regime $N\ll k$.
We consider operators that may be heavy in that their scaling dimensions may scale in $N$,
but not in $k$ which is larger. In this setup, one can ignore the contributions 
from the so-called magnetic monopole operators \cite{Borokhov:2002ib}. 
The latter operators are defined by giving singular boundary conditions 
near the operator insertion point $x$, with nonzero magnetic flux on the
small $S^2$ which surrounds $x$. The Gauss law $k\star F_\mu\sim J_\mu$ 
of Chern-Simons-matter theory, with the gauge current $J_\mu$, demands that 
such operators with quantized flux $\int_{S^2}F$ are dressed by order $k$ quanta 
of matter fields. So the scaling dimensions of the monopole 
operators scale in $k$, which we can ignore in our setup.

In our constructions of new BPS operators in this section, the index 
of these operators will provide useful guidance. So we explain the index and the 
useful formula to compute it \cite{Bhattacharya:2008zy,Bhattacharya:2008bja,Kim:2009wb}. 
The index of the $\mathcal{N}=6$ SCFT is defined by
\begin{equation}\label{index}
  Z(x,y_1,y_2)={\rm Tr}\left[(-1)^F x^{E+J}y_1^{F_1}y_2^{F_2}\right]\ ,
\end{equation}
where $J$ is the angular momentum and $F_{1,2}$ are the 
Cartans of $SU(2)\times SU(2)\subset SU(4)_R$ in $OSp(4|2)$ which commutes with our 
$Q,Q^\dag$. The trace is taken over the Hilbert space of local gauge-invariant operators. 
We note that $\mathcal{N}=6$ SCFTs also have a $U(1)$ flavor symmetry 
\cite{Bashkirov:2011fr,Cordova:2016emh}, whose fugacity may further refine the index. 
However, in the ABJ theory, this is realized as a topological $U(1)$ symmetry carried by the 
magnetic monopole operators.
that decouple in this section's setup $N\ll k$, $E\ll k$.
Since the charges appearing in the trace commute with $Q,Q^\dag$, 
pairs of operators which do not preserve $Q,Q^\dag$ cancel by $(-1)^F$. 
When the monopole operators are decoupled at $E\ll k$, the index is independent 
of $k$ \cite{Kim:2009wb} and thus of the coupling $\lambda=\frac{N}{k}$. 
In this case, one finds the following expression for the 
index \cite{Bhattacharya:2008bja}:
\begin{eqnarray} \label{index-formula}
  \hspace*{-1cm}Z \!\!&=&\!\! \frac{1}{N!}
  \int_0^{2\pi}\!\! \frac{d^N\alpha}{(2\pi)^N}\prod_{a\neq b}(1-e^{i\alpha_{ab}})
  \exp\left[\sum_{n=1}^\infty\frac{1}{n}\left(\frac{x^{\frac{n}{2}}}{1-x^{2n}}(y_1^n+y_1^{-n})
  -\frac{x^{\frac{3n}{2}}}{1-x^{2n}}(y_2^n+y_2^{-n})\right)\sum_{a=1}^Ne^{in\alpha_a}\right.\nonumber\\
  &&\hspace{2cm}\left.
  +\sum_{n=1}^\infty\frac{1}{n}\left(\frac{x^{\frac{n}{2}}}{1-x^{2n}}(y_2^n+y_2^{-n})
  -\frac{x^{\frac{3n}{2}}}{1-x^{2n}}(y_1^n+y_1^{-n})\right)\sum_{a=1}^Ne^{-in\alpha_a}\right]\ .
\end{eqnarray}
Since the index (\ref{index-formula}) is independent of  
$\lambda=\frac{N}{k}$, it can be understood in various ways. It may be understood as the index over 
the free BPS states. Alternatively, one can regard it as the index over the 2-loop BPS states.
Equivalently, it is the index over the cohomology classes with respect to the classical
supercharge $Q$ of (\ref{Q-free-BPS}) acting on the free BPS letters.  

We also add the following two comments on the validity of the monopole-free index 
(\ref{index-formula}). First, although in this section we shall study the cohomologies in 
the regime $N\ll k$, the expression (\ref{index-formula}) is valid as long as $E<E_0\sim k$, 
where $E_0$ is the lightest gauge-invariant monopole operator which is proportional to $k$.
Second, in Section 3, 
we shall study this index in the large $N$ 't Hooft limit, keeping $\lambda=\frac{N}{k}$ 
fixed, at high temperature $\beta\equiv -\log x \sim N^{-1}\ll 1$. Although in this regime 
there will be many magnetic monopole operators, demanding corrections of (\ref{index-formula}), 
whether the monopoles will affect the dominant saddle point is a subtler matter. 
This issue will be also discussed in Section 3 (before the start of Section 3.1).

One can also consider the BPS partition function, which depends on $\lambda$. 
We can define it as a 1-parameter generalization of the index: 
\begin{equation}\label{partition}
  Z(x,y_{1,2},y)={\rm Tr}\left[x^{E+J}y_1^{F_1}y_2^{F_2}y^R\right]
  ={\rm Tr}\left[(-1)^F(x^{\frac{1}{2}}y)^R(-x)^{2J}y_1^{F_1}y_2^{F_2}\right]\ .
\end{equation}
Unlike for the index, the trace is taken over the local BPS operators only.
Unrefining $(x^{\frac{1}{2}},y)\rightarrow (ix^{\frac{1}{2}},-i)$,
one recovers the index (\ref{index}). 
As noted above, the quantum number $R$ can be regarded as the letter number
when acting on the free BPS fields.
Since $Q$ increases the letter number by $2$, $[R,Q]=2Q\neq 0$
and $y^R$ does not commute with $Q,Q^\dag$. So this partition function is not protected by 
supersymmetry and depends on $\lambda$. We computed this partition function
(up to a certain order in $x$) by counting the cohomologies of $Q$: 
See Appendix \ref{sec:appA.1} for the outline of the calculations 
and the results. 

There will be three different classes of cohomologies that we study in this paper:
\begin{enumerate}
\item single- and multi-particle states of BPS gravitons, 
\item multi-particle BPS bound states that contain the non-BPS higher spin particles,
\item and new heavy states which become $Q$-closed due to the finiteness of $N$.
\end{enumerate}
The class 1 is simply given by the products of (\ref{BPS-B1}) and their 
conformal descendants within $OSp(4|2)$, with $\partial$ replaced by $D$. 
We will first count them and subtract their contributions $Z_{N,\textrm{grav}}$
to the index (\ref{index}) or the BPS partition function (\ref{partition}).
From the subtracted partition functions $Z_N-Z_{N,\textrm{grav}}$ or
$Z_\infty - Z_{\infty,\textrm{grav}}$, one can notice the charges of the new
cohomologies in the classes 2 and 3.
This information will guide us to detect and construct the representatives
of these new cohomologies.
Similar strategy was taken in \cite{Choi:2023znd,Choi:2023vdm}
to construct analogous cohomologies in the 4d $\mathcal{N}=4$ Super-Yang-Mills theory.

The distinction between the classes 2 and 3 is one of the key messages of this section.
At first sight, cohomologies in the class 2 seem to be monotone in a sense that
their versions with higher values of $N$ are no longer $Q$-closed,
but at the same time they do not belong to the class 1 of BPS gravitons.
As it will become clear in this section, these are cohomologies that are to be fortuitous
with respect to $U(N')$ trace relations, where we have taken $N'=1$
from the $U(N)_k\times U(N^\prime)_{-k}$ ABJ theory to define the ABJ vector model.
On the other hand, cohomologies in the class 3 are fortuitous with respect to $U(N)$ trace relations,
and thus they are genuinely heavy cohomologies within the $N'=1$ ABJ vector model.

The three subsections of this section will roughly correspond to the three classes of cohomologies
described above.
In section \ref{sec:2.1} we discuss multi-trace operators built from the single-trace BPS
graviton operators (\ref{BPS-B1}). We also discuss higher spin particles
that become anomalous with nonzero interaction.
In section \ref{sec:2.2} we introduce the concept of the higher spin BPS bounds
corresponding to class 2 and construct an explicit example thereof.
Then in section \ref{sec:2.3} we construct the first example of class 3 cohomologies at $N=2$.
In Section \ref{sec:3}, we shall also make a large $N$ approximation of the index
(at suitably scaled chemical potential) and study possible phases of these BPS states.

\subsection{BPS gravitons and anomalous higher spin particles}\label{sec:2.1}

We first explain what happen to the free single-trace BPS operators with nonzero 
interaction, $\frac{1}{k}\neq 0$.
The right hand sides of the $Q$ transformation (\ref{Q-free-BPS}) on the
free BPS fields are now nonzero, and many of them lift to the non-BPS sector.

We first discuss the graviton multiplet $B_1[0]_1^{(0,1,1)}$ that contains
the single-trace operators (\ref{BPS-B1}).
All these operators can be obtained by acting the $OSp(4|2)$ supercharges
$Q_{ia}$ on the primary $u_{ia}=q_i\cdot\tilde{q}_a$:
\begin{eqnarray}
  Q_{ia}u_{jb}&=&\epsilon_{ij}\tilde{v}_{ab}-\epsilon_{ab}v_{ij}
  +{\textstyle \frac{1}{2}}\epsilon_{ij}\epsilon_{ab}v~,
  \nonumber\\
  Q_{ia}v_{jk}&=&-\epsilon_{i(j}\left[D\tilde{q}_a\cdot q_{k)}+\psi_{k)}\cdot\tilde\psi_a\right]
  =-{\textstyle \frac{1}{2}}\epsilon_{i(j}w_{k)a}
  -{\textstyle \frac{1}{2}}\epsilon_{i(j}\partial u_{k)a}~,
  \nonumber\\
  Q_{ia}\tilde{v}_{bc}&=&-{\textstyle \frac{1}{2}}\epsilon_{a(b|}w_{i|c)}+
  {\textstyle \frac{1}{2}}\epsilon_{a(b|}\partial u_{i|c)}~,
  \nonumber\\
  Q_{ia}v&=&-\partial u_{ia}~,
  \nonumber\\
  Q_{ia}w_{jb}&=&-{\textstyle \frac{1}{2}}\epsilon_{ij}\epsilon_{ab}x
  -\epsilon_{ij}\partial\tilde{v}_{ab}-\epsilon_{ab}\partial v_{ij}~.
\end{eqnarray}
The definitions of $u_{ia}$, $v_{ij}$, $\tilde{v}_{ab}$, $v$ are the same as (\ref{BPS-B1}), 
and here we define 
\begin{eqnarray}\label{graviton-w-x}
  w_{ia}&=&D\tilde{q}_a\cdot q_i-\tilde{q}_a\cdot Dq_i+2\psi_i\cdot\tilde\psi_a~, \nonumber\\
  x&=&3Dq^i\cdot \psi_i-q^i\cdot D\psi_i+3D\tilde{q}^a\cdot\tilde\psi_a
  -\tilde{q}^a\cdot D\tilde\psi_a~.
\end{eqnarray}
which are different from (\ref{BPS-B1}) by covariantizing the derivative $\partial\rightarrow D$ 
and suitably adding the conformal descendants. All the other single-trace 
graviton states are obtained by acting $\partial^j$ on these, becoming the conformal descendants.
As already explained, the multiplet $B_1[0]_1^{(0,1,1)}$ which contains these operators 
is absolutely protected, so these operators remain $Q$-closed even after turning on the 
cubic terms in (\ref{Q-free-BPS}). One can readily show this explicitly.

Taking products of the single-trace operators of the previous paragraph,
one obtains multi-particle graviton states.
With interactions of the ABJ theory, many multi-trace operators become $Q$-exact.
When an operator $O$ becomes $Q$-exact, i.e. $O=Q\Lambda$,
$O$ belongs to the trivial cohomology.
(Physically, the superpartner pair $(\Lambda,O)$ lifts to the non-BPS sector.)

In our system, $Q$-exact multi-graviton operators can appear for two reasons. 
First, this happens by the multi-trace interactions in the ABJ vector model. 
To see this, note that the $U(N)_k\times U(N^\prime)_{-k}$ ABJ theory
has the superpotential 
\begin{equation}\label{superpotential}
  W(q_i,\tilde{q}_a)\sim {\rm tr}\left(\epsilon^{ij}\epsilon^{ab}
  q_i\tilde{q}_a q_j\tilde{q}_b\right)~,
\end{equation}
in 3d $\mathcal{N}=2$ language, where $q_i$ and $\tilde{q}_a$ are $N\times N^\prime$ 
and $N^\prime \times N$ matrices, respectively.
At $N^\prime=1$, this superpotential is factorized into a double-trace of the form 
$W\sim \epsilon^{ij}\epsilon^{ab}(q_i\cdot\tilde{q}_a)(q_j\cdot\tilde{q}_b)$. 
So in the vector CS model, the interaction does not preserve the trace number.
In our problem, the $Q$-transformations of (\ref{Q-free-BPS}) have
inner products on the right hand sides. So certain combinations of multi-graviton 
operators can be $Q\Lambda$ where $\Lambda$ has one less trace number.

For example, consider the multi-trace operators of the primaries 
$u_{ia}=q_i\cdot\tilde{q}_a$. The $n$-particle states are given by linear combinations of 
\begin{equation}\label{utothen}
  u_{i_1a_1}\cdots u_{i_n a_n}
  =(q_{i_1}\cdot\tilde{q}_{a_1}) \cdots (q_{i_n}\cdot\tilde{q}_{a_n})~.
\end{equation}
In the interacting theory, some combination of these operators can be $Q$-exact 
from the interacting $Q$-transformations (\ref{Q-free-BPS}), especially from 
\begin{equation}
  Q\psi_i=\tilde{q}^au_{ia}=\epsilon^{ab}u_{i[a}\tilde{q}_{b]}~, \qquad 
  Q\tilde\psi_a=q^iu_{ia}=\epsilon^{ij}q_{[j}u_{i]a}\ .
\end{equation}
If any pair of $SU(2)$ indices is antisymmetrized in (\ref{utothen}),  
i.e. $u_{[i|a}u_{|j]b}$ or $u_{i[a|}u_{j|b]}$, it is $Q$-exact.  
Thus, the only nontrivial cohomologies are 
those with $i_1,\cdots,i_n$ and $a_1,\cdots, a_n$ symmetrized,
\begin{equation}\label{multi-u-sym}
  {u^{(i_1}}_{(a_1}\cdots {u^{i_n)}}_{a_n)}\ .
\end{equation}
The counting problem of these scalar multi-trace primaries is the same as that in
the $N=1$ theory. This is because the positions of all $q$'s and $\tilde{q}$'s
in (\ref{multi-u-sym}) are irrelevant, so they behave as numbers rather than vectors. 
This counting rule can also be phrased as the counting based on quantizing the moduli space. 
Including all the other gravitons, the nontrivial polynomials of (\ref{BPS-B1}), 
(\ref{graviton-w-x}) are also reduced in the interacting theory by the $Q$-transformation 
(\ref{Q-free-BPS}). (Some examples will be provided below.) Unfortunately, we are not
aware of a simple method to count all the BPS multi-gravitons: for instance, 
we find that using only the light field components on the generic point of the moduli space 
yields a wrong counting.\footnote{In 4d $U(N)$ $\mathcal{N}=4$ Yang-Mills theory, 
the moduli space counting of gravitons was successfully employed in  
\cite{Choi:2023znd,Choi:2023vdm,deMelloKoch:2024pcs,Gadde:2025yoa}. 
The same approach may fail in the ABJ vector model due to the singularity of the 
moduli space $\mathbb{C}^4/\mathbb{Z}_k$, but a good understanding is lacking. 
This counting scheme also fails in other models \cite{Chang:2025mqp,Choi:2025pqr,JHC}.}
We resort to a brute force counting on a computer.
The reduction of the multi-trace BPS states in the interacting theory that we just explained 
applies to arbitrary $N$. That is, even at large $N$ and low energy $\sim\mathcal{O}(1)$, 
the BPS multi-gravitons do not behave like an ideal gas at $\lambda\neq 0$. 
Note that this is different from the multi-gravitons of the weakly-coupled 
string theories in AdS, say on AdS$_5\times S^5$. There, cohomologies of multi-gravitons 
at low energy do behave like an ideal gas in that all multi-particle 
states are present. It is the multi-trace 
nature of the interactions in the vector model which breaks such ideal gas properties.

There is a second way in which multi-gravitons may be $Q$-exact. 
This may happen when the size of the operators scales in $N$, 
due to various relations of heavy multi-trace operators. To explain this, 
first consider the rank $n$ $U(N)$ tensors 
\begin{equation}
  V_1^{p_1}\cdots V_n^{p_n}~, \qquad (W_1)_{q_1}\cdots(W_n)_{q_n}~,
\end{equation}
where $p_i,q_i=1,\cdots,N$ are $U(N)$ fundamental/anti-fundamental indices, respectively. 
If $n>N$, the complete antisymmetrization of $p_1,\cdots p_n$ or 
$q_1,\cdots q_n$ must be zero, 
\begin{equation}\label{vector-anti}
  V_1^{[p_1}\cdots V_n^{p_n]}=0\ \ ,\ \ (W_1)_{[q_1}\cdots(W_n)_{q_n]}=0\ .
\end{equation}
So the following gauge invariant operator  
\begin{equation}\label{vector-trace-relation}
  V_1^{p_1}\cdots V_n^{p_n}(W_1)_{[p_1}\cdots(W_n)_{p_n]}\sim 
  \sum_{\rho\in S_n}(-1)^{{\rm sgn}(\rho)}(V_1\cdot W_{\rho(1)})\cdots (V_n\cdot W_{\rho(n)})~,
\end{equation}
must be zero if $n>N$. In other words, some polynomials of single-trace operators 
are zero when $N$ is smaller than the trace number $n$.
More generally, relations like (\ref{vector-trace-relation}) can be found from the linear 
combinations of the form 
\begin{equation}\label{vector-trace-relation-2}
  \sum_{\rho\in S_n}\chi_R(\rho)(V_1\cdot W_{\rho(1)})\cdots (V_n\cdot W_{\rho(n)})~,
\end{equation}
where $R$ is a representation of the symmetric group $S_n$, associated with a Young diagram 
with $n$ boxes, and $\chi_R(\sigma)$ is its character. (\ref{vector-trace-relation-2}) is zero if 
the Young diagram for $R$ has more than $N$ rows. (For instance, see \cite{deMelloKoch:2024sdf} for 
a review.) In our cohomology problem, it may happen that a large multi-graviton cohomology 
can be written as $Q\Lambda$ plus various operators of the form 
(\ref{vector-trace-relation-2}). If this happens, such a cohomology 
is trivial if $N$ is smaller than the row number of the Young diagram $R$. 
The size of such operators scales in $N$.

This mechanism has an analogue in AdS string theory. With $N\times N$ matrix fields 
$M_i$, the number of multi-graviton cohomologies reduces relative to the naive count,
due to the relations 
\begin{equation}\label{matrix-trace-relation}
  \sum_{\rho\in S_n}\chi_R(\rho)
  {(M_1)^{p_1}}_{\rho(p_1)}\cdots {(M_n)^{p_n}}_{\rho(p_n)}=0
\end{equation}
when the Young diagram for $R$ has more than $N$ rows. 
In the bulk, the heavy gravitons with reduced degrees of freedom are called 
giant gravitons \cite{McGreevy:2000cw}. The bulk picture for the 
heavy BPS multi-gravitons is unclear in the higher spin gravity. See Section \ref{sec:2.2} 
for interesting giant graviton like phenomena, and Sections \ref{sec:3.3} and \ref{sec:4} 
for further comments.

Counting of the multi-graviton cohomologies at different $N$ and charges, 
subject to both aforementioned reduction mechanisms, is explained in Appendix \ref{sec:appA}. 
The results are summarized as the 2-loop BPS partition function $Z_{N,\textrm{grav}}(x,y_{1,2},y)$
of (\ref{partition}) but with ${\rm Tr}$ restricted to gravitons only.

Now we consider other single-trace free BPS operators in the interacting 
theory. It turns out that all the other single-trace operators become non-BPS. 
This can be easily understood by recalling the multiplet contents of the free BPS operators, 
explained earlier in this section. Apart from the BPS graviton operators which are in 
the absolutely protected multiplet, other single-trace BPS states in the free theory are in 
the multiplets $A_2[0]_1^{(0,0,0)}$ or $A_1[2s]_{s+1}^{(0,0,0)}$ (with $s=1,2,\cdots$) 
which contain higher spin currents. In the interacting theory, these currents are no longer 
conserved \cite{Maldacena:2011jn,Chang:2012kt}. So their multiplets combine with other 
multi-trace multiplets and become anomalous. Again one can concretely check 
from (\ref{Q-free-BPS}) that they are not $Q$-closed, not representing 
nontrivial cohomologies. For example, the operator $q^i\cdot\psi_i+\tilde{q}^a\cdot\tilde\psi_a$ 
is a free BPS operator which belongs to the multiplet $A_2[0]_1^{(0,0,0)}$. 
From (\ref{Q-free-BPS}),
\begin{equation}
  Q(q^i\cdot \psi_i+\tilde{q}^a\cdot\tilde\psi_a)=2u^{ia}u_{ia}\ ,
\end{equation}
so it forms a non-BPS pair with a double-trace graviton.

Similar lifts of single-trace free BPS operators happen in the AdS/CFT models 
of superstring theory. For instance, in 4d $\mathcal{N}=4$ Yang-Mills theory, the single-trace 
operators are classified into protected Kaluza-Klein graviton multiplets and the rest. Only the 
graviton multiplets are protected, while the others acquire nonzero anomalous dimensions already 
at the leading 1-loop level $\sim \mathcal{O}(g_{\rm YM}^2)$. At strong coupling, 
$\lambda\equiv Ng_{\rm YM}^2\gg 1$, we expect them to acquire large anomalous dimensions 
$\sim \lambda^{\frac{1}{4}}$ and to be dual to the oscillation modes of fundamental 
strings in $AdS_5\times S^5$. That is, the `zero modes' of the string corresponding to gravitons 
are BPS while other typical oscillations are non-BPS. In higher spin gravity, the tower of 
higher-spin currents are somewhat analogous to the tower of string oscillating modes, 
which also become anomalous. 

So far, we have discussed how the single-trace operators in the
higher-spin current multiplets become anomalous at $\lambda\neq 0$.
One can further discuss the multi-trace BPS operators made of all 
the single-trace free BPS operators, including gravitons and higher-spin particles. 
These operators will be discussed in the next two subsections.

\subsection{BPS bounds of anomalous higher spin particles}\label{sec:2.2}

In 4d $\mathcal{N}=4$ Yang-Mills theory, it is by now well known that there exist 
multi-trace (multi-particle, loosely speaking) BPS operators whose
single-trace (single-particle) constituents are non-BPS in general.
Although some single-trace partons are not $Q$-closed,
$Q$ acting on the whole operator can be a linear combination 
of the trace relations of the forms (\ref{matrix-trace-relation}) and vanish. 
They are necessarily heavy operators, since trace relations
require more than $N$ fields. Such operators that become $Q$-closed by  
trace relations are called fortuitous cohomologies \cite{Chang:2024zqi}. 
(See also \cite{Chang:2024lxt,Chen:2025sum}.)
They are being studied to better understand the BPS black hole microstates in 
$AdS_5\times S^5$.

We study similar phenomena in the ABJ vector model.
It is helpful to consider a generalized setup of the
$U(N)_k\times U(N^\prime)_{-k}$ ABJ theory, at least conceptually. 
Now there are two possible classes of relations.
If the operator contains more than $N$ letters, 
there may appear relations due to the identities like (\ref{vector-anti}) 
(understanding that the $U(N^\prime)$ indices are implicit in (\ref{vector-anti})). 
Similarly, if it has more than $N^\prime$ letters, identities similar to (\ref{vector-anti}) 
for the $U(N^\prime)$ indices may yield relations. So $Q$ acting on 
multi-trace operators can be zero by two different classes of trace relations.
Each class starts to apply above the threshold $\sim N$ and $\sim N^\prime$, respectively. 
So there are two notions of fortuity, each with their own energy threshold. 
In the ABJM limit $N^\prime=N$, the two thresholds will merge. In the regime 
$1\ll N^\prime \ll N$, there will be three hierarchies of states with two well-separated
thresholds.

We study the extreme limit of this phenomena at $N^\prime=1$. Since $U(N^\prime)$
trace relations have an order $1$ threshold in this case,
new multi-trace cohomologies appear at low energy even in the large $N$ limit.
At $N^\prime=1$, applying (\ref{vector-anti}) for $U(N^\prime)$ implies a trivial identity  
${(V_1)^{p_1}}_{[q_1^\prime}{(V_2)^{p_2}}_{q_2^\prime]}=0$ between $U(N)$ vectors $V_1$, $V_2$, 
where $p_1,p_2=1,\cdots,N$ and $q_1^\prime,q_2^\prime=1$ are respectively the $U(N)$ and 
formal $U(1)$ indices. For instance, for two identical bosonic vectors $V_1=V_2\equiv V$, 
this trivial identity can be rephrased as 
\begin{equation}\label{Nprime-1-relation}
  V^{[p_1}V^{p_2]}={V^{[p_1}}_{1}{V^{p_2]}}_{1}={V^{p_1}}_{[1}{V^{p_2}}_{1]}=0\ .
\end{equation}
(Had $V$ been fermionic, $V^{\{p_1}V^{p_2\}}=0$.)
So the vanishing skew-symmetrization of two identical vectors can be 
understood as an $N^\prime=1$ trace relation. As we will explain below, 
many multi-trace cohomologies can be constructed using (\ref{Nprime-1-relation}). 
Cohomologies constructed from such $N^\prime=1$ 
trace relations are studied in this subsection. Those constructed with $U(N)$ 
trace relations will be studied in the next subsection. 

We first present an infinite class of multi-trace operators. 
We claim that they contain anomalous higher spin operators and become BPS by 
the $N^\prime=1$ relations of the form (\ref{Nprime-1-relation}).  
Consider the following rank $r$ ($\geq 2$) antisymmetric representations of $U(N)$, 
\begin{equation}\label{skew-rank-r}
  (q^j\wedge q_j\wedge \tilde\psi_{a_1}\wedge\cdots \wedge \tilde\psi_{a_{r-2}})\ ,\qquad 
  (\tilde{q}^b\wedge \tilde{q}_b\wedge \psi_{i_1}\wedge\cdots \wedge \psi_{i_{r-2}})\ ,
\end{equation}
where we use the wedge notation to denote
$[V_1\wedge \cdots \wedge V_r]^{p_1\cdots p_r}=r!V_1^{[p_1}\cdots V_r^{p_r]}$, etc. 
These operators are nonzero when $N\geq r$. They are trivially $Q$-closed at $r=2$ since 
there are no fermions, so we turn to those with $r\geq 3$.
One can show that they are $Q$-closed by (\ref{Nprime-1-relation}). 
Acting $Q$ on the first one of (\ref{skew-rank-r}), $Q$ applies to one of 
the fermions as $Q\tilde\psi_a=q^i u_{ia}$. Then one obtains a skew-symmetric 
product containing $q^j\wedge q_j\wedge q^i\wedge \cdots$.
Since one of the two scalars in $q^j\wedge q_j$ is identical to $q^i$,
this expression vanishes by (\ref{Nprime-1-relation}). 
The second one of (\ref{skew-rank-r}) is also $Q$-closed for the same reason.
So the following gauge-invariant operators are $Q$-closed,
\begin{equation}\label{non-grav-mono}
  O^{(r)}_{a_1\cdots a_{r-2},i_1\cdots i_{r-2}}\equiv 
  (q^j\wedge q_j\wedge \tilde\psi_{a_1}\wedge\cdots \wedge \tilde\psi_{a_{r-2}})\cdot
  (\tilde{q}^b\wedge \tilde{q}_b\wedge \psi_{i_1}\wedge\cdots \wedge \psi_{i_{r-2}})\ ,
\end{equation}
where $\cdot$ between a pair of rank $r$ tensors denotes $\frac{1}{r!}$ times pairwise 
index contractions. This operator exists for $N\geq r$: otherwise, it vanishes due to 
the $U(N)$ relation (\ref{vector-anti}). (\ref{non-grav-mono}) transforms in the $r-1$ dimensional 
(i.e. spin $\frac{r-2}{2}$) representation of both $SU(2)$ global symmetries. 
It remains to be seen whether (\ref{non-grav-mono}) is $Q$-exact or not, 
and also whether its single-trace contents contain gravitons only or have higher 
spin particles as well. 

We first discuss the exceptional cases at $r=2,3$. At $r=2$, (\ref{non-grav-mono}) 
becomes $u^{jb}u_{jb}=\frac{1}{2}Q(q^i\cdot\psi_i+\tilde{q}^a\cdot\tilde\psi_a)$. 
So this cohomology is trivial. At $r=3$, (\ref{non-grav-mono}) can be written as
\begin{eqnarray}\label{r=3-graviton}
  3\,(q^j\wedge q_j\wedge \tilde\psi_a)\cdot (\tilde{q}^b\wedge \tilde{q}_b\wedge \psi_i)
  &=&u_{ia}u^{jb}w_{jb}-8u^{jb}v_{ij}\tilde{v}_{ab}\\
  &&-Q\left[4(D\q^b \cdot q_i)\tilde{v}_{ab}
  - 2(D\q_a \cdot q_i)(\q^b \cdot \psit_b) + 4{u_i}^b(D\q_{(a} \cdot \psit_{b)}) 
  \right.\nn\\
  &&\hspace{1cm} + \frac14 u_{ia}D(\psi_j \cdot q^j + \q^b \cdot \psit_b)
\left.+ 4(\q^b \cdot \psit_b)(\psi_i \cdot \psit_a) \right]\ .
  \nonumber
\end{eqnarray}
The operator (\ref{non-grav-mono}) is cohomologous to multi-gravitons,
namely the first line of \eqref{r=3-graviton}, at $r=3$.\footnote{The
graviton operator on the right hand side of (\ref{r=3-graviton}) 
is $Q$-exact at $N\leq 2$ since the left hand side vanishes. 
This is an example of reduced multi-graviton states due to trace relations, 
as explained in Section \ref{sec:2.1}}  
So at $r=2,3$, we do not find new cohomologies beyond gravitons.

Then for higher $r\geq 4$ and large enough $N$, we have checked that (\ref{non-grav-mono})
for $r=4,5,6,7$ are not $Q$-exact and also not cohomologous to any multi-gravitons. 
These have been checked rather brutally on a computer by consturcting all
BPS cohomologies and quotienting graviton cohomologies at $N=\infty$.
We conjecture that (\ref{non-grav-mono}) is neither $Q$-exact nor multi-graviton 
for all $r\geq 4$. The discussions below show that various features of the index and 
the BPS partition function can be naturally understood based on this conjecture.

First, we look for a sign of the operators (\ref{non-grav-mono}) in the index.
We compute $Z_N-Z_\infty$ for $N=1,2,\cdots,7$ and find that 
\begin{equation}\label{index-difference}
  Z_N(x,y_{1,2})-Z_\infty(x,y_{1,2}) = -x^{3N-1}\chi_N\hat\chi_N+\cdots~
   \leftrightarrow \quad
  \frac{Z_N(x,y_{1,2})}{Z_\infty(x,y_{1,2})}
  = 1-x^{3N-1}\chi_N\hat\chi_N+\cdots \ .
\end{equation}
$\chi_n(y_1)$ and $\hat\chi_n(y_2)$ are respectively the characters of the two $SU(2)$ symmetries 
for the dimension $n$ representations.
Note that if the operator (\ref{non-grav-mono}) exists as a nontrivial cohomology, 
it contributes to the index $+x^{3r-4}\chi_{r-1}\hat\chi_{r-1}$ for $N\geq r\geq 3$. 
Since the operator 
(\ref{non-grav-mono}) does not exist due to the trace relation for $r >N$, 
its contribution should be present in $Z_\infty$ but not in $Z_{N<r}$. The right hand sides
of (\ref{index-difference}) precisely measures the lightest absent operator at 
$r=N+1$ in the $U(N)$ theory. We interpret this as an indirect evidence for the presence 
of (\ref{non-grav-mono}) in $Z_\infty$.
(At $N=1,2$, the lightest absent operators $O^{(N)}$ are multi-gravitons.)
In fact this term in the index was the original motivation for us to construct the 
operators (\ref{non-grav-mono}). Here note that, in 4d $U(N)$ $\mathcal{N}=4$ 
Yang-Mills theory, such lightest absent operator in the index is the maximal giant graviton 
operator made of $N+1$ scalars. It is curious that such lightest excluded 
operators in the ABJ vector model are typically not multi-gravitons.

We have further evidence for the higher spin BPS bounds,
both of the type (\ref{non-grav-mono}) and beyond.
As we outline in Appendix \ref{sec:appA}, we have separately counted all cohomologies
and the multi-graviton cohomologies.
The results are summarized in the BPS partition function $Z_N(x,y_{1,2},y)$  
in Appendix \ref{sec:appA}.
We quote part of \eqref{partition-fortuity} and \eqref{Z-infty-non-grav} here:
\begin{eqnarray}\label{Z-Zgrav-4andinfty}
  Z_4-Z_{4,\textrm{grav}}&=&x^8 y^8 \chi_3\hat\chi_3+\cdots~,
  \nonumber\\
  Z_\infty-Z_{\infty,\textrm{grav}}&=&\left[
  y^8\left(x^8\chi_3\hat\chi_3+\mathcal{O}(x^{15})\right)+y^{10}\left(x^{10}\chi_3\hat\chi_3
  +x^{11}(3\chi_4\hat\chi_4+\chi_4\hat\chi_2+\chi_2\hat\chi_4)+\mathcal{O}(x^{12})\right)
  \right.\nonumber\\
  &&\left.+y^{12}\left(2x^{13}\chi_4\hat\chi_4+x^{14}(6\chi_5\hat\chi_5+\cdots)
  +\mathcal{O}(x^{15})\right)\right]\chi_{\textrm{desc}}\ ,
\end{eqnarray}
where in the second equation, we have factored out the contributions $\chi_{\textrm{desc}}$ 
from the superconformal descendants:
\begin{equation}
\chi_{\rm desc} = \frac{\prod_{\pm} (1 + xy_1^{\pm 1})(1 + xy_2^{\pm 1})}{1-x^2}~.
\end{equation}
The contribution of (\ref{non-grav-mono}) 
to the partition function is $x^{3r-4}y^{2r}\chi_{r-1}\hat\chi_{r-1}$ for $N\geq r$,
and since they are not gravitons for $r\geq 4$, we expect this contribution
in \eqref{Z-Zgrav-4andinfty}.
For $N=4$ and $N=\infty$, the first terms are given by $x^8y^8\chi_3\hat\chi_3$
meeting the expectation, so we interpret them as coming from $O^{(4)}$ of (\ref{non-grav-mono}).  
There are also terms $3x^{11}y^{10}\chi_4\hat\chi_4$ and $6x^{14}y^{12}\chi_5\hat\chi_5$
which we may interpret as partly coming from $O^{(5)}$ and $O^{(6)}$, respectively. 
In particular, in the non-graviton index $Z_\infty-Z_{\infty,\textrm{grav}}$,
every term is an evidence for a higher spin BPS state beyond (\ref{non-grav-mono}). 
(Since $N=\infty$, none of them can come from $U(N)$ fortuity.)
See Appendix \ref{sec:appA} for the explicit constructions of some of these cohomologies
based on dressing (\ref{non-grav-mono}) by gravitons.

One can find more evidences that such multi-particle BPS bounds are abundant in 
the low energy spectrum. We present one from the large $N$ index at fixed `temperature' $\beta^{-1}$ 
(related to the fugacity by $x=e^{-\beta}$). In Appendix \ref{sec:appA}, the following expression 
for the large $N$ index $Z_\infty$ is derived:
\begin{eqnarray}\label{Z-infty}
  Z_\infty(x,y_{1,2})&=&\exp\left[\sum_{n=1}^\infty\frac{1}{n}
  \frac{x^{\frac{n}{2}}(y_1^n+y_1^{-n})-x^{\frac{3n}{2}}(y_2^n+y_2^{-n})}{1-x^{2n}}
  \frac{x^{\frac{n}{2}}(y_2^n+y_2^{-n})-x^{\frac{3n}{2}}(y_1^n+y_1^{-n})}{1-x^{2n}}\right]
  \nonumber\\
  &=&\exp\left[\sum_{n=1}^\infty\frac{1}{n}f_{\rm F}(x^n,y_{1,2}^n)f_{\rm A}(x^n,y_{1,2}^n)\right]\ .
\end{eqnarray}
$f_{\rm F}$ and $f_{\rm A}$ denote fundamental/anti-fundamental letter indices, respectively.
Since each of them counts the column/row $N$-vector fields acted on by derivatives, 
of schematic forms $\partial^{j_1}V$ and $\partial^{j_2}W$,
(\ref{Z-infty}) says that the index acquires contributions from 
the bilinears $\partial^{j_2}W\cdot \partial^{j_1}V$ 
and their multi-traces. These bilinears can be combined 
into the form $\partial^{j}(W\cdot \partial^{k}V+\cdots)$, which are the operators in the 
higher spin current multiplets including the conformal descendants. Of course there are 
many cancellations due to the minus signs in $f_{\rm F}$ and $f_{\rm A}$, but we find 
that the typical contributions to $Z_\infty$ come from non-graviton states.

To be concrete about the last claim, we study the high temperature behaviors of 
this index, defined as follows. In the partition function without the $(-1)^F$ insertion,
the high temperature limit is given by $x=e^{-\beta}\rightarrow 1^{-}$. However, 
with minus signs appearing in the index, one should take a limit in which 
the `free energy' $\log Z_\infty$ diverges the fastest. To simplify the discussions, 
let us turn off the extra fugacities $y_{1,2}$ to the values which preserve 
the $SU(2)\times SU(2)$ symmetries. It turns out that one can take $y_1=y_2=1$. 
(See Section \ref{sec:3} for further explanation about this point.)
The index $Z_\infty$ in this setup is given by
\begin{equation}\label{Z-infty-unrefined}
  Z_\infty=\exp\left[\sum_{n=1}^\infty\frac{1}{n}\frac{4x^n}{(1+x^n)^2}\right]\ .
\end{equation}
We want to take the limit in which $\log Z_\infty$ diverges the fastest. One motivation 
for this is that we want to go to the regime in the $x$ space whose Legendre transformation 
yields the maximal indicial entropy. Some terms in the exponent of (\ref{Z-infty-unrefined}) 
will diverge if $x^{n_0}\rightarrow -1$ for some positive integer $n_0$. In this limit, 
the divergent terms in (\ref{Z-infty-unrefined}) are for $n$ given by $n_0$ times odd
integers. One finds that the fastest divergence happens at $n_0=1$: at other $n_0$'s, one finds 
$\left[\log Z_\infty\right]_{n_0}=\frac{1}{n_0}\left[\log Z_\infty\right]_{n_0=1}$. 
So we define $\beta$ by $x\equiv -e^{-\beta}$ and take the `high temperature' limit 
$\beta\rightarrow 0$, regarding 
$\beta^{-1}$ formally as the temperature conjugate to the `energy' $j= E+J$.
The nature of this high temperature limit is that we first take $N$ large an then 
take $\beta$ small: $N^{-1}\ll |\beta|\ll 1$. In Section \ref{sec:3}, we shall consider 
the large $N$ and high temperature limits either in the opposite order or as a simultaneous 
scaling limit, to unveil more interesting physics.

The high temperature free energy is given by
\begin{equation}\label{Z-infty-high-T}
  \log Z_\infty\approx -\frac{4}{\beta^2}\sum_{n=\textrm{odd}}\frac{1}{n^3}
  =-\frac{7\zeta(3)}{2\beta^2}\ .
\end{equation}
The indicial entropy at large $j=E+J$ (conjugate to $\beta$) is given by the 
Legendre transformation, the large charge saddle point approximation of 
the Laplace transformation, which extremizes 
\begin{equation}
  S(j,\beta)=\log Z_\infty +\beta j\approx-\frac{7\zeta(3)}{2\beta^2}+\beta j
\end{equation}
in $\beta$. There are three solutions of the Legendre transformation, 
\begin{equation}\label{Z-infty-legendre-root}
  \beta=e^{\pm \frac{\pi i}{3}}\left(\frac{7\zeta(3)}{j}\right)^{\frac{1}{3}}\ \ ,\ 
  -\left(\frac{7\zeta(3)}{j}\right)^{\frac{1}{3}}\ .
\end{equation}
The last real negative solution is unphysical: among others, it violates the physical 
requirement ${\rm Re}(\beta)>0$. The other two saddles are complex and appear in conjugate pairs. 
Adding their contributions to the Laplace transformation, one obtains the macroscopic 
indicial degeneracies including the possible fluctuations of overall signs
(See \cite{Agarwal:2020zwm} or discussion around \eqref{degeneracy-oscillate}).
At these complex $\beta$, $\log Z_\infty$ has the positive real part.
This means that, despite the negative coefficient of (\ref{Z-infty-high-T}) 
which would have yielded $|Z_\infty|\ll 1$ at real $\beta$,
$|Z_\infty|\gg 1$ at suitably complex $\beta$.

For our purpose now, to see the cotributions from the higher spin multi-particles,
note that the free energy (\ref{Z-infty-high-T}) diverges quadratically 
in $T\sim \beta^{-1}$, i.e. $\log Z_\infty\sim \beta^{-2}$. Our BPS operators include 
only one derivative $D\equiv D_{1+i2}$ among the three of them on $\mathbb{R}^3$. 
Had a partition function been acquiring contributions from finite species of particles, 
it should have diverged as $\log Z\sim \beta^{-1}$ in the high temperature limit. 
So the high temperature limit of the graviton free energy cannot diverge faster than 
$\beta^{-1}$. The quadratic divergence like (\ref{Z-infty-high-T}) 
is possible only when the contribution comes from infinitely many particle species, 
for instance like $W\cdot \partial^k V+\cdots$ for all $k\geq 0$ as we asserted below (\ref{Z-infty}). 
Similar studies are made for the partition function of the vector model in the free limit
\cite{Shenker:2011zf}, in which one finds $\log Z\sim T^4$ for higher spin particles instead 
of $\log Z\sim T^2$ for finite particle species. This can be understood as volume$^2\sim (T^2V)^2$
contribution to the free energy from the bilocal higher spin fields. 
In our index, $T^2$ may be regarded as the square of the `holomorphic volume' probed by 
$D_{1+i2}$.

The BPS states of this subsection became $Q$-closed by $N^\prime=1$ trace relations. 
Since the $U(N^\prime)$ fortuitous cohomologies are not stable against changing $N^\prime$,  
the BPS states of this subsection are no longer BPS for the ABJ theories with higher $N^\prime$. 
In fact, the thresholds for these $U(N^\prime)$
fortuitous cohomologies will increase in $N^\prime$. So if we move 
from the region $\lambda_B=\frac{N^\prime}{N}\ll 1$ of higher spin gravity 
to that $\lambda_B\approx 1$ for string theory, these operators will become heavy  
and indistinguishable from the typical black hole states which are $U(N)$ fortuitous. 
Changing the viewpoint around, some of the black hole fortuitous states at $N^\prime=N$ will 
get lighter as $N^\prime$ is reduced, eventually coming down to low energies 
when $N^\prime\sim \mathcal{O}(1)$. In this sense, we view our higher spin BPS 
bounds as a low energy remnant of some black hole states in the higher spin gravity regime. 
In Section \ref{sec:3}, we will study the large $N$ phases of this system. We will find saddles which 
qualitatively behave like large black holes, but none that behaves like small black holes. 
Instead, we find that the low temperature saddles acquire richer structures due to 
the higher spin BPS bounds studied in this section. So we view the higher spin bounds 
as a remnant of the small black hole states remaining in the higher spin gravity regime $N^\prime=1$.

The BPS bounds of this subsection are formed in the weakly coupled regime of the higher spin gravity. 
So it may be possible to address these objects from the bulk Vasiliev 
theory.\footnote{We thank Chi-Ming Chang for pointing out and discussing this question.}

\subsection{A new heavy cohomology at $N=2$}\label{sec:2.3}

In this subsection, we present a cohomology 
which becomes $Q$-closed by $U(N)$ trace relations. 
Although such $U(N)$ fortuitous cohomologies are expected to exist,
it is not easy to construct concrete examples.
One reason is that, among the set of free BPS operators, the fortuitous 
cohomologies are relatively sparse. Another reason is that their existence depends on 
trace relations which are specific to each gauge group, so there are no universal 
frameworks available to study them (so far). So very conservatively, we find 
it important to first illustrate their existence in the ABJ vector model.
We present the first example of fortuitous cohomology
with the smallest gauge group $N=2$.

The sign of the smallest $U(2)$ fortuitous cohomology can be detected by studying the 
cohomology counting that we summarize in Appendix \ref{sec:appA}.
To certain charges (i.e. to certain order of $x$),
we counted all $N=2$ cohomologies as well as the $N=2$ multi-graviton cohomologies.
Subtracting the two, the first few terms of the BPS non-graviton partition function are  
\begin{equation}\label{N=2-fortuitous-Z}
  Z_2(x,y_{1,2},y)-Z_{2,\textrm{grav}}(x,y_{1,2},y)
  =x^8y^6+x^9y^6\chi_2\hat\chi_2+x^{10}y^6(2+\chi_3+\hat\chi_3)+\cdots\ .
\end{equation}
So from the leading term $x^8y^6$, 
one expects a fermionic non-graviton cohomology with $E+J=8$ and $R=6$.  
It is a femionic state, since $J=\frac{E}{2}-\frac{R}{4}=\frac{5}{2}$.
From (\ref{N=2-fortuitous-Z}) alone, it is unclear whether this operator is
fortuitous at $N=2$ or not.
Here note that we did a similar calculation at $N=3$, finding that
$Z_3(x,y_{1,2},y)-Z_{3,\textrm{grav}}(x,y_{1,2},y)$ starts from an order
higher than $x^8$: see Appendix \ref{sec:appA}. 
So we expect to find an $N=2$ fortuitous cohomology at this order.

We present a representative of this cohomology: 
\begin{eqnarray}
    O&=&(\psi_i\cdot q^i)(\psi_j\cdot\psit_a)(\psi^j\cdot\psit^a) 
    + 2(\psi_i\cdot q^i)(\psi_j\cdot Dq_k)(\psi^j\cdot q^k) \nonumber\\
    &&-2 (\psi_i\cdot Dq^j)(\psi_j\cdot q^k)(\psi_k\cdot q^i)
    -2 (\q^a\cdot Dq^i)(\psi_i\cdot q^j)(\psi_j\cdot \psit_a) \ .
\end{eqnarray}
This can also be written as
\begin{equation}\label{N=2-fortuitous}
  O=(\psi_i\cdot q^i)(\psi_j\cdot\psit_a)(\psi^j\cdot\psit^a) 
    + 2(\psi_i\cdot q^i)(\psi_j\cdot Dq_k)v^{jk}
    -2 (\psi_i\cdot Dq^j){v_j}^k{v_k}^i
    -2 (\q^a\cdot Dq^i)(\psi_i\cdot q^j)(\psi_j\cdot \psit_a)\ ,
\end{equation}
rewriting some single-traces as gravitons \eqref{BPS-B1}.
We checked that $O$ is $Q$-closed, not $Q$-exact and
not cohomologous to a graviton.
See Appendix \ref{sec:appA.2} for the outline of these calculations.

It is illustrative to see how this operator becomes $Q$-closed by using $N=2$ 
trace relations.
After some calculations, one obtains
\begin{equation}
    QO = \frac12 w^{ia}(q^k\wedge q_k \wedge \psit_a)\cdot(\q^c\wedge \q_c \wedge \psi_i) 
    - v^{ij}(q^k\wedge q_k \wedge\psit_a)\cdot(\q^a\wedge\psi_i\wedge\psi_j) 
    -v^{ij}(q^k\wedge q_k\wedge Dq_i)\cdot(\q^a\wedge\q_a\wedge \psi_j) \;.
\end{equation}
(Recall that $\cdot$ on a pair of rank $r$ tensors means $\frac{1}{r!}$ times 
the full pairwise index contractions.) The right hand side vanishes at $N=2$ 
since all terms involve rank $3$ antisymmetric tensors.
These terms neither vanish nor mutually cancel for $N\geq 3$, 
showing that it is a $U(2)$ fortuitous cohomology.

\section{The large $N$ phases of the index}\label{sec:3}

In this section we study the large $N$ saddle points of the integral 
(\ref{index-formula}) for the index. We will also study the BPS phases represented 
by these saddles and the transition between them.
This index may be regarded as counting either the free BPS states, the 2-loop 
BPS states of Section \ref{sec:2}, or abstractly the interacting BPS states below the 
monopole operator threshold.\footnote{We comment on the effect of monopole operators
before start of section \ref{sec:highTsaddle}.}
Since the free BPS states undergo big cancellations between the 
superpartner pairs, it is better to view the index as counting the interacting BPS states. 
(For instance, the BPS phases deduced by the index is 
very different from that of the free partition function: see Section \ref{sec:3.3}.) 
In principle, there may also be extra accidental cancellations between states 
which are not superpartners. So if there are fine-tuned cancellations even at 
macroscopic charges, the index will substantially underestimate the BPS entropy.  
In AdS superstring theories, the index over the Kaluza-Klein supergravitons has such 
cancellations \cite{Kinney:2005ej} while the black hole index does not 
\cite{Cabo-Bizet:2018ehj,Choi:2018hmj,Benini:2018ywd}. Since the large extra cancellations 
without clear reasons are unnatural, we will assume that the indicial entropy of the ABJ theory 
represents the correct BPS entropy  
at the leading order in large charges.\footnote{As for the KK graviton towers of AdS string theory, 
perhaps the reason for the fine cancelation is their origin from the 10d/11d supermultiplet with 
$32$ supersymmetry. On the other hand, the index  over 
the higher spin particles of Section \ref{sec:2} do not seem to suffer from such big cancellations 
since it respects the kinematic (holomorphic volume)$^2$ structure 
$\log Z_\infty\sim \frac{1}{\beta^2}$ at high temperature.}
Perhaps the fortuity of heavy cohomologies causes irregularities of the spectrum and 
disallows fine-tuned cancellations.

At order $1$ fugacity $x$ (i.e. $|x|$ not close to $1$),
the large $N$ index $Z_\infty(x,y_{1,2})$ is computed in Appendix \ref{sec:appA.1}. As shown in Section \ref{sec:2.2}, 
it captures gravitons and the higher spin BPS bounds at low energy, 
where no trace relation is in effect.
Also, it does not capture the $U(N)$ fortuitous states. 
So it lacks interesting finite $N$ information on the heavy operators. 
The large $N$ eigenvalue distribution which yields $Z_\infty$ 
is the uniform distribution on the circle $|e^{i\alpha_a}|=1$, the confining saddle point
\cite{Sundborg:1999ue,Aharony:2003sx}.

In vector models, more interesting large $N$ saddles appear when the
temperature-like chemical potential scales in $N$ in the following way.
Again for simplicity we turn off $y_1=y_2=1$ at the $SO(4)=SU(2)\times SU(2)$ symmetric point. 
(There are four $SU(2)\times SU(2)$ invariant points $y_1=\pm 1$, $y_2=\pm 1$, but 
the others are related to $y_1=y_2=1$ by suitable phase shifts of $x$ 
and/or $e^{i\alpha_a}$'s.) In this case, the integrand 
of (\ref{index-formula}) is given by the exponential of 
\begin{equation}\label{eff-action}
  -S(\{\alpha\})=\sum_{a\neq b}\log(1-e^{i(\alpha_a-\alpha_b)})
  +2\sum_{a=1}^N\sum_{n=1}^\infty\frac{1}{n}\frac{x^{\frac{n}{2}}}{1+x^{n}}
  (e^{in\alpha_a}+e^{-in\alpha_a})\ .
\end{equation}
In this effective action, the first and second sums respectively have $N^2$ and $N$ terms 
which cannot balance each other to yield nontrivial large $N$ saddles at fixed $|x|<1$: 
the first term will dominate and yield the uniform distribution. To have nontrivial saddles, 
$x$ should scale in $N$ so that the second term has an extra divergent factor,
i.e. the denominator of $\frac{x^{\frac{n}{2}}}{1+x^n}$ should be close to $0$ for some $n$.
For this to happen starting from the `largest' term $n=1$,
one should take $x=-e^{-\beta}$ with small (complex) $\beta$.
With this scaling, all terms labeled by odd $n$'s acquire 
extra large factors $\frac{x^{\frac{n}{2}}}{1+x^n}\approx \frac{(-1)^{\frac{n}{2}}}{n\beta}$.
So nontrivial saddles will appear in the large $N$ and high temperature
scaling limit with fixed $\gamma\equiv\beta N$. 
(Other scalings would presumably yield subleading saddles:
for instance this is clearly true in the Cardy limit, (\ref{Cardy-free}).) 
Note here that \cite{Shenker:2011zf,Jain:2013py} studied the 
large $N$ partition functions of all local operators in the vector models. 
There, nontrivial saddles appear with $\beta\sim N^{-\frac{1}{2}}\ll 1$ scaling.  
In Section \ref{sec:3.3}, we will compare our results for the index with those for the partition 
function in the literature.

In this scaling limit, one should further choose a value between 
$x^{\frac{1}{2}}\approx (-1)^{\frac{1}{2}}=\pm i$ because it appears in (\ref{eff-action}). 
The two choices yield the the two effective actions from (\ref{eff-action}), 
\begin{eqnarray}\label{eff-action-pm}
  \hspace*{-0.5cm}-S_\pm(\{\alpha\})&\approx& \sum_{a\neq b}\log(1-e^{i(\alpha_a-\alpha_b)})
  \pm\frac{2i}{\beta}\sum_{a=1}^N\sum_{n=\textrm{odd}}
  \frac{(-1)^{\frac{n-1}{2}}}{n^2}(e^{in\alpha_a}+e^{-in\alpha_a})\\
  &=&\sum_{a\neq b}\log(1\!-\!e^{i(\alpha_a-\alpha_b)})
  \pm\frac{1}{\beta}\sum_{a=1}^N\left[{\rm Li}_2\left(ie^{i\alpha_a}\right)
  \!-\!{\rm Li}_2\left(-ie^{i\alpha_a}\right)\!+\!{\rm Li}_2\left(ie^{-i\alpha_a}\right)
  \!-\!{\rm Li}_2\left(-ie^{-i\alpha_a}\right)\right]\ .\nonumber
\end{eqnarray}
However, one finds that 
\begin{equation}\label{pm-relation}
  S_-(\beta,\{\alpha\})=S_+(\beta,\{\alpha+\pi\})\ .
\end{equation}
Since $\alpha_a$'s are integration variables, (\ref{pm-relation})
implies that the two choices $x^{\frac{1}{2}}\approx \pm i$ yield identical 
matrix integrals. Also note the relation that involves complex conjugate of $\beta$,
\begin{equation}\label{eff-action-conjugation}
  S_+(\beta,\{\alpha\})^\ast=S_-(\beta^\ast,\{\alpha^\ast\})
  =S_+(\beta^\ast,\{\alpha^\ast+\pi\})\ .
\end{equation}
We shall mostly use $S_+$ for the computations.

Before proceeding, we comment on an interpretation of the two dual descriptions $S_\pm$. 
They are complex functions in the sense that the coefficients are complex, i.e. 
the factor $i$ on the second term. This is related to the fact that the large $N$ saddle 
point calculation of the indicial entropy uses the complex chemical potential. 
The reason for this is as follows \cite{Agarwal:2020zwm}. The microcanonical 
index $\Omega(j)$ at fixed charge $j$ is obtained by expanding the grand canonical
partition function $Z(x)$ in the chemical potential $x$:
\begin{equation}
  Z(x)=\sum_{j}\Omega(j) x^j\ .
\end{equation}
Equivalently, $\Omega(j)$ can be obtained from $Z(x)$ by the Laplace transformation:
\begin{equation}\label{laplace}
  \Omega(j)=\frac{1}{2\pi i}\oint \frac{dx}{x} x^{-j}Z(x)\ .
\end{equation}
$\Omega(j)$ is an integer-valued function of quantized $j$. It increases very quickly in $j$, 
but with alternating signs depending on whether bosons or fermions dominate \cite{Murthy:2020scj,Agarwal:2020zwm}. 
On the other hand, the saddle point calculation of the indicial entropy involves 
the large charge approximation of the integral (\ref{laplace}). 
As a result, one obtains a continuous function of $j$, $\Omega_\star(j)\sim e^{S_\star(j)}$, 
with the discreteness of $\Omega(j)$ and $j$ obscured. 
With a complex effective action like (\ref{eff-action-pm}), $S_\star(j)$ will be a complex 
function of real $j$, which by itself does not even represent the coarse-grained degeneracy. 
The coarse-grained index should be a real oscillating function.
Instead, the real oscillating function is obtained from a complex conjugate pair of saddles
$x_\star$, $x_\star^\ast$ for (\ref{laplace}), 
\begin{equation}\label{degeneracy-oscillate}
  \Omega(j)\sim e^{S_\star(j)}+e^{S_\star(j)^\ast}\sim 
  e^{{\rm Re}[S_\star(j)]}\cos[{\rm Im}(S_\star(j))]\ .
\end{equation}
${\rm Re}[S_\star(j)]$ provides the leading entropy and leads to the enveloping function,
while the cosine function represents the oscillating signs.

We will approximately compute $Z(\beta)$ in the integrand of (\ref{laplace})
using the large $N$ saddle point approximation for $\alpha_a$'s.
In this setup, the pair $x_\star, x_\star^\ast$ can appear in two possible ways.
First, they may appear from a definite real function $\log Z(\beta)$ whose 
Legendre transformation has a pair of complex roots.
An example is the uniform confining distribution for $Z_\infty$ that is self-conjugate.
Second, they may come from two different complex functions $\log Z_\pm(\beta)$
(i.e. two distinct large $N$ saddles)
that are approximations of $\log Z(x)$ in different regions of $x$,
whose respective Legendre transformations yield the complex conjugate pairs
$S_\star(j)$ and $S_\star(j)^\ast$. 
In our setup, two different complex background values of the chemical potential,
namely $x^{\frac12} \approx \pm i$, lead to the pair of effective actions $S_\pm$
that are conjugate to each other by (\ref{eff-action-conjugation}).
Given a saddle $\{\alpha\}$ for $S_+$ at $\beta$,
a conjugate saddle $\{\alpha^\ast\}$ can be found from $S_-$ at $\beta^\ast$.
Both saddles contribute to the integral in \eqref{laplace} and
play the role of $S_\star(j)$ and $S_\star(j)^\ast$ in \eqref{degeneracy-oscillate}.

Note that the exact $Z(\beta)$ is real by definition, since
all coefficients $\Omega(j)$ are real. The complexity of $\log Z_\pm$ may appear only 
due to the large $N$ saddle point approximation that specifies particular complex background 
values of the chemical potential, such as $x^{\frac{1}{2}}\approx \pm i$ here. 
Pairs of conjugate saddles play important roles for computing 
the black hole entropy from the index \cite{Cabo-Bizet:2018ehj,Choi:2018hmj,Benini:2018ywd},  
which will also be the case in our ABJ vector model.

The external potential of (\ref{eff-action-pm}), 
the second term consisting of the ${\rm Li}_2(\pm ie^{\pm i\alpha_a})$ functions, 
is singular at $\alpha_a=\pm\frac{\pi}{2}$ because the dilogarithm ${\rm Li}_2(x)$
has a cusp at $x=1$. The potential function itself is finite, but the force 
$\sim {\rm Li}_1(\pm ie^{\pm i\alpha_a})=-\log(1-e^{\pm i(\alpha_a\pm\frac{\pi}{2})})$ 
given by its $\alpha_a$ derivative diverges there. Since $\alpha_a=\pm\frac{\pi}{2}$ are 
on the original integration contour, one should clarify the origin 
of this singularity to understand the calculations using such a singular potential. 
Each term of the potential arises from the infinite sum of the form
\begin{equation}\label{div-pot-1}
  \sum_{n=0}^\infty \log (1-x^{l+2n}e^{\pm i\alpha_a})\ ,
\end{equation}
where $l=\frac{1}{2}$ for bosons and $\frac{3}{2}$ for fermions. 
When  $x^{\frac{1}{2}}=\pm ie^{-\beta/2}$ 
with $\beta\rightarrow 0$, (\ref{div-pot-1}) becomes
\begin{equation}\label{div-pot-2}
  \sum_{n=0}^\infty\log(1-e^{-M-2n\beta})
  \sim \frac{1}{\beta}\int_{M}^\infty dE~\log(1-e^{-E})\sim \frac{1}{\beta}{\rm Li}_2(e^{-M})
\end{equation}
where $M=\pm i(\alpha_a\pm\frac{\pi}{2})$ is interpreted as an effective mass. 
The integral has a singularity $M\log M$ when $M\rightarrow 0$: 
the potential is finite but the force $\log M$ diverges there. The sum (\ref{div-pot-1}) or the integral (\ref{div-pot-2}) is formally that of the $D_{\rm eff}=1+1$ dimensional field of mass 
$M$ at high temperature. More generally, for such a field in $D$ spacetime dimension, 
one finds 
\begin{equation}\label{div-pot-3}
  \frac{1}{\beta^{D-1}}\int_{M}^\infty dE~ E^{D-2}\log(1-e^{-E})\sim 
  \frac{M^{D-1}\log M}{\beta^{D-1}}\ \ \ \ \textrm{when }\ M\rightarrow 0\ .
\end{equation}
For instance, this is the behavior of the partition function on $S^{D-1}\times S^1$. 
For the index on $S^{D-1}\times S^1$, $D$ in the expression above is replaced by the effective 
spacetime dimension $D_{\rm eff}=\lfloor \frac{D}{2}\rfloor+1$. For larger $D$ or $D_{\rm eff}$, 
the effective potential $\sim M^{D-1}\log M$ is less singular since the IR divergence is milder 
in higher dimensions as we remove the IR regulator $M\rightarrow 0$. For the partition
function of the $D=2+1$ vector model \cite{Shenker:2011zf,Jain:2013py}, both 
the potential $\sim M^2\log M$ and the force $\sim M\log M$ on the eigenvalue $\alpha_a$ are finite 
in the massless limit. (Since interactions induce nonzero thermal mass \cite{Jain:2013py}, 
the massless limit can be reached only in the free theory.) On the other hand, our index 
with $D_{\rm eff}=2$ suffers from more violent IR divergence.

From (\ref{div-pot-1}), the divergence is caused by the accumulation of infinitely many 
singularities at $\alpha_a=\pm\frac{\pi}{2}\pm 2ni\beta$ for $\beta\rightarrow 0$ 
(all four sign choices possible). 
We discuss the implications of these singularities in the saddle point 
approximation of the integral.  One deforms the original contour to the 
steepest descent contour for calculations. The saddle points that we will find in this section are 
all away from $\alpha_a=\pm\frac{\pi}{2}$, locally free of the singularities. 
Furthermore, during the contour deformation, one should add the extra residue 
contributions if the contour crosses the poles of the integrand. To be definite, let us choose 
the effective action $S_+$, for which $x^{\frac{1}{2}}\approx i$. The singularities in 
(\ref{div-pot-1}) caused by bosons are poles of the integrand.
The poles near $\alpha_a=\pm\frac{\pi}{2}$ are at $\alpha_a=\pm(\frac{\pi}{2}+2n\beta i)$, 
which are all $\sim \pm(\frac{\pi}{2}+i\epsilon)$ in the $\beta\rightarrow 0$ limit. 
In other words, the poles approach the limiting points on the original contour (real $\alpha_a$) 
from one side rather than pinching it, i.e. approach $+\frac{\pi}{2}$ from above  
and $-\frac{\pi}{2}$ from below. So if the deformation towards the 
steepest descent contour happens in the direction avoiding these accumulating poles, 
there will be no issue of the extra residue contributions.

Deciding the steepest descent contour is beyond our scope. 
As is often the practice, we will assume that our saddle points are on the steepest 
descent contour. However, we will see in this section and Appendix \ref{sec:appB} that the complex 
eigenvalue distributions at the saddle points are distributed below $+\frac{\pi}{2}$ 
and above $-\frac{\pi}{2}$ (see Fig. \ref{1-cut}), which we think may be a sign that 
the steepest descent contour avoids the poles accumulating in the $\beta\rightarrow 0$ limit.

The large $N$ saddle point approximation with (\ref{eff-action-pm}) 
is studied in the continuum approximation. The eigenvalues $\alpha_a$ are densely 
distributed along a curve $\theta(s)$ labeled by a real parameter $s$,
on the complex plane for $\alpha_a$ which we call the $\theta$-plane. 
The distribution may be along one segment of a curve, or many disconnected segments. 
We call these segments `cuts.' The cuts are called $C_i$, where $i=1,\cdots,\#(\textrm{cuts})$. 
The eigenvalue distribution on the cut is specified by the density function $\rho(s)$, $\frac{1}{N}$
times the number density of eigenvalues, constrained by
\begin{equation}\label{rho-constraints}
  \sum_{i}\int_{C_i}ds\rho(s)=1\ ,\ \ \rho(s)\geq 0\ .
\end{equation}
To find the saddle point solution, one should determine $C=\cup_i C_i$ 
as well as the density function on the cuts which extremizes the following 
continuum effective action ($\gamma \equiv \beta N$)
\begin{eqnarray}\label{eff-action-continuum}
  -\frac{S_\pm}{N^2}&=&\int\int ds ds^\prime \rho(s)\rho(s^\prime)
  \log(1-e^{i(\alpha(s)-\alpha(s^\prime))})\\
  &&
  \pm\frac{1}{\gamma}\int ds \rho(s)\left[{\rm Li}_2\left(ie^{i\alpha(s)}\right)
  \!-\!{\rm Li}_2\left(-ie^{i\alpha(s)}\right)\!+\!{\rm Li}_2\left(ie^{-i\alpha(s)}\right)
  \!-\!{\rm Li}_2\left(-ie^{-i\alpha(s)}\right)\right]~, \nonumber
\end{eqnarray}
subject to the constraints (\ref{rho-constraints}).

The effective action (\ref{eff-action-continuum}) has a 2-body interaction (first term) and 
a background potential (second term). When the background potential is a real function of 
$\alpha_a$, the cuts $C_i$ can also be taken on the real axis. 
When the potential is furthermore a finite polynomial of $e^{\pm i\alpha_a}$, the saddle point 
solutions have been studied systematically: see for instance \cite{Jurkiewicz:1982iz}. 
These studies are extended to the case in which the potential is a general real function, i.e. 
an infinite series of $e^{\pm i\alpha_a}$, where a formal infinite series solution for 
$\rho(s)$ is obtained \cite{Aharony:2003sx}. 
For complex potentials, one should also determine the cuts $C_i$ on the complex plane. 
This can be done as follows \cite{David:1990sk,Copetti:2020dil,Choi:2021lbk}. (We follow 
the notations and setups of \cite{Choi:2021lbk}.) 
By explicitly computing the formal solution of \cite{Aharony:2003sx}, one first 
obtains the `bulk density function' $\rho(\theta)$ for the eigenvalue distribution, 
which is locally a holomorphic function. ($\rho(\theta)$ 
suffers from branch point singularities at certain points, as will be explained below.)
The cuts $C_i$ are then locally determined from $\rho(\theta)$ by finding the curves for which
$\rho(\theta(s))d\theta(s)$ is real and positive. If such cuts globally exist, the 
distribution $\rho(s)$ is given by the pullback $\rho(s)ds=\rho(\theta(s))d\theta(s)$.
For given $\rho(\theta)$, whether the cuts $C_i$ exist or not depends on situations.
As we explain below, one can determine $\rho(\theta)$ of our interest analytically, 
while the cuts $\cup_i C_i$ are determined only numerically. With analytic knowledge of $\rho(\theta)$, 
one can sometimes compute physical quantities like the free energy $\log Z$ analytically 
at the saddle points, without knowing the analytic expressions for the cuts.

We will need to study the 1-cut and 2-cut saddle point solutions to understand 
the BPS phases of the ABJ vector model. We will first study the 1-cut solutions for $S_+$
that are centered around $\theta=0$ and reflection symmetric in $\theta\to -\theta$.
(This is a symmetry of the effective action, which we impose on the solutions.)
For given complex $\gamma=N\beta$, its bulk $\rho(\theta)$ is given by
\begin{eqnarray}\label{1-cut-summary}
  \rho(\theta)&=&\frac{2i}{\pi\gamma}\tan^{-1}\left[
  \frac{\sqrt{\sin^2\frac{\theta_0}{2}-\sin^2\frac{\theta}{2}}}
  {\cos\frac{\theta}{2}\sqrt{\cos\theta_0}}\right]=
  \frac{2i}{\pi \gamma}\tan^{-1}\sqrt{\frac{\cos\theta-\cos\theta_0}{\cos\theta_0(1+\cos\theta)}}
  \nonumber\\
  \gamma&=&i\left(\pi-4\tan^{-1}\sqrt{\cos\theta_0}\right)
  \ \longleftrightarrow\ \cos\theta_0=\left(\frac{1+ie^{\frac{\gamma}{2}}}{i+e^{\frac{\gamma}{2}}}\right)^2\ ,
\end{eqnarray}
where $\pm\theta_0$ are the endpoints of the cut $C\equiv C_1$.
This result is derived in Appendix \ref{sec:app1cut}. This saddle point is not self-conjugate:  
its conjugate saddle will be another 1-cut solution centered around $\pi$. The functions in 
(\ref{1-cut-summary}) suffer from singularities/ambiguities of branch points and branch cuts. 
The branch cut choices are merely conventions locally, but global monodromies around log 
(i.e. $\tan^{-1}$) branch points should be specified in particular manners for (\ref{1-cut-summary}) 
to describe the saddle points correctly. See Section \ref{sec:highTsaddle} and 
Appendix \ref{sec:app1cut} for details.

The 2-cut solutions have the first reflection-symmetric cut $C_1$ around $\theta=0$,
and the second reflection-symmetric 
cut $C_2$ around $\theta=\pi$. The bulk density function is given by
\begin{eqnarray}\label{2-cut-summary}
  \rho(\theta)&=&\frac{1}{\pi\gamma}
  \left[\tanh^{-1}\frac{\sqrt{(c_\theta-c_1)(c_\theta-c_2)}}{c_\theta-i\sqrt{-c_1c_2}}
  -\tanh^{-1}\frac{\sqrt{(c_\theta-c_1)(c_\theta-c_2)}}{c_\theta+i\sqrt{-c_1c_2}}\right]
  \nonumber\\
  \gamma&=&i\left(\pi-4\tan^{-1}\sqrt{\frac{c_1}{-c_2}}\right)\ \ \longleftrightarrow\ \ 
  \frac{\cos\theta_1}{-\cos\theta_2}=\left(\frac{1+ie^{\frac{\gamma}{2}}}
  {i+e^{\frac{\gamma}{2}}}\right)^2\ ,
\end{eqnarray}
where $c_\theta\equiv \cos\theta$, $c_{1,2}\equiv \cos\theta_{1,2}$. The cuts $C_1$, $C_2$ are 
respectively intervals between $(-\theta_1,\theta_1)$ and $(2\pi-\theta_2,\theta_2)$.  
Note that, setting $c_1>0$ and $c_2<0$ for instance, one can rewrite 
\begin{equation}\label{2-cut-C1}
  \rho(\theta)=\left\{\begin{array}{ll}
    \frac{2i}{\pi\gamma}\tan^{-1}\sqrt{\frac{-c_2}{c_1}\frac{c_\theta-c_1}{c_\theta-c_2}}&
    \textrm{if }\ \theta\in C_1\\
    -\frac{2i}{\pi\gamma}\tan^{-1}\sqrt{\frac{c_1}{-c_2}
  \frac{c_2-c_\theta}{c_1-c_\theta}}&
  \textrm{if }\ \theta\in C_2
  \end{array}
  \right.
\end{equation}
by choosing the log branches carefully. So the 2-cut solution 
(\ref{2-cut-C1}) reduces to the 1-cut solution (\ref{1-cut-summary}) centered at $\theta=0$ 
when $c_2=-1$. When $c_1=1$, it reduces to a 
1-cut solution centered at $\theta=\pi$, conjugate to
(\ref{1-cut-summary}). Again this bulk function has branch cut ambiguities, 
whose determination is explained in Section \ref{sec:lowTsaddle} and Appendix \ref{sec:app2cut}.

At given complex $\gamma$, the second equation of (\ref{2-cut-summary}) only fixes
two real parameters among the four real (two complex) $\theta_{1,2}$. 
The extra 2 real parameters are fixed as follows. 
First note that the bulk function $\rho(\theta)$ like (\ref{2-cut-summary}) is obtained
by solving some part of the saddle point equations assuming that the 
cut $C_1\cup C_2$ exists, determined by the local condition $\rho(\theta)d\theta=\textrm{real}>0$. 
The last assumption is violated unless we tune one of the remaining 2 real parameters. 
After this tuning, the saddle point equation is fully solved with  
$C_1\cup C_2$ determined, but still with the last real parameter unfixed. This parameter is the 
`filling fraction' of the 2-cut solution. Namely, there is a 1-parameter family of saddle point 
solutions labeled by $\nu\equiv\int_{C_{1}} d\theta \rho(\theta)$ 
satisfying $0\leq\nu\leq 1$. $\nu$ is $\frac{1}{N}$ times the number of eigenvalues
on the first cut. One usually maximizes $\log Z$ with respect to $\nu$ to find the 
dominant contribution. This issue is quite subtle for complex saddles, which 
will be explained in Section \ref{sec:lowTsaddle}.

To better motivate the studies of the one- and two-cut saddle points, 
it is helpful to first understand the extreme high and low temperature limits. 
Recall that the large $N$ limit already involved a high temperature scaling: 
$N\gg 1$, $|\beta|\ll 1$ with $\gamma\equiv N\beta$ fixed.
The low temperature limit in this setup refers to taking the second limit $|\gamma|\gg 1$,
so that $N^{-1}\ll |\beta|\ll 1$. One can alternatively approach 
this region by changing the order of limits: first take $N\gg 1$ with $\beta$ fixed, 
and then take $|\beta|\ll 1$.
We have already taken the latter approach in (\ref{Z-infty-high-T}) to obtain
$\log Z\sim -\frac{7\zeta(3)}{2\beta^2}$. 
From the viewpoint of the former order of limits, one can rewrite it as 
$\log Z\sim -\frac{7\zeta(3)N^2}{2\gamma^2}$, consistent with the
$\log Z\sim N^2f(\gamma)$ scaling of (\ref{eff-action-continuum}).
As we will explain in Section \ref{sec:lowTsaddle}, this behavior will
demand the low temperature phase to be described by the 2-cut saddles (\ref{2-cut-summary}):
neither gapless nor 1-cut distributions will exhibit this behavior.

As for the high temperature limit, we now discuss 
the `Cardy limit' defined by taking $|\beta|\ll 1$ first with $N$ fixed. After this limit, 
one can then take $N\gg 1$ to study the region $|\beta|\ll N^{-1}\ll 1$. 
We will show shortly that $\log Z\propto \frac{N}{\beta}$ in this region, implying 
that $\log Z$ sees $\mathcal{O}(N)$ species of particles. 
Alternatively in the large $N$ scaling limit with $\gamma$ fixed, one can approach the same 
region by taking $|\gamma|\ll 1$. The Cardy free energy in this viewpoint can be 
written as $\log Z\propto \frac{N^2}{\gamma}$, again taking the form of $N^2f(\gamma)$. 
As we will explain in Section \ref{sec:highTsaddle}, the Cardy regime will appear as the high temperature 
limit $\gamma\rightarrow 0$ of the one-cut saddles (\ref{1-cut-summary}).

We study the Cardy limit $\beta\ll 1$ in detail. 
Now the second term of the effective action (\ref{eff-action-pm}) proportional to $\frac{N}{\beta}$
is much larger than the first term. So ignoring the first term, $S_+$ is given by 
\begin{eqnarray}\label{eff-Cardy}
  \hspace*{-0.5cm}-S_+(\{\alpha\})&\sim&
  \frac{1}{\beta}\sum_{a=1}^N\left[{\rm Li}_2\left(ie^{i\alpha_a}\right)
  \!-\!{\rm Li}_2\left(-ie^{i\alpha_a}\right)\!+\!{\rm Li}_2\left(ie^{-i\alpha_a}\right)
  \!-\!{\rm Li}_2\left(-ie^{-i\alpha_a}\right)\right]\ ,
\end{eqnarray}
in which different $\alpha_a$'s decouple. 
The saddle point equation for each eigenvalue is given by
\begin{equation}
  1=\frac{(1-iz)^2(1+iz^{-1})^2}{(1-iz^{-1})^2(1+iz)^2}=
  \left(\frac{2-i(z-z^{-1})}{2+i(z-z^{-1})}\right)^2\ \ \ \textrm{where}\ z\equiv e^{i\alpha_a}\ .
\end{equation}
Its solutions are $z=\pm 1$, or $\alpha_a=0,\pi$. An eigenvalue at $e^{i\alpha_a}=\pm 1$  
contributes to (\ref{eff-Cardy}) as
\begin{equation}
  \log Z~\leftarrow~ \pm\frac{2}{\beta}\left[{\rm Li}_2(i)-{\rm Li}_2(-i)\right]
  ~=~\pm\frac{4iG}{\beta}
\end{equation}
respectively, where ${\rm Li}_2(\pm i)=\pm i G-\frac{\pi^2}{48}$ and
$G=\sum_{n=0}^\infty\frac{(-1)^n}{(2n+1)^2}\approx 0.916$ is the Catalan's constant.
If $0\leq N_1\leq N$ eigenvalues are at $\alpha_a=0$ and the remaining $N-N_1$ of them are 
at $\alpha_a=\pi$, the net Cardy free energy is given by 
\begin{equation}\label{Cardy-free}
  \log Z\sim \frac{4iGN(2\nu-1)}{\beta}
\end{equation}
where $\nu\equiv \frac{N_1}{N}\in [0,1]$ is the filling fraction of eigenvalues at $\alpha=0$. 
So we have found $N+1$ distinct Cardy saddles, labeled by discrete $\nu$.

At fixed $\nu$, the entropy in the Cardy limit is obtained by extremizing
\begin{equation}\label{entropy-Cardy}
  S(\beta,\nu)=\frac{4iGN(2\nu-1)}{\beta}+\beta j
\end{equation}
in $\beta$, where $j= E+J$ is fixed. 
The solution for $\beta$ satisfying ${\rm Re}(\beta)>0$ is given by
\begin{equation}
  \beta_\star=
  \left\{\begin{array}{ll}
     \sqrt{\frac{4GN(2\nu-1)}{j}}e^{i\frac{\pi}{4}}&\textrm{if } \frac{1}{2}<\nu\leq 1\\
     \sqrt{\frac{4GN(1-2\nu)}{j}}e^{-i\frac{\pi}{4}}&\textrm{if } 0\leq\nu <\frac{1}{2}
  \end{array}
  \right.\ .
\end{equation}
The entropy is given by ${\rm Re}[S(\beta_\star)]$ from (\ref{entropy-Cardy}), 
as already explained. So one obtains
\begin{equation}
  S(j,\nu)={\rm Re}\left[4\sqrt{GN|2\nu-1|j}e^{\pm i\frac{\pi}{4}}\right]
  =2\sqrt{2GN|2\nu-1|j}\ .
\end{equation}
Note that this entropy is maximal at $\nu=1$ and $0$, 
\begin{equation}
  S(j,1)=S(j,0)=2\sqrt{2GNj}\ ,
\end{equation}
and minimal at $\nu=\frac{1}{2}$, $S(j,\frac{1}{2})\sim 0$. The maximal saddles 
$\nu=1,0$ in the microcanonical ensemble have one cut. Note that the 
the single cut saddles at $\nu=1,0$ are the mutually conjugate ones, related by the 
$\pi$ shifts of the eigenvalues $\alpha_a$. The contribution of this pair is actually 
what ensures the real oscillating degeneracy (\ref{degeneracy-oscillate}).

One can also select the maximal saddle in the grand canonical ensemble, 
at fixed complex $\beta$, arriving at the same conclusion $\nu=1$ or $0$. Since this 
is a special case of selecting the filling fraction of 2-cut saddles, and also since we would 
like to suggest a more natural prescription for the grand 
canonical calculation below, we postpone the discussion to Sections \ref{sec:highTsaddle} and \ref{sec:lowTsaddle}.

In the meantime, let us comment on the effect of magnetic monopoles on the studies of this section.
At typical $\gamma = N\beta \sim \mathcal{O}(1)$ that we shall take throughout this section,
the thermal expectation value of the energy is at order $E \sim N^3$,
see e.g. below \eqref{entropy-function}. 
On the other hand, since the threshold of the monopole operators is $E \sim k$, 
one might think that our analysis ignoring monopoles would be valid only when 
$N^3\ll k$, i.e. $\lambda=\frac{N}{k}\sim N^{-2}\ll 1$. However, although we use 
the index formula ignoring the monopoles whose strict validity is $E<k$, there are reasons 
to believe that the large $N$ high temperature thermodynamics in this setup can be trusted 
till $\lambda$ at order $1$ below a threshold.

In various studies, it turns out that the dominant effect of the monopole operators 
in large $N$ thermodynamics is reducing the states rather than increasing them, which 
needs to be discussed more carefully than the naive threshold level analysis.
Roughly speaking, inserting monopole operators in a gauge-invariant operator tends to
reduce the effective gauge symmetry from $U(N)$ to the
one unbroken by the GNO charges of the monopoles.
The elementary fields which minimally couple to these GNO charges acquire
nonzero effective masses, and their contribution to the (deconfined) free energy is suppressed. 
For instance, in the ABJM theory, it was shown that the monopole condensation 
reduces the $N^2$ light fields down to an effective $\sim N^{\frac{3}{2}}$ `near-diagonal' fields, where the latter acquire no bigger than $\mathcal{O}(1)$ mass from the monopoles \cite{Choi:2019zpz}.
So the deconfined phase of the ABJM theory experiences partial Higgsing/confinement 
from the monopole condensation/proliferation.

Rather similar observations have been made in the (not necessarily supersymmetric)
vector-Chern-Simons model \cite{Aharony:2012ns,Jain:2013py}.
For definiteness, consider the ordinary partition function of such theories on $S^2\times S^1$.
In this case, the renormalized inverse temperature is 
$\gamma = N^{\frac{1}{2}} \beta \sim \mathcal{O}(1)$
and the typical thermal energy is at order $E\sim N^{\frac{5}{2}}$.
At such a high temperature, it was argued \cite{Jain:2013py} that the monopole charges
appearing in the letter partition function can be neglected,
so that the sum over the GNO charges yields
\begin{equation}
  \prod_a \sum_{m_a=-\infty}^\infty e^{ik\sum_a m_a \alpha_a}
  \, \sim \, \prod_a \delta(k \alpha_a )\ .
\end{equation}
Here the delta functions $\delta(k\alpha_a)$ are $2\pi$-periodic on a circle,
meaning that the holonomy variables to be integrated over are peaked at
discrete values $\alpha_a=\frac{2\pi n_a}{k}$, where $n_a=0,\cdots,k-1$.
Combined with the Haar measure of the integral $\prod_{a<b}\sin^2\frac{\alpha_a-\alpha_b}{2}$,
this renders an incompressible nature to the eigenvalues $\alpha_a$:
no more than one eigenvalue can be located at a discrete site.
So the effect of monopoles is to make the large $N$ eigenvalue distribution `capped,' constraining it by 
\begin{equation}\label{monopole-cap-onset}
  \rho(\theta)\leq \frac{k}{2\pi N}=\frac{1}{2\pi \lambda}\ . 
\end{equation}  
Since the 
eigenvalue distribution can be less compressed by the capping phenomenon, 
the high temperature growth of the free energy is tamed and so is the entropy growth.
At higher $k$ (smaller $\lambda$), the onset of the capped phase will be delayed to smaller $\gamma$
(higher temperature).

In typical eigenvalue distributions obtained by ignoring the effects of monopoles,
$\rho(\theta) \sim \mathcal{O}(1)$ for $\gamma \sim \mathcal{O}(1)$.
Thus, in the picture of \cite{Jain:2013py} summarized in the previous paragraph,
the onset of the monopole effects (\ref{monopole-cap-onset}) 
happens at finite $\lambda$ (i.e. $k \sim N$). At such an onset point, 
there are already many magnetic monopole operators at energy $N^{\frac{5}{2}} \gg k$, 
but due to the arguments above, their dominant effect is believed to be reducing the states 
via (\ref{monopole-cap-onset}).
Of course, the capping effect will always be important in the extreme high-temperature limit, 
i.e. large-$j$ and small-$|\gamma|$ limit in section \ref{sec:highTsaddle}, 
where $\rho(\theta)$ blows up. If $\lambda$ is smaller, our results in this section 
ignoring such monopole effects will be reliable up to higher temperatures.

For the index, the fact that the eigenvalue distribution is complex technically
complicates the story. However, along the line of discussions made for indices
in Section 4.2 of \cite{Choi:2019zpz}, we expect similar arguments to apply to
our ABJ vector-Chern-Simons theory. In other words, we expect that a complex capped 
distribution would be replacing the saddle points that we have discussed in this paper. 
It will be interesting to study this systematically in the future.

\subsection{High temperature saddles and a threshold}\label{sec:highTsaddle}

In this subsection we study the 1-cut large $N$ saddle point solutions summarized by the bulk 
function (\ref{1-cut-summary}). To complete the construction of the solutions, one should 
determine the eigenvalue cut $C$ which ends on $\pm\theta_0(\gamma)$ given by (\ref{1-cut-summary}). 
$\rho(\theta)d\theta$ must be real and positive along $C$.
This condition is nontrivial because, although the condition of real $\rho(\theta)d\theta$ 
can always determine $C$ incrementally from an initial point, it is not guaranteed that such 
a curve that starts at $-\theta_0$ ends on $+\theta_0$. We examine this problem
mostly numerically, except in certain limits.

We start by explaining the branch point structures and the related branch cut 
conventions of the bulk function (\ref{1-cut-summary}) on the $\theta$-plane. Since $C$ is 
determined by integrating $\rho(\theta)$, a key requirement for the convention 
is that $C$ does not intersect the branch cuts of $\rho(\theta)$.

First, from the argument 
\begin{equation}\label{arctan-argument}
  x\equiv\frac{\sqrt{\sin^2\frac{\theta_0}{2}-\sin^2\frac{\theta}{2}}}
  {\cos\frac{\theta}{2}\sqrt{\cos\theta_0}}
\end{equation}
of $\tan^{-1}$ in (\ref{1-cut-summary}), one finds square-root branch points  
at $\theta=\pm\theta_0$ where the numerator vanishes.
Two branch cuts start from these branch points and move outwards to infinity: 
see \cite{Choi:2021lbk} (in particular Fig. 1) for examples.
We would like to take the region containing $C-\{\pm\theta_0\}$ to be free of the branch cuts
for the bulk function $\rho(\theta)$.
Since our $C$ always passes through $\theta=0$, we set the square root branch such that 
$\sqrt{\sin^2\frac{\theta_0}{2}-\sin^2\frac{\theta}{2}}\stackrel{~\theta=0~}{\longrightarrow} +\sin\frac{\theta_0}{2}$.
This choice is then continued to a region which contains $C-\{\pm\theta_0\}$,
making $\rho(\theta)$ holomorphic there.

Although our main interest here is the branch structures of $\rho(\theta)$ on the 
$\theta$-plane, there is also a square-root branch issue for $\theta_0$ coming from the 
denominator $\sqrt{\cos\theta_0(\gamma)}$ of (\ref{arctan-argument}). We comment 
on it here before proceeding. As we change the chemical potential $\gamma$, and thus $\theta_0$, 
the saddle points will change continuously within a given phase. We will study the family of 
one-cut saddles which contains the high temperature Cardy limit.
In this limit, $\theta_0(\gamma\rightarrow 0)\rightarrow 0$ 
and we choose the $\sqrt{\cos\theta_0}\rightarrow +1$ branch. From this point, 
we will continuously change $\gamma$ and $\theta_0(\gamma)$ along a particular curve 
on the complex $\gamma$ or $\theta$ plane (e.g. determined by the Legendre transformation of 
$\log Z(\gamma)$ at various real charge $j$). Depending 
on how this curve goes around the branch point $\theta_0=\frac{\pi}{2}$, we continue the 
function $\sqrt{\cos\theta_0}$ continuously along this curve.\footnote{One may wonder 
if there are other classes of 1-cut saddles elsewhere on the complex 
$\gamma$-plane, disconnected to the high temperature Cardy regime. We did 
not find any, but we do not claim that our study was comprehensive.}

\begin{figure}[t]
	\centering
	\includegraphics[width=0.6\textwidth]{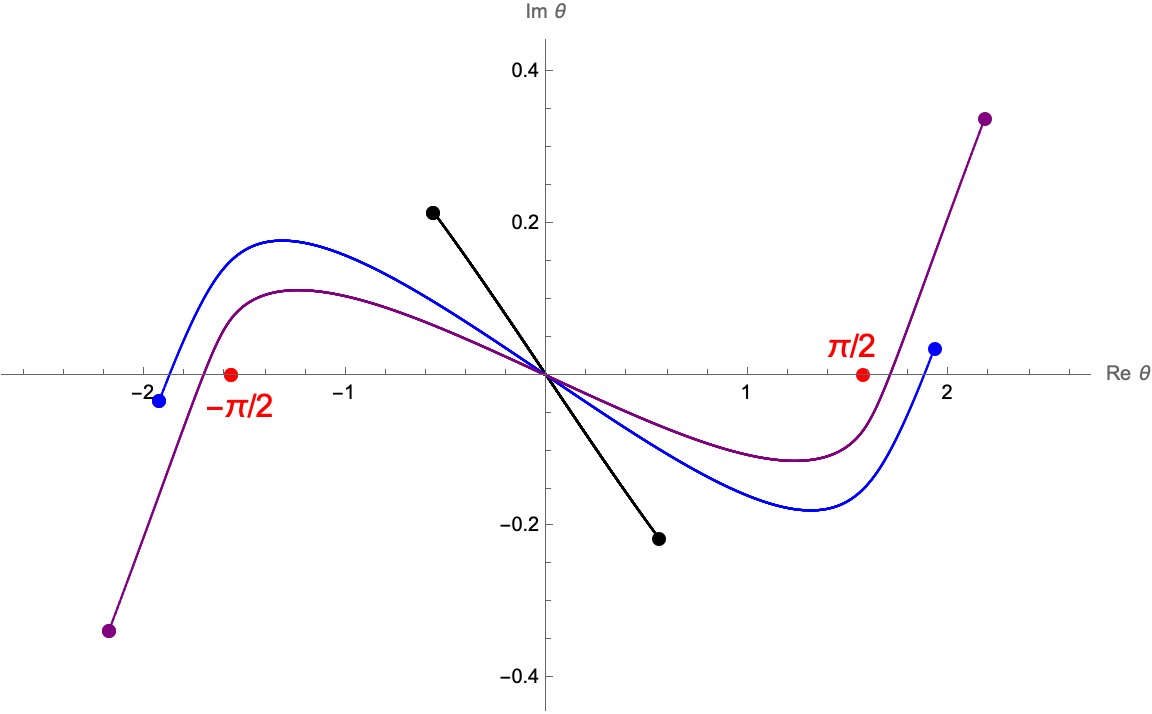}
	\caption{The cut $C$ for the single cut saddles at various values of 
    $\gamma(j)$: $j=100$ (black), $j=0.13$ (blue), $j=0.05$ (purple). The red 
    dots are the branch points $\theta=\pm\frac{\pi}{2}$. ($\gamma(j)$ is determined 
    by Legendre transformation at fixed charge $j$: see below for explanations.)}
	\label{1-cut}
\end{figure}

Now we explain more unusual branch points for $\rho(\theta)$ at $\theta=\pm\frac{\pi}{2}$. 
This singularity originates from the singular external potential in the $\beta\rightarrow 0$ limit 
explained earlier. At these points, (\ref{arctan-argument}) approaches 
$x=\sqrt{-1}=\pm i$, at which $\tan^{-1}x$ diverges. (The choice between $\pm$ depends 
on the square-root branch choices explained in the previous two paragraphs.) Since 
\begin{equation}\label{arctan-log}
  \tan^{-1}x=\frac{i}{2}\log(1-ix)-\frac{i}{2}\log(1+ix)\ ,
\end{equation}
the divergence of $\rho(\theta)$ is logarithmic, $\propto \log(\theta\mp\frac{\pi}{2})$. 
These singularities create branch cuts, which we again align to not cross $C$. 
The local shape of the branch cut of course depends on the convention. However, 
the monodromy for these cuts is not a matter of convention but is determined
while deriving (\ref{1-cut-summary}). For instance, suppose $\theta_0$ is large enough, 
located on the right side of $\theta=\frac{\pi}{2}$ (like the blue or purple curves 
of Fig. \ref{1-cut}). Depending on whether the cut $C$ connects $\theta=0$ and $\theta_0$ clockwise 
or anti-clockwise, the branch choice for $\rho(\theta)$ around 
$\theta=\frac{\pi}{2}$ should differ by 
a monodromy because the branch cut should avoid $C$. In other words, the log branch choice is related 
to the orientation of the cut $C$ around $\theta=\frac{\pi}{2}$. As we explain in Appendix \ref{sec:app1cut}, 
around Fig. \ref{fig:1cut}, $C$ should go around $\theta=\frac{\pi}{2}$ anti-clockwise and   
the log branch cuts have to be aligned to avoid such $C$.
The numerically determined $C$'s all satisfy this, as illustrated in Fig. \ref{1-cut}.

More concretely, we can again prescribe the branch sheet choices for the two log 
functions of (\ref{arctan-log}) by specifying them in a limit. 
Since we will demand the continuity in $\theta_0$ as explained above, 
we consider the Cardy limit $\theta_0\rightarrow 0$, 
$\gamma\approx \frac{i\theta_0^2}{2} \rightarrow 0$ (from (\ref{1-cut-summary})). 
In this limit, one finds 
\begin{equation}
  x= \frac{\sqrt{\sin^2\frac{\theta_0}{2}-\sin^2\frac{\theta}{2}}}
  {\cos\frac{\theta}{2}\sqrt{\cos\theta_0}}\approx 
  \frac{1}{2}\sqrt{\theta_0^2-\theta^2}\ ,
\end{equation}
where on the second step we used the anticipated fact that $\theta$ 
is also very small if $\theta_0$ is (i.e. $C$ is a very short segment: 
this can be easily justified below). At this small $x$, we select the branches 
for the two log functions in (\ref{arctan-log}) such that $\log(1\mp ix)\approx \mp ix$. 
This yields
\begin{equation}\label{rho-Cardy}
  \rho(\theta)=\frac{2i}{\pi \gamma}\tan^{-1}(x)
  \approx\frac{2}{\pi\theta_0^2}\sqrt{\theta_0^2-\theta^2}
\end{equation}
which correctly integrates to $\int_{-\theta_0}^{\theta_0}d\theta\rho(\theta)=1$ 
along the short cut $C$. Continuously changing
$\theta$ and $\theta_0$ from this Cardy regime, 
turning anti-clockwise around $\frac{\pi}{2}$ as we asserted in the previous paragraph,
one is led to pick definite branch sheets for the log functions.

Now we explain how to determine $C$. We start from $\theta=-\theta_0$ and determine 
the curve $C$ incrementally by demanding $\rho(\theta)d\theta$ to be real and
positive.\footnote{Practically, since $C$ is symmetric in $\theta\rightarrow -\theta$,
it always passes through $\theta=0$. So it suffices to start from $\theta = 0$
and determine only half of $C$ between $\theta=0$ and $\theta_0$.}
If this curve indeed ends at $\theta=\theta_0$, the saddle point equation is completely 
solved and we have found a solution. If this curve does not end on $\theta_0$, then for that 
$\theta_0(\gamma)$ the one-cut saddle does not exist. If one can find an explicit expression 
for $s(\theta)=\int_0^{\theta}d\theta^\prime \rho(\theta^\prime)$
with $\rho(\theta)$ given by (\ref{1-cut-summary}), this problem becomes easy 
to solve because $C$ will be the segment of the curve ${\rm Im}[s(\theta)]=0$ stretched 
between $\pm \theta_0$. 
Unfortunately, we failed to obtain a closed form expression for $s(\theta)$ with  
(\ref{1-cut-summary}). So we construct $C$ numerically. 
The simple method is to discretize the parameter $s\in [0,1]$,
i.e. $s_i=\frac{i}{N}$ for $i=1,\cdots,N$ with a large $N$, and integrate 
the condition $\rho(\theta(s))d\theta(s)=\textrm{real}>0$ discretely. 
If we take the number of steps $N$ to be the eigenvalue number, one may regard 
$i$ as labeling the eigenvalues, in which case 
$\rho(\theta_i)\Delta\theta_i=\frac{1}{N}$. 
Starting from $\theta_1=-\theta_0$, one can determine $\theta_i$ iteratively from 
\begin{equation}\label{cut-iteration}
  \theta_i=\theta_{i-1}+\Delta \theta_{i-1}=\theta_{i-1}+\frac{1}{N\rho(\theta_{i-1})}\ .
\end{equation}
In practice, we use an improved two-step method by determining $\theta_i$ using the above,
then taking the average of $\rho(\theta_{i-1})$ and $\rho(\theta_i)$ to recalculate $\theta_i$.
This is summarized as
\begin{equation}\label{cut-iteration-improved}
  \begin{cases}
  \tilde\theta_i = \theta_{i-1}+\frac{1}{N\rho(\theta_{i-1})}~, \\
  \theta_i = \theta_{i-1}+\frac{2}{N \left( \rho(\theta_{i-1}) + \rho(\tilde\theta_i) \right)}~.
  \end{cases}
\end{equation}
If $\theta_N$ determined this way indeed agrees with $\theta_0$, 
it signals that we have finally constructed a saddle point solution.
Fig. \ref{1-cut} shows such cuts for certain
values of $\theta_0$. (Our selections of $\theta_0$ in this figure are explained below.)

In some limits, one can analytically determine $C$. For instance, in the Cardy limit 
$|\gamma|\ll 1$, recall that $\rho(\theta)$ is approximately given by (\ref{rho-Cardy}).
With any complex number $\theta_0$, aligning $d\theta$ parallel to $\theta_0$ 
on the complex plane renders $\rho(\theta)d\theta$ real and positive. The solution for 
the cut $C$ is a straight interval, for instance parametrized as 
$\theta(s)=2s\theta_0$ with $-\frac{1}{2}<s<\frac{1}{2}$. On this cut, 
(\ref{rho-Cardy}) yields the Wigner semicircle distribution. 
We have also determined $C$ semi-analytically in the opposite limit $|\gamma|\gg 1$, 
in the sense of computing $s(\theta)=\int_0^\theta d\theta^\prime \rho(\theta^\prime)$ 
analytically but plotting the curve ${\rm Im}[s(\theta)]=0$ numerically. 
In this case, from the second line of (\ref{1-cut-summary}), one finds $\theta_0\approx \pi$. 
The cut $C$ connects $\pm \theta_0 \approx \pm \pi$ while staying close to the real axis,
but passing slightly below the singularity at $\theta=\frac{\pi}{2}$ and
slightly above the singularity at $\theta=-\frac{\pi}{2}$.
This is consistent with our assertion below (\ref{arctan-log}) that $C$ should
go around $\theta=\frac{\pi}{2}$ anti-clockwise.

We did not scan the entire $\gamma$ plane to see which domain 
hosts consistent $C$, not even numerically (e.g. by discretizing the plane into a fine grid). 
Rather, we focus on the curve $\gamma(j)$ on the complex 
$\gamma$ plane which is conjugate under Legendre transformation to a real positive 
charge $j\sim E+J$. That is, we are not interested in general complex temperature $\gamma^{-1}$ 
for its own sake, but only in those values which admit micro-canonical/grand-canonical 
duality. As discussed in \cite{Choi:2018vbz,Choi:2021lbk}, we interpret other points on the 
$\gamma$-plane as suffering from coarse-grained cancellations of the nearby indices 
$\Omega(j)$ and thereby misrepresenting the large $N$ BPS phases.

To determine $\gamma(j)$, one should somehow know the free energy $\log Z(\gamma)$ 
(\ref{eff-action-continuum}) for the saddle point solution, which is an integral along 
the cut $C$. Then one extremizes 
\begin{equation}\label{entropy-function}
  \frac{S(\gamma,j)}{N^2}=\frac{\log Z(\gamma)}{N^2}+\gamma j
\end{equation}
in $\gamma$ to find $\gamma(j)$, where $j\equiv\frac{E+J}{N^3}$. (In our 
scaling large $N$ limit, the charge scales like $N^3$. We redefine $j$ 
with this $N^3$ scaling from now on.) It is possible to compute $\log Z$ on the 
saddle points before fully knowing it, i.e. without knowing $C$ yet. 
Since the bulk function $\rho(\theta)$ is free of branch cuts in a 
region containing $C$, the integral (\ref{eff-action-continuum}) 
can be promoted to a bulk integral 
\begin{eqnarray}\label{eff-action-bulk}
  -\frac{S_\pm}{N^2}&=&\int_{-\theta_0}^{\theta_0}\int_{-\theta_0}^{\theta_0} 
  d\theta d\theta^\prime \rho(\theta)\rho(\theta^\prime)
  \log(1-e^{i(\theta-\theta^\prime))})\\
  &&
  \pm\frac{1}{\gamma}\int_{-\theta_0}^{\theta_0} 
  d\theta \rho(\theta)\left[{\rm Li}_2\left(ie^{i\theta}\right)
  \!-\!{\rm Li}_2\left(-ie^{i\theta}\right)\!+\!{\rm Li}_2\left(ie^{-i\theta}\right)
  \!-\!{\rm Li}_2\left(-ie^{-i\theta}\right)\right]\ .\nonumber
\end{eqnarray}
The integral can be performed on any curve ending on $\pm\theta_0$, not necessarily 
on $C$, as long as the two can be deformed into each other without crossing
the branch points $\theta=\pm\frac{\pi}{2}$. 
As explained above, it suffices to fix the curve between $\theta=0$ to $\theta_0$. 
After the curve starts at $\theta=0$, the curve reaches $\theta_0$ following a `short'
path (i.e. not going around $\frac{\pi}{2}$) if $\theta_0$ is not too far away. 
If $\theta_0$ is large, located on the right side of 
the branch point $\frac{\pi}{2}$, the curve goes around the branch point 
anti-clockwise. On such a curve, plugging (\ref{1-cut-summary}) into 
(\ref{eff-action-bulk}), we compute the integrals and obtain
\begin{equation}\label{free-1-cut}
  \frac{1}{N^2}\log Z(\gamma)=-\frac{\pi^2}{4\gamma}
  +\frac{1}{\gamma^2}\left[\frac{7}{4}\zeta(3)+\frac{\pi^3 i}{4}
  +8{\rm Li}_3(-ie^{-\frac{\gamma}{2}})-{\rm Li}_3(e^{-\gamma})\right]\ .
\end{equation}
See Appendix \ref{sec:app1cut} around (\ref{app:1cutlogZ}) for its derivation. As a small 
check of this formula, note that its small $\gamma$ expansion is given by
\begin{equation}
  \frac{1}{N^2}\log Z=\frac{4iG}{\gamma}+\frac{1}{2}\log(\gamma/2) 
  -\frac{\pi i}{4}-\frac{3}{4}+\mathcal{O}(\gamma)\ ,
\end{equation}
whose leading term $\frac{4iG}{\gamma}$ agrees with the Cardy free energy 
(\ref{Cardy-free}) at $\nu=1$.

\begin{figure}[t]
	\centering
	\includegraphics[width=0.45\textwidth]{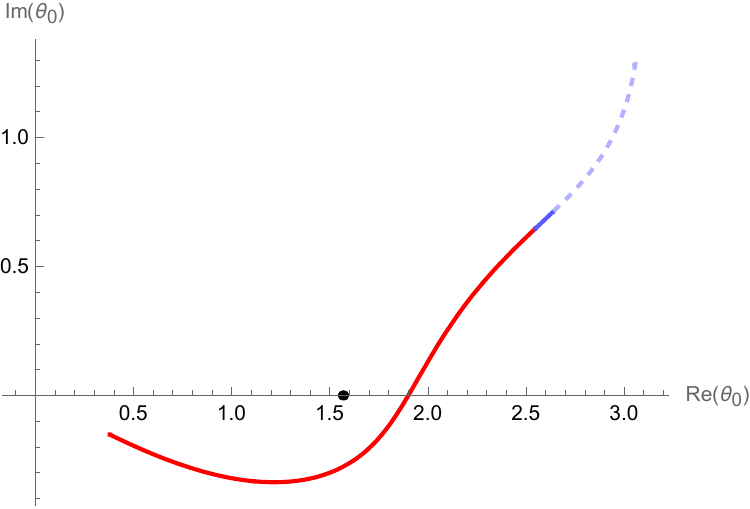}
    \hspace{1cm}
    \includegraphics[width=0.45\textwidth]{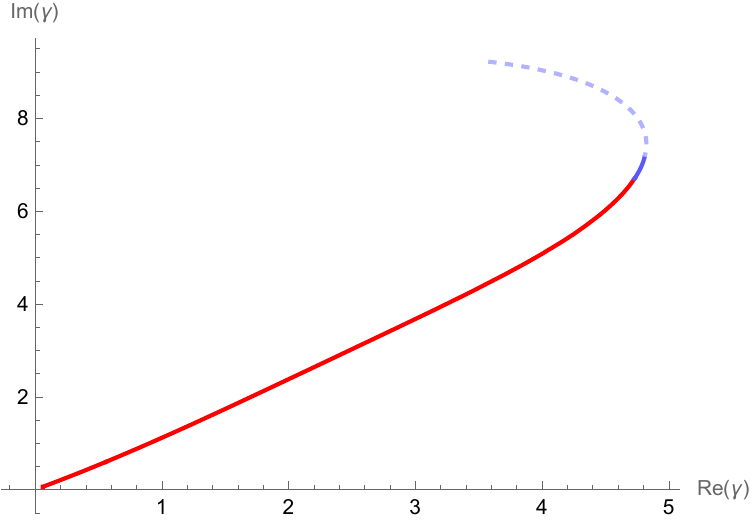}
	\caption{Plots of $\theta_0(j)$ (left) and $\gamma(j)$ (right). 
    The red curve is for $j_c<j<500$; 
    solid blue for $j_0<j<j_c$; dashed light blue for 
    $0<j<j_0$ ($j_c=0.017674$, $j_0=0.013924$). For $j<j_c$, the cut $C$ does not exist but we have shown the 
    formal results using the function (\ref{free-1-cut}).
    The black dot on the left figure is the branch point $\theta_0=\frac{\pi}{2}$.}
	\label{theta0-legendre}
\end{figure}

With (\ref{free-1-cut}), we numerically extremize (\ref{entropy-function}) in $\gamma$ 
at various $j>0$.
The resulting curve $\gamma(j)$, or $\theta_0(j)$, is shown in Fig. \ref{theta0-legendre}.
Different parts of the curves are distinguished by solid red, solid blue, and dashed light blue, 
whose meaning we explain now. If one takes the function $\log Z(\gamma)$ given by (\ref{free-1-cut}) 
and extremize (\ref{entropy-function}) for $j>0$, one obtains the entire curve shown in 
Fig. \ref{theta0-legendre}. (The red curve on the left end extrapolates to $\theta_0=0$  
for $j\rightarrow \infty$.) However, one should check if the cuts $C$ that would lead to
(\ref{free-1-cut}) indeed exist at those values of $\gamma(j)$.
With the iteration method explained 
around (\ref{cut-iteration}), one finds that $C$ exists only for $\gamma$'s on the red part 
of the curve. See Fig. \ref{1-cut} for the shapes of the cuts on 
this part of the curve. The right ends of the red curves in Fig. \ref{theta0-legendre} 
correspond to the charge $j_c\approx 0.017674$, at $\gamma(j_c)\approx 4.73+6.70 i$. 
For $j<j_c$, the cut $C$ (and thus the 1-cut saddle) does not exist.
We will explain this phenomenon in more detail below.

\begin{figure}[t]
	\centering
    \includegraphics[width=0.5\textwidth]{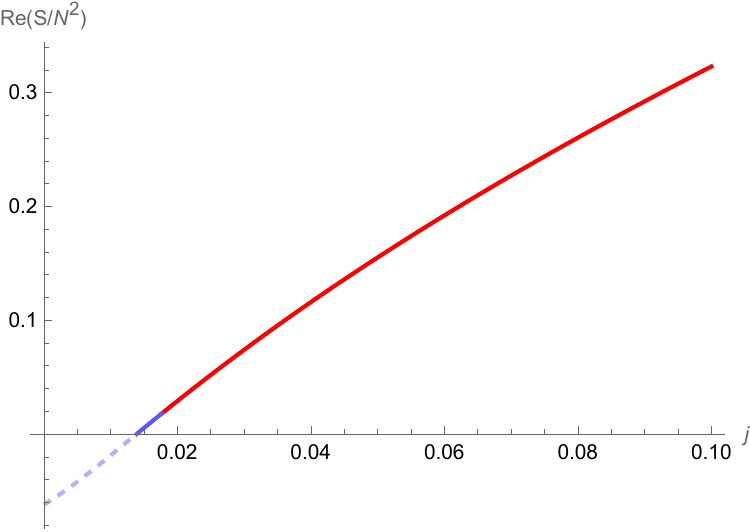}
	\caption{Plot of $\frac{1}{N^2}{\rm Re}[S(j)]$: colors/dash 
    of the curve denote the same ranges as in Fig. \ref{theta0-legendre}.}
	\label{entropy-1-cut-plot}
\end{figure}

If one `formally' continues to use (\ref{free-1-cut}) and Legendre transform, 
being blind to whether there exist such saddle points or not, one obtains 
the other part of the curve. For $j_0<j<j_c$ where $j_0\approx 0.013924$, 
the indicial entropy 
${\rm Re}[S(j)]$ from (\ref{entropy-function}) is positive: see Fig. \ref{entropy-1-cut-plot}.
This region is shown in solid blue curves. So had there been saddle points whose free energy 
is given by (\ref{free-1-cut}), it would have represented macroscopic entropy for $j>j_0$.
Below $j_0$, one finds ${\rm Re}[S(j)]<0$ and it cannot represent an ensemble
with large entropy even if the saddle point existed.
This region is shown in dashed light blue curves.
Although there are no saddles for $j<j_c$, we show these formal results as a mathematical
property of the function (\ref{free-1-cut}), and also to trigger some speculations below.

\label{paragraph-coarsegrained}
We also study the saddle point free energy $\log Z(\gamma)$ as a function of temperature. 
This function will be important for understanding the grand canonical 
phase transition, after we study another set of saddles in Section \ref{sec:lowTsaddle}.
To determine the dominant phase, one should pick the saddle with largest $|Z(\gamma)|$, 
i.e. largest ${\rm Re}[\log Z(\gamma)]$. We stress that, when discussing the competition 
between different saddles, we do \textit{not} compare them at the same complex value 
of $\gamma$. Rather, we will consider the thermodynamics only on the curve $\gamma(j)$ 
which admits micro-canonical/grand-canonical duality. The interpretation of this curve 
in the grand canonical ensemble is as follows. We  
regard $T^{-1}\equiv{\rm Re}[\gamma(j)]$ as relating the real chemical potential and the charge, 
changing the ensemble, while $\varphi(T)\equiv{\rm Im}[\gamma(j(T))]$ at fixed $T$ is regarded 
as optimally tuning the phase of fugacity to obstruct the coarse-grained cancellations of 
nearby $\Omega(j)$ in the index. Away from the curve $\gamma(j)$, 
the coarse-grained formal entropy will under-estimate $\Omega(j)$
and misrepresent the BPS phases.
At $\varphi=0$, the under-estimation results in an apparent absence of
the deconfinement phase transition in the BPS sector \cite{Kinney:2005ej}.
Similarly, at general nonzero $\varphi\neq \varphi(T)$, deconfinement transition
is visible but at delayed higher temperatures 
than the one at the optimal $\varphi(T)$ \cite{Copetti:2020dil,Choi:2021lbk}.
$\varphi(T)$ can be determined purely within the grand canonical ensemble by noting 
that the imaginary part of the extremization of (\ref{entropy-function}) is given by
${\rm Im}[\frac{\log Z(\gamma)}{\partial\gamma}]=0$, relating $T$ and $\varphi$ without 
referring to any $j$. To summarize, we regard the grand canonical ensemble of the index 
as labeled by real $T$, in 1-to-1 map 
to the micro-canonical ensemble. The chosen $\varphi_i(T)$ depends on the 
saddle point, which we label by $i$. The dominant phase at fixed $T$ is determined 
by comparing ${\rm Re}[\log Z_i(T^{-1}+i\varphi_i(T))]$. 

\begin{figure}[t]
	\centering
	\includegraphics[width=0.45\textwidth]{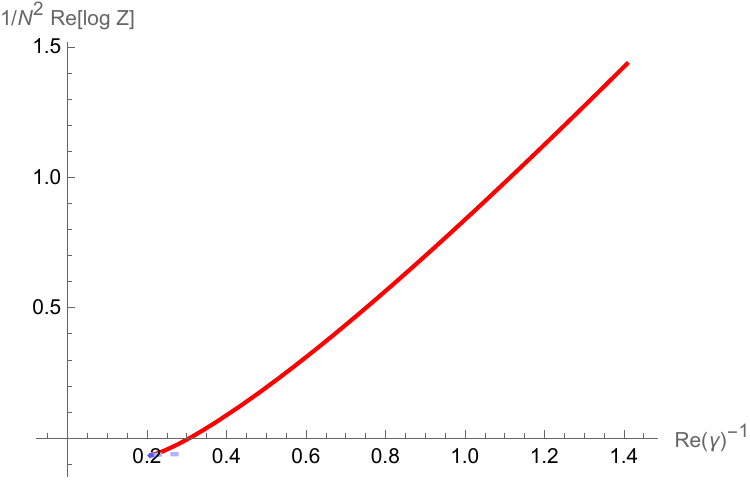}
    \hspace{1cm}
    \includegraphics[width=0.45\textwidth]{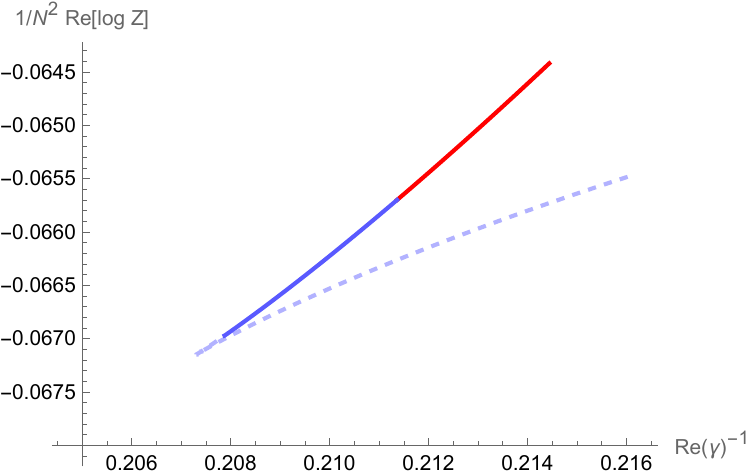}
	\caption{(Left) Plot of the `real temperature' ${\rm Re}(\gamma)^{-1}$ vs. the free energy 
    $\frac{1}{N^2}{\rm Re}(\log Z)$. (Right) Zoom-in to the cusp region $0.007<j<0.02$. 
    (Colors/dash mean the same as in Fig. \ref{theta0-legendre}.)}
	\label{free-plot}
\end{figure}

We plot ${\rm Re}[\log Z]$ of (\ref{free-1-cut}) as a function of
temperature $T={\rm Re}[\gamma(j)]^{-1}$ in Fig. \ref{free-plot}.
The red part of the curve has saddle points, and those in other colors are
formal results without the corresponding saddles.
On the right panel of Fig. \ref{free-plot}, we zoom into the cusp region of the left.

Now we explain how the $1$-cut saddle point disappears at $j=j_c$. 
As $j$ decreases, $\theta_0(j)$ moves around the branch point $\theta=\frac{\pi}{2}$
anti-clockwise as shown on the left figure of Fig. \ref{theta0-legendre}.
The corresponding cut $C$ also goes around the branch point anti-clockwise, 
see Fig. \ref{1-cut}. As $j$ approaches $j_c\approx 0.017674$ from above, 
part of $C$ approaches very close to $\theta=\frac{\pi}{2}$ from the right: 
see Fig. \ref{fig:1cutx}(b) in Appendix \ref{sec:app1cut}.
At $j=j_c$, the cut $C$ touches the branch point and the ansatz (\ref{1-cut-summary})
breaks down. Continuous change of $C$ across $\theta=\frac{\pi}{2}$ is
not guaranteed to yield solutions, and in fact forbidden because the cut $C$
has to go around $\theta=\frac{\pi}{2}$ anti-clockwise as already explained.
Blindly trying to get the clockwise solution by choosing the clockwise branch sheet
for (\ref{1-cut-summary}) and iterating with (\ref{cut-iteration}),
one indeed does not get the cut which ends on $\theta_0$.

Mathematically, one may view this as a kind of wall-crossing phenomenon at $j=j_c$, beyond 
which the solution disappears. From the viewpoint of our original matrix integral problem 
(\ref{index-formula}) at small but finite $\beta$, the force in the region close to
$\theta=\frac{\pi}{2}$ becomes large due to an infinite tower of light 
degrees of freedom, as explained around (\ref{div-pot-2}) and (\ref{div-pot-3}).
The large $N$ saddles near $j=j_c$ will suffer from large fluctuations of these modes. 
As explained in the paragraph below the one containing (\ref{div-pot-3}),
the poles of the integral (\ref{index-formula}) are accumulated on the other side
of $\theta=\frac{\pi}{2}$ than $C$. If $C$ approaches $\theta=\frac{\pi}{2}$ and
tries to `cross' it, so will the steepest descent contour. 
Such a contour deformation will require an extra contribution to the index from
a large number of residues: the number of residues to add will increase
as $C$ moves farther from $\theta=\frac{\pi}{2}$ after crossing it.
They may be important to understand the true quantum fate of this saddle for $j<j_c$.
It is possible that, collecting the contributions from these residues, 
the free energy (\ref{free-1-cut}) may continue to express a phase below this threshold.

We have to find a new class of saddles in the range $0<j<j_c$, or 
$0<T<T_c$, for a consistent picture of the large $N$ BPS phases. 
Note that for $j=\frac{E+J}{N^3}\ll 1$, the large $N$ index $Z_\infty$ of Section \ref{sec:2} 
is obtained from a uniform confining eigenvalue distribution. So we would like to find a 
new class of saddles that has the uniform distribution as its limit, since the saddles
discussed in this subsection fail to do so.
In the next subsection, we will present new saddles that (partly) do this job.
In particular, in the grand canonical ensemble with fixed 
$T\equiv {\rm Re}[\gamma(j)]^{-1}$, we will show in Section \ref{sec:lowTsaddle} that the subtleties of 
the 1-cut saddles around $j=j_c$ can be bypassed. This is because, as one reduces $T$, 
the 1-cut phase undergoes a phase transition to the new low temperature phase 
at a critical temperature higher than $T_c\equiv {\rm Re}[\gamma(j_c)]^{-1}$.

We discuss the physical implications of the 1-cut saddles,
in particular the physics regarding the small black hole branch,
in the remainder of this subsection.

We first compare our 1-cut saddles with the black holes in AdS$_{D\geq 4}$ Einstein gravity. 
The latter system has black holes at arbitrarily small charge as long as it is macroscopic. 
Those are called the small black holes, which have negative specific heat.
The energy (mass) $E$ of an AdS black hole is an increasing function of 
temperature $T$ when it is larger than a critical mass, while being a decreasing function 
below it. The BPS AdS black holes have an analogous feature
between an energy like charge (call it $j$) and its inverse chemical potential 
(which we keep calling $T$). As for our one-cut saddles in the ABJ vector model, the susceptibility
$\frac{d j(T)}{dT}$ is always positive. So one may interpret these 1-cut saddles as the vector 
model analogue of large black holes. One can also interpret these saddles as describing the 
deconfined phase, since at very high temperature $\gamma\rightarrow 0$ one finds
$\log Z\sim \frac{N^2}{\gamma}=\frac{N}{\beta}$,
similar to the contributions from $N$ liberated quarks. 
In Einstein gravity, as we reduce the energy, the large black hole branch
terminates semi-classically by switching to the small black hole branch. 
On the other hand, in our vector model, the fate of the large black hole like branch 
at low charges is unclear due to the large quantum fluctuations of the light matters.

It is somewhat curious to find that the analytic function (\ref{free-1-cut}) 
formally `knows' the small black hole like branch. If for instance the infinitely many 
residue contributions near $\theta=\frac{\pi}{2}$ retain the free energy (\ref{free-1-cut}) 
beyond the apparent threshold, its Legendre transformation may look like exhibiting a branch 
with negative susceptibility. That is, on the right hand side of Fig. \ref{theta0-legendre}, 
the dashed blue part of the curve shows a decreasing function $T(j)$ in $j$.
Note however that there is another, statistical, obstruction against extending
these saddles to the small black hole like region.
As shown in Fig. \ref{entropy-1-cut-plot}, the `entropy' ${\rm Re}[S(j)]$
in this region is negative. 
So even if there are saddles with free energy (\ref{free-1-cut}), the region with negative 
susceptibility is subdominant in the microcanonical ensemble, not representing macroscopic 
entropy. This may be implying that the vector model does not have enough degrees of freedom to make 
small black holes.

The presence or absence of the small BPS black hole branch may also be understood from the 
different combinatoric natures of the matrix and vector trace relations. 
Strictly free BPS entropy always shows positive specific heat. 
(It may be infinity at the Hagedorn temperature, but 
not negative.) The entropy $S_{\rm free}(j)$ 
of the free theory is thus concave, $\frac{d^2S_{\rm free}(j)}{dj^2}\leq 0$,
from the positivity of susceptibility. With interaction, most of the free single-trace
BPS states except gravitons are lifted, and their multi-traces remain non-BPS
until trace relations make some of them $Q$-closed.
Therefore, the shape of the function $S(j)$ in the interacting theory depends on the
energy scales at which various multi-trace operators re-enter the BPS sector
thanks to trace relations. The more delayed their re-entrance is, the sharper
the increase of $S(j)$ could be at higher $j$.
If the entropy increase is sharp enough to have convex $S(j)$, $\frac{d^2S(j)}{dj^2}>0$,
in some energy range, the susceptibility
will be negative. With matrices, trace relations start to appear at energy scales of 
order $j\sim N$. One has to wait until even higher energies till a substantial 
number of multi-trace $Q$-closed operators appear by trace relations.
On the other hand, we have seen in Section \ref{sec:2} that multi-traces of non-BPS operators
can become $Q$-closed already at $\mathcal{O}(1)$ energies by vector trace relations.
Earlier re-entrance to the BPS sector at $E\sim \mathcal{O}(1)$ and
the deconfinement at $E\sim \mathcal{O}(N^3)$ exhibits a big energy range,
which may cause a milder growth of $S(j)$ and the absence of the small black hole like region
in the vector model.

Although these considerations are speculative, we think they
will be relevant when we consider the family of 
ABJ theories with increasing $N^\prime$. As one increases $N^\prime$,
appearance of the trace relations between the rectangular matrices
will be delayed to higher energies because the threshold for 
the $U(N^\prime)$ trace relations grows in $N^\prime$. In particular, 
the multi-trace BPS bounds studied in Section \ref{sec:2.2}
will start to form at higher energies. This effect, and also that there are more 
degrees of freedom at larger $N^\prime$, will make $S(j)$ increase more sharply
in some energy range, eventually forming a small black hole branch as 
$N^\prime$ increases towards $N$. For instance, if one increases $N^\prime$ together 
with decreasing $k$ to reach the regime of the type IIA gravity dual, there clearly exist 
small black holes. It will be interesting to see, at least in the weakly-coupled setup 
at $N\ll k$, the $N^\prime$ dependence of the BPS thermodynamics.

Reversing the viewpoint, one can start from the matrix theory with $N^\prime=N$ 
at small $k\ll N$ with a type IIA dual, and then reduce $N^\prime$ together with 
increasing $k$ to reach our weakly-coupled vector model regime. The small black holes 
will disappear, but some of their heavy microstates will descend down 
to low energies because the threshold of the $U(N^\prime)$ trace relations is lowered.
In this sense, we are tempted to view the multi-trace BPS bounds of Section \ref{sec:2.2} as 
the `quantum low energy remnant' of the small black hole states left in 
the higher spin gravity.

We observe that some features of our 1-cut saddles are similar to 
the BTZ black holes. BTZ black holes exist above a threshold $E_0=\frac{c}{12}$  
where $c$ is the central charge of the dual CFT, and also, they always have positive
specific heat. Here we note that the CFT$_2$ dual to AdS$_3$ gravity may be viewed 
as a kind of large $N$ vector model.\footnote{We thank Robert de Mello Koch for the suggestion.} For instance, the CFT on $N_1$ D1-branes and $N_5$ 
D5-branes is described by the sigma model on $(T^4)^N/S_N$, 
where $N=N_1N_5$. The permutation $S_N$ is a gauge symmetry 
of this theory, which might be (at least morally) understood as coming from the UV system of 
$N$ D1-branes on $1$ D5-brane. The gauge symmetry of the latter system is $U(N)$, 
of which $S_N$ is a subgroup, acting on the $N\times 1$ vector-like open string 
modes. It would be interesting to see if the supercharge cohomology problem of this model
\cite{Chang:2025rqy} has any vector-like features. Of course, we should also 
stress that many features of the BPS states and the saddles are quite
different between our vector model and the sigma model. One difference is that ours 
have finite entropy at $j=j_c$, while the threshold BTZ black hole has zero entropy 
(analogous to the point $j=j_0$ in our model). This is because nontrivial large $N$ saddles 
in our model appear in the scaling limit $\beta\sim N^{-1}\ll 1$, 
causing the large quantum fluctuations at $\theta=\frac{\pi}{2}$ to disturb the classical 
saddle before its entropy vanishes. $U(N)$ gauge singlet constraint is stronger than the $S_N$ 
constraint, allowing the deconfined phase only at very high temperature 
which in turn causes large quantum effects. On the other hand, 
the deconfinement temperature $\sim\mathcal{O}(1)$ is much lower for the sigma model 
because the permutation gauge invariance is easier to 
locally overcome. (Matrix models with permutation gauging also have much lower transition 
temperature than those with $U(N)$ gauging \cite{OConnor:2024udv}.)

\subsection{Low temperature saddles and the phase transition}\label{sec:lowTsaddle}

Recall that in the previous subsection, we found 1-cut saddle points only above 
a critical charge $j_c$, or equivalently above a critical temperature. In this subsection, 
we study another class of saddle points which we claim dominate at low temperatures.

Since the large $N$ confining saddle point at $\mathcal{O}(1)$ temperature is the
uniform distribution on the unit circle, we naturally seek gapless 
non-uniform saddle points at the low temperature part (large $|\gamma|$)
of our scaling limit (large $N$ with fixed $\gamma \equiv N \beta$). 
In fact in many matrix integrals, one finds such saddles at low temperature. 
This is the case for the Gross-Witten-Wadia (GWW) model 
\cite{Gross:1980he,Wadia:1980cp}, and also for the 
partition functions of the 3d vector models \cite{Shenker:2011zf,Jain:2013py}. 
Even with complex effective action, gapless distributions on a complex `cut' $C$
(which is circular now) may exist and dominate at low temperature.
For instance, see \cite{Copetti:2020dil} for such a case in the complex GWW model.
The general form of the gapless density function is given by 
\begin{equation}\label{rho-gapless}
  \rho(\theta)=\frac{1}{2\pi}\left[1+\sum_{n=1}^\infty a_n (e^{in\theta}+e^{-in\theta})\right]~,
\end{equation}
when the external potential $V(\theta)$ is given by
\begin{equation}
  -V(\theta)=N\sum_{n=1}^\infty\frac{a_n}{n}(e^{in\theta}+e^{-in\theta})\ .
\end{equation}
For the complex GWW model, $a_1\equiv \frac{g}{2}$ is the complex parameter 
and all other $a_n$'s are zero. From the bulk function
$\rho(\theta)=\frac{1}{2\pi}[1+g\cos\theta]$ 
and
$s(\theta)=\int_0^\theta d\theta^\prime \rho(\theta^\prime)=\frac{1}{2\pi}[\theta+g\sin\theta]$, the condition ${\rm Im}[s(\theta)]=0$ 
admits gapless $C$ for certain complex $g$.
However, for our matrix model (\ref{index-formula}), such gapless saddles cannot 
be found in the scaling limit. From (\ref{eff-action-pm}), one obtains $a_n=\frac{2i^n}{n\gamma}$
for odd $n$ and $0$ for even $n$. The infinite sum (\ref{rho-gapless})
in the scaling limit converges only on the real axis, so we sum it for
real $\theta$ and then try to continue it to the complex plane.\footnote{With
general potential (\ref{eff-action}) before taking the scaling limit, the sum converges for 
$|{\rm Im}(\theta)|<\frac{1}{2}{\rm Re}(\beta)$.}
From (\ref{rho-gapless}), one obtains
\begin{equation}\label{rho-gapless-vector}
  \rho(\theta)=\frac{1}{2\pi}\left[1+\frac{1}{\gamma}\log\left(\frac{(1+ie^{i\theta})(1+ie^{-i\theta})}
  {(1-ie^{i\theta})(1-ie^{-i\theta})}\right)\right]
  =\frac{1}{2\pi}\left[1+\frac{1}{\gamma}\log(-1)\right]\ .
\end{equation}
Considering possible log branch choices,
this renders $\rho(\theta)$ piecewise (complex) constant.
No matter how one chooses the branches, one can never obtain a gapless $C$ 
from (\ref{rho-gapless-vector}).

Having failed to find gapless saddles at finite $\gamma$, 
one may then ask if the 1-cut saddles of (\ref{1-cut-summary}) asymptotes to the 
uniform gapless distribution as we take $\gamma\rightarrow\infty$.
According to the relation $\theta_0(\gamma)$ of (\ref{1-cut-summary}),
one can reach $\theta_0\rightarrow\pi$ asymptotically
as ${\rm Re}(\gamma)\rightarrow\infty$.
However, from the studies of Section \ref{sec:highTsaddle}, we already know 
that this limit cannot be reached with a definite micro-canonical dual,
since the latter terminates at a lower bound.
It is clear from their free energy (\ref{free-1-cut})
why the 1-cut saddles cannot be continued to arbitrarily low temperature.
Expanding (\ref{free-1-cut}) in large $\gamma$,
with ${\rm Re}(\gamma)\gg 1$, one obtains 
\begin{equation}
  \frac{1}{N^2}\log Z=-\frac{\pi^2}{4\gamma}+\frac{7\zeta(3)+\pi^3i}{4\gamma^2}
  +\mathcal{O}(e^{-\frac{\gamma}{2}})\ .
\end{equation}
The leading term $-\frac{\pi^2}{4\gamma}$ disagrees with the 
expected behavior $\frac{1}{N^2}\log Z_\infty\sim -\frac{7\zeta(3)}{2\gamma^2}$.
Furthermore, Legendre transformation of this leading term, obtained by extremizing
$-\frac{\pi^2}{4\gamma}+j\gamma$, leads to $\gamma=\pm \frac{\pi i}{2\sqrt{j}}$
which violates the assumption ${\rm Re}(\gamma)\gg 1$. 
Therefore, the 1-cut saddles (\ref{1-cut-summary}) cannot describe 
the BPS phase at large ${\rm Re}(\gamma)$ in the large $N$ scaling limit.

This led us to search for 2-cut eigenvalue distributions for the low temperature phase. 
Just to give a rough idea first, at very low temperature 
${\rm Re}(\gamma)\gg 1$, the two-cut distribution will be such that
the cuts are almost entirely along the real axis of the complex plane for $\theta$,
i.e. it will be a small deformation of the uniform confining saddle.
However, due to the strong external force near $\theta=\pm\frac{\pi}{2}$
as explained around (\ref{div-pot-2}),
the eigenvalues will be repelled from these two points and two small gaps will form there.

\begin{figure}[t]
	\centering
	\includegraphics[width=\textwidth]{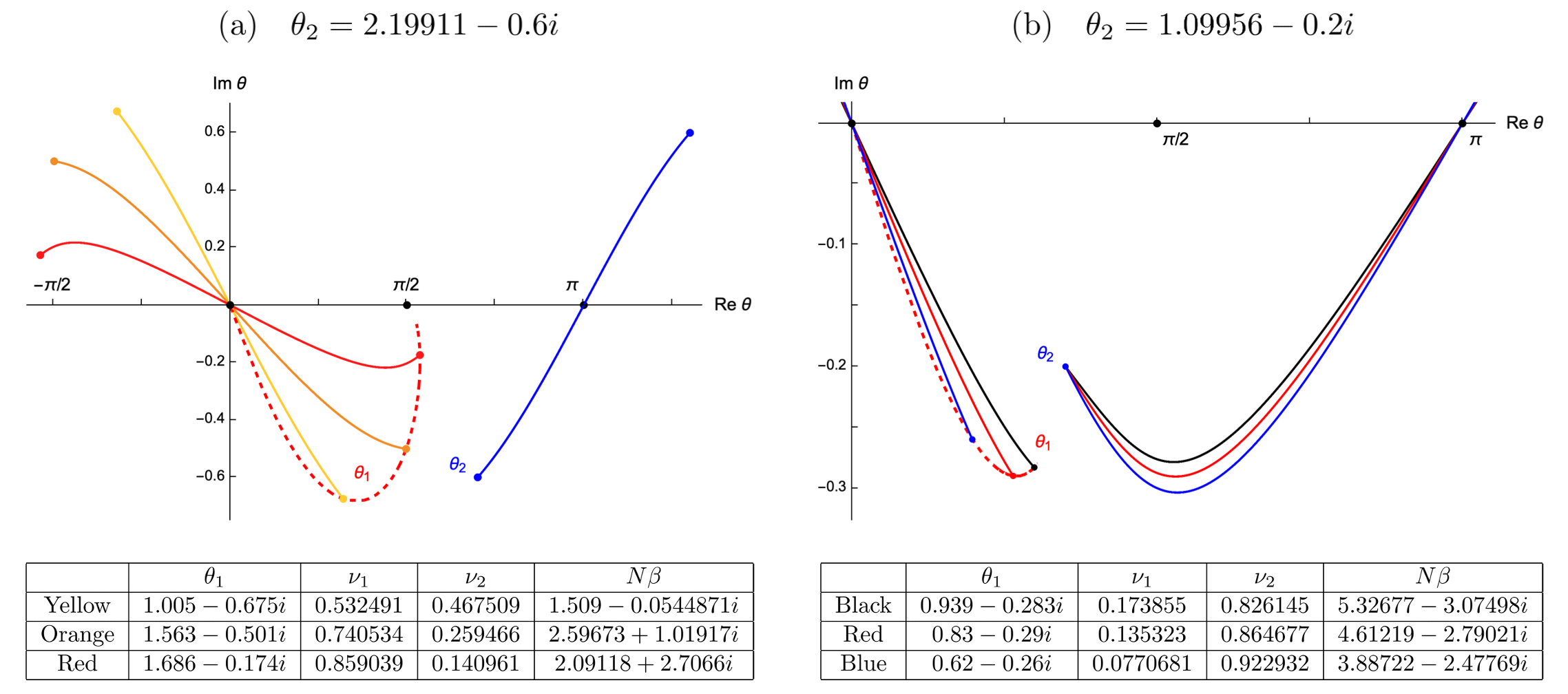}
	\caption{Examples of the double cut $C=C_1\cup C_2$. At fixed $\theta_2$, the dashed 
    curve shows those $\theta_1$ which admit the saddle point solutions. The tables show 
    the filling fractions $\nu_1\equiv \nu$, $\nu_2\equiv 1-\nu$ for the cuts and $\gamma=N\beta$. 
    (On the left figure, $C_2$ for the three chosen $\theta_1$ are almost degenerate. 
    On the right figure, only parts of $C_1$ and $C_2$ are shown.)}
	\label{2-cut-plot}
\end{figure}

Following similar computational strategies to the 1-cut case,
we computed the infinite series in the formal solution for $\rho(\theta)$ to obtain
the 2-cut bulk function (\ref{2-cut-summary}): see Appendix \ref{sec:app2cut} for some details.
From the second line of (\ref{2-cut-summary}), 
one complex (two real) parameter between $\theta_1,\theta_2$ is left unfixed 
at given $\gamma$. As sketched below (\ref{2-cut-summary}), we must
tune one of these two real parameters to obtain a consistent cut $C=C_1\cup C_2$: 
otherwise the integral $\int_{C_i}\rho(\theta)d\theta$ along each cut yields
a complex number, obstructing the existence of physical saddles.
See Fig. \ref{2-cut-plot} for the numerically determined cuts after the tunings.
The 2-cut saddles are labeled by a complex $\gamma$ and one extra real parameter 
$\nu$ defined by 
\begin{equation}
  \nu \equiv \int_{C_1}d\theta \rho(\theta)\ ,\ \ 0\leq \nu\leq 1\ .
\end{equation}
$\nu$ is $\frac{1}{N}$ times the number of eigenvalues on the first cut.
The 2-cut saddles are labeled by $\nu=0,\frac{1}{N},\frac{2}{N},\cdots,1$. 
The 1-cut solutions of Section \ref{sec:highTsaddle} are special cases with $\nu=1$ or $0$.

$Z$ receives contributions from 
these saddles, 
\begin{equation}\label{filling-fraction-sum}
  Z(\beta)\leftarrow \sum_{\nu}Z(\gamma,\nu)
  =\sum_{\nu}\exp\left[N^2f(\gamma,\nu)+f_1(\gamma,\nu)+O(N^{-2})\right]\ .
\end{equation}
To find the dominant contribution in (\ref{filling-fraction-sum}), 
one has to choose $\nu$ which maximizes the sum in (\ref{filling-fraction-sum}), 
i.e. the one with maximal 
\begin{equation}\label{maximize-nu}
  |Z(\gamma,\nu)|=e^{{\rm Re}[\log Z(\gamma,\nu)]}\sim e^{N^2{\rm Re}[f(\gamma,\nu)]}\ .
\end{equation}
Naive maximization of the function ${\rm Re}[f]\sim f(\gamma,\nu)+f^\ast(\gamma^\ast,\nu)$ 
yields a real non-holomorphic solution $\nu_\star(\gamma,\gamma^\ast)$ unless maximized at 
the edges $\nu=1$ or $0$.
This leads to the phenomenon of holomorphic anomaly, best known in  
topological string theories \cite{Bershadsky:1993ta} but also known in 
matrix models \cite{Bonnet:2000dz,Eynard:2007hf,Eynard:2008yb,Eynard:2008he}. 
Approximating a holomorphic function $Z(\beta)$ by a non-holomorphic expression $f(\gamma,\nu_\star(\gamma,\gamma^\ast))$ is nonsensical.

This puzzle is resolved by remembering that $\nu$ is discrete.
The true maximum $\nu_c$ is $\frac{1}{N}$ times an integer, close to 
the continuous function $\nu_\star$ but not quite the same. 
(B.10) of \cite{Bonnet:2000dz} provides the correct saddle point estimate reflecting 
the discreteness of $\nu$, which is given by\footnote{We correct 
$2\pi F_0^{\prime\prime}(x_c)\rightarrow \frac{F_0^{\prime\prime}(x_c)}{2\pi}$
in the formula of \cite{Bonnet:2000dz} (which is $-\frac{f^{\prime\prime}(\nu_\star)}{2\pi}$ 
in our notation).} 
\begin{equation}\label{saddle-nu-osc}
  \log Z(\gamma)\sim N^2 f(\gamma,\nu_c)+\frac{\pi i}{\tau}u_c^2+\log[\theta_3(u_c|\tau)]
  -\frac{1}{2}\log[{\textstyle -\frac{f^{\prime\prime}(\nu_\star)}{2\pi}}]+f_1(\nu_\star)+
  O(N^{-2})\ .
\end{equation}
The quantities appearing in this expression are given by
\begin{equation}\label{saddle-nu-def}
  \nu_c=\frac{\lfloor N\nu_\star\rfloor}{N}\ ,\ \ 
  u_c=[N\nu_\star]-\tau[N\Delta_\star]\ ,\ \ \tau=-\frac{2\pi i}{f^{\prime\prime}(\nu_\star)}
  \ ,\ \ \Delta_\star=-\frac{1}{2\pi}{\rm Im}[f^\prime(\nu_\star)]
\end{equation}
where primes denote $\nu$ derivatives, $\nu_\star$ satisfies ${\rm Re}[f^\prime(\nu_\star)]=0$, 
$[u]\equiv u-\lfloor u\rfloor$ is the fractional part of $u$, and $\theta_3$ is a Jacobi 
theta function. In (\ref{saddle-nu-osc}), the first term carries a factor of $N^2$ which naively 
makes it the dominant classical term, while the other terms are apparently subdominant. 
As long as one does not take $\gamma$ derivatives, 
this naive estimate is correct.
In particular,  since the value of $\nu_c$ is 
very close to $\nu_\star$, the first term of (\ref{saddle-nu-osc}) is approximately the 
same as the naive expression that we obtained above. However, this naive estimate becomes wrong 
if one takes sufficiently many $\gamma$ derivatives. 
First note that $\nu_c$ is a piecewise constant function in a domain,
so its $\gamma$ derivative vanishes.
$\gamma$ derivative on $u_c$ will yield a factor of $N$ because $u_c$ given by
(\ref{saddle-nu-def}) is a fast oscillating function with a steep slope of order $N$.
So for observables which contain two derivatives of $\gamma$, the second and third
terms are proportional to $N^2$ and cannot be neglected compared to the first.
In this sense, the $\frac{1}{N}$ expansion is `non-universal.'

The apparent $\gamma^\ast$ dependence of (\ref{saddle-nu-osc}), through $\nu_\star$,
cancels between various terms \cite{Eynard:2008yb,Eynard:2008he},
ensuring the background independence.
To see this concretely in the first few terms of (\ref{saddle-nu-osc}),
first recall that $\nu_c$ is a piecewise constant function in a domain.
This makes the $\gamma^\ast$ derivative of the first term vanish, at the $N^2$ order.
Then, as for the second and third terms, it may appear that $\gamma^\ast$
derivative on $u_c$ will yield a factor of $N$ because of its fast oscillation,
yielding terms at the $N^1$ order. 
However, one can check that $\frac{\partial u_c}{\partial \gamma^\ast}=\mathcal{O}(1)$
after cancellations, implying that the $\gamma^\ast$ derivative vanishes at the $N^1$ order.
These arguments can be continued to higher orders.

For our purpose of studying the large $N$ thermodynamics, we will at most take one derivative 
of $\log Z(\gamma)$ in $\gamma$ for the Legendre transformation. The first term of 
(\ref{saddle-nu-osc}) will remain dominant for these calculations. (However, the susceptibility is 
two-derivative, subject to large non-universal fluctuations. Also, the order of 
phase transitions higher than two seems to suffer from this issue.)
As mentioned in the previous paragraph, $\nu_c$ is a piecewise constant function, 
so both its $\gamma$ and $\gamma^\ast$ derivatives are 
zero. Therefore, we use $\log Z(\gamma)\sim N^2f(\gamma,\nu_c)$ as our leading holomorphic 
free energy, and $\nu_c\approx\nu_\star(\gamma,\gamma^\ast)$ can be inserted only after 
the $\gamma$ derivative is taken.

Along the spirit of using the index only at those $\gamma=\gamma_R+i\gamma_I$ 
without coarse-grained cancellations (see page \pageref{paragraph-coarsegrained}),
we tune $\gamma_I$ as a function of $\gamma_R=T^{-1}$ by demanding 
${\rm Im}[\frac{\partial}{\partial\gamma}\log Z(\gamma)]=0$. 
If we select the maximal $\nu_c\approx \nu_\star(\gamma,\gamma^\ast)$ first and 
then tune $\gamma_I$, one obtains the condition
\begin{equation}
  0=\partial_\gamma f(\gamma,\nu_c)-\partial_{\gamma^\ast}f^\ast(\gamma^\ast,\nu_c)
  =\frac{1}{2}\partial_R(f-f^\ast)
  -\frac{i}{2}\partial_I(f+f^\ast)\ \leftrightarrow \ 
  \partial_R(f-f^\ast)=i\partial_I(f+f^\ast)\ ,
\end{equation}
where $\partial_R\equiv\frac{\partial}{\partial\gamma_R}$ and 
$\partial_I\equiv\frac{\partial}{\partial\gamma_I}$ do not act on $\nu_c$.
Also, from the holomorphy of $f$, $\partial_{\gamma^\ast}f=0$,  
one also finds the conditions $\partial_I f=i\partial_Rf$ and 
$\partial_If^\ast=-i\partial_R f^\ast$. Combing these conditions, 
\begin{equation}\label{tune-legendre}
  \partial_R(f(\gamma,\nu_c)-f^\ast(\gamma^\ast,\nu_c))=0\ ,\ 
  \partial_I(f(\gamma,\nu_c)+f^\ast(\gamma^\ast,\nu_c))=0
\end{equation}
are satisfied at the optimal $\gamma_I(\gamma_R)$. 

Alternatively, one may try to tune $\gamma_I$ first on (\ref{filling-fraction-sum}) before selecting the 
maximal term $\nu_c$, by first Legendre transforming each term in (\ref{filling-fraction-sum}) 
and then finding the maximal $\nu$. One can easily check that changing the order of 
tuning $\gamma_I$ and maximizing in $\nu$ yields the same final result. If we follow the 
order just stated, we first tune $\gamma_I$ for each 
$Z(\gamma,\nu)$, obtaining the optimal $\gamma_I(\gamma_R,\nu)$ which depends on $\nu$. 
Then to find the maximal $\nu$ with $\gamma$ restricted, one should maximize 
\begin{equation}
  f(\gamma_R+i\gamma_I(\gamma_R,\nu),\nu)+
  f^\ast(\gamma_R-i\gamma_I(\gamma_R,\nu),\nu)\ .
\end{equation}
We maximize this in continuous real $\nu$, to find the coarse-grained non-holomorphic 
maximum analogous to $\nu_\star$ above. Fixing $\gamma_R$ and taking $\nu$ derivative, 
one obtains
\begin{equation}\label{maximal-order}
  0=\partial_\nu(f+f^\ast)+\partial_I(f+f^\ast)
  \left.\frac{\partial}{\partial\nu}\right|_{\gamma_R}\gamma_I(\gamma_R,\nu)\ ,
\end{equation}
where $\partial_\nu$, $\partial_R$, $\partial_{I}$ denote 
derivatives before inserting $\gamma_I(\gamma_R,\nu)$. Since $\gamma_I$ 
appearing in $f,f^\ast$ are already fixed to satisfy 
${\rm Im}[\frac{\partial}{\partial\gamma}\log Z]=0$ for a given $\nu$, 
it satisfies $\partial_I(f(\gamma,\nu)+f^\ast(\gamma^\ast,\nu))=0$ 
by following the same arguments which led to (\ref{tune-legendre}). 
So the second term of (\ref{maximal-order}) is zero, yielding the 
equation $\partial_\nu(f+f^\ast)=0$
which is the same as the equation for the coarse-grained $\nu_c\approx\nu_\star$.
Therefore, no matter whether one tunes $\gamma_I$ first or maximizes in $\nu$ first, 
one arrives at the same expressions for $\gamma_I,\nu$ as functions of $\gamma_R=T^{-1}$ if one remembers that $\nu_c$ is piecewise constant.

Although we can construct the 2-cut saddle points numerically at various selected complex 
values of $\gamma$ and $\nu$ as illustrated in Fig. \ref{2-cut-plot}, 
again we are unable to determine the cut $C=C_1\cup C_2$ analytically. 
What makes the situation worse than the 1-cut case is that we are also unable to
obtain the general expression for the saddle point free energy such as (\ref{free-1-cut}), 
by evaluating the integral (\ref{eff-action-bulk}) with (\ref{2-cut-summary}). 
So the studies of the 2-cut saddles will be somewhat limited below.

We first study the 2-cut saddle points at very low temperatures. That is, 
we consider (\ref{2-cut-summary}) at ${\rm Re}(\gamma)\gg 1$ (then $|\gamma|\gg 1$ follows). 
In this case, we can perform perturbative expansion in $\gamma^{-1}$
to systematically approximate the saddle point solutions,
and further maximize in $\nu$ to find the free energy.
As explained in detail in Appendix \ref{sec:app2cut}, let us parametrize 
$c_1\equiv c_0(1-i\epsilon)$ and $c_2=c_0(1+i\epsilon)$, 
where $c_{1,2}\equiv\cos\theta_{1,2}$. If the 2-cut ansatz is the correct one at low 
temperature, it will asymptote to the uniform gapless distribution on the unit circle.  
This demands that the two gaps asymptotically close in the $|\gamma|\rightarrow \infty$
limit, $\theta_1=\theta_2$, which will be realized as the small $\epsilon$ limit in 
the parametrization above.
So with foresight, let us first expand various quantities in small $\epsilon$.
Expanding the second line of (\ref{2-cut-summary}) 
in small $\epsilon$ (and large $|\gamma|$), one obtains 
$\gamma=2\log(2/\epsilon)+\mathcal{O}(\epsilon^2)$, or 
$\epsilon\approx 2e^{-\frac{\gamma}{2}}$. 
Therefore, $\gamma$ indeed becomes large at small $\epsilon$. 
The bulk function (\ref{2-cut-summary}) can be approximated as
\begin{equation}\label{rho-2-cut-lowT}
  \rho(\theta)=\frac{1}{2\pi}+\frac{1}{2\pi\gamma}
  \log\left[-\frac{(\cos\theta-c_0)^2}{\cos^2\theta}\right]+\mathcal{O}(\epsilon^2)~,
\end{equation}
as shown in (\ref{app:2cutlowTrho}). Note that all the ignored terms 
are nonperturbatively (exponentially) suppressed in small $\gamma^{-1}$. 
The leading term $\rho(\theta)\approx \frac{1}{2\pi}$ leads to the cut,
determined by real positive $\rho(\theta)d\theta$,
that is the gapless line $-\pi<\theta<\pi$ on the real axis, 
with uniform eigenvalue density. The leading value of 
$\theta_1\approx\theta_2$ is $\frac{\pi}{2}$, making the two cuts $C_1$, $C_2$ to 
meet asymptotically.

We would like to study the effects of the second term of (\ref{rho-2-cut-lowT}), 
perturbatively in small $\frac{1}{|\gamma|}$. 
We want to determine: the small gap, i.e. deviations of $\theta_{1,2}$ 
away from $\frac{\pi}{2}$; the filling fraction $\nu_\star$ that gives the dominant saddle;
and the leading free energy $\log Z$. 
We solve these problems by following the procedures outlined earlier in 
this section. The leading gap is determined from the results shown in the previous paragraph, 
$c_2-c_1\approx 2ic_0 \epsilon\approx 4ic_0e^{-\frac{\gamma}{2}}$, once we compute $c_0$ in 
terms of $\gamma$. After the calculations explained in Appendix \ref{sec:app2cut}, one obtains
\begin{eqnarray}
c_0 &=& \frac{i\pi^2}{\gamma^\ast} \cdot \left[ \frac12 -
\frac{1}{{\rm Re}\,(\gamma)} \cdot \left( 1 + \log \frac{4|\gamma|}{\pi^2} \right) \right]
- \frac{\pi^5}{4c_\mu |\gamma|^2} \cdot \frac{{\rm Im}\,(\gamma)}{{\rm Re}\,(\gamma)}
+ O(|\gamma|)^{-3}~, \nonumber\\
\nu_\star &=& \frac12 + \frac{i\pi}{2\gamma}
- \frac{c_0}{\pi} - \frac{ic_0}{\gamma} - \frac{2c_0}{\pi \gamma} \left( 1 + \log\frac{2}{c_0} \right)
- \frac{c_0^3}{6\pi} + O(c_0^4)~,
\end{eqnarray}
where $c_\mu\approx 9.8696$ is a constant whose exact expression is given by (\ref{app:cmu}). 
The expressions are non-holomorphic in $\gamma$, reflecting the holomorphic anomaly
discussed earlier in this subsection. 
We insert the expression for $c_0$ on the first into the second line to obtain 
\begin{equation}
  \nu_\star(\gamma,\gamma^\ast)=\frac{1}{2}+\frac{\pi \, {\rm Im}(\gamma)}{|\gamma|^2}+
  \frac{\pi}{|\gamma|^2}\left[\frac{{\rm Im}(\gamma)}{{\rm Re}(\gamma)}
  \left(\frac{\pi^3}{4c_\mu}-1-\log|\gamma|\right)+
  \frac{i}{2}\log\left(\frac{\gamma}{\gamma^\ast}\right)\right]+\cdots\ .
\end{equation}
We also compute the leading order free energy, (see \eqref{app:2cutlogZ}) 
\begin{equation}\label{2-cut-low}
  \frac{\log Z(\gamma,\nu_\star)}{N^2}=-\frac{7\zeta(3)}{2\gamma^2}+\mathcal{O}(|\gamma|^{-3})\ .
\end{equation}
Non-holomorphicity is not visible at the leading order. 
The leading term agrees with $\log Z_\infty$ that we computed in the regime 
$N^{-1}\ll \beta\ll 1$, providing the correct low temperature limit.
This supports our assertion that the low temperature phase of the index is 
described by the 2-cut distributions \eqref{2-cut-summary}.

Beyond the approximation $|\gamma| \gg 1$,
we could not analytically compute the free energy $\log Z(\gamma,\nu)$ for the general 
2-cut function (\ref{2-cut-summary}). This makes it hard to find $\varphi(T)={\rm Im}(\gamma)$ 
as a function of $T={\rm Re}(\gamma)^{-1}$ from ${\rm Im}[\partial_\gamma\log Z(\gamma)]=0$, 
because we do not know the analytic expression of the latter. This further makes it hard 
to compare the 1-cut and 2-cut saddles and determine the dominant phase and their transition. 
Note however that the threshold ($j=j_c$) 1-cut saddle appears around $T\sim 0.2$,
below which our 2-cut saddle is the only saddle known to us, with none to compete against.

We can however do the following calculation to constrain the phase transition temperature between 
our 1-cut and 2-cut saddles. We consider the configuration in which $N-1$ eigenvalues 
form the 1-cut distribution of (\ref{1-cut-summary}), centered around $\theta=0$,
while the last eigenvalue is located at $\theta=\pi$.
From the viewpoint of 2-cut distribution, it corresponds to the largest non-trivial
filling fraction $\nu=1-\frac{1}{N}$. However, it is better to view this configuration as 
an eigenvalue instanton correction to the 1-cut distribution. The last eigenvalue can be 
treated as a probe in the 1-cut background. 
One can compute the $\mathcal{O}(N^1)$ subleading correction to $\log Z$ by studying the 
`chemical potential' of the probe eigenvalue,
\begin{eqnarray}\label{filling-integral}
  \mu_1&=&2\int_{C_1}d\theta^\prime \rho(\theta^\prime)
  \log\left(4\sin^2\frac{\theta^\prime-\theta}{2}\right)+
  \frac{1}{\gamma}\left[{\rm Li}_2(ie^{i\theta})-{\rm Li}_2(-ie^{i\theta})+
  {\rm Li}_2(ie^{-i\theta})-{\rm Li}_2(-ie^{-i\theta})\right]~, \nonumber\\
  \mu_2&=&2\int_{C_1}d\theta^\prime \rho(\theta^\prime)
  \log\left(4\sin^2\frac{\theta^\prime-\pi}{2}\right)+
  \frac{2}{\gamma}\left[{\rm Li}_2(-i)-{\rm Li}_2(i)\right]\ ,
\end{eqnarray}
where $C_1$ is the curve for the 1-cut saddle point.
$\mu_1$ is $\frac{1}{N}$ times the contribution to $\log Z$ (\ref{eff-action-continuum})
from the probe eigenvalue that is placed at $\theta\in C_1$. 
Since $\rho(\theta^\prime)$ satisfies the saddle point equation, 
$\mu_1$ cannot depend on $\theta$ if it is on $C_1$.
Therefore, we will insert $\theta=0$ and evaluate 
\begin{equation}\label{filling-integral-2}
  \mu_1=2\int_{C_1}d\theta^\prime \rho(\theta^\prime)
  \log\left(4\sin^2\frac{\theta^\prime}{2}\right)+
  \frac{2}{\gamma}\left[{\rm Li}_2(i)-{\rm Li}_2(-i)\right]\ .
\end{equation}
$\mu_2$ is $\frac{1}{N}$ times the contribution to $\log Z$
from the probe eigenvalue when it is placed at $\theta=\pi$.
Evaluations using the techniques of Appendix \ref{sec:appB}, we obtain
\begin{eqnarray}\label{instanton-action}
  \mu_1&=&\frac{1}{\gamma}\left[{\rm Li}_2(e^{-\gamma})-4\,{\rm Li}_2(-ie^{-\frac{\gamma}{2}})
  -\frac{\pi^2}{4}\right]\ ,\\
  \mu_2&=&\frac{1}{\gamma}\left[-4\,{\rm Li}_2(-ie^{-\frac{\gamma}{2}})
  +8\,{\rm Li}_2(-e^{-\frac{\gamma}{4}})+8\,{\rm Li}_2(-ie^{-\frac{\gamma}{4}})
  +\frac{3\pi^2}{4}\right]\ .
  \nonumber
\end{eqnarray}
Since each chemical potential represents the free energy `cost' for placing
the probe eigenvalue at either location,
the sign of ${\rm Re}(\mu_1-\mu_2)$ determines whether it is thermodynamically favored
to absorb the probe eigenvalue in $C_1$ while preserving its shape,
or to create a second cut at $\pi$.

\begin{figure}[t]
	\centering
	\includegraphics[width=0.6\textwidth]{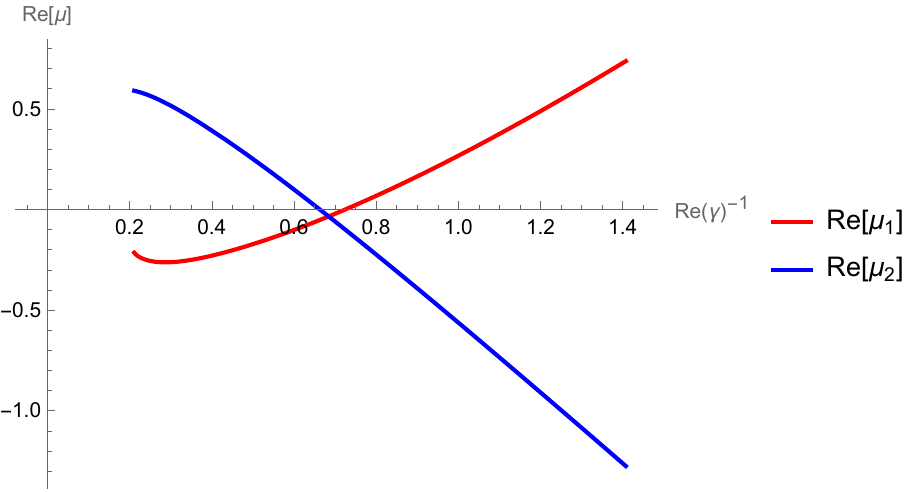}
	\caption{Plots of ${\rm Re}[\mu_1]$ and ${\rm Re}[\mu_2]$. The two curves 
    cross at  $T_0\sim 0.68$. Plots are shown only for $j>j_c$, or $T>T_c\approx 0.21$, 
    where the 1-cut saddles exist.}
	\label{mu1-mu2}
\end{figure}

Fig. \ref{mu1-mu2} plots ${\rm Re}(\mu_1)$ and ${\rm Re}(\mu_2)$ 
as functions of $T$, where $\gamma=T^{-1}+i\varphi_1(T)$.
$\varphi_1(T)$ is the optimal value of ${\rm Im}(\gamma)$ given $T$,
determined by ${\rm Im}[\partial_\gamma\log Z_{\textrm{1-cut}}]=0$ at the 1-cut saddle.
$\varphi(T)$ depends on the filling fraction if a large number of eigenvalues 
is moved from one cut to the other, but since we are moving only one eigenvalue, 
$\Delta\nu=\frac{1}{N}$, the deviation of $\varphi(T)$ from $\varphi_1(T)$ is negligible.
If ${\rm Re}(\mu_1)>{\rm Re}(\mu_2)$, the 1-cut saddle is preferred
against creating a new small cut at $\theta=\pi$. If ${\rm Re}(\mu_2)>{\rm Re}(\mu_1)$, 
creating the new cut is preferred.

In Fig. \ref{mu1-mu2}, one finds ${\rm Re}(\mu_1)<{\rm Re}(\mu_2)$ for $T<T_0\approx 0.68$, 
meaning that the 1-cut saddle cannot be dominant in this region.
On the other hand for $T>T_0$, one finds ${\rm Re}(\mu_1)>{\rm Re}(\mu_2)$ which means that 
the 1-cut saddle is more dominant than the 2-cut saddle with the infinitesimal second cut.
However, our calculations do not rule out more dominant 2-cut saddles with finite
nonzero $\nu_2$. If there exist 2-cut saddles at $T=T_0$ with already finite filling fraction 
with larger ${\rm Re}[\log Z]$, the phase transition will happen 
at a temperature higher than $T_0$.

Eventually, at sufficiently high temperature, we expect the 1-cut saddles to be dominant. 
This can be easily seen by studying the extreme high temperature Cardy limit. 
In this limit, the most dominant saddles are given by the 1-cut distributions, at $\nu=1$ or $0$, 
because the numerator of the free energy (\ref{Cardy-free}) has the maximal
absolute value at these values. Therefore, we expect that $T_0\sim 0.68$ of 
Fig. \ref{mu1-mu2} is a lower bound for the deconfinement phase transition.

\label{T0-transition}
Given these constraints, there are two natural scenarios of the phase transition.
First, if the critical temperature is higher than $T_0$, the transition will be of 
first order between the two distinct saddles.
Second, if the transition happens at $T_0$, the second cut 
(say $C_2$) of the 2-cut phase will gradually shrink as $T$ increases towards $T_0$, 
merging with the 1-cut saddle at $T=T_0$. This type of transition was studied in
\cite{Jurkiewicz:1982iz} in a simple model, where the transition is of second order. 
Had there been no issue of non-universality explained around (\ref{saddle-nu-def}), we would 
have also naturally expected the transition in our second scenario to be of second order. 
However, the fast oscillating might spoil the standard considerations after two derivatives. 
It seems quite clear that such a transition will be no smoother than 
second order.

To determine the phase transitions, at least between our 
1-cut and 2-cut saddles, one should compute the classical $\log Z$ for the 2-cut 
exactly and go through the maximization procedures discussed in this section. 
As these include interesting issues like the holomorphic anomaly and background independence 
of quantum gravity, we think it will be a valuable exercise. In particular, recall that in the low
temperature 2-cut phase, the second derivative of the free energy $\log Z$ in $\gamma$ suffers from 
non-universal contributions, from the fast oscillating terms of (\ref{saddle-nu-osc}). 
Among others, this may affect the computation of the susceptibility/specific heat. The `average'
susceptibility computed from the leading term of (\ref{2-cut-low}) is positive, but we could not 
compute the oscillating part even in this limit. The large fluctuations would mean that the
thermodynamic reactions of the system do not exhibit uniform semi-classical behaviors, perhaps
highlighting the subtle natures of the BPS sector of the higher spin gravity.

We close this subsection by discussing the connections between the
low/high temperature phases introduced in this section,
and the microstate contents explored in Section \ref{sec:2}.
Since the $N$ quark degrees of freedom are visible at extreme high temperature,
the 1-cut saddles are naturally regarded as describing the deconfined phase.
The dominant cohomologies are presumably the $U(N)$ fortuitous cohomologies.
As for the low temperature 2-cut phase, we have seen 
at very low temperature $\frac{1}{|\gamma|}\ll 1$ that the free energy 
$\log Z\approx -\frac{7N^2\zeta(3)}{2\gamma^2}=-\frac{7\zeta(3)}{2\beta^2}$ can be 
accounted for by the multi-particle higher spin BPS states in the strict large $N$ limit. 
At higher temperature, the $U(N)$ trace relations will reduce the number of
independent states among these, imposing a giant graviton like exclusion principle. 
A natural possibility is that the 2-cut phase is dominated by such 
reduced states alone, and the phase transition is the point where the new $U(N)$ fortuitous 
states start to affect the large $N$ thermodynamics. It will be again helpful to know 
the semi-classical 2-cut free energy exactly, to better address this question. 
(See Sections \ref{sec:3.3} and \ref{sec:4} for further comments on this issue.)

\subsection{Comparison to the large $N$ partition functions}\label{sec:3.3}

We compare our results in this section to the closely related studies in the literature. 
There are many works on the large $N$ vector Chern-Simons holography: see for instance 
\cite{Shenker:2011zf,Aharony:2011jz,Giombi:2011kc,Maldacena:2011jn,Maldacena:2012sf,Banerjee:2012gh,Chang:2012kt,Aharony:2012nh,Jain:2012qi,Yokoyama:2012fa,Banerjee:2012aj,Gur-Ari:2012lgt,Aharony:2012ns,Jain:2013py}. In this subsection, we focus on
\cite{Shenker:2011zf} and \cite{Jain:2013py} for comparison, which studied 
the $S^2\times S^1$ partition function of the vector Chern-Simons theory, respectively 
in the free limit and at nonzero 't Hooft coupling $\lambda\neq 0$.

\cite{Jain:2013py} studied the partition functions of large $N$ vector models at nonzero interactions.
The large $N$ matrix model at high temperature $T\sim \sqrt{N}$ scaling is given by
\begin{equation}\label{partition-matrix}
  Z(T)\sim \int [dU]e^{-T^2V_2v(U)}\ ,
\end{equation}
where $V_2$ is the volume of the spatial 2-sphere and $U$ is a unitary matrix whose eigenvalues 
are $e^{i\alpha_a}$, $a=1,\cdots,N$. \cite{Jain:2013py} computed the external potential 
\begin{equation}
  v(U)\sim \sum_{a=1}^Nv(e^{i\alpha_a})=N\int d\alpha\rho(\alpha)v(e^{i\alpha})~,
\end{equation}
in various theories. To be definite, we consider the theory of a scalar 
in the fundamental representation of $U(N)$, with CS 't Hooft coupling 
$\lambda=\frac{N}{k}$ and a sixth order potential of the form $\lambda_6\phi^6$. 
Apart from an $\alpha$-independent constant, the potential is given by \cite{Jain:2013py} 
\begin{eqnarray}\label{potential-interacting}
  v(e^{i\alpha})
  &=&-\frac{1}{2\pi}\int_\sigma^\infty dy
  \, y \left({\rm Li}_1(e^{-y+i\alpha})+{\rm Li}_1(e^{-y-i\alpha})\right)\\
  &=&-\frac{1}{2\pi}\left[{\rm Li}_3(e^{-\sigma+i\alpha})+{\rm Li}_3(e^{-\sigma-i\alpha})
  +\sigma\left( {\rm Li}_2(e^{-\sigma+i\alpha})+{\rm Li}_2(e^{-\sigma-i\alpha})
  \right)\right]~.\nonumber
\end{eqnarray}
$\sigma$ is a parameter appearing in the 
thermal mass $\Sigma=\sigma T^2$ of the scalar field determined by
\begin{equation}\label{thermal-mass}
  \sigma=-\frac{1}{2}\sqrt{\frac{\lambda_6}{8\pi^2}+\lambda^2}
  \int_{-\pi}^\pi d\alpha\rho(\alpha)\left[\log(2\sinh{\textstyle \frac{\sigma-i\alpha}{2}})
  +\log(2\sinh{\textstyle \frac{\sigma+i\alpha}{2}})\right]\ .
\end{equation}
Note that the overall factor of $T^2$ in (\ref{partition-matrix})
and the potential take the form of (\ref{div-pot-3}) at $D=3$. 
In the free limit, $\lambda,\lambda_6\rightarrow 0$, one finds 
$\sigma=0$ from (\ref{thermal-mass}) and the potential (\ref{potential-interacting})
reduces to 
\begin{equation}\label{potential-free}
  v(e^{i\alpha})\rightarrow -\frac{1}{2\pi}
  \left[{\rm Li}_3(e^{i\alpha})+{\rm Li}_3(e^{-i\alpha})\right]\ .
\end{equation}
This is the matrix model potential of the free partition function
\cite{Shenker:2011zf}.\footnote{
The temperatures of \cite{Shenker:2011zf,Jain:2013py} are related by
$(T\textrm{ of \cite{Shenker:2011zf}})=r(T\textrm{ of \cite{Jain:2013py}})$, 
where $r$ is the radius of $S^2$. 
Multiplying an extra $4\pi$ to (\ref{potential-free}) which comes from $V_2=4\pi r^2$ of 
(\ref{partition-matrix}), one obtains the potential of \cite{Shenker:2011zf} at $N_f=1$.}

To compare with these problems, recall that the matrix model for the index 
studied in this paper takes the form of 
\begin{equation}\label{Z-as-v}
  Z(\beta)\sim \int [dU]e^{-\frac{1}{\beta}v(U)}~,
\end{equation}
with $v(U)=\sum_a v(e^{i\alpha_a})$ and 
\begin{equation}\label{potential-index}
  v(e^{i\alpha})=-\left[{\rm Li}_2(ie^{i\alpha})-{\rm Li}_2(-ie^{i\alpha})
  +{\rm Li}_2(ie^{-i\alpha})-{\rm Li}_2(-ie^{-i\alpha})\right]~,
\end{equation}
from (\ref{eff-action-pm}). The potential takes the form of (\ref{div-pot-3}) 
at effective spacetime dimension $D=2$, because the BPS operators carry only one 
(holomorphic) derivative.

We first compare the interacting partition function with the potential
(\ref{potential-interacting}) and the index with (\ref{potential-index}).
The interacting partition function depends on the nonzero thermal mass parameter $\sigma$,
which keeps the integration contour (real $\alpha$) free of singularities
even in the $\beta\sim N^{-\frac{1}{2}} \ll 1$ limit.
The potential is furthermore a real function, so one naturally stays
on this contour while finding the large $N$ saddle points.
On the other hand, the potential (\ref{potential-index}) for the index 
suffers from a singularity on the original contour of real $\alpha$,
specifically at $\alpha=\pm\frac{\pi}{2}$, as we take $\beta\sim N^{-1} \ll 1$.
What saves our setup is that the potential is complex,
demanding the saddle point solutions to deviate from the original integration contour.
In fact, all solutions that we have found in this work
have their cuts away from the singular points $\alpha=\pm\frac{\pi}{2}$.

Now consider the free limit of the partition function, whose potential
(\ref{potential-free}) is singular at $\alpha=0$.
However, this singularity is milder than those in (\ref{potential-index}) for the index 
at $\alpha=\pm\frac{\pi}{2}$.
The function ${\rm Li}_3(e^{\pm i\alpha})$ in (\ref{potential-free}) is 
finite up to the first derivative at $\alpha=0$, i.e. ${\rm Li}_3(1)=\zeta(3)$ and 
$\left.\frac{\partial{\rm Li}_3(e^{\pm i\alpha})}{\partial\alpha}\right|_{\alpha=0}
=\pm\frac{\pi i}{6}$, meaning that both the potential and the force are finite there. 
So one obtains continuous solution for $\rho(\theta)$ across this singular point, 
as explored in \cite{Shenker:2011zf}. On the other hand, the potential
${\rm Li}_2(\pm ie^{\pm i\alpha})$ for our index has divergent first derivatives 
at $\alpha=\pm\frac{\pi}{2}$, disallowing the large $N$ saddle points across these points. 
This was the key technical reason for the existence of a threshold of our 1-cut saddles,
as well as for the appearance of 2-cut (as opposed to gapless) distributions
at low temperatures.
The different natures of the singularities in the potential lead to 
different phase structures between the partition function and the index. 
The absence/mildness of the singularity for the partition function 
rendered the relatively smoother third order 
phase transition between the gapless and the 1-cut gapped saddles. The transition 
for the index should be more singular, as we discussed in 
Section \ref{sec:lowTsaddle}.

Physically, the different phase transition structures of the large $N$ index and 
the partition functions may be understood as follows.
We first compare the free partition function (\ref{potential-free}) of \cite{Shenker:2011zf}
and the index (\ref{potential-index}).
The free partition function counts all the higher spin current multiplets while 
the index only counts those protected against interactions. 
In the former, as we go to higher energies (temperature), the only possible 
finite $N$ effect is to have fewer states by trace relations. The phase transition 
which creates a gap in $\rho(\theta)$ was interpreted in \cite{Shenker:2011zf} as reflecting 
such reduction of states.\footnote{The gap (interval with 
$\rho(\theta)=0$) implies many trace relations between the Fourier coefficients 
$\rho_n\equiv\frac{1}{N}{\rm tr}(U^n)$ of $\rho(\theta)$ near the saddle point. $\rho(\theta)$ 
can also be interpreted as the Fermi surface on the phase space for $\rho$-$\theta$, 
after reformulating the problem \cite{Dutta:2007ws}. Having the bottom $\rho=0$ of the Fermi sea exposed means that giant graviton like exclusion principle affects the states \cite{Lin:2004nb}.}
On the other hand, since our index captures fortuitous states,
trace relations can result in more states counted by the index as the energy increases.
Incidentally, the phase transition of the index annihilates the second cut 
rather than creating a gap, which is presumably more singular. We interpret the 
more singular phase transition in the index as a consequence of the fortuitous states. 
Similar phenomenon is observed in the matrix field theories. The free partition 
function exhibits marginally first order phase transition by creating a gap
\cite{Sundborg:1999ue,Aharony:2003sx}, while the index (affected by fortuity) 
undergoes a proper first order phase transition \cite{Copetti:2020dil,Choi:2021lbk} 
which is in a sense more singular.

The interacting large $N$ partition functions of \cite{Jain:2013py} 
seem to exhibit qualitatively similar phase structures to the free partition function, 
among others having similar matrix model potentials 
and undergoing third order phase transitions.
(This is modulo an interesting new effect of the `capped' saddles \cite{Jain:2013py}.)
To better understand this, 
first note that the anomalous dimensions $\Delta E$ of the higher spin particles are 
suppressed by $\frac{1}{N}$. So the large $N$ interacting partition function counts the 
anomalous operators with almost the same weight as in the free theory, 
$e^{-\beta E}=e^{-\beta E_{\rm free}+\Delta E}\approx e^{-\beta E_{\rm free}}$, 
even at finite $\lambda$. This naturally explains that the interaction does not 
affect the qualitative structures of the phase transition.

We also briefly discuss the free BPS partition function, 
counting the multi-trace operators made of (\ref{single-trace-boson-BPS}) and 
BPS derivatives $\partial$ on them. It is given by
the following matrix integral,
\begin{equation}
  \int [dU]\exp\left[2\sum_{a=1}^N\sum_{n=1}^\infty\frac{1}{n}\frac{x^{\frac{n}{2}}
  +(-1)^{n-1}x^{\frac{3n}{2}}}
  {1-x^{2n}}(e^{in\alpha_a}+e^{-in\alpha_a})\right]\ .
\end{equation}
We keep only one fugacity $x$ for simplicity, and removed $(-1)^F$.
We take the scaling limit $\beta\sim N^{-1}\ll 1$ (where $x=e^{-\beta}$)
for nontrivial large $N$ saddles, and the matrix integral is written
in the form \eqref{Z-as-v} with a potential
\begin{equation}\label{BPS-partition}
  v(e^{i\alpha})=-{\rm Li}_2(e^{i\alpha})-{\rm Li}_2(e^{-i\alpha})
  +{\rm Li}_2(-e^{i\alpha})+{\rm Li}_2(-e^{-i\alpha})\ .
\end{equation}
This potential can be rewritten as  
\begin{equation}\label{BPS-partition-potential}
  v=-\frac{\pi^2}{2}+\pi|\alpha|\ \ \ \textrm{for }\ -\pi<\alpha<\pi\ ,
\end{equation}
and is $2\pi$ periodic in $\alpha$. The forces at $\alpha=0,\pi$ are finite but 
discontinuous. With this potential, one may look for the gapless distribution 
at fixed $\gamma=N\beta$. One finds that $\rho(\theta)$ for the gapless saddle, given by 
(\ref{rho-gapless}), always violates $\rho(\theta)\geq 0$ near $\theta=\pi$, thus failing 
to exist. This is because the potential (\ref{BPS-partition-potential}) has 
a cusp at $\alpha=\pi$, whose force repels eigenvalues away from this point in both directions.
We expect that the dominant large $N$ saddles 
at finite $\gamma$ are always 1-cut distributions, with a gap around $\theta=\pi$. 
The gap would close only at $\gamma=\infty$. This is similar to our low temperature 
2-cut saddles of the index. In both cases, the gaps are always created at finite $\gamma$, 
meaning that both free energies see the reduction of states by trace relations 
(applying the interpretation of \cite{Shenker:2011zf}). Since there is no fortuity 
in the free spectrum, it is natural that there are no further phase 
transitions of (\ref{BPS-partition}) at higher temperatures.

Reduction of states by trace relation starts to happen at energies of order $N$. 
In partition functions and the indices, this affects the terms which are powers of 
$e^{-N\beta}$. This factor is finite in the index \eqref{potential-index} and in the free BPS partition function 
(\ref{BPS-partition}) in their large $N$ scaling limits, in which we keep $\gamma=N\beta$ 
fixed. So the gap should exist in the large $N$ saddle for arbitrary $\gamma$, 
interpreting \cite{Shenker:2011zf} the gap as trace relations reducing the states. 
This explains why the low temperature saddles for the index should be 2-cuts rather than 
gapless, and also supports our expectation in the previous paragraph. On the other hand, 
the full partition functions of the vector model \cite{Shenker:2011zf,Jain:2013py}
are studied in the scaling limit with $N\beta^2$ fixed. In this setup, the factor 
$e^{-N\beta}$ is very small, meaning that the trace relations are not visible 
unless the large entropic factor overcomes this energy suppression factor.
So the gap can be created only at small enough $N\beta^2$ with large enough entropy, 
as is the case \cite{Shenker:2011zf,Jain:2013py}. See Section \ref{sec:4} for further comments on 
the trace relations and the giant graviton like effects in these free energies.

\cite{Shenker:2011zf} and \cite{Jain:2013py} explicitly constructed the gapless saddles 
at low temperature, but not the gapped high temperature saddles. 
Since we obtained analytic expressions for the gapped saddles for the index 
in this paper, one may wonder if similar exact gapped solutions can be obtained for 
the partition functions of the vector models. In fact, this is possible. 
In Appendix \ref{sec:appC}, we use the techniques developed in Appendix \ref{sec:appB} 
to construct the high temperature gapped saddles of the 
free partition function. Similar calculations should be possible for 
the interacting partition functions \cite{Jain:2013py} and the free 
BPS partition function (\ref{BPS-partition}).

\section{Conclusion and discussions}\label{sec:4}

In this paper, we studied the BPS states of the ABJ vector Chern-Simons theory at weak-coupling and 
also explored their BPS phases from the index. First, by studying the $Q$-cohomologies for 
the 2-loop BPS states, we found low energy multi-trace/multi-particle BPS states with $U(1)$ 
trace relations that we call `BPS bounds',
and also a heavy BPS operator with $U(N)$ trace relations. We then studied 
the large $N$ high temperature scaling limit of the index which counts these BPS operators 
and constructed saddle point solutions. The low temperature 
phase is described by novel 2-cut eigenvalue distributions, while the high temperature 
phase is described by 1-cut distributions. We made a preliminary study of the  
phase transition and found a sign that the transition is either of first or second order. 
Comparing our results with phase transitions of the partition functions, 
we discussed possible roles of the fortuitous states.

At 2-loop level, most of the free BPS single-trace operators are anomalous, 
except for those in the spin $2$ graviton multiplet. At the multi-trace level,
many operators that contain non-BPS single-trace operators re-enter the BPS sector.
Unlike in the matrix QFT in which 
such re-entrance happens at energies scaling in $N$, it happens from low energies in 
the vector model. We explicitly constructed (\ref{non-grav-mono}),
of which the simplest are $4$ particle states.
It would be interesting to see if such effects can be computed directly 
from the supersymmetric Vasiliev theory.

In string theory, the transitions between small black holes and excited strings (and branes) 
are discussed in \cite{Bowick:1985af,Susskind:1993ws,Horowitz:1996nw,Giveon:2006pr}. Similar 
transitions between the small black holes and the classical solutions for the string condensates 
\cite{Horowitz:1997jc}  are studied in \cite{Chen:2021dsw}. The multi-particle
bounds of the higher spin particles we found could be a BPS higher spin theory analogue 
of such excited strings/branes at weak bulk coupling $\lambda_B=\frac{N^\prime}{N}\ll 1$. 
At $N^\prime=1$, we found no large $N$ phases behaving like small black holes 
but only these higher spin bound states. As $\lambda_B$ increases, it has been suggested that 
the higher spin particles combine to form fundamental strings \cite{Chang:2012kt}, and they could 
also be the partons of branes. So it is possible that our higher spin BPS bounds are 
primordial remnants of these strings/branes at weak coupling 
after the black hole/string transition. It would be interesting to study the spectrum at 
$N^\prime>1$ with these issues in mind.

We note that, shortly after this article appeared, an instance of pairwise lifting of
BPS cohomologies due to higher order corrections to the supercharge has been reported
in the $\frac{1}{16}$-BPS spectrum of 4d $\mathcal{N}=4$ Yang-Mills theory with $SO(7)$ gauge group
\cite{Gadde:2025yoa,Chang:2025mqp,Choi:2025bhi}.
We cannot rule out the possibility that similar lifting of BPS cohomologies occur
in the ABJ model that we study in this article.
The coupling dependence is more tricky in this model because even though we take $k \to \infty$,
the Chern-Simons level $k$ is still an integer and thus the 't Hooft coupling $\lambda = \frac{N}{k}$ is discrete.
While a non-renormalization conjecture was made in a similar context for the ABJM theory
in \cite{Belin:2025hsg}, we leave the full discussion on this issue for future studies.

To study the large $N$ BPS phases of the vector model from the index, 
we took full advantage of the solvability of the large $N$ matrix model with 
an external potential to obtain semi-analytic expressions for the saddle point solutions. 
This allowed us to derive certain classes of large $N$ saddles without any guess. 
Various physical aspects of our saddles are novel, which include (among others): termination of the 
high temperature branch of saddles at a threshold charge; dominance of the two-cut 
saddles at low temperature; subtleties of holomorphic anomaly and background 
independence. Since many subtle quantum aspects seem to appear in the $\frac{1}{N}$ 
expansion of this model, we find it is worth further studies.

We showed that these studies can be extended to the 
partition functions of large $N$ vector models. We made concrete calculations for 
the free partition function in Appendix \ref{sec:appC}, and we feel that they can be generalized to
interacting partition functions. It will also be interesting to go beyond the weak-coupling 
limit $\lambda\ll 1$ and study the physics of magnetic monopole operators given by the 
`capped' eigenvalue distributions \cite{Aharony:2012ns,Jain:2013py}.

It will be interesting to see if the technical advances in this paper can 
be applied to the $Sp(N)$ vector model for the de Sitter higher spin gravity  
\cite{Anninos:2011ui}, e.g. along the lines of \cite{Anninos:2012ft}.

We studied the scaling limit of the index, $\beta\sim N^{-1}\rightarrow 0$ with $\gamma=N\beta$ 
fixed. This scaling retains interesting terms which are powers of $e^{-\gamma}=e^{-N\beta}$ in 
the fugacity expansion. In other words, the fugacity expansion 
structures are partly unspoiled for heavy states at $E\sim N$.
In supersymmetric QFT's of matrices, like 4d $\mathcal{N}=4$ Yang-Mills, 
these terms are studied as the giant graviton expansion 
\cite{Imamura:2021ytr,Gaiotto:2021xce,Murthy:2022ien}. 
Although the meaning of `giant gravitons' is unclear in higher spin gravity, 
in field theory it simply means the finite $N$ effects on the spectrum 
of the $U(N)$ trace relations. 
In the $\mathcal{N}=4$ Yang-Mills theory, the leading large $N$ free energy does not 
keep such power series in $e^{-N\beta}$ because $\beta$ is kept fixed. However, 
free energies like (\ref{free-1-cut}) 
or the instanton actions like (\ref{instanton-action}) contain nontrivial 
series in $e^{-N\beta}$. This should provide useful information on the finite 
$N$ effects in the vector models.

We illustrate the origin of such terms in a simple model. Consider the half-BPS partition 
function of the 4d $\mathcal{N}=4$ Yang-Mills theory with $U(N)$ gauge group,
\begin{equation}
  \log Z(\beta)=-\sum_{n=1}^N\log(1-e^{-n\beta})\ ,
\end{equation}
in the scaling limit $\beta\sim N^{-1}\rightarrow 0$. 
We can approximate the sum by an integral over $x=n\beta$, 
whose error is suppressed by $\beta\sim N^{-1}$ and thus ignored in the leading term.
One obtains
\begin{equation}\label{half-scaling}
  \log Z\sim -\frac{1}{\beta}\int_{0}^{N\beta} dx\log(1-e^{-x})
  =\frac{1}{\beta}\left[{\rm Li}_2(1)-{\rm Li}_2(e^{-N\beta})\right]
  =\frac{\pi^2}{6\beta}-\frac{1}{\beta}\sum_{n=1}^\infty\frac{1}{n^2}e^{-nN\beta}\ .
\end{equation}
The first term $\frac{\pi^2}{6\beta}$ is the free energy of the half-BPS Kaluza-Klein gravitons,
while the other terms are finite $N$ effects on the heavy states. 
The negativity of the latter terms implies the subtractions of null states 
from the naive KK graviton spectrum. The factor $-\frac{1}{n^2\beta}$ comes from 
$-\frac{e^{-nN\beta}}{n(1-e^{-n\beta})}\rightarrow -\frac{e^{-nN\beta}}{n^2\beta}$ in the 
scaling limit. The $i$'th term in the expansion $-\frac{e^{-nN\beta}}{1-e^{-nN\beta}}=
-\sum_{i=0}^\infty e^{-n(N+i)\beta}$ subtracts the redundant half-BPS operator 
${\rm tr}(Z^{N+i})$. More generally, a function 
$\frac{1}{\beta^{D}}{\rm Li}_{D+1}(e^{-N\beta})$ may
represent a tower of states with $D$ dimensional momenta.
(\ref{free-1-cut}) and (\ref{instanton-action}) have ${\rm Li}_3$ functions, which might 
be alluding to $D=2$.

Coming back to the ABJ vector model, the series in $e^{-\gamma}$ in the 1-cut free energy 
(\ref{free-1-cut}) should be the 
finite $N$ effects in the `black hole like' sector, either subtracting the null states or 
adding fortuitous states. Since each saddle point is complex, the spectral 
interpretation may be partly restored after adding contributions from the pair of conjugate 
saddles. (In particular, the $-i$ factor in ${\rm Li}_3(-ie^{-\frac{\gamma}{2}})$ 
obstructs simple spectral interpretation.) It will be very interesting to
extract information on the finite $N$ spectrum from the expression 
(\ref{free-1-cut}). It will also be interesting to analytically compute 
the full free energy of the low temperature two-cut phase, which we 
did not manage to do in this paper, and learn the patterns of the 
finite $N$ effects.

The derivation of AdS/CFT from the vector model has been discussed in the 
literature (e.g. see \cite{Das:2003vw}), using the collective fields given by the gauge 
invariant bilinears of the vector field. At high energy, 
$E\gtrsim N$, these bulk fields should be redundant due to trace relations. The patterns of 
trace relations and the giant graviton like exclusion principle that one may extract from 
our studies could shed lights on the correct bulk variables at high energies.

\vskip 0.5cm

\hspace*{-0.8cm} {\bf\large Acknowledgements}
\vskip 0.2cm

\hspace*{-0.75cm} 
We thank Chi-Ming Chang, Yiming Chen, JaeHyeok Choi, Sunjin Choi, Robert de Mello Koch, 
Minkyoo Kim, Shiraz Minwalla, Stephen Shenker and Xi Yin for helpful discussions. 
This work is supported in part by the NRF grant 2021R1A2C2012350 (SK, JL, HO),
FWO projects G094523N and G003523N, KU Leuven project C16/25/010
and the FWO Junior Postdoctoral Fellowship 1274626N (SL).

\appendix

\section{Counting and constructing operators}\label{sec:appA}

In this appendix, we first derive the large $N$ index $Z_\infty$ with finite `temperature',
which counts the higher spin BPS states at low energy.
Then we present results for the cohomology counting at $N = 2, 3, 4, \infty$.
Finally, we explain the construction of a fortuitous cohomology at $N = 2$.

\subsection{Index and BPS partition function}\label{sec:appA.1}

We analyze the index (\ref{index-formula}) in the large $N$ limit
with the temperature kept at order $1$.
Following \cite{Sundborg:1999ue,Aharony:2003sx}, we introduce 
$\rho_n = \frac1N \sum_{i=1}^N e^{-in\alpha_i}$,
which is the $n$-th Fourier mode of the eigenvalue density $\rho(\alpha)$.
In the large $N$ limit, each $\rho_n$ can be treated as an independent variable.
The integrand in (\ref{index-formula}) is given by
\begin{align}
    \exp \biggl[\, \sum_{n=1}^\infty \frac{1}{n}
    \biggl( 
    -N^2\rho_n \rho_{-n} &+ N \left(\frac{x^\frac{n}{2}}{1-x^{2n}}(y_1^n+y_1^{-n})-\frac{x^\frac{3n}{2}}{1-x^{2n}}(y_2^n+y_2^{-n}) \right) \rho_n  \nn\\
    &+ N \left(\frac{x^\frac{n}{2}}{1-x^{2n}}(y_2^n+y_2^{-n})-\frac{x^\frac{3n}{2}}{1-x^{2n}}(y_1^n+y_1^{-n}) \right) \rho_{-n} \biggr)
    \biggr]\ .
\end{align}
The Gaussian integral over $\rho_n,\rho_{-n}$  yields
\begin{equation}
  Z_{\infty}(x,y_1,y_2)= \exp\left[\sum_{n=1}^\infty\frac{1}{n}
  \left(\frac{x^{\frac{n}{2}}}{1-x^{2n}}\chi_2(y_1^n)
  -\frac{x^{\frac{3n}{2}}}{1-x^{2n}}\hat\chi_2(y_2^n)\right)
  \left(\frac{x^{\frac{n}{2}}}{1-x^{2n}}\hat\chi_2(y_2^n)
  -\frac{x^{\frac{3n}{2}}}{1-x^{2n}}\chi_2(y_1^n)\right)\right]
\end{equation}
where $\chi_m$ and $\hat\chi_m$ denote characters of the dimension $m$ representation
of the two $SU(2)$'s.

Now we address the counting of supercharge cohomologies.
The counting proceeds as follows:
\begin{enumerate}
    \item In each sector with fixed charges, we construct all independent operators. 
    \item We then act $Q$ to extract all $Q$-closed operators.
    With the remaining non-$Q$-closed operators, we construct a basis of
    $Q$-exact operators in the `next' sector that has $Y$ increased by two units
    (because $Y[Q]=2$) and all other charges unchanged.
    \item Repeating the two steps above in all charge sectors up to certain orders,
    we count the number of cohomologies in each charge sector by
    subtracting the number of $Q$-exact operators from the number of $Q$-closed operators.
    \item Similarly, we construct all graviton operators in each charge sector
    and count them modulo $Q$-exact operators, yielding the number of graviton cohomologies.
    \item The number of all cohomologies minus the number of graviton cohomologies
    gives the number of non-graviton cohomologies in each sector.
\end{enumerate}

We counted all cohomologies for $N=2,3,4$ up to orders $x^{11}$, $x^{10}$, and $x^9$,
respectively, and for $N = \infty$ up to order $x^{14}$. 
We present the result of this counting in the form of the BPS partition function (\ref{partition}). 
The BPS partition functions only over the multi-gravitons
(obtained from step 4 above) at $N=2,3,4$ are
\vspace{-0.6cm}

%
{\allowdisplaybreaks \footnotesize
\begin{align}
    Z_{2,{\rm grav}} =& 1 + x y^2 \chi_2 \hat\chi_2 + x^2 \bigl(y^2 (1 + \chi_3 + \hat\chi_3) +y^4 \chi_3 \hat\chi_3 \bigr) + x^3 \bigl( 2 y^2 \chi_2 \hat\chi_2  - y^4 (2 \chi_2 \hat\chi_2 + \chi_2 \hat\chi_4 + \chi_4 \hat\chi_2) + y^6 \chi_4 \hat\chi_4 \bigr) \nn\\
    &+ x^4 \bigl( y^2 (2 + \chi_3 + \hat\chi_3) + y^4 (1 + 3 \chi_3 + 3 \hat\chi_3 + 3 \chi_3 \hat\chi_3) + y^6 (2 \chi_3 \hat\chi_3 + \chi_3 \hat\chi_5 + \chi_5 \hat\chi_3) + y^8 \chi_5 \hat\chi_5  \bigr) \nn\\
    &+ x^5 \bigl( 2 y^2 \chi_2 \hat\chi_2 + y^4 (8 \chi_2 \hat\chi_2 + 3 \chi_4 \hat\chi_2 + 3 \chi_2 \hat\chi_4) + y^6 (\chi_2 \hat\chi_2 + 3 \chi_2 \hat\chi_4 + 3 \chi_4 \hat\chi_2 + 3 \chi_4 \hat\chi_4) \nn\\
    &\qquad + y^8 (2 \chi_4 \hat\chi_4 + \chi_4 \hat\chi_6 + \chi_6 \hat\chi_4) + y^{10} \chi_6 \hat\chi_6 \bigr) \nn\\
    &+ x^6 \bigl( y^2 (2 + \chi_3 + \hat\chi_3) + y^4 (7 + 6 \chi_3 + 6 \hat\chi_3 + \chi_5 + \hat\chi_5 + 7 \chi_3 \hat\chi_3 ) \nn\\
    &\qquad + y^6 (1+ 3 \chi_3 + 3 \hat\chi_3 + \chi_5 + \hat\chi_5 + 10 \chi_3 \hat\chi_3  + 3 \chi_3 \hat\chi_5 + 3 \chi_5 \hat\chi_3 ) + y^8 (\chi_3 \hat\chi_3 + 3 \chi_3 \hat\chi_5 + 3 \chi_5 \hat\chi_3 + 3 \chi_5 \hat\chi_5) \nn\\
    &\qquad + y^{10} (2 \chi_5 \hat\chi_5 + \chi_5 \hat\chi_7 + \chi_7 \hat\chi_5) + y^{12} \chi_7 \hat\chi_7 \bigr) \nn\\
    &+ x^7 \bigl( 2 y^2 \chi_2 \hat\chi_2 + y^4 (16 \chi_2 \hat\chi_2 + 5 \chi_2 \hat\chi_4 + 5 \chi_4 \hat\chi_2) + y^6 (13 \chi_2 \hat\chi_2 + 12 \chi_2 \hat\chi_4 + 12 \chi_4 \hat\chi_2 + \chi_2 \hat\chi_6 + \chi_6 \hat\chi_2 + 9 \chi_4 \hat\chi_4) \nn\\
    &\qquad + y^8 (3 \chi_2 \hat\chi_4 + 3 \chi_4 \hat\chi_2 + \chi_2 \hat\chi_6 + \chi_6 \hat\chi_2 + 10 \chi_4 \hat\chi_4 + 3 \chi_4 \hat\chi_6 + 3 \chi_6 \hat\chi_4) \nn\\
    &\qquad + y^{10} (\chi_4 \hat\chi_4 + 3 \chi_4 \hat\chi_6 + 3 \chi_6 \hat\chi_4 + 3 \chi_6 \hat\chi_6) + y^{12} (2 \chi_6 \hat\chi_6 + \chi_6 \hat\chi_8 + \chi_8 \hat\chi_6) + y^{14} \chi_8 \hat\chi_8 \bigr) \nn\\
    &+ x^8 \bigl( y^2 (2 + \chi_3 + \hat\chi_3) + y^4 (9 + 12 \chi_3 + 12 \hat\chi_3 + \chi_5 + \hat\chi_5 + 9 \chi_3 \hat\chi_3) \nn\\
    &\qquad + y^6 (7 + 17 \chi_3 + 17 \hat\chi_3 + 5 \chi_5 + 5 \hat\chi_5 + 31 \chi_3 \hat\chi_3 + 9 \chi_3 \hat\chi_5 + 9 \chi_5 \hat\chi_3) \nn\\
    &\qquad + y^8 (\chi_3 + \hat\chi_3 + 3 \chi_5 + 3 \hat\chi_5 + 12 \chi_3 \hat\chi_3 + 12 \chi_3 \hat\chi_5 + 12 \chi_5 \hat\chi_3 + \chi_3 \hat\chi_7 + \chi_7 \hat\chi_3  + 9 \chi_5 \hat\chi_5 ) \nn\\
    &\qquad + y^{10} (3 \chi_3 \hat\chi_5 + 3 \chi_5 \hat\chi_3  + \chi_3 \hat\chi_7 + \chi_7 \hat\chi_3 + 10 \chi_5 \hat\chi_5 + 3 \chi_5 \hat\chi_7 + 3 \chi_7 \hat\chi_5 ) + y^{12} (\chi_5 \hat\chi_5 + 3 \chi_5 \hat\chi_7 + 3 \chi_7 \hat\chi_5 + 3 \chi_7 \hat\chi_7) \nn\\
    &\qquad + y^{14} (2 \chi_7 \hat\chi_7 + \chi_7 \hat\chi_9 + \chi_9 \hat\chi_7) + y^{16} \chi_9 \hat\chi_9 \bigr)  \nn\\
    &+ x^9 \bigl( 2 y^2 \chi_2 \hat\chi_2 + y^4 (24 \chi_2 \hat\chi_2 + 7 \chi_2 \hat\chi_4 + 7 \chi_4 \hat\chi_2) + y^6 (46 \chi_2 \hat\chi_2 + 32 \chi_2 \hat\chi_4 + 32 \chi_4 \hat\chi_2 + 3 \chi_2 \hat\chi_6 + 3 \chi_6 \hat\chi_2 + 19 \chi_4 \hat\chi_4 ) \nn\\
    &\qquad + y^8 (7 \chi_2 \hat\chi_2 + 19 \chi_2 \hat\chi_4 + 19 \chi_4 \hat\chi_2 + 6 \chi_2 \hat\chi_6 + 6 \chi_6 \hat\chi_2  + 34 \chi_4 \hat\chi_4 + 9 \chi_4 \hat\chi_6 + 9 \chi_6 \hat\chi_4) \nn\\
    &\qquad + y^{10} (\chi_2 \hat\chi_4 + \chi_4 \hat\chi_2 + 3 \chi_2 \hat\chi_6 + 3 \chi_6 \hat\chi_2 +  12 \chi_4 \hat\chi_4 + 12 \chi_4 \hat\chi_6 + 12 \chi_6 \hat\chi_4 + \chi_4 \hat\chi_8 + \chi_8 \hat\chi_4 + 9 \chi_6 \hat\chi_6) \nn\\
    &\qquad + y^{12} (3 \chi_4 \hat\chi_6 + 3 \chi_6 \hat\chi_4 + \chi_4 \hat\chi_8 + \chi_8 \hat\chi_4 + 10 \chi_6 \hat\chi_6 + 3 \chi_6 \hat\chi_8 + 3 \chi_8 \hat\chi_6) + y^{14} (\chi_6 \hat\chi_6 + 3 \chi_6 \hat\chi_8 + 3 \chi_8 \hat\chi_6 + 3 \chi_8 \hat\chi_8) \nn\\
    &\qquad + y^{16} (2 \chi_8 \hat\chi_8 + \chi_8 \hat\chi_{10} + \chi_{10} \hat\chi_8) + y^{18} \chi_{10} \hat\chi_{10} \bigr) \nn\\
    &+ x^{10} \bigl(y^2 (2 + \chi_3 + \hat\chi_3) + y^4 (17 + 15 \chi_3 + 15 \hat\chi_3 + 2 \chi_5 + 2 \hat\chi_5 + 13 \chi_3 \hat\chi_3) \nn\\
    &\qquad+ y^6 (27 + 43 \chi_3 + 43 \hat\chi_3 + 13 \chi_5 + 13 \hat\chi_5 + 69 \chi_3 \hat\chi_3 + 16 \chi_3 \hat\chi_5 + 16 \chi_5 \hat\chi_3) \nn\\
    &\qquad + y^8 (2 + 16 \chi_3 + 16 \hat\chi_3 + 12 \chi_5 + 12 \hat\chi_5 + \chi_7 + \hat\chi_7 + 52 \chi_3 \hat\chi_3 + 40 \chi_3 \hat\chi_5 + 40 \chi_5 \hat\chi_3 + 3 \chi_3 \hat\chi_7 + 3 \chi_7 \hat\chi_3 + 22 \chi_5 \hat\chi_5 ) \nn\\
    &\qquad + y^{10} (3 \chi_5 + 3\hat\chi_5 +\chi_7 + \hat\chi_7 + 6 \chi_3 \hat\chi_3 + 19 \chi_3 \hat\chi_5 + 19 \chi_5 \hat\chi_3 + 6 \chi_3 \hat\chi_7 + 6 \chi_7 \hat\chi_3 + 34 \chi_5 \hat\chi_5 + 9 \chi_5 \hat\chi_7 + 9 \chi_7 \hat\chi_5) \nn\\
    &\qquad+ y^{12} (\chi_3 \hat\chi_5 + \chi_5 \hat\chi_3 + 3 \chi_3 \hat\chi_7 + 3 \chi_7 \hat\chi_3 +  12 \chi_5 \hat\chi_5 + 12 \chi_5 \hat\chi_7 + 12 \chi_7 \hat\chi_5 + \chi_5 \hat\chi_9 + \chi_9 \hat\chi_5  + 9 \chi_7 \hat\chi_7) \nn\\ 
    &\qquad + y^{14} (3 \chi_5 \hat\chi_7 + 3 \chi_7 \hat\chi_5 + \chi_5 \hat\chi_9 + \chi_9 \hat\chi_5 + 10 \chi_7 \hat\chi_7 + 3 \chi_7 \hat\chi_9 + 3 \chi_9 \hat\chi_7) + y^{16} (\chi_7 \hat\chi_7 + 3 \chi_7 \hat\chi_9 + 3 \chi_9 \hat\chi_7 + 3 \chi_9 \hat\chi_9) \nn\\
    &\qquad + y^{18} (2 \chi_9 \hat\chi_9 + \chi_9 \hat\chi_{11} + \chi_{11} \hat\chi_9 ) + y^{20} \chi_{11} \hat\chi_{11} \bigr) \nn\\
    &+ x^{11} \bigl( 2 y^2 \chi_2 \hat\chi_2 + y^4 (32 \chi_2 \hat\chi_2 + 9 \chi_2 \hat\chi_4 + 9 \chi_4 \hat\chi_2) + y^6 (102 \chi_2 \hat\chi_2 + 64 \chi_2 \hat\chi_4 + 64 \chi_4 \hat\chi_2 + 6 \chi_2 \hat\chi_6 + 6 \chi_6 \hat\chi_2 + 31 \chi_4 \hat\chi_4) \nn\\
    &\qquad+ y^8 (48 \chi_2 \hat\chi_2 + 72 \chi_2 \hat\chi_4 + 72 \chi_4 \hat\chi_2 + 20 \chi_2 \hat\chi_6 + 20 \chi_6 \hat\chi_2  + 92 \chi_4 \hat\chi_4 + 22 \chi_4 \hat\chi_6 + 22 \chi_6 \hat\chi_4) \nn\\
    &\qquad+ y^{10} (\chi_2 \hat\chi_2 + 15 \chi_2 \hat\chi_4 + 15 \chi_4 \hat\chi_2 + 15 \chi_2 \hat\chi_6 + 15 \chi_6 \hat\chi_2 + \chi_2 \hat\chi_8 + \chi_8 \hat\chi_2 + 51 \chi_4 \hat\chi_4 \nn\\
    &\qquad + 40 \chi_4 \hat\chi_6 + 40 \chi_6 \hat\chi_4 + 3 \chi_4 \hat\chi_8 + 3 \chi_8 \hat\chi_4  + 22 \chi_6 \hat\chi_6) \nn\\
    &\qquad+ y^{12} (3 \chi_2 \hat\chi_6\! +3 \chi_6 \hat\chi_2\! + \chi_2 \hat\chi_8\!+ \chi_8 \hat\chi_2\! + 6 \chi_4 \hat\chi_4\! + 19 \chi_4 \hat\chi_6\! + 19 \chi_6 \hat\chi_4\! + 6 \chi_4 \hat\chi_8\! + 6 \chi_8 \hat\chi_4\!  + 34 \chi_6 \hat\chi_6\! + 9 \chi_6 \hat\chi_8\! + 9 \chi_8 \hat\chi_6) \nn\\
    &\qquad+ y^{14} (\chi_4 \hat\chi_6 + \chi_6 \hat\chi_4 + 3 \chi_4 \hat\chi_8 + 3 \chi_8 \hat\chi_4 +  12 \chi_6 \hat\chi_6 + 12 \chi_6 \hat\chi_8 + 12 \chi_8 \hat\chi_6 + \chi_6 \hat\chi_{10} + \chi_{10} \hat\chi_6  + 9 \chi_8 \hat\chi_8) \nn\\
    &\qquad + y^{16} (3 \chi_6 \hat\chi_8 + 3 \chi_8 \hat\chi_6 + \chi_6 \hat\chi_{10} + \chi_{10} \hat\chi_6  + 10 \chi_8 \hat\chi_8 + 3 \chi_{10} \hat\chi_8  + 3 \chi_8 \hat\chi_{10}) \nn\\
    &\qquad + y^{18} (\chi_8 \hat\chi_8 + 3 \chi_8 \hat\chi_{10} + 3 \chi_{10} \hat\chi_8 + 3 \chi_{10} \hat\chi_{10}) + y^{20} (2 \chi_{10} \hat\chi_{10} + \chi_{10} \hat\chi_{12} + \chi_{12} \hat\chi_{10}) \bigr) + y^{22} \chi_{12} \hat\chi_{12} +O(x^{12})   \;, \nn\\
    Z_{3,{\rm grav}} =& 1 + x y^2 \chi_2 \hat\chi_2 + x^2 \bigl(y^2 (1 + \chi_3 + \hat\chi_3) +y^4 \chi_3 \hat\chi_3 \bigr) + x^3 \bigl( 2 y^2 \chi_2 \hat\chi_2  - y^4 (2 \chi_2 \hat\chi_2 + \chi_2 \hat\chi_4 + \chi_4 \hat\chi_2) + y^6 \chi_4 \hat\chi_4 \bigr) \nn\\
    &+ x^4 \bigl(y^2 (2 + \chi_3 + \hat\chi_3) + y^4 (1 + 3 \chi_3 + 3 \hat\chi_3 + 3 \chi_3 \hat\chi_3) + y^6 (2 \chi_3 \hat\chi_3 + \chi_3 \hat\chi_5 + \chi_5 \hat\chi_3) + y^8 \chi_5 \hat\chi_5  \bigr) \nn\\
    &+ x^5 \bigl( 2 y^2 \chi_2 \hat\chi_2 + y^4 (8 \chi_2 \hat\chi_2 + 3 \chi_4 \hat\chi_2 + 3 \chi_2 \hat\chi_4) + y^6 (2\chi_2 \hat\chi_2 + 3 \chi_2 \hat\chi_4 + 3 \chi_4 \hat\chi_2 + 3 \chi_4 \hat\chi_4) \nn\\
    &\qquad + y^8 (2 \chi_4 \hat\chi_4 + \chi_4 \hat\chi_6 + \chi_6 \hat\chi_4) + y^{10} \chi_6 \hat\chi_6 \bigr) \nn\\
    &+ x^6 \bigl(y^2 (2 + \chi_3 + \hat\chi_3) + y^4 (7 + 6 \chi_3 + 6 \hat\chi_3 + \chi_5 + \hat\chi_5 + 7 \chi_3 \hat\chi_3 ) \nn\\
    &\qquad + y^6 (2+ 4 \chi_3 + 4 \hat\chi_3 + \chi_5 + \hat\chi_5 + 11 \chi_3 \hat\chi_3  + 3 \chi_3 \hat\chi_5 + 3 \chi_5 \hat\chi_3 ) + y^8 (\chi_3 \hat\chi_3 + 3 \chi_3 \hat\chi_5 + 3 \chi_5 \hat\chi_3 + 3 \chi_5 \hat\chi_5) \nn\\
    &\qquad + y^{10} (2 \chi_5 \hat\chi_5 + \chi_5 \hat\chi_7 + \chi_7 \hat\chi_5) + y^{12} \chi_7 \hat\chi_7 \bigr) \nn\\
    &+ x^7 \bigl( 2 y^2 \chi_2 \hat\chi_2 + y^4 (16 \chi_2 \hat\chi_2 + 5 \chi_2 \hat\chi_4 + 5 \chi_4 \hat\chi_2) + y^6 (16 \chi_2 \hat\chi_2 + 13 \chi_2 \hat\chi_4 + 13 \chi_4 \hat\chi_2 + \chi_2 \hat\chi_6 + \chi_6 \hat\chi_2 + 9 \chi_4 \hat\chi_4) \nn\\
    &\qquad + y^8 (\chi_2\hat\chi_2 + 3 \chi_2 \hat\chi_4 + 3 \chi_4 \hat\chi_2 + \chi_2 \hat\chi_6 + \chi_6 \hat\chi_2 + 10 \chi_4 \hat\chi_4 + 3 \chi_4 \hat\chi_6 + 3 \chi_6 \hat\chi_4) \nn\\
    &\qquad + y^{10} (\chi_4 \hat\chi_4 + 3 \chi_4 \hat\chi_6 + 3 \chi_6 \hat\chi_4 + 3 \chi_6 \hat\chi_6) + y^{12} (2 \chi_6 \hat\chi_6 + \chi_6 \hat\chi_8 + \chi_8 \hat\chi_6) + y^{14} \chi_8 \hat\chi_8 \bigr) \nn\\
    &+ x^8 \bigl(y^2 (2 + \chi_3 + \hat\chi_3) + y^4 (9 + 12 \chi_3 + 12 \hat\chi_3 + \chi_5 + \hat\chi_5 + 9 \chi_3 \hat\chi_3) \nn\\
    &\qquad + y^6 (9 + 19 \chi_3 + 19 \hat\chi_3 + 5 \chi_5 + 5 \hat\chi_5 + 33 \chi_3 \hat\chi_3 + 9 \chi_3 \hat\chi_5 + 9 \chi_5 \hat\chi_3) \nn\\
    &\qquad + y^8 (1 + 3\chi_3 + 3\hat\chi_3 + 3 \chi_5 + 3 \hat\chi_5 + 15 \chi_3 \hat\chi_3 + 12 \chi_3 \hat\chi_5 + 12 \chi_5 \hat\chi_3 + \chi_3 \hat\chi_7 + \chi_7 \hat\chi_3  + 9 \chi_5 \hat\chi_5 ) \nn\\
    &\qquad + y^{10} (3 \chi_3 \hat\chi_5 + 3 \chi_5 \hat\chi_3  + \chi_3 \hat\chi_7 + \chi_7 \hat\chi_3 + 10 \chi_5 \hat\chi_5 + 3 \chi_5 \hat\chi_7 + 3 \chi_7 \hat\chi_5 ) + y^{12} (\chi_5 \hat\chi_5 + 3 \chi_5 \hat\chi_7 + 3 \chi_7 \hat\chi_5 + 3 \chi_7 \hat\chi_7) \nn\\
    &\qquad + y^{14} (2 \chi_7 \hat\chi_7 + \chi_7 \hat\chi_9 + \chi_9 \hat\chi_7) + y^{16} \chi_9 \hat\chi_9 \bigr)  \nn\\
    &+ x^9 \bigl( 2 y^2 \chi_2 \hat\chi_2 + y^4 (24 \chi_2 \hat\chi_2 + 7 \chi_2 \hat\chi_4 + 7 \chi_4 \hat\chi_2) + y^6 (50 \chi_2 \hat\chi_2 + 33 \chi_2 \hat\chi_4 + 33 \chi_4 \hat\chi_2 + 3 \chi_2 \hat\chi_6 + 3 \chi_6 \hat\chi_2 + 19 \chi_4 \hat\chi_4 ) \nn\\
    &\qquad + y^8 (16 \chi_2 \hat\chi_2 + 25 \chi_2 \hat\chi_4 + 25 \chi_4 \hat\chi_2 + 6 \chi_2 \hat\chi_6 + 6 \chi_6 \hat\chi_2  + 36 \chi_4 \hat\chi_4 + 9 \chi_4 \hat\chi_6 + 9 \chi_6 \hat\chi_4) \nn\\
    &\qquad + y^{10} (\chi_2 \hat\chi_4 + \chi_4 \hat\chi_2 + 3 \chi_2 \hat\chi_6 + 3 \chi_6 \hat\chi_2 +  12 \chi_4 \hat\chi_4 + 12 \chi_4 \hat\chi_6 + 12 \chi_6 \hat\chi_4 + \chi_4 \hat\chi_8 + \chi_8 \hat\chi_4 + 9 \chi_6 \hat\chi_6) \nn\\
    &\qquad + y^{12} (3 \chi_4 \hat\chi_6 + 3 \chi_6 \hat\chi_4 + \chi_4 \hat\chi_8 + \chi_8 \hat\chi_4 + 10 \chi_6 \hat\chi_6 + 3 \chi_6 \hat\chi_8 + 3 \chi_8 \hat\chi_6) + y^{14} (\chi_6 \hat\chi_6 + 3 \chi_6 \hat\chi_8 + 3 \chi_8 \hat\chi_6 + 3 \chi_8 \hat\chi_8) \nn\\
    &\qquad + y^{16} (2 \chi_8 \hat\chi_8 + \chi_8 \hat\chi_{10} + \chi_{10} \hat\chi_8) + y^{18} \chi_{10} \hat\chi_{10} \bigr) \nn\\
    &+ x^{10} \bigl( y^2 (2 + \chi_3 + \hat\chi_3) + y^4 (17 + 15 \chi_3 + 15 \hat\chi_3 + 2 \chi_5 + 2 \hat\chi_5 + 13 \chi_3 \hat\chi_3) \nn\\
    &\qquad+ y^6 (29 + 45 \chi_3 + 45 \hat\chi_3 + 13 \chi_5 + 13 \hat\chi_5 + 71 \chi_3 \hat\chi_3 + 16 \chi_3 \hat\chi_5 + 16 \chi_5 \hat\chi_3) \nn\\
    &\qquad + y^8 (10 + 27 \chi_3 + 27 \hat\chi_3 + 14 \chi_5 + 14 \hat\chi_5 + \chi_7 + \hat\chi_7 + 68 \chi_3 \hat\chi_3 + 43 \chi_3 \hat\chi_5 + 43 \chi_5 \hat\chi_3 + 3 \chi_3 \hat\chi_7 + 3 \chi_7 \hat\chi_3 + 22 \chi_5 \hat\chi_5 ) \nn\\
    &\qquad + y^{10} (1+ 3 \chi_5 + 3\hat\chi_5 +\chi_7 + \hat\chi_7 + 7 \chi_3 \hat\chi_3 + 19 \chi_3 \hat\chi_5 + 19 \chi_5 \hat\chi_3 + 6 \chi_3 \hat\chi_7 + 6 \chi_7 \hat\chi_3 + 34 \chi_5 \hat\chi_5 + 9 \chi_5 \hat\chi_7 + 9 \chi_7 \hat\chi_5) \nn\\
    &\qquad+ y^{12} (\chi_3 \hat\chi_5 + \chi_5 \hat\chi_3 + 3 \chi_3 \hat\chi_7 + 3 \chi_7 \hat\chi_3 +  12 \chi_5 \hat\chi_5 + 12 \chi_5 \hat\chi_7 + 12 \chi_7 \hat\chi_5 + \chi_5 \hat\chi_9 + \chi_9 \hat\chi_5  + 9 \chi_7 \hat\chi_7) \nn\\ 
    &\qquad + y^{14} (3 \chi_5 \hat\chi_7 + 3 \chi_7 \hat\chi_5 + \chi_5 \hat\chi_9 + \chi_9 \hat\chi_5 + 10 \chi_7 \hat\chi_7 + 3 \chi_7 \hat\chi_9 + 3 \chi_9 \hat\chi_7) + y^{16} (\chi_7 \hat\chi_7 + 3 \chi_7 \hat\chi_9 + 3 \chi_9 \hat\chi_7 + 3 \chi_9 \hat\chi_9) \nn\\
    &\qquad + y^{18} (2 \chi_9 \hat\chi_9 + \chi_9 \hat\chi_{11} + \chi_{11} \hat\chi_9 ) + y^{20} \chi_{11} \hat\chi_{11} \bigr) +O(x^{11})   \;, \nn\\
    Z_{4,{\rm grav}} =& 1 + x y^2 \chi_2 \hat\chi_2 + x^2 \bigl(y^2 (1 + \chi_3 + \hat\chi_3) +y^4 \chi_3 \hat\chi_3 \bigr) + x^3 \bigl( 2 y^2 \chi_2 \hat\chi_2  + y^4 (2 \chi_2 \hat\chi_2 + \chi_2 \hat\chi_4 + \chi_4 \hat\chi_2) + y^6 \chi_4 \hat\chi_4 \bigr) \nn\\
    &+ x^4 \bigl(y^2 (2 + \chi_3 + \hat\chi_3) + y^4 (1 + 3 \chi_3 + 3 \hat\chi_3 + 3 \chi_3 \hat\chi_3) + y^6 (2 \chi_3 \hat\chi_3 + \chi_3 \hat\chi_5 + \chi_5 \hat\chi_3) + y^8 \chi_5 \hat\chi_5  \bigr) \nn\\
    &+ x^5 \bigl( 2 y^2 \chi_2 \hat\chi_2 + y^4 (8 \chi_2 \hat\chi_2 + 3 \chi_4 \hat\chi_2 + 3 \chi_2 \hat\chi_4) + y^6 (2\chi_2 \hat\chi_2 + 3 \chi_2 \hat\chi_4 + 3 \chi_4 \hat\chi_2 + 3 \chi_4 \hat\chi_4) \nn\\
    &\qquad + y^8 (2 \chi_4 \hat\chi_4 + \chi_4 \hat\chi_6 + \chi_6 \hat\chi_4) + y^{10} \chi_6 \hat\chi_6 \bigr) \nn\\
    &+ x^6 \bigl(y^2 (2 + \chi_3 + \hat\chi_3) + y^4 (7 + 6 \chi_3 + 6 \hat\chi_3 + \chi_5 + \hat\chi_5 + 7 \chi_3 \hat\chi_3 ) \nn\\
    &\qquad + y^6 (2+ 4 \chi_3 + 4 \hat\chi_3 + \chi_5 + \hat\chi_5 + 11 \chi_3 \hat\chi_3  + 3 \chi_3 \hat\chi_5 + 3 \chi_5 \hat\chi_3 ) + y^8 (\chi_3 \hat\chi_3 + 3 \chi_3 \hat\chi_5 + 3 \chi_5 \hat\chi_3 + 3 \chi_5 \hat\chi_5) \nn\\
    &\qquad + y^{10} (2 \chi_5 \hat\chi_5 + \chi_5 \hat\chi_7 + \chi_7 \hat\chi_5) + y^{12} \chi_7 \hat\chi_7 \bigr) \nn\\
    &+ x^7 \bigl( 2 y^2 \chi_2 \hat\chi_2 + y^4 (16 \chi_2 \hat\chi_2 + 5 \chi_2 \hat\chi_4 + 5 \chi_4 \hat\chi_2) + y^6 (16 \chi_2 \hat\chi_2 + 13 \chi_2 \hat\chi_4 + 13 \chi_4 \hat\chi_2 + \chi_2 \hat\chi_6 + \chi_6 \hat\chi_2 + 9 \chi_4 \hat\chi_4) \nn\\
    &\qquad + y^8 (\chi_2\hat\chi_2 + 3 \chi_2 \hat\chi_4 + 3 \chi_4 \hat\chi_2 + \chi_2 \hat\chi_6 + \chi_6 \hat\chi_2 + 10 \chi_4 \hat\chi_4 + 3 \chi_4 \hat\chi_6 + 3 \chi_6 \hat\chi_4) \nn\\
    &\qquad + y^{10} (\chi_4 \hat\chi_4 + 3 \chi_4 \hat\chi_6 + 3 \chi_6 \hat\chi_4 + 3 \chi_6 \hat\chi_6) + y^{12} (2 \chi_6 \hat\chi_6 + \chi_6 \hat\chi_8 + \chi_8 \hat\chi_6) + y^{14} \chi_8 \hat\chi_8 \bigr) \nn\\
    &+ x^8 \bigl(y^2 (2 + \chi_3 + \hat\chi_3) + y^4 (9 + 12 \chi_3 + 12 \hat\chi_3 + \chi_5 + \hat\chi_5 + 9 \chi_3 \hat\chi_3) \nn\\
    &\qquad + y^6 (9 + 19 \chi_3 + 19 \hat\chi_3 + 5 \chi_5 + 5 \hat\chi_5 + 33 \chi_3 \hat\chi_3 + 9 \chi_3 \hat\chi_5 + 9 \chi_5 \hat\chi_3) \nn\\
    &\qquad + y^8 (1 + 3\chi_3 + 3\hat\chi_3 + 3 \chi_5 + 3 \hat\chi_5 + 15 \chi_3 \hat\chi_3 + 12 \chi_3 \hat\chi_5 + 12 \chi_5 \hat\chi_3 + \chi_3 \hat\chi_7 + \chi_7 \hat\chi_3  + 9 \chi_5 \hat\chi_5 ) \nn\\
    &\qquad + y^{10} (3 \chi_3 \hat\chi_5 + 3 \chi_5 \hat\chi_3  + \chi_3 \hat\chi_7 + \chi_7 \hat\chi_3 + 10 \chi_5 \hat\chi_5 + 3 \chi_5 \hat\chi_7 + 3 \chi_7 \hat\chi_5 ) + y^{12} (\chi_5 \hat\chi_5 + 3 \chi_5 \hat\chi_7 + 3 \chi_7 \hat\chi_5 + 3 \chi_7 \hat\chi_7) \nn\\
    &\qquad + y^{14} (2 \chi_7 \hat\chi_7 + \chi_7 \hat\chi_9 + \chi_9 \hat\chi_7) + y^{16} \chi_9 \hat\chi_9 \bigr)  \nn\\
    &+ x^9 \bigl( 2 y^2 \chi_2 \hat\chi_2 + y^4 (24 \chi_2 \hat\chi_2 + 7 \chi_2 \hat\chi_4 + 7 \chi_4 \hat\chi_2) + y^6 (50 \chi_2 \hat\chi_2 + 33 \chi_2 \hat\chi_4 + 33 \chi_4 \hat\chi_2 + 3 \chi_2 \hat\chi_6 + 3 \chi_6 \hat\chi_2 + 19 \chi_4 \hat\chi_4 ) \nn\\
    &\qquad + y^8 (16 \chi_2 \hat\chi_2 + 25 \chi_2 \hat\chi_4 + 25 \chi_4 \hat\chi_2 + 6 \chi_2 \hat\chi_6 + 6 \chi_6 \hat\chi_2  + 36 \chi_4 \hat\chi_4 + 9 \chi_4 \hat\chi_6 + 9 \chi_6 \hat\chi_4) \nn\\
    &\qquad + y^{10} (\chi_2 \hat\chi_4 + \chi_4 \hat\chi_2 + 3 \chi_2 \hat\chi_6 + 3 \chi_6 \hat\chi_2 +  12 \chi_4 \hat\chi_4 + 12 \chi_4 \hat\chi_6 + 12 \chi_6 \hat\chi_4 + \chi_4 \hat\chi_8 + \chi_8 \hat\chi_4 + 9 \chi_6 \hat\chi_6) \nn\\
    &\qquad + y^{12} (3 \chi_4 \hat\chi_6 + 3 \chi_6 \hat\chi_4 + \chi_4 \hat\chi_8 + \chi_8 \hat\chi_4 + 10 \chi_6 \hat\chi_6 + 3 \chi_6 \hat\chi_8 + 3 \chi_8 \hat\chi_6) + y^{14} (\chi_6 \hat\chi_6 + 3 \chi_6 \hat\chi_8 + 3 \chi_8 \hat\chi_6 + 3 \chi_8 \hat\chi_8) \nn\\
    &\qquad + y^{16} (2 \chi_8 \hat\chi_8 + \chi_8 \hat\chi_{10} + \chi_{10} \hat\chi_8) + y^{18} \chi_{10} \hat\chi_{10} \bigr) +O(x^{10})   \;.
\end{align}
}
The $y_{1,2}$ dependence is encoded in the character $\chi_n \hat\chi_m$ of the 
$n \times m$-dimensional representation of $SU(2)\times SU(2)$.
We also present the BPS partition function for non-graviton cohomologies for $N=2,3,4$,
up to $x^{11},x^{10},x^9$, respectively,
\begin{align} \label{partition-fortuity}
  Z_2 - Z_{2,{\rm grav}} ~=~& y^6 x^8 + y^6x^9\chi_2\hat\chi_2 + y^6x^{10}(2+\chi_3+\hat\chi_3) + x^{11} (3y^6\chi_2\hat\chi_2 +y^8\chi_2\hat\chi_2) + O(x^{12}) \nn\\
  ~=~& \left[y^6 (x^8 + x^{10} + O(x^{12})) + y^8 (x^{11} \chi_2\hat\chi_2 + O(x^{12}))\right] \chi_{\rm desc} \;, \nn\\
  Z_3-Z_{3,{\rm grav}} ~=~& O(x^{11}) \;, \nn\\
  Z_4-Z_{4,{\rm grav}} ~=~& \chi_3\hat\chi_3 y^8 x^8 + (\chi_2\hat\chi_2+\chi_2\hat\chi_4+\chi_4\hat\chi_2+\chi_4\hat\chi_4)y^8 x^9 +O(x^{10}) \nn\\
  ~=~& \left[\chi_3\hat\chi_3 y^8 x^8 +O(x^{10})\right]\chi_{\rm desc} \;,
\end{align}
where
\begin{equation}\label{chi-desc}
\chi_{\rm desc} = \frac{\prod_{\pm} (1 + xy_1^{\pm 1})(1 + xy_2^{\pm 1})}{1-x^2}~,
\end{equation}
encodes contributions from all descendants given a (superconformal) primary.
Not all $OSp(6|4)$ multiplets share the same descendant structure
but there are exceptions when the primary has small quantum numbers,
see \cite{Cordova:2016emh} for details.
However, such non-generic multiplets do not appear in the partition functions
that we compute, so the descendant contribution can be simply factored out
by \eqref{chi-desc}.

We counted $N=\infty$ cohomologies in a similar manner.
$N = \infty$ is taken into account by treating each row-column contraction
(i.e. single-trace) as the elementary variables,
since the absence of $U(N)$ trace relations guarantee their independence.
The BPS partition function for non-graviton cohomologies at large $N$ is computed
till $x^{14}$, and presented following similar notation to \eqref{partition-fortuity}
\begin{eqnarray}\label{Z-infty-non-grav}
Z_\infty - Z_{\infty,{\rm grav}} \!\!&=&\!\!
\bigg[ y^8 \left( x^8 \chi_3 \hat\chi_3 + O(x^{15}) \right) \\
&&\!\! + y^{10} \big(
x^{10} \chi_3 \hat\chi_3
+ x^{11} \left( 3\chi_4 \hat\chi_4 + \chi_4 \hat\chi_2 + \chi_2 \hat\chi_4 \right)
+ x^{12} \left( \chi_5 \hat\chi_3 + \chi_3 \hat\chi_5
+ \chi_3 \hat\chi_3 + \chi_3 + \hat\chi_3 \right)
\nn\\
&&\!\! \qquad + x^{13} \left( 2\chi_4 \hat\chi_4 + 3\chi_4 \hat\chi_2
+ 3\chi_2 \hat\chi_4 + 3\chi_2 \hat\chi_2 \right)
\nn\\
&&\!\! \qquad + x^{14} \left( \chi_5 \hat\chi_3 + \chi_3 \hat\chi_5 + 5\chi_3 \hat\chi_3
+ 2\chi_3 + 2 \hat\chi_3 + 1 \right) + O(x^{15}) \big) \nn\\
&&\!\! + y^{12} \left( 2x^{13}\chi_4 \hat\chi_4 +
x^{14} \left( 6\chi_5 \hat\chi_5 + 3\chi_5 \hat\chi_3 + 3\chi_3 \hat\chi_5
+ 5\chi_3 \hat\chi_3 + \chi_5 + \hat\chi_5 \right) + O(x^{15}) \right)
\bigg]\chi_{\rm desc} ~. \nn
\end{eqnarray}
Contents of this partition function must represent multi-trace operators
that contain non-BPS single-trace operators in the higher spin current multiplets
rather than in the graviton multiplet,
but that are BPS due to the `$N^\prime=1$ fortuity.'
Throughout this paper, those are referred to as the `BPS bound states of higher spin particles',
or as its shortened versions.

As explained in the main body of this paper,
\eqref{non-grav-mono} with $r \geq 4$ provides a class of such operators.
For the simplest case $r=4$, the operator is explicitly written as
\begin{equation}\label{app:r=4NGM}
    O^{(4)}_{a_1a_2,i_1i_2} \equiv
    (q^j\wedge q_j\wedge \tilde\psi_{a_1}\wedge \tilde\psi_{a_2})\cdot
    (\tilde{q}^b\wedge \tilde{q}_b\wedge \psi_{i_1}\wedge \psi_{i_2})~.
\end{equation}
This operator vanishes by $U(N)$ trace relation if $N \leq 3$,
but for larger $N$ it is a nontrivial non-graviton cohomology.
Therefore, it also accounts for the first term in $Z_4-Z_{4,{\rm grav}}$
in \eqref{partition-fortuity}.
We have also checked that $r=5,6,7$ versions of (\ref{non-grav-mono}) are all
new non-graviton cohomologies.

Going further, we can account for the next term in \eqref{Z-infty-non-grav}
at the $x^{10}y^{10}$ order, by multiplying $v$-type gravitons on \eqref{app:r=4NGM}.
In general, an operator in the class \eqref{non-grav-mono} dressed (multiplied)
by graviton operators are clearly a new example of $Q$-closed operators.
However, it can represent a new non-graviton cohomology that appears in
\eqref{Z-infty-non-grav} only if it is not cohomologous to any graviton (nor to 0).
We consider the 3 candidates with correct $SU(2) \times SU(2)$ representation,
\begin{equation}
v \cdot O^{(4)}_{a_1a_2,i_1i_2}~, \qquad
{v_{(i_1|}}^{l} \cdot O^{(4)}_{a_1a_2,l |i_2)}~, \qquad
{v_{(a_1|}}^{b} \cdot O^{(4)}_{b |a_2),i_1i_2}~.
\end{equation}
Of these, the first candidate and the sum of the second and the third
turn out to be graviton cohomologies.
Any other combination of the second and the third candidates (e.g. just either one)
is a non-graviton cohomology that accounts for the second term in
(\ref{Z-infty-non-grav}) at the $x^{10}y^{10}$ order.
Similarly, we find that graviton dressings of \eqref{app:r=4NGM} generally yield
non-graviton cohomologies except for a few simplest cases.
Examples that we have explicitly confirmed to be non-graviton cohomologies are,
in non-decreasing order of the fugacity $x$,
\begin{itemize}
\item One combination of the ${v_{(i_1|}}^{l} \cdot O^{(4)}_{a_1a_2,l |i_2)}$
and ${v_{(a_1|}}^{b} \cdot O^{(4)}_{b |a_2),i_1i_2}$, as just explained.

\item All products of $O^{(4)}_{a_1a_2,l |i_2)}$ and $w_{ia}$.

\item All products of $O^{(4)}_{a_1a_2,l |i_2)}$ and $D u_{ia}$,
except the one with irrep $\chi_2 \hat\chi_2$ of $SU(2) \times SU(2)$.

\item All products of $O^{(4)}_{a_1a_2,l |i_2)}$ and $Dv_{ij}$, $Dv_{ab}$, $Dv$ or $x$.

\item All products of $O^{(4)}_{a_1a_2,l |i_2)}$ and $DDu_{ia}$ or $Dw_{ia}$.
\end{itemize}
Note that these do not fully explain the non-graviton partition function
(\ref{Z-infty-non-grav}), even within the order computed.
For example, the second and the third bullet points would contribute
$2 \chi_4 \hat\chi_4 + 2 \chi_4 \hat\chi_2 + 2 \chi_2 \hat\chi_4 + \chi_2 \hat\chi_2$
to the $y^{10}x^{11}$ order in \eqref{Z-infty-non-grav},
already exceeding what is written there.
We expect that some of such graviton dressings are descendants of the earlier primary,
namely the one explained in the first bullet point,
so their contribution is absorbed in the $\chi_{\rm desc}$ factor.
Based on these observations, we expect that the spectrum of the BPS bound states
are much richer than just those that belong to the class \eqref{non-grav-mono}.

\subsection{Constructing an $N=2$ fortuitous cohomology}\label{sec:appA.2}

In this subsection we construct a new `heavy' cohomology for $N=2$ with
$E+J=8$, $Y=6$ and $F_1=F_2=0$ (or $R_2=R_3=0$ in the notation of \cite{Cordova:2016emh}),
which belongs to the $A_1[4]_5^{(2,0,0)}$ multiplet. 
While counting cohomologies, we have already enumerated the basis operators,
$Q$-closed operators, $Q$-exact operators, and graviton cohomologies in the given charge sector. 
The counting shows that there are $8$ cohomologies, of which $7$ are gravitons. 
In the following, we construct all $8$ cohomologies and identify one that is not cohomologous to gravitons.

Operators with $E+J=8$, $Y=6$ and $R_2=R_3=0$ take the form
(in terms of the number of constituent letters) of either
$q\psi^3\psit^2$, $\q\psi^2\psit^3$, $Dq^3\psi^3, D\q^3\psit^3$, $Dq^2\q\psi^2\psit$, $Dq\q^2\psi\psit^2$, $D^2q^3\q^2\psi$, or $D^2q^2\q^3\psit$. The overcomplete list of these operators is
{\allowdisplaybreaks
\begin{align}
    q\psi^3\psit^2 :\ &(\psi_i\cdot q^i)(\psi_j\cdot\psit_b)(\psi^j\cdot\psit^b) \;, \nn\\
    Dq^3\psi^3 :\ & (\psi_i\cdot q^i)(D\psi_j\cdot q_k)(\psi^j\cdot q^k) \;, (\psi_i\cdot q^i)(\psi_j\cdot Dq_k)(\psi^j\cdot q^k) \;, \nn\\
    &(D\psi_i\cdot q^j)(\psi_j\cdot q^k)(\psi_k\cdot q^i) \;, (\psi_i\cdot Dq^j)(\psi_j\cdot q^k)(\psi_k\cdot q^i) \;, \nn\\
    Dq^2\q\psi^2\psit :\ & (D\q^a\cdot q^i)(\psi_j\cdot q^j)(\psi_i\cdot\psit_a) \;, (\q^a\cdot Dq^i)(\psi_j\cdot q^j)(\psi_i\cdot\psit_a) \;, (\q^a\cdot q^i)(D\psi_j\cdot q^j)(\psi_i\cdot\psit_a) \;, \nn\\
    &(\q^a\cdot q^i)(\psi_j\cdot Dq^j)(\psi_i\cdot\psit_a) \;, (\q^a\cdot q^i)(\psi_j\cdot q^j)(D\psi_i\cdot\psit_a) \;, (\q^a\cdot q^i)(\psi_j\cdot q^j)(\psi_i\cdot D\psit_a) \;, \nn\\
    &(D\q^a\cdot q^i)(\psi_i\cdot q^j)(\psi_j\cdot \psit_a) \;, (\q^a\cdot Dq^i)(\psi_i\cdot q^j)(\psi_j\cdot \psit_a) \;, (\q^a\cdot q^i)(D\psi_i\cdot q^j)(\psi_j\cdot \psit_a) \;, \nn\\
    &(\q^a\cdot q^i)(\psi_i\cdot Dq^j)(\psi_j\cdot \psit_a) \;, (\q^a\cdot q^i)(\psi_i\cdot q^j)(D\psi_j\cdot \psit_a) \;, (\q^a\cdot q^i)(\psi_i\cdot q^j)(\psi_j\cdot D\psit_a) \;, \nn\\
    &(\q^a\cdot \psit_a)(Dq^i\cdot\psi_i)(q^j\cdot\psi_j) \;, (\q^a\cdot \psit_a)(q^i\cdot D\psi_i)(q^j\cdot\psi_j) \;,\nn\\
    &(\q^a\cdot \psit_a)(Dq^i\cdot\psi_j)(q^j\cdot\psi_i) \;, (\q^a\cdot \psit_a)(q^i\cdot D\psi_j)(q^j\cdot\psi_i) \;, \nn\\
    D^2q^3\q^2\psi :\ & (D^2\psi_i\cdot q^j)(\q^a\cdot q^i)(\q_a\cdot q_j) \;,\ (\psi_i\cdot D^2q^j)(\q^a\cdot q^i)(\q_a\cdot q_j) \;,\ (\psi_i\cdot q^j)(D^2\q^a\cdot q^i)(\q_a\cdot q_j) \;,\nn\\
    &(\psi_i\cdot q^j)(\q^a\cdot D^2q^i)(\q_a\cdot q_j) \;,\ (\psi_i\cdot q^j)(\q^a\cdot q^i)(D^2\q_a\cdot q_j) \;,\ (\psi_i\cdot q^j)(\q^a\cdot q^i)(\q_a\cdot D^2q_j) \;,\nn\\
    &(D\psi_i\cdot Dq^j)(\q^a\cdot q^i)(\q_a\cdot q_j) \;,\ (D\psi_i\cdot q^j)(D\q^a\cdot q^i)(\q_a\cdot q_j) \;,\ (D\psi_i\cdot q^j)(\q^a\cdot Dq^i)(\q_a\cdot q_j) \;,\nn\\
    &(D\psi_i\cdot q^j)(\q^a\cdot q^i)(D\q_a\cdot q_j) \;,\ (D\psi_i\cdot q^j)(\q^a\cdot q^i)(\q_a\cdot Dq_j) \;,\ (\psi_i\cdot Dq^j)(D\q^a\cdot q^i)(\q_a\cdot q_j) \;,\nn\\
    &(\psi_i\cdot Dq^j)(\q^a\cdot Dq^i)(\q_a\cdot q_j) \;,\ (\psi_i\cdot Dq^j)(\q^a\cdot q^i)(D\q_a\cdot q_j) \;,\ (\psi_i\cdot Dq^j)(\q^a\cdot q^i)(\q_a\cdot Dq_j) \;,\nn\\
    &(\psi_i\cdot q^j)(D\q^a\cdot Dq^i)(\q_a\cdot q_j) \;,\ (\psi_i\cdot q^j)(D\q^a\cdot q^i)(D\q_a\cdot q_j) \;,\ (\psi_i\cdot q^j)(D\q^a\cdot q^i)(\q_a\cdot Dq_j) \;,\nn\\
    &(\psi_i\cdot q^j)(\q^a\cdot Dq^i)(D\q_a\cdot q_j) \;,\ (\psi_i\cdot q^j)(\q^a\cdot Dq^i)(\q_a\cdot Dq_j) \;,\ (\psi_i\cdot q^j)(\q^a\cdot q^i)(D\q_a\cdot Dq_j) \;,\nn\\
    &(D^2\psi_i\cdot q^i)(\q^a\cdot q^j)(\q_a\cdot q_j) \;,\ (\psi_i\cdot D^2q^i)(\q^a\cdot q^j)(\q_a\cdot q_j) \;,\ (\psi_i\cdot q^i)(D^2\q^a\cdot q^j)(\q_a\cdot q_j) \;,\nn\\
    &(\psi_i\cdot q^i)(\q^a\cdot D^2q^j)(\q_a\cdot q_j) \;,\ (D\psi_i\cdot Dq^i)(\q^a\cdot q^j)(\q_a\cdot q_j) \;,\ (D\psi_i\cdot q^i)(D\q^a\cdot q^j)(\q_a\cdot q_j) \;,\nn\\
    &(D\psi_i\cdot q^i)(\q^a\cdot Dq^j)(\q_a\cdot q_j) \;,\ (\psi_i\cdot Dq^i)(D\q^a\cdot q^j)(\q_a\cdot q_j) \;,\ (\psi_i\cdot Dq^i)(\q^a\cdot Dq^j)(\q_a\cdot q_j) \;,\nn\\
    &(\psi_i\cdot q^i)(D\q^a\cdot Dq^j)(\q_a\cdot q_j) \;,\ (\psi_i\cdot q^i)(D\q^a\cdot q^j)(D\q_a\cdot q_j) \;,\ (\psi_i\cdot q^i)(D\q^a\cdot q^j)(\q_a\cdot Dq_j) \;,\nn\\
    &(\psi_i\cdot q^i)(\q^a\cdot Dq^j)(\q_a\cdot Dq_j) \;. 
\end{align}
}
and those which can be obtained from the above by exchanging $q \leftrightarrow \tilde{q}$ and $\psi \leftrightarrow \tilde{\psi}$. 
Among these operators, $72$ are independent, and after the action of $Q$,
$31$ remain independent, meaning that there are $41$ $Q$-closed operators.

The $Q$-exact operators in the charge sector of our interest 
are constructed by acting $Q$ on operators in the `previous' charge sector,
with $E+J=8$, $Y=4$ and $R_2=R_3=0$.
The operators belong to either one of $D\psi^2\psit^2$, $D^2q^2\psi^2$, $D^2\q^2\psit^2$, $D^2q\q\psi\psit$, $D^3q^2\q^2$. 
Explicitly,
{\allowdisplaybreaks
\begin{align}
    D\psi^2\psit^2 :& \
    (D\psi_i\cdot\psit_a)(\psi^i\cdot\psit^a) \;, (\psi_i\cdot D\psit_a)(\psi^i\cdot\psit^a) \;, \nn\\
    D^2q^2\psi^2 :& \
    (D^2\psi_j\cdot q^i)(\psi_i\cdot q^j) \;, (\psi_j\cdot D^2q^i)(\psi_i\cdot q^j) \;, (D\psi_j\cdot Dq^i)(\psi_i\cdot q^j) \;, (D\psi_j\cdot q^i)(\psi_i\cdot Dq^j) \;, \nn\\
    &\ (D^2\psi_i\cdot q^i)(\psi_j\cdot q^j) \;, (\psi_i\cdot D^2q^i)(\psi_j\cdot q^j) \;, (D\psi_i\cdot Dq^i)(\psi_j\cdot q^j) \;, (D\psi_i\cdot q^i)(\psi_j\cdot Dq^j) \;, \nn\\
    D^2\q^2\psit^2 : & \
    (D^2\q^a\cdot\psit_b)(\q^b\cdot\psit_a) \;, (\q^a\cdot D^2\psit_b)(\q^b\cdot\psit_a) \;, (D\q^a\cdot D\psit_b)(\q^b\cdot\psit_a) \;, (D\q^a\cdot \psit_b)(\q^b\cdot D\psit_a) \;, \nn\\
    &\ (D^2\q^a\cdot\psit_a)(\q^b\cdot\psit_b) \;, (\q^a\cdot D^2\psit_a)(\q^b\cdot\psit_b) \;, (D\q^a\cdot D\psit_a)(\q^b\cdot\psit_b) \;, (D\q^a\cdot\psit_a)(\q^b\cdot D\psit_b) \;, \nn\\
    D^2q\q\psi\psit :& \
    (D^2\q^a\cdot q^i)(\psi_i\cdot\psit_a) \;, (\q^a\cdot D^2q^i)(\psi_i\cdot\psit_a) \;, (\q^a\cdot q^i)(D^2\psi_i\cdot\psit_a) \;, (\q^a\cdot q^i)(\psi_i\cdot D^2\psit_a) \;, \nn\\
    &\ (D\q^a\cdot Dq^i)(\psi_i\cdot\psit_a) \;, (D\q^a\cdot q^i)(D\psi_i\cdot\psit_a) \;, (D\q^a\cdot q^i)(\psi_i\cdot D\psit_a) \;, (\q^a\cdot Dq^i)(D\psi_i\cdot\psit_a) \;, \nn\\
    &\ (\q^a\cdot Dq^i)(\psi_i\cdot D\psit_a) \;, (\q^a\cdot q^i)(D\psi_i\cdot D\psit_a) \;, \nn\\
    &\ (D^2\psi_i\cdot q^i)(\q^a\cdot\psit_a) \;, (\psi_i\cdot D^2q^i)(\q^a\cdot\psit_a) \;, (\psi_i\cdot q^i)(D^2\q^a\cdot\psit_a) \;, (\psi_i\cdot q^i)(\q^a\cdot D^2\psit_a) \;,\nn\\
    &\ (D\psi_i\cdot Dq^i)(\q^a\cdot\psit_a) \;, (D\psi_i\cdot q^i)(D\q^a\cdot\psit_a) \;, (D\psi_i\cdot q^i)(\q^a\cdot D\psit_a) \;, (\psi_i\cdot Dq^i)(D\q^a\cdot\psit_a) \;,\nn\\
    &\ (\psi_i\cdot Dq^i)(\q^a\cdot D\psit_a) \;, (\psi_i\cdot q^i)(D\q^a\cdot D\psit_a) \;, \nn\\
    D^3q^2\q^2 :&\ 
    (D^3\q^a\cdot q^i)(\q_a\cdot q_i) \;, (\q^a\cdot D^3q^i)(\q_a\cdot q_i) \;, (D^2\q^a\cdot Dq^i)(\q_a\cdot q_i) \;, (D\q^a\cdot D^2q^i)(\q_a\cdot q_i) \;, \nn\\
    &\ (D^2\q^a\cdot q^i)(D\q_a\cdot q_i) \;, (D^2\q^a\cdot q^i)(\q_a\cdot Dq_i) \;, (\q^a\cdot D^2q^i)(D\q_a\cdot q_i) \;, (\q^a\cdot D^2q^i)(\q_a\cdot Dq_i) \;, \nn\\
    &\ (D\q^a\cdot Dq^i)(D\q_a\cdot q_i) \;, (D\q^a\cdot Dq^i)(\q_a\cdot Dq_i) \;.
\end{align}
}
After acting $Q$, we confirmed that there are $33$ independent operators.
They form the basis of $Q$-exact operators in the original charge sector with $Y=6$.
Thus, we have $8$ cohomologies in the latter sector as mentioned above.

One can also construct gravitons in the given charge sector
\begin{align}
    &Du_{ia} u^{ia} x \;,\ u_{ia}u^{ia}Dx \;,\ D^2u_{ia}u^{ia}v \;,\ Du_{ia}Du^{ia}v \;,\ Du_{ia}u^{ia}Dv \;,\ u_{ia}u^{ia}D^2v \;, \nn\\
    &D^2{u^i}_a u^{ja} v_{ij} \;,\ D{u^i}_a u^{ja} Dv_{ij} \;,\ D^2{u_i}^a u^{ib} v_{ab} \;,\ D{u_i}^a u^{ib} Dv_{ab} \;,\ u_{ia}w^{ia}x \;,\ \nn\\
    & Du_{ia}w^{ia}v \;,\ u_{ia}Dw^{ia}v \;,\ u_{ia}w^{ia}Dv \;,\ w_{ia}w^{ia}v \;,\ \nn\\
    &D{u^i}_a w^{ja} v_{ij} \;,\ {u^i}_a Dw^{ja} v_{ij} \;,\ {u^i}_a w^{ja} Dv_{ij} \;,\ D{u_i}^a w^{ib} v_{ab} \;,\  {u_i}^a Dw^{ib} v_{ab} \;,\  {u_i}^a w^{ib} Dv_{ab} \;,\ \nn\\
    &Dv_{ij} v^{ij}v\;,\ Dv_{ab}v^{ab}v \;,\ Dv_{ij} {v^i}_k v^{jk} \;,\ Dv_{ab}{v^a}_cv^{bc} \;.
\end{align}
With the basis of $Q$-exact operators obtained above,
we verified that there are $7$ graviton cohomologies. 
Thus, we have $1$ non-graviton cohomology whose representative is given as
\begin{align}
    O =&\  (\psi_i\cdot q^i)(\psi_j\cdot\psit_a)(\psi^j\cdot\psit^a) 
    + 2(\psi_i\cdot q^i)(\psi_j\cdot Dq_k)(\psi^j\cdot q^k) \nn\\
    &+2 (D\psi_i\cdot q^j)(\psi_j\cdot q^k)(\psi_k\cdot q^i)
    -2 (\q^a\cdot Dq^i)(\psi_i\cdot q^j)(\psi_j\cdot \psit_a) \;.
\end{align}
The representative of a non-graviton cohomology is not unique,
in a sense that any $Q$-exact and/or graviton operator may be added.
An alternative representative which may be useful, is
\begin{align}
    O'=&\ (\q^a\cdot\psit_a)(\psi_i\cdot\psit_b)(\psi^i\cdot\psit^b) 
    +2(\q^a\cdot\psit_a)(D\q_b\cdot\psit_c)(\q^b\cdot\psit^c) \nn\\
    &-2(D\q^c\cdot\psit_a)(\q^a\cdot\psit_b)(\q^b\cdot\psit_c) 
    +2(D\q^a\cdot q^i)(\q^b\cdot\psit_a)(\psi_i\cdot\psit_b)\;.
\end{align}
We showed that this cohomology is $N=2$ fortuitous. Numerically, we checked that 
it is not $Q$-closed for $N=3,4$. Analytically, this follows by carefully rearranging $QO$,
\begin{align}
    QO = & \
    \frac12 w^{ia}(q^k\wedge q_k \wedge \psit_a)(\q^c\wedge q_c \wedge \psi_i) 
    - v^{ij}(q^k\wedge q_k \wedge\psit_a)(\q^a\wedge\psi_i\wedge\psi_j) \nn\\
    &-v^{ij}(q^k\wedge q_k\wedge Dq_i)(\q^a\wedge\q_a\wedge \psi_j) \;, 
\end{align}
which vanishes for $N=2$ but is nontrivial for larger $N$.

So far, we constructed the first fortuitous cohomology $O$ for $N=2$,
that accounts for the leading term in the first line of (\ref{partition-fortuity}). 
Its superconformal descendants, obtained by acting $Q_{ia}$ and $D$ on it, are obviously fortuitous.
One can further ask whether multiplying gravitons to the fortuitous cohomology $O$ is fortuitous or not.
We show that $u_{ia}O$ and $v_{ij}O, v_{ab}O, vO$ are cohomologous to gravitons. 
If we choose a `better' representative of $O$, those product operators become $Q$-exact, meaning 
that a carefully chosen representative $O$ does not admit those graviton `hairs'.
On the other hand, $O$ admits $w_{ia}, x$ graviton hair, i.e. $w_{ia}O$ and $xO$ are 
fortuitous cohomologies.
$w_{ia}O$ accounts for the primary factor $x^{11}y^8\chi_2\hat\chi_2$
on the second line of (\ref{partition-fortuity}).
Although we have not performed comprehensive counting of all charge sectors in $x^{12}$,
which is why $Z_2 - Z_{2,{\rm grav}}$ in (\ref{partition-fortuity}) was truncated at this order,
we have analyzed the specific charge sector that contains $xO$
to confirm that $xO$ is a fortuitous cohomology.

\section{Matrix model calculations}\label{sec:appB}

In this appendix, we present solutions of the unitary matrix model relevant for Section \ref{sec:3}.
We only briefly review the standard procedure for solving the general class of models
to arrive at the answer quickly.
More details can be found for example in \cite{Jurkiewicz:1982iz,Aharony:2003sx},
and we mainly refer to Appendix A of \cite{Choi:2021lbk} for maximal coherence.
We then apply the one-cut and two-cut solutions of the general model
to our specific model to obtain expressions for respective solutions.

The matrix model of our interest is described by the following unitary matrix integral
with input parameters $g_n$:
\begin{eqnarray}
Z &=& \int [dU] \exp \left[ \sum_{n=1}^\infty \frac{g_n}{n}
(\tr U^n + \tr {U^\dagger}^n) \right]~.
\end{eqnarray}
In the present work, we are interested in a specific model described by
\begin{equation}\label{app:gn}
g_n = \frac{i^n - (-i)^n}{n\beta} =
\begin{cases} \frac{2 i^n}{n\beta}~, & (n: \text{ odd}) \\ 0~. & (n: \text{ even}) \end{cases}
\end{equation}
We will restrict to such a model later, but for now we leave $g_n$'s as general parameters.
The matrix integral can be interpreted as an integral over $N$ eigenvalues
with the Haar measure. Eigenvalues of the unitary matrices lie on the unit circle,
parametrized by $e^{i \theta}$ with $\theta \in [0, 2\pi)$.
In the large-$N$ limit, the eigenvalue configuration is well approximated by a
continuous distribution of $\theta$ throughout the (periodic) interval $[0, 2\pi)$,
described by the density function $\rho(\theta)$
that is normalized as $\int_0^{2\pi} \rho(\theta) d\theta = 1$,
so that the displacement between two adjacent $\theta$'s is $\frac{1}{N \rho(\theta)}$.
Then the matrix model is a path integral whose effective action is a functional of $\rho$,
\begin{eqnarray}\label{app:ZandS}
Z \!\!&=&\!\! \int [d \rho] e^{-S[\rho(\theta)]}~, \\
- \frac{S[\rho(\theta)]}{N^2} \!\!&=&\!\!
\iint_0^{2\pi} d\theta_1 d\theta_2
\log \left[ 1-e^{i (\theta_1 - \theta_2)}\right] \rho(\theta_1) \rho(\theta_2)
+ \frac{1}{N} \int_0^{2\pi} d\theta
\left( \sum_{n=1}^\infty \frac{g_n}{n} (e^{in\theta} + e^{-in\theta} ) \right) \rho(\theta)~. \nn
\end{eqnarray}

When the input parameters $g_n$'s are all real, such as in the case of
\cite{Gross:1980he,Wadia:1980cp,Jurkiewicz:1982iz},
the path integral is evaluated using the saddle point approximation,
i.e. to find the saddle eigenvalue distribution $\rho(\theta)$ that minimizes the effective action.
However, as we will be interested in the model \eqref{app:gn},
we shall more generally study the matrix model with complex coefficients.
Then we must allow contour deformations of each eigenvalue integral
and find a complex saddle where the effective action is extremized as a complex function.
In the complex saddle, the eigenvalues may be scattered around the complex plane.
We nevertheless assume that they are distributed only along a one-real-dimensional curve,
or a set of disjoint such curves $\calC = \calC_1 \cup \calC_2 \cup \cdots$ in the complex plane.
This assumption allows the standard solution \cite{Jurkiewicz:1982iz,Aharony:2003sx}
for real matrix models to be readily generalized.
The density function $\rho(\theta)$ is defined on the cut ($\theta \in \calC$) by the condition
that $N \cdot \int \rho(\theta) d\theta$ along any segment of the cut gives
the number of eigenvalues on that segment.
It follows from this definition that $\int_{\calC} \rho(\theta) d\theta = 1$ along the entire cut.
We expect $\rho(\theta)$ to be a holomorphic function of $\theta$,
although its value outside $\calC$ is irrelevant for the eigenvalue distribution.

We now evaluate the matrix integral \eqref{app:ZandS} by finding the complex saddle,
i.e. the eigenvalue distribution that extremizes the complex function $S[\rho(\theta)]$.
It is useful to change basis via $z = e^{i\theta}$.
The density function is easily translated according to the principle that
$\rho(\theta)d\theta$ and $\rho(z)dz$ represent the same coordinate-independent quantity,
namely the number of eigenvalues.
\begin{equation}
\rho(\theta) d\theta = \rho(z) dz
\qquad \leftrightarrow \qquad
\rho(\theta) = i z \rho(z)~.
\end{equation}
In this basis, the effective action is rewritten as
\begin{eqnarray}
- \frac{S[\rho(z)]}{N^2} \!\!&=&\!\!
\frac12 \iint_\calC dz_1 dz_2
\log \left[ -\frac{(z_1-z_2)^2}{z_1z_2}\right] \rho(z_1) \rho(z_2)
+ \frac{1}{N} \sum_{n=1}^\infty \frac{g_n}{n} \cdot
\int_\calC dz (z^n+z^{-n}) \rho(z)~. \nn
\end{eqnarray}
For $\rho(z)$ to extremize the action, the action must not change under infinitesimal
displacement of each eigenvalue. This condition is equivalent to the chemical potential
\begin{equation}
\mu(z) \equiv \frac{\delta}{\delta \rho(z)} S[\rho(z)]~,
\end{equation}
being constant along a continuous cut.\footnote{
Whether it must also be equal between disjoint cuts, calls for a separate discussion
because it corresponds to extremizing the action under changing the filling fraction of each cut,
which is not a continuous deformation. This will be discussed later in subsection \ref{sec:app2cut}.}
This leads to what is often referred to as the force-free equation:
\begin{eqnarray}\label{app:FFE}
\int_\calC dz' \rho(z') {\cal P} \frac{2}{z-z'} - \frac{1}{z}
+ \frac{1}{N} \sum_n g_n \cdot \frac{z^n-z^{-n}}{z} = 0~, ~~~~~ (\forall z \in \calC)
\end{eqnarray}
where ${\cal P}$ indicates the principal value.

The standard treatment of this equation is to define an auxiliary function
\begin{equation}\label{app:defy}
y(z) \equiv - \int_\calC dz' \rho(z') \frac{2}{z-z'} + \frac{1}{z}
- \frac{1}{N} \sum_n g_n \cdot \frac{z^n-z^{-n}}{z}~.
\end{equation}
$y(z)$ is well-defined for any $z \notin \calC$,
but it has branch cuts along $\calC$ such that
\begin{equation}\label{app:rhofromy}
y(z + i\epsilon) - y(z - i\epsilon) = 4 \pi i \rho(z) ~, ~~~~~ (\forall z \in \calC)
\end{equation}
and \eqref{app:FFE} manifests that ${\cal P} y(z) = 0$ for $z \in \calC$.
So it is crucial to locate the branch points/cuts of $y(z)$ and evaluate the function in vicinity,
in order to obtain the saddle $\rho(z)$.
One can show that $y(z)$ satisfies (for details see Appendix A of \cite{Choi:2021lbk})
\begin{eqnarray}\label{app:eqnfory}
z^2 y^2(z) &=& \left( 1- \frac{1}{N} \sum_n g_n (z^n-z^{-n}) \right)^2 + 4z\rho_1 \\
&& + \frac{4}{N} \sum_n g_n \left( z^n \rho_0 + z^{n-1} \rho_1 + \cdots + z^2 \rho_{n-2}
+ z \rho_{n+1} + \rho_{n} + \cdots + z^{-n+1} \rho_1 \right)~, \nn
\end{eqnarray}
where we used the Fourier modes of $\rho(z)$
\begin{equation}
\rho_n = \rho_{-n} = \int_\calC dz \rho(z) z^n~.
\end{equation}
The first equality holds because we assume the symmetry
$\rho(\theta) = \rho(-\theta)~\leftrightarrow~ z\rho(z) = \frac{1}{z}\rho\left(\frac{1}{z}\right)$
of the saddle based on that of the model.
The square root branch cut that arises from \eqref{app:eqnfory} should coincide with $\calC$.
In particular, the branch points where the RHS vanishes,
define the endpoints of (each disjoint piece of) the cut.
Starting from each endpoint, one can repetitively add the complex number $\frac{1}{N \rho(z)}$
to locate subsequent eigenvalues until it reaches another endpoint.
This will determine the precise shape of the complex eigenvalue cut.

For a more concrete argument, we momentarily suppose that there are only a finite number $p$
of non-zero $g_n$, i.e. $g_{p+1} = g_{p+2} = \cdots = 0$,
so the sums over $n$ in \eqref{app:eqnfory} run from $n=1$ to $p$ only.
We will later take $p \to \infty$.
Then the powers of $z$ on the RHS of \eqref{app:eqnfory} range from $z^{-2p}$ to $z^{2p}$,
so it is a polynomial (times an overall $z^{-2p}$) in $z$ of degree $4p$.

For an $m$-cut saddle where $\calC$ consists of $m$ disjoint pieces of continuous curves,
the RHS of \eqref{app:eqnfory} must have $2m$ single roots where the cuts start or end.
Then for the remaining $4p-2m$ roots to not cause $y(z)$ to have additional branch cuts,
they must be double roots (or roots with even multiplicity)
so that cuts appear and vanish immediately.
Moreover, one can derive from the definition \eqref{app:defy} that $zy(z)$ is odd under
$z \to \frac{1}{z}$,
\begin{equation}\label{app:ysymmetry}
\frac{1}{z} y\left(\frac{1}{z}\right) = -zy(z)~,
\end{equation}
and thus $(zy(z))^2$ is even.
This property is naturally connected to the aforementioned symmetry
$\frac{1}{z} \rho\left(\frac{1}{z}\right) = z\rho(z)$ via \eqref{app:rhofromy},
the extra minus sign in \eqref{app:ysymmetry} arising from the fact that
$z \leftrightarrow \frac{1}{z}$ exchanges the ``above'' and the ``below'' of $z \in \calC$.
Therefore, roots of the RHS of \eqref{app:eqnfory} always come in pairs of $(z, \frac{1}{z})$.
Such a pair of single roots naturally mark the two endpoints of each cut
as the shape of each cut must be symmetric under $z \leftrightarrow \frac{1}{z}$.
On the other hand, the double roots must also come in pairs.
When $m$ is odd, this is only possible if one (or an odd number) of the double roots
is either of $z = \pm 1$, the fixed point of the exchange $z \leftrightarrow \frac{1}{z}$.

According to the arguments given so far, we may now require that
\begin{eqnarray}\label{app:y1cut}
\eqref{app:eqnfory} &\propto&
\frac{(z-a_1)(z-a_1^{-1})}{z}
\cdot \left( \prod_{i=1}^{p-1} \frac{(z-d_i)^2(z-d_i^{-1})^2}{z^2} \right)
\cdot \frac{(z+1)^2}{z}~,
\end{eqnarray}
for a one-cut saddle.
$a_1^{\pm 1}$ indicate the symmetric endpoints of the only cut,
and $p-1$ $d_i$'s parametrize the pairs of double roots.
Note that we have chosen the fixed point $-1$ for the unpaired double root,
because we will let the only cut to pass through $\theta=0 \leftrightarrow z=1$.\footnote{
One could otherwise choose that the cut to pass through $\theta=\pi \leftrightarrow z=-1$ instead.
We do not treat them separately since they yield the
complex conjugate saddle, see discussion around \eqref{pm-relation}.}
For a two-cut saddle, we can similarly write
\begin{eqnarray}\label{app:y2cut}
\eqref{app:eqnfory} &\propto&
\frac{(z-a_1)(z-a_1^{-1})(z-a_2)(z-a_2^{-1})}{z^2}\cdot
\left( \prod_{i=1}^{p-1} \frac{(z-d_i)^2(z-d_i^{-1})^2}{z^2} \right)~,
\end{eqnarray}
where $z=a_1^{\pm 1}$ and $z=a_2^{\pm 1}$ are the endpoints of each cut,
and the unpaired double root is not needed for the two-cut saddle.

\eqref{app:y1cut} or \eqref{app:y2cut} on its own is sufficient to determine all coefficients
$a_{1,2}$ and $d_i$, given the input $g_n$ of the model.
They are in fact overconstraining for $a_{1,2}$ and $d_i$,
so $\rho_n$'s are also determined by these equations.
For example, imposing the $z \to z^{-1}$ symmetry that is required by symmetry and
is apparent from \eqref{app:y1cut} or \eqref{app:y2cut}, on the right hand side of \eqref{app:eqnfory},
and equating the $z^p$ and the $z^{-p}$ coefficients (recall that the sum over $n$ runs up to $p$),
it gives $\rho_0 = 1$, the overall normalization condition.
Equivalently, $\rho$ can be determined via \eqref{app:rhofromy} given $y(z)$.
The endpoints of the cut(s) and $\rho$ together fully determine the saddle eigenvalue distribution.
We present the specific expressions, of one- and two-cut saddles separately,
in subsequent subsections.

\subsection{One-cut saddles}\label{sec:app1cut}

We study one-cut saddles with $\calC = (-\theta_1, \theta_1)$ that pass through $\theta = 0$.
\eqref{app:y1cut} can be written equivalently as
\begin{eqnarray}\label{app:y1cut2}
\eqref{app:eqnfory} &=& \frac{(z-a_1)(z-a_1^{-1})}{z} \cdot
\left( \sum_{n=1}^p Q_n \cdot \frac{z^{n-\frac12} + z^{-n+\frac12}}{2} \right)^2~,
\end{eqnarray}
where $Q_n$'s simply replace $d_n$'s as unknown coefficients.
To be more precise, $Q_1, \cdots, Q_{p-1}$ replace the same number of $d_n$'s
and then $Q_p$ is introduced to eliminate the proportionality sign in favor of an equality.
We simultaneously expand the RHS of \eqref{app:eqnfory} and the RHS of \eqref{app:y1cut2}
around $z=0$, the leading order being $z^{-2p}$, and compare coefficients to determine $Q_n$.
For the first $p+1$ order, that is until $z^{-p}$, the second line of \eqref{app:eqnfory}
does not enter, so all $Q_n$ can be written purely in terms of $g_n$ without $\rho_n$.
Taking $p \to \infty$ after this step, we have
\begin{equation}\label{app:1cutQgen}
Q_n = \sum_{k=0}^\infty \frac{2g_{n+k}}{N} P_k (c_1)~,
\end{equation}
where $P_k$ are Legendre polynomials and
\begin{equation}
a_1 = e^{i \theta_1}~, \qquad c_1 = \cos \theta_1
\qquad \Rightarrow \qquad c_1 = \frac{a_1+a_1^{-1}}{2}~,
\end{equation}
all relate to the endpoints of the cut.
After determining $p$ coefficients $Q_n$'s,
one equation from the first $p+1$ order of \eqref{app:y1cut2} still remains.
This puts a constraint that is ultimately equivalent to the normalization condition
$\rho_0 = 1$. Again taking the $p \to \infty$ limit, the constraint is
\begin{equation}\label{app:1cutnormgen}
\sum_{k=0}^\infty \frac{2g_k}{N} (P_{k-1} (c_1) - P_k (c_1)) = 2~.
\end{equation}
This can be understood to determine $c_1$.
\eqref{app:1cutQgen} and \eqref{app:1cutnormgen} are results well known from \cite{Aharony:2003sx}.

We now apply the general solution to our specific model with input parameters \eqref{app:gn}.
First, we examine \eqref{app:1cutnormgen} that determines the endpoint $c_1$
via normalization of $\rho (\alpha)$.
It becomes
\begin{equation}\label{app:1cutnorm1}
N\beta = \sum_{n=1}^\infty \frac{i^n - (-i)^n}{n} (P_{n-1}(c_1) - P_n(c_1))~,
\end{equation}
One needs to be cautious with the infinite sum on the right hand side, however.
A careful analytic continuation must be performed to avoid branch cuts.
We first write \eqref{app:1cutnorm1} as an integral over an auxiliary variable $t$:
\begin{equation}\label{app:1cutnorm1.5}
N\beta = \left[ \sum_{n=1}^\infty \frac{t^n}{n} (P_{n-1}(c_1) - P_n(c_1)) \right]_{t=-i}^{t=i}
= \int_\mathcal{T} \left( \sum_{n=1}^\infty t^{n-1} (P_{n-1}(c_1) - P_n(c_1)) \right) dt~,
\end{equation}
where $\mathcal{T}$ is a contour that starts at $-i$ and ends at $i$.
Using the generating function of Legendre polynomials
\begin{equation}\label{app:Legendregenfn}
\sum_{n=0}^\infty P_n(x) t^n = \frac{1}{\sqrt{1-2xt+t^2}}~,
\end{equation}
we can rewrite and even naively evaluate the integral,
\begin{eqnarray}\label{app:1cutnorm2}
N\beta &=& \int_\mathcal{T} \left( \frac{1-\frac{1}{t}}{\sqrt{1-2c_1t+t^2}} + \frac{1}{t} \right) dt \nn\\
&=& \left[ \log \frac{t(1-t+\sqrt{1-2c_1t+t^2})}{(1+t-\sqrt{1-2c_1t+t^2})(c_1-t+\sqrt{1-2c_1t+t^2})}
\right]_{t=-i}^{t=i} + 2\pi i n~.
\end{eqnarray}
However, there are two ambiguities in the last expression.
First, the $\log$ may always be added by any multiples of $2\pi i$ as we explicitly wrote with $n$.
Second, there can be sign choices for the square roots $\sqrt{1-2c_1t+t^2}$
because in general the expression inside the square root is complex.

Both ambiguities can be and should be fixed by a careful choice of the contour $\mathcal{T}$.
Recall that the square roots originate from the generating function for
Legendre polynomials \eqref{app:Legendregenfn}.
There, the sign choice for the square root is completely unambiguous at $t=0$,
and it is indeed an expansion of \eqref{app:Legendregenfn} around $t = 0$
that defines the Legendre polynomials.
Thus, one must design the contour $\mathcal{T}$ for $t$ and the branch cuts of $\sqrt{1-2c_1t+t^2}$,
so that the contour continuously connects $-i$ to $i$ via 0,
without crossing the branch cuts.
In this way, the branches for the square root at both endpoints of $\mathcal{T}$,
namely at $t = \pm i$, are defined unambiguously.
Moreover, tracking along the contour $\mathcal{T}$ the complex phase of the expression
inside the $\log$ in \eqref{app:1cutnorm2} will unambiguously determine $n$.
Thus, we have a principled way of fixing all branch cut ambiguities in \eqref{app:1cutnorm2}.
Further restrictions for $\mathcal{T}$ will come shortly from determining $\rho$.
For cases of interest in this paper, the correct choices give
\begin{equation}\label{app:1cutsqrtchoice}
\sqrt{1-2c_1 (\pm i)+(\pm i)^2} = \sqrt{\mp 2ic_1} = (1 \mp i) \sqrt{c_1}~,
\end{equation}
for the square root branches at the endpoints of $\mathcal{T}$,
and \eqref{app:1cutnorm2} can be taken as
\begin{equation}\label{app:1cutnorm}
N\beta = i\pi - 4i \tan^{-1} \sqrt{c_1}~.
\end{equation}

Next we study the density function $\rho(\theta)$.
Recall that $\calC$ is defined as the square root branch cut of $zy(z)$,
which means that $zy(z)$ flips its sign across $\calC$.
On one side of $\calC$, say for $z+i\eps$, we have from \eqref{app:y1cut2}
(recall that $z=e^{i\theta}$)
\begin{eqnarray}
zy(z+i\eps) &=& \sqrt{2} \cdot \sqrt{\cos\theta - c_1} \cdot
\sum_{n=1}^\infty Q_n \cos \left[ \left(n-{\textstyle\frac12}\right) \theta \right]~.
\end{eqnarray}
It then follows that (recall that $\rho(\theta) = iz\rho(z)$)
\begin{eqnarray}\label{app:1cutrho1.5}
\rho(\theta)
&=& \frac{zy(z+i\epsilon)}{2\pi} 
= \frac{\sqrt{\cos\theta - c_1}}{\sqrt{2}\,\pi} \cdot
\sum_{n=1}^\infty Q_n \cos \left[ \left(n-{\textstyle\frac12}\right) \theta \right] \nn\\
&=& \frac{\sqrt{\cos \theta - c_1}}{\sqrt{2} \, \pi N\beta} \cdot
\sum_{n=1}^\infty  \sum_{k=0}^\infty \left(
\frac{ \left( i^{n+k} - (-i)^{n+k} \right) \left(e^{i(n-\frac12)\theta} + e^{-i(n-\frac12)\theta}\right)}{n+k}
\cdot P_k (c_1) \right) \nn\\
&=& \frac{\sqrt{\cos\theta - c_1}}{\sqrt{2} \, \pi N \beta} \cdot 2 \cos\frac{\theta}{2} \cdot
\int_\mathcal{T} \frac{1-t}{(1-2t \cos\theta + t^2) \sqrt{1-2c_1t+t^2}} dt~.
\end{eqnarray}
For the last equality we similarly used the \eqref{app:Legendregenfn} and wrote as a
contour integral from $-i$ to $i$.
The contour $\mathcal{T}$ must coincide with that used in \eqref{app:1cutnorm2},
to ensure that \eqref{app:1cutnorm2} is equivalent to normalization of $\rho$.
As we have explained, $\mathcal{T}$ must be chosen such that it connects $-i$ to $i$
continuously via $0$, and the branch for the square root factor $\sqrt{1-2c_1t+t^2}$
will be determined so that the branch cut is not crossed while following $\mathcal{T}$.
However, the pole due to $\frac{1}{1-2t \cos\theta + t^2}$ adds an extra constraint on the
choice of $\mathcal{T}$; the pole should be avoided while following the contour.
Note that \eqref{app:1cutnorm1.5} needs to be evaluated for \emph{all} $\theta \in \calC$.
Thus, the contour must avoid the pole for \emph{all} $\theta \in \calC$,
otherwise $\rho(\theta)$ before and after encountering the pole will be discontinuous.
To summarize, there must be no combination of $t \in \mathcal{T}$ and $\theta \in \calC$
where $1-2t \cos\theta + t^2=0 ~\leftrightarrow~ t = e^{\pm i \theta}$.
Graphically, this means that when $\mathcal{T}$ is drawn on the complex plane for $t$,
it must not intersect with the eigenvalue cut $\calC$ drawn on the complex plane for $e^{i\theta}$.\footnote{
Since the cut is symmetric under $\theta \to - \theta$,
$t = e^{\pm i \theta}$ for some $t \in \mathcal{T}$ and $\theta \in \calC$
with both signs are equivalent statements.}
One practical difficulty with this constraint on $\mathcal{T}$ is that the precise shape
of the cut $\calC$ is determined only after $\rho(\theta)$ has been properly evaluated,
which requires one to determine $\mathcal{T}$ first.
However, this difficulty can be overcome by estimating a rough shape of the cut to draw
$\mathcal{T}$, evaluating $\rho$ based on this choice, and confirming that the cut
indeed does not intersect with the $\mathcal{T}$ chosen.
In practice, the only choice that matters at the stage of rough estimation of the cut is
whether the cut will pass $\theta = \frac{\pi}{2} ~\leftrightarrow~ e^{i\theta} = i$ above or below it.

Naively evaluating the integral in \eqref{app:1cutrho1.5}, one obtains
\begin{equation}\label{app:1cutrho}
\rho(\theta) = \frac{1}{2\pi N\beta} \cdot 4i \tan^{-1} \sqrt{\frac{\cos\theta - c_1}{c_1 (1+\cos\theta)}}~.
\end{equation}
However, this expression has many ambiguities.
Not only the sign of the square root is ambiguous, but the $\tan^{-1}$ function
is always ambiguous under addition of any multiple of $\pi$.
Which multiple of $\pi$ should be added to the standard branch of $\tan^{-1}$
may even differ between different values of $\theta$.
It is possible and sometimes more practical to fix these ambiguities empirically.
That is, one can add $n \pi$ with suitable $n$ to the $\tan^{-1}$ function and
choose the branch for the square root by trial and error for each $\theta$,
to avoid discontinuity in $\rho(\theta)$ along the eigenvalue cut $\calC$
and ensure that the cut that started at one endpoint indeed ends at the other endpoint.
The procedure explained in the last several paragraphs provides a principled way
to choose the correct branches, rather than by trials and errors.

\begin{figure}[t]
\centering
\hspace{-0.02\textwidth}
\begin{subfigure}[t]{0.38\textwidth}
\includegraphics[width=\textwidth]{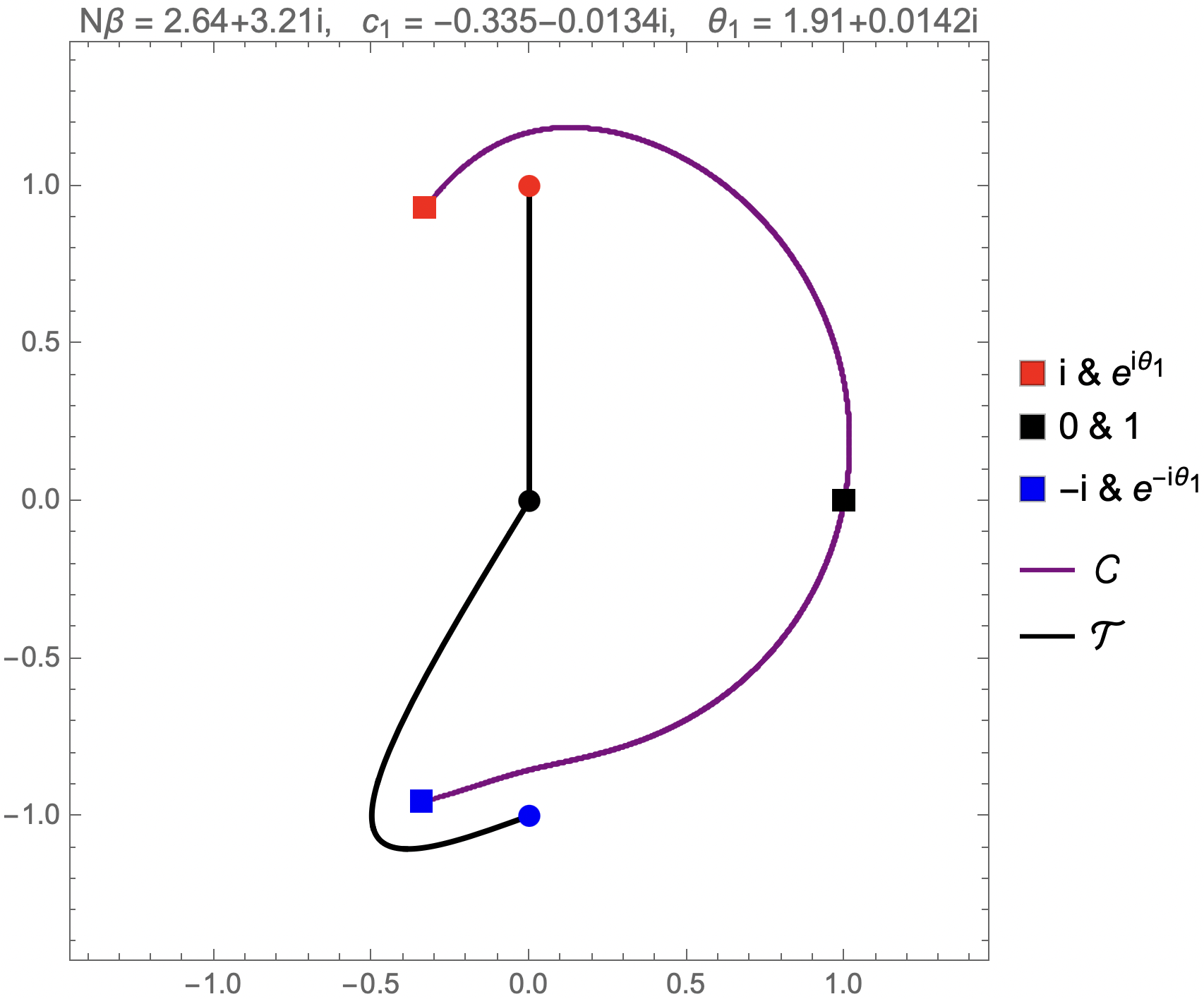}
\caption{}
\end{subfigure}
\hspace{0.02\textwidth}
\begin{subfigure}[t]{0.28\textwidth}
\includegraphics[width=\textwidth]{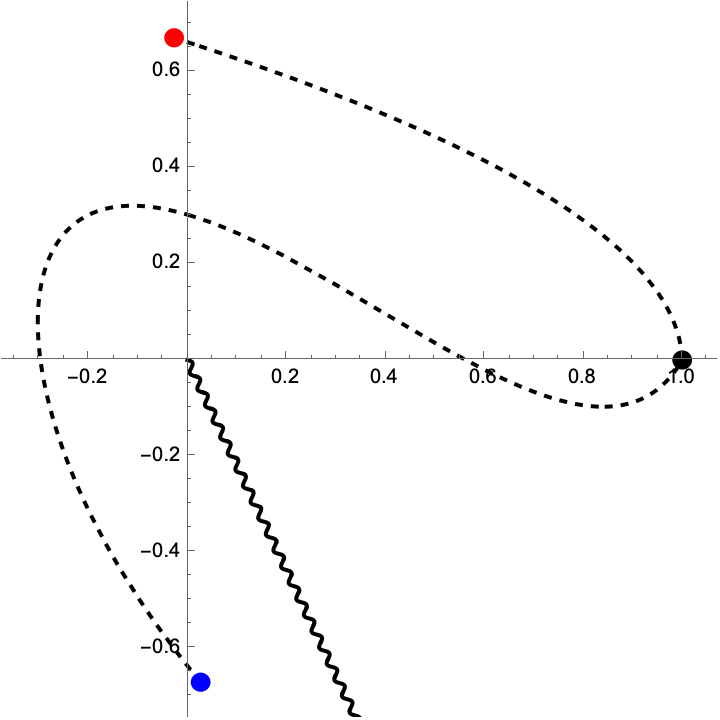}
\caption{}
\end{subfigure}
\hspace{0.02\textwidth}
\begin{subfigure}[t]{0.28\textwidth}
\includegraphics[width=\textwidth]{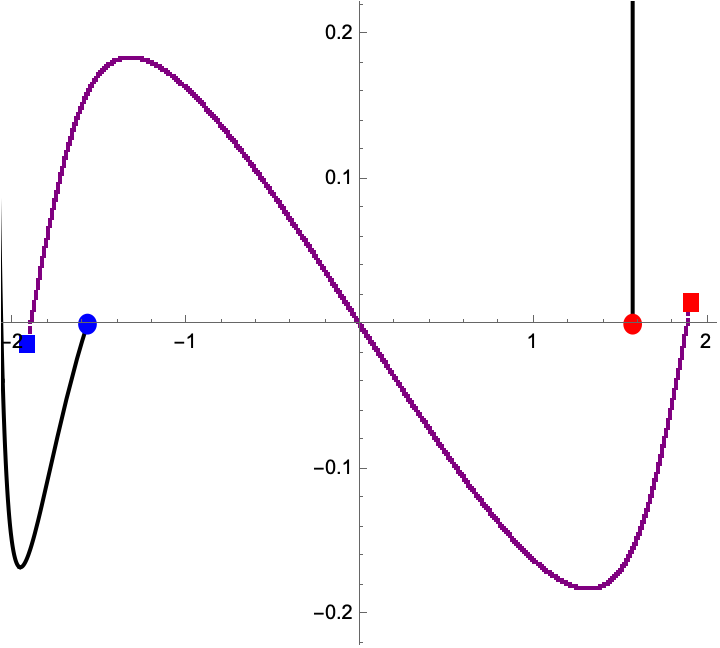}
\caption{}
\end{subfigure}
\hspace{-0.02\textwidth}
\caption{\label{fig:1cut}
Graphics for determining the correct branches.
(a) Given the endpoints of the cut (squares), a putative eigenvalue cut $\calC$ (purple curve) is first drawn on the $e^{i\theta}$ plane.
The contour $\mathcal{T}$ (black curve) connecting the dots is then drawn,
such that it does not intersect with $\calC$.
(b) $1-2c_1t+t^2$ for $t \in \mathcal{T}$ is drawn on the complex plane (black dashed curve).
Its square root branch cut (wavy line) should be chosen so that it is not crossed
for $t \in \mathcal{T}$.
(c) With the branches chosen, $\rho(\theta)$ is computed and thus the actual cut
$\calC$ is obtained numerically for $N=4001$.
The result is plotted in purple curve on the complex $\theta$ plane
(and on the $e^{i\theta}$ plane with hindsight on (a)).
The black curve corresponds to that in (a), and it marks where the bulk function
$\rho(\theta)$ suffers branch cut discontinuities that emanate from $\pm\pi/2$.}
\end{figure}

Let us illustrate the one-cut saddle and the rather abstract procedure for determining
the branches with an example.
We consider the matrix model with an input $N\beta = 2.6435 + 3.2112\,i$,
which approximately corresponds to $j=0.14$ upon Legendre transformation,
following discussion in Section \ref{sec:highTsaddle}.
According to \eqref{app:1cutnorm}, this corresponds to the endpoint parameters
$c_1 = -0.33518 - 0.013398\,i ~\leftrightarrow~\theta_1 = 1.9126 + 0.014220\,i$.
The two endpoints on the $e^{i\theta}$ plane, namely $e^{\pm i \theta_1}$,
are marked in Figure \ref{fig:1cut}(a) with red and blue squares, respectively.
The black square marks $1 = e^{i \cdot 0}$ that the contour is expected to pass by symmetry.
It is reasonable to presume that the eigenvalue cut $\calC$ drawn on the $e^{i\theta}$ plane
will roughly look like the purple curve.
Then, recall that we must draw a contour $\mathcal{T}$ on the same complex plane,
that connects $-i$ (blue dot) to $0$ (black dot) to $i$ (red dot),
without intersecting with $\calC$ (purple curve).
A natural choice is the black curve, whose exact shape is not important as
continuous deformations thereof lead to identical results.

Then we move on to determine the square root branch cut for $\sqrt{1-2c_1t+t^2}$
such that the branch cut is not encountered for $t \in \mathcal{T}$.
On Figure \ref{fig:1cut}(b), the values of $1-2c_1t+t^2$ along $t \in \mathcal{T}$
are plotted on the complex plane.
At $t = -i$ (blue dot), the phase is ${\rm arg}\,(2 i c_1) \approx 4.75$.
Following $\mathcal{T}$, it decreases at $t=0$ (black dot) to 0
and increases back at $t=i$ (red dot) to ${\rm arg}\,(-2 i c_1) \approx 1.61$.
The square root branch cut can be avoided by placing it at ${\rm arg}\,z = \pi+2$,
as described by the wavy line,
thus allowing the phases to take values in $(-\pi+2, \pi+2]$.
So for example, $\sqrt{2ic_1}$ will be on the second quadrant
even though $2ic_1$ (blue dot) lies on the fourth quadrant.
This justifies the choice \eqref{app:1cutsqrtchoice} and thus the formula \eqref{app:1cutnorm}
(which involves tracking the phase of the logarithm to ensure that correct $n$ has been chosen)
that we have already used to determined $c_1$ and $\theta_1$ from $N\beta$.
Furthermore, with suitable branch choices in \eqref{app:1cutrho} determined
in the principled way from \eqref{app:1cutrho1.5},
$\rho(\theta)$ can be evaluated at any given $\theta \in \calC$.
Then, starting from the midpoint $\theta = 0$ we can find subsequent eigenvalues
towards both directions by adding (or subtracting) $\frac{1}{N\rho(\theta)}$ each time.
We compute the eigenvalues numerically for $N=4001$,
where we chose an odd number so that the eigenvalue in exactly the middle of the cut is $0$.
The resulting eigenvalue cut is the purple curve in Figure \ref{fig:1cut}(a)
that we have already drawn with hindsight.
On Figure \ref{fig:1cut}(c), the same eigenvalue cut is drawn on the $\theta$ plane
as opposed to the $e^{i\theta}$ plane on Figure \ref{fig:1cut}(a).
On the same Figure \ref{fig:1cut}(c), the analogue of the black curve $\mathcal{T}$
in Figure \ref{fig:1cut}(a) is also drawn.
As it is obvious from the last line of \eqref{app:1cutrho1.5},
$\rho(\theta)$ is discontinuous when $t = e^{i\theta}$ for some $t \in \mathcal{T}$.
So the black curve in \ref{fig:1cut}(c) is where the density $\rho(\theta)$,
as a complex function of eigenvalues, suffers branch cut discontinuities.
This reemphasizes why $\mathcal{T}$ had to be chosen so that it does not
intersect with the (putative) eigenvalue cut.

\begin{figure}[t]
\centering
\begin{subfigure}[t]{0.38\textwidth}
\includegraphics[width=\textwidth]{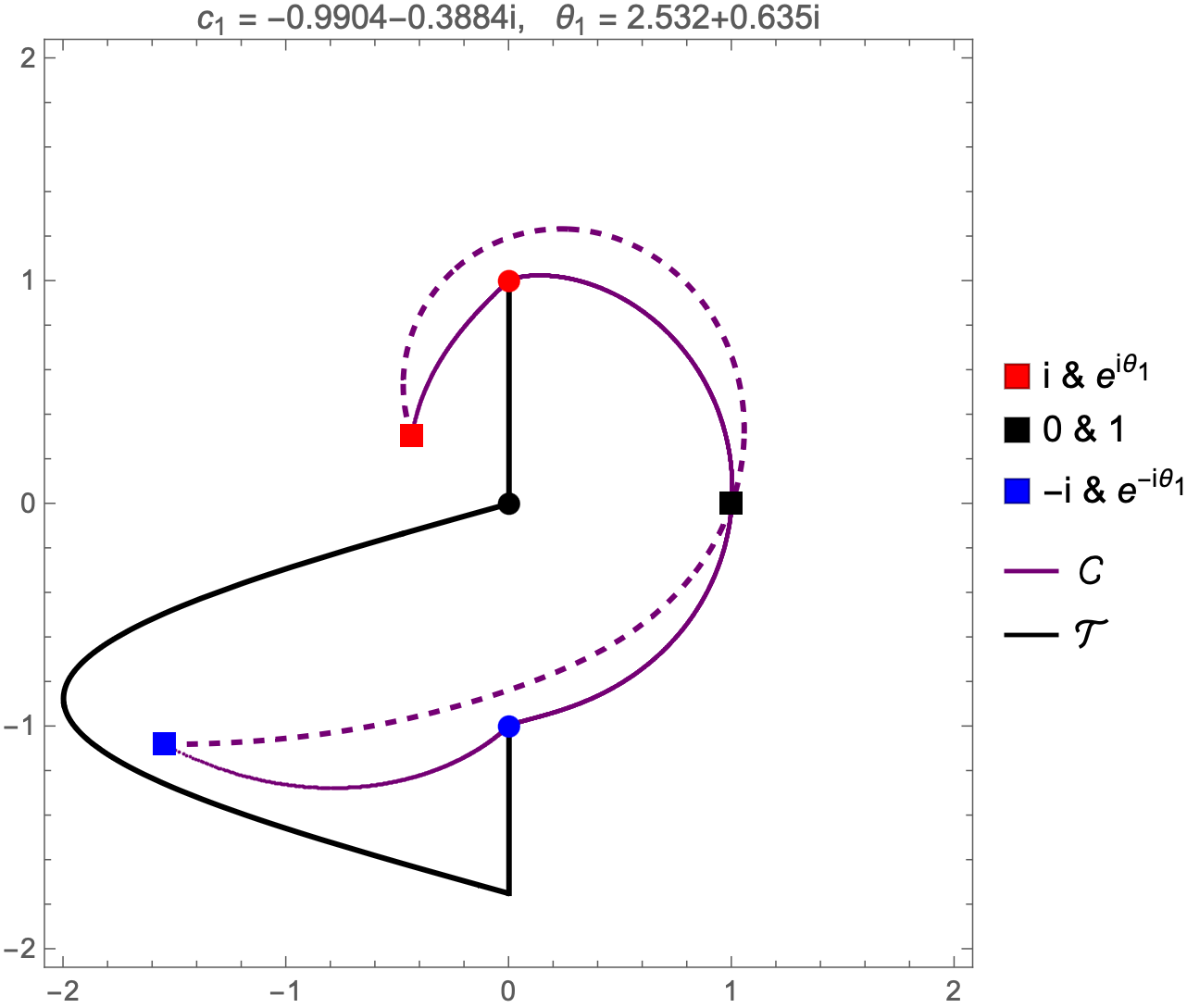}
\caption{}
\end{subfigure}
\hspace{0.02\textwidth}
\begin{subfigure}[t]{0.29\textwidth}
\includegraphics[width=\textwidth]{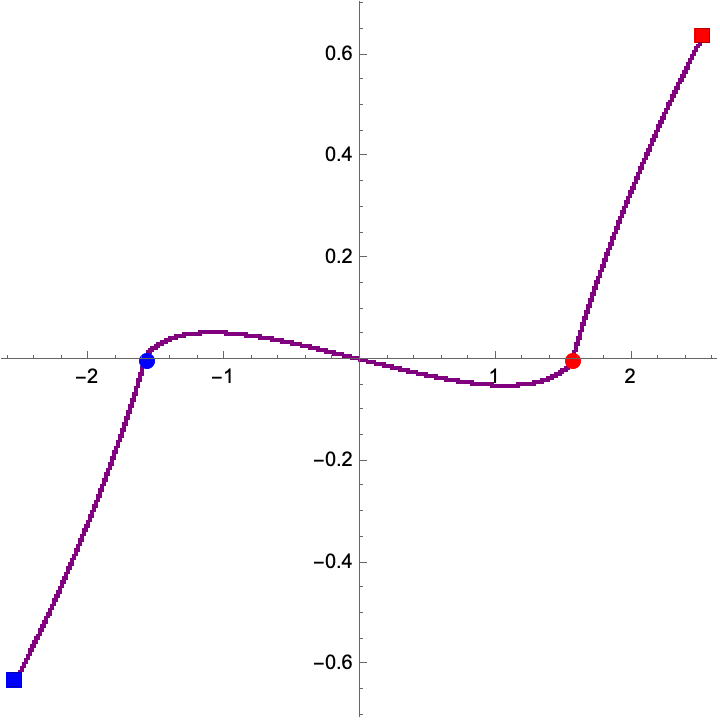}
\caption{}
\end{subfigure}
\medskip

\begin{subfigure}[t]{0.31\textwidth}
\includegraphics[width=\textwidth]{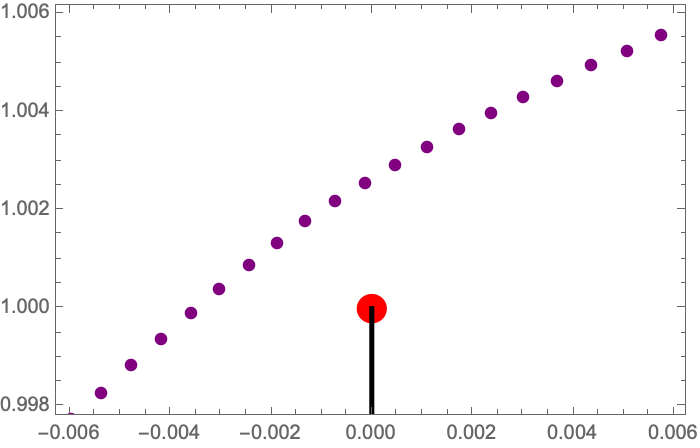}
\caption{}
\end{subfigure}
\hspace{0.02\textwidth}
\begin{subfigure}[t]{0.31\textwidth}
\includegraphics[width=\textwidth]{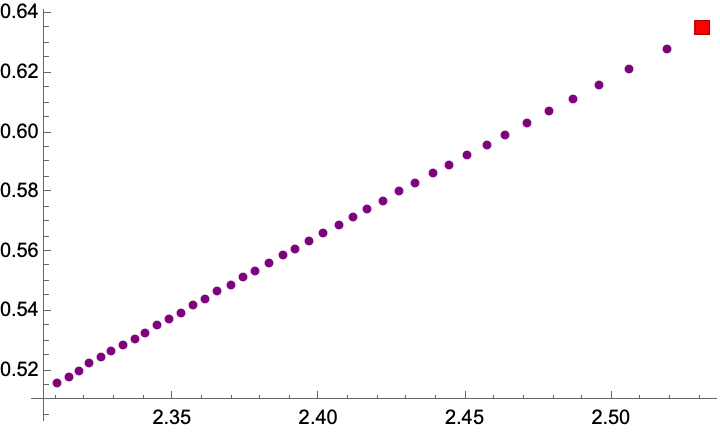}
\caption{}
\end{subfigure}
\hspace{0.02\textwidth}
\begin{subfigure}[t]{0.31\textwidth}
\includegraphics[width=\textwidth]{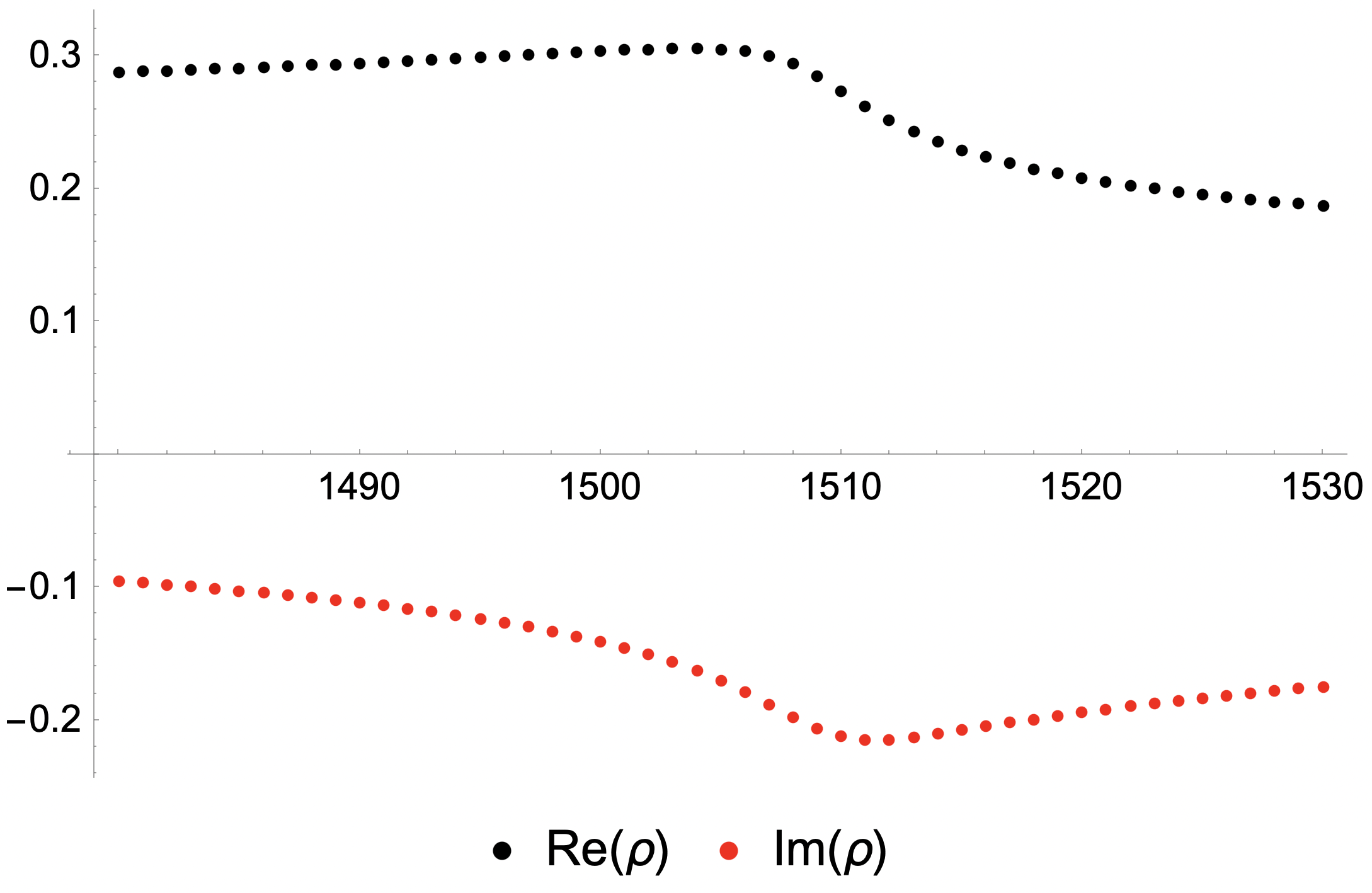}
\caption{}
\end{subfigure}
\caption{\label{fig:1cutx}
The one-cut saddle for $N\beta = 4.7081 + 6.6131\,i$ ($\leftrightarrow~ j=0.0185$),
close to the wall-crossing phenomenon at $j_c\approx 0.017674$.
(a) We first assume the shape of the eigenvalue cut $\calC$ (dashed purple curve) on the $e^{i\theta}$ plane
and draw $\mathcal{T}$ (black curve) that avoids the cut.
The actual eigenvalue cut is found numerically with $N=4001$ (solid purple curve)
(b) and also drawn on the $\theta$-plane.
(c) Zooming into the red dot in (a) on the $e^{i\theta}$ plane,
the actual cut barely avoids intersecting with $\mathcal{T}$.
(d) Zooming into the red square in (b) on the $\theta$ plane,
the actual cut safely ends at the expected endpoint.
(e) Real and imaginary parts of $\rho(\theta)$ for eigenvalues close to $\pi/2$
change somewhat rapidly, albeit continuous.
The horizontal axis enumerates the eigenvalues sequentially starting from $0$.
}
\end{figure}

\begin{figure}[t]
\centering
\begin{subfigure}[t]{0.4\textwidth}
\includegraphics[width=\textwidth]{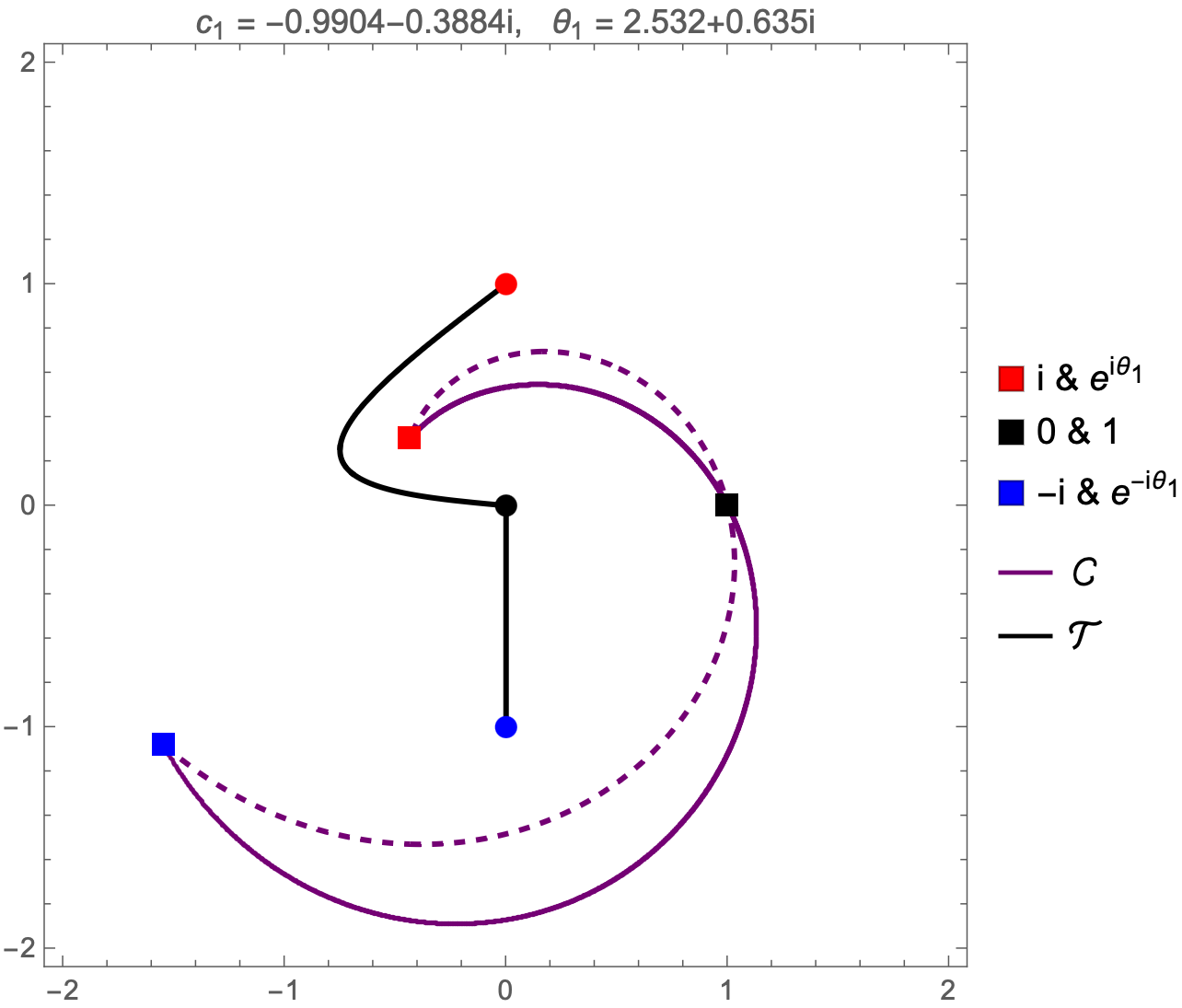}
\caption{}
\end{subfigure}
\hspace{0.02\textwidth}
\begin{subfigure}[t]{0.4\textwidth}
\includegraphics[width=\textwidth]{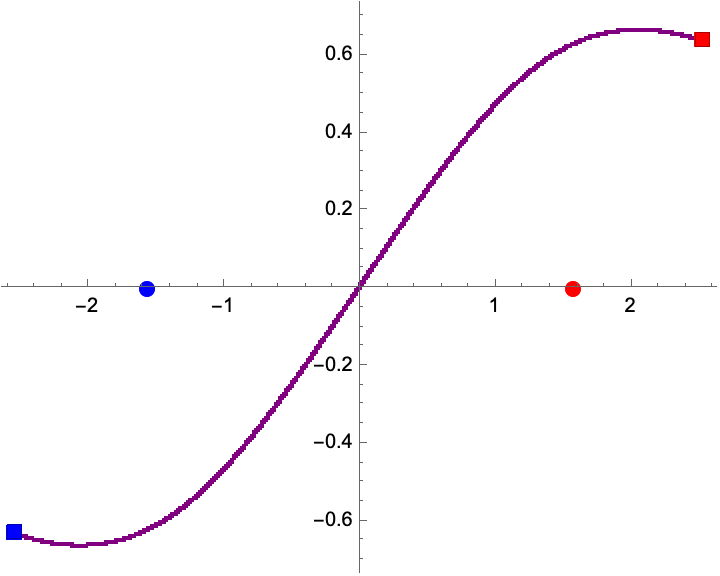}
\caption{}
\end{subfigure}
\caption{\label{fig:1cuty}
An one-cut saddle for $N\beta = -(4.7081 + 6.6131\,i)+2\pi i$.
(a) The putative (dashed) and the actual (solid) eigenvalue cut $\calC$ (purple curves)
obtained numerically for $N=4001$,
and the choice of the contour $\mathcal{T}$ (black curve).
(b) The actual eigenvalue cut drawn on the $\theta$ plane.}
\end{figure}

At this point, we can also demonstrate how certain value of $N\beta$ may lead to
absence of a one-cut saddle, relevant to the wall-crossing phenomenon at
$j=j_c \approx 0.017674$ discussed in section \ref{sec:highTsaddle}.
(This marks the point in Figure \ref{theta0-legendre} where the red line turns to blue.)
For this purpose, we consider the model with an input $N\beta = 4.7081 + 6.6131\,i$,
which corresponds to $j = 0.0185$ upon Legendre transformation,
meaning that this is a point on the red curve in Figure \ref{theta0-legendre}
close to where it becomes blue.
According to \eqref{app:1cutnorm}, the endpoint parameter is equal to
$c_1 = -0.99036 - 0.38843\,i ~\leftrightarrow~ \theta_0 = 2.53164 + 0.63463\,i$.
The first step, which is the only step that is not algorithmically straightforward,
is to presume a rough shape of $\calC$ and then to draw $\mathcal{T}$ that does not intersect.
In Figure \ref{fig:1cutx}(a) and Figure \ref{fig:1cuty}(a),
we marked the endpoints by red and blue squares as we did in Figure \ref{fig:1cut}(a).
We also marked $\pm i$ and $0$ that $\mathcal{T}$ must connect, by dots.
For the eigenvalue cut $\calC$ that connects the squares,
there are essentially two discrete options:
to pass above the red and blue dots (dashed purple curve in Figure \ref{fig:1cutx}(a))
or to pass below both dots (dashed purple curve in Figure \ref{fig:1cuty}(a)).
It is not possible to pass above one and below another because $\calC$ is symmetric in $\theta \to -\theta$.

Let us consider the first scenario, depicted in Figure \ref{fig:1cutx}(a).
This is qualitatively similar to the situation in Figure \ref{fig:1cut}.
The contour $\mathcal{T}$ is drawn, taking a big detour around the blue square.
We denote this contour by the black curve.
With this choice of $\mathcal{T}$ and the suitable square root branch that follows,
one can check that proper evaluation of \eqref{app:1cutnorm2} confirms
the relation \eqref{app:1cutnorm} between $N\beta$ and $c_1$
that we obtained by taking the branch choice \eqref{app:1cutsqrtchoice}.
Then one can also evaluate $\rho(\theta)$ properly from \eqref{app:1cutrho1.5},
and construct the eigenvalue cut numerically.
We do it for $N=4001$ and plot the eigenvalue cut on Figure \ref{fig:1cutx}(b).
It turns out that the eigenvalue cut thus obtained passes barely above
the blue and the red dots in Figure \ref{fig:1cutx}(a)
(equivalently, below the red dot and above the blue dot in Figure \ref{fig:1cutx}(b))
as we have assumed when drawing the putative cut as the dashed purple curve.
Figure \ref{fig:1cutx}(c) shows the eigenvalue cut zoomed into the red dot.
In other words, the sequence of eigenvalues obtained recursively from $\theta = 0$,
comes very close to intersecting with $\mathcal{T}$ (solid black curve) near $\pm i$.
The latter are the points where the external potential is singular.
As a result, $\rho(\theta)$ along $\theta \in \calC$ starts to develop a kink at this point
although it is still continuous for this case, see Figure \ref{fig:1cutx}(e).
When the sequence of $N$ eigenvalues is completed, it indeed ends up
at the expected endpoints $e^{\pm i\theta_1}$, see Figure \ref{fig:1cutx}(d).
The last statement is equivalent to $\int_\calC \rho(\theta) d\theta \neq 1$.
Therefore, we have justified a one-cut saddle for $N\beta = 4.7081 + 6.6131\,i$,
but we also observe that when extended further, the cut $\calC$ will intersect with $\mathcal{T}$,
thus causing a discontinuity in $\rho(\theta)$, and cease to yield a justifiable one-cut saddle.
(For example, if one insists on the discontinuous $\rho(\theta)$ to complete the cut of
$N$ eigenvalues, it does not end at the expected endpoints $e^{\pm i\theta_1}$.)

Let us also consider the second scenario, where the presumed shape of $\calC$ is
the dashed purple curve in Figure \ref{fig:1cuty}(a).
The contour $\mathcal{T}$ must take a detour around the red square instead of the blue square,
resulting in what we have plotted as the black curve.
However, with this choice of $\mathcal{T}$ and with the suitable square root branch that follows,
\eqref{app:1cutnorm2} results in $N\beta = -(4.7081 + 6.6131\,i)+2\pi i$,
instead of $N\beta = 4.7081 + 6.6131\,i$.
In other words, \eqref{app:1cutnorm2} leads to \eqref{app:1cutnorm} but with the non-standard
branch choices for the square root and for the $\tan^{-1}$ function there.
So the contour $\mathcal{T}$ leads to a consistent one-cut saddle,
although for a different value of $N\beta$.
Evaluating $\rho(\theta)$ properly via \eqref{app:1cutrho1.5} and constructing the cut
numerically for $N=4001$, we obtain the eigenvalue cut drawn with solid purple curves in
Figure \ref{fig:1cuty}(a) and (b).
This is a valid one-cut saddle for a different value of $N\beta$ from what we have aimed for,
but this value has ${\rm Re}(N\beta) < 0$ so it has no thermodynamic implications.

To exhaust all other scenarios for a given input $N\beta$,
we can examine all values of $c_1$ that may yield the desired $N\beta$
via \eqref{app:1cutnorm} under some choice of branches.
Then for each such $c_1$, we repeat the procedure described above and see if
i) it yields a viable solution,
ii) and if the branch choice principled in \eqref{app:1cutnorm2} is indeed what gives the desired $N\beta$.
Although at first sight, there can be infinitely many values of $c_1$ that may
lead to a given $N\beta$ due to ambiguity of $\tan^{-1}$ by addition of $n \pi$,
such infinitely many cases are not realized for one-cut saddles,
so we may treat only a finite number of options.
This is in contrast to the two-cut saddle, see discussions around footnote \ref{footnoteonk}.
In this way, we convincingly conclude the uniqueness of the saddles that we construct,
such as the one described in Figure \ref{fig:1cutx} for $N\beta = 4.7081 + 6.6131\,i$,
and similarly that no one-cut saddle exists for certain values of $N\beta$,
for instance for those beyond the wall-crossing phenomenon
(on the blue part of the curve in Figure \ref{theta0-legendre}).

With the eigenvalue distribution determined, one can evaluate the free energy of the one-cut saddle.
This involves evaluating the integral formula \eqref{app:ZandS}
for $\rho(\theta)$ given by \eqref{app:1cutrho} and the contour $\calC$ determined thereby.
We do not fully lay out the long and brutal computations but only present a few key intermediate steps.

To start with, we define the chemical potential $\mu(\theta)$
which must be constant along the cut (i.e. for $\theta \in \calC$), as
\begin{equation}\label{app:defmu}
\mu(\theta)
\equiv \frac{\delta}{\delta \rho(\theta)} \left[ -\frac{S[\rho(\theta)]}{N^2} \right] 
= \int_{\calC} d\theta' \rho(\theta') \log\left(4\sin^2\frac{\theta-\theta'}{2}\right)
+ \frac{1}{N} \sum_n \frac{g_n}{n} (e^{in\theta} + e^{-in\theta})~.
\end{equation}
Because it is constant ($\mu = \mu(\theta)$),
we can use it to simplify the effective action \eqref{app:ZandS},
\begin{equation}
  -\frac{S[\rho(\theta)]}{N^2} = \mu -\log 2 + \frac{S_2[\rho(\theta)]}{N^2}~,
\end{equation}
where $S_2$ refers to the two-body interaction term of the effective action,
\begin{eqnarray}\label{app:logZincosmodes}
- \frac{S_2[\rho(\theta)]}{N^2} \!\!&=&\!\!
\iint_\calC d\theta_a d\theta_b \rho(\theta_1) \rho(\theta_2)
\log\left|\sin\frac{\theta_{ab}}{2}\right| \\
\!\!&=&\!\!\! \sum_{n_1,n_2=1}^\infty \!\! \frac{Q_{n_1} Q_{n_2}}{2\pi^2} \iint_\calC d\theta_a d\theta_b
  \sqrt{\cos\theta_a \!-\! c_1} \sqrt{\cos\theta_b \!-\! c_1}
   \cos (n_1\!-\!{\textstyle \frac12}) \theta_a \cos (n_2\!-\!{\textstyle \frac12}) \theta_b
  \log\left|\sin\frac{\theta_{ab}}{2}\right|~.\nn
\end{eqnarray}
For the second equality we used $\rho(\theta)$ in the form of the second line in \eqref{app:1cutrho1.5}.

The constant value of $\mu$ is evaluated at $\theta = 0$, and it is
\begin{eqnarray}\label{app:muresult}
  \mu &=&
  \frac{2}{N\beta} \sum_\pm \left[
  {\rm Li}_2\left({\textstyle \frac{e^{\pm\frac{\pi i}{4}}(1+i\sqrt{c_1})}{\sqrt{2}}}\right)
  -{\rm Li}_2\left({\textstyle \frac{e^{\pm\frac{\pi i}{4}}(1-i\sqrt{c_1})}{\sqrt{2}}}\right)
  \right] \nonumber\\
  &=& \frac{1}{N\beta}\left[{\rm Li}_2(e^{-N\beta})-4\,{\rm Li}_2(-ie^{-\frac{N\beta}{2}})
  -\frac{\pi^2}{4}\right]~,
\end{eqnarray}
where ${\rm Li}_n$ is the polylogarithm and the two lines are related via \eqref{app:1cutnorm}.
For evaluation of the double integral \eqref{app:logZincosmodes},
the following table of integrals is useful:
\begin{eqnarray}\label{app:integrals2}
\mathcal{I} (s,t) \!\!&=&\!\! \sum_{l,m=1}^\infty s^lt^m
\iint_{-\theta_1}^{\theta_1} d\theta_a d\theta_b \log \left| \sin \frac{\theta_{ab}}{2} \right|
\sqrt{\cos\theta_a - c_1} \sqrt{\cos\theta_b - c_1}
\cos(l-{\textstyle \frac12}) \theta_a \cos(m-{\textstyle \frac12}) \theta_b \nonumber \\
\!\!&=&\!\! \frac{\pi^2}{2} \sqrt{1-2sc_1+s^2}\sqrt{1-2tc_1+t^2} \cdot \log
\frac{2 \left( (1-t)\sqrt{1-2sc_1+s^2}+(1-s)\sqrt{1-2tc_1+t^2} \right)}
{\left(\sqrt{1-2sc_1+s^2}+(1-s)\right) \left(\sqrt{1-2tc_1+t^2}+(1-t)\right)} \nonumber \\
&& +\frac{\pi^2}{2} \left(\sqrt{1-2sc_1+s^2}-(1-s)\right) \left(\sqrt{1-2tc_1+t^2}-(1-t)\right)
\cdot \log \frac{\sin \frac{\theta_1}{2}}{2}~.
\end{eqnarray}
This formula can be understood as giving an integral for every order of auxiliary variables $s$ and $t$,
but the generating function itself is more useful for our purpose
because upon substituting \eqref{app:1cutQgen} for $Q_n$, we have
\begin{equation}
- \frac{S_2[\rho(\theta)]}{N^2} =
\frac{2}{(\pi N\beta)^2} \cdot
\iint_\mathcal{T} dsdt \frac{s^{-1} t^{-1}}{\sqrt{1-2c_1s+s^2}\sqrt{1-2c_1t+t^2}} \cdot \mathcal{I}(s,t)~.
\end{equation}
Here the integration is along the contour $\mathcal{T}$ from $-i$ to $i$,
for the same reason as when it was introduced around \eqref{app:1cutnorm1.5}.
The last integral can be performed by treating $c_1$ as a variable to
differentiate in $c_1$, perform the integral and integrate back $c_1$ to its fixed value.
As a result, we obtain
\begin{eqnarray}\label{app:S2}
  && - \frac{S_2[\rho(\theta)]}{N^2} + \log 2\\
  &&= \frac{1}{N\beta} \left[ {\rm Li}_2 (e^{-N\beta}) - 4\,{\rm Li}_2 (-ie^{-\frac{N\beta}{2}}) \right]
  - \frac{1}{(N\beta)^2} \left[ \frac74 \zeta(3) + \frac{i\pi^3}{4}
  + 8\,{\rm Li}_3 (-ie^{-\frac{N\beta}{2}}) - {\rm Li}_3 (e^{-N\beta}) \right]~. \nn
\end{eqnarray}
Combining with \eqref{app:muresult} and simplifying some of the polylogarithms,
we arrive at the final result for the free energy of the one-cut saddle,
\begin{equation}\label{app:1cutlogZ}
- \frac{S[\rho(\theta)]}{N^2} =
-\frac{\pi^2}{4N\beta}+\frac{1}{(N\beta)^2}\left[\frac{7}{4}\zeta(3)+\frac{\pi^3 i}{4}
  +8\,{\rm Li}_3(-ie^{-\frac{N\beta}{2}})-{\rm Li}_3(e^{-N\beta})\right]~,
\end{equation}
where $\zeta(3) = {\rm Li}_3 (1) \approx 1.202$.
This formula for the free energy as well as the constancy of the chemical potential \eqref{app:defmu}
have been checked numerically for the saddles discussed in Section \ref{sec:highTsaddle}
including the examples displayed in this subsection,
by evaluating them as discrete summations over eigenvalue distributions with finite $N = O(10^4)$.

\subsection{Two-cut saddles}\label{sec:app2cut}

We now study two-cut saddles with the cut $\calC = \calC_1 \cup \calC_2$ where
$\calC_1 = (-\theta_1, \theta_1)$, $\calC_2 = (\theta_2, 2\pi - \theta_2)$.
$\calC_1$ passes through $\theta = 0$ and $\calC_2$ through $\theta = \pi$,
and both are reflection symmetric.
\eqref{app:y2cut} can be written equivalently as
\begin{eqnarray}\label{app:y2cut2}
\eqref{app:eqnfory} &=&
\frac{(z-a_1)(z-a_1^{-1})}{z} \cdot \frac{(z-a_2)(z-a_2^{-1})}{z} \cdot
\left( \sum_{n=0}^{p-1} Q_n \cdot \frac{z^{n} + z^{-n}}{2} \right)^2~.
\end{eqnarray}
Similarly as in \eqref{app:y1cut2}, $p$ parameters $Q_n$'s replace
$p-1$ parameters $d_n$'s and turns the proportionality sign into an equality.
Again expanding both sides in small $z$ and comparing the first $p+1$ orders,
we can determine all $p$ parameters $Q_n$'s and still one constraint remains.
Taking $p \to \infty$ in the formulae for $Q_n$ thus obtained,
\begin{eqnarray}\label{app:2cutQgen}
Q_n &=& \sum_{n_1,n_2=0}^\infty \frac{2g_{n+1+n_1+n_2}}{N} P_{n_1} (c_1) P_{n_2} (c_2)~,
\qquad (n \geq 1) \nn\\
Q_0 &=& \sum_{n_1,n_2=0}^\infty \frac{g_{1+n_1+n_2}}{N} P_{n_1} (c_1) P_{n_2} (c_2)~.
\end{eqnarray}
$Q_0$ acquired an exceptional factor of $\frac12$ because of the
obvious $z$-series structure of the terms in the parentheses in \eqref{app:y2cut2}.
Here,
\begin{equation}
a_{1,2} = e^{i \theta_{1,2}}~, \qquad
c_{1,2} = \cos \theta_{1,2} = \frac{a_{1,2}+a_{1,2}^{-1}}{2}~,
\end{equation}
all relate to the endpoints of the cuts.

The one remaining constraint requires that $Q_1$ must also satisfy
\begin{equation}\label{app:2cutQ1}
Q_1 = 2 + \sum_{\substack{n_1,n_2 \geq 0 \\ n_1+n_2 \geq 1}} \frac{2g_{n_1+n_2}}{N}
P_{n_1} (c_1) P_{n_2} (c_2)~.
\end{equation}
Its compatibility with \eqref{app:2cutQgen} is ultimately equivalent to
the normalization $\int_{\calC} \rho(\alpha) d\alpha = 1$
and imposes a constraint between $c_1$ and $c_2$.
Note that in this normalization condition, the integral is over $\calC = \calC_1 \cup \calC_2$,
namely over both pieces of the eigenvalue cut.

We now apply the general solution to our specific model with input parameters \eqref{app:gn}.
The $i^n - (-i)^n$ structure inside $g_n$ is suited for turning the sums over Legendre polynomials
into its closed-form generating function \eqref{app:Legendregenfn}.
Thus similarly to what was done in Appendix \ref{sec:app1cut},
we can rewrite \eqref{app:2cutQgen} as an integral along a contour $\mathcal{T}$
that connects $-i$ to $i$ via $0$:
\begin{eqnarray}\label{app:2cutQgenintegral}
Q_n &=& \frac{2}{N\beta} \int_{\mathcal{T}}
\frac{t^n dt}{\sqrt{1-2tc_1 + t^2} \cdot \sqrt{1-2tc_2 + t^2}}~.
\qquad (n \geq 1)
\end{eqnarray}
For $Q_0$ simply put in $n=0$ and multiply by $\frac12$.
In this formula, the square root branches should be chosen such that the square roots
take the standard branch $\sqrt{1} = 1$ at $t=0$, and are continuous along $\mathcal{T}$.
It is straightforward to write also \eqref{app:2cutQ1} as such an integral.

We examine the compatibility condition between \eqref{app:2cutQgen} and \eqref{app:2cutQ1},
that represents the normalization of $\rho (\alpha)$.
Using the integral formula, it reads
\begin{equation}\label{app:2cutnorm}
N\beta =
\int_{\mathcal{T}} \frac{t-\frac{1}{t}}{\sqrt{1-2tc_1 + t^2} \cdot \sqrt{1-2tc_2 + t^2}} dt 
= -2 \log \frac{i(\sqrt{c_1} - \sqrt{c_2})}{\sqrt{c_1} + \sqrt{c_2}}~.
\end{equation}
%
Similar comments to Appendix \ref{sec:app1cut} regarding the appropriate choice of $\mathcal{T}$ for
a principled fixing of branch cut ambiguities would follow.
That is, $\mathcal{T}$ must not intersect with the eigenvalue cut $\calC$ in its $e^{i\theta}$ plane.
Along with the requirement that the square roots be continuous along $\mathcal{T}$,
the first line of \eqref{app:2cutnorm} is free of branch cut ambiguities.
This will determine which branch and sheet to take for the expression in the second line.
We have discussed this way of fixing branches in detail for 1-cut saddles in Appendix \ref{sec:app1cut}.
However, for two-cut saddles in this subsection, we shall avoid discussing such complication
and instead work with branch cut choices confirmed empirically and numerically.
For example, the way the second line of \eqref{app:2cutnorm} is written is such that 
the standard branch thereof gives correct formula for examples to be discussed later in this subsection.

Next we study the density function $\rho(\theta)$.
From \eqref{app:rhofromy}, \eqref{app:y2cut2} and \eqref{app:2cutQgenintegral},
we obtain (recall that $z = e^{i\theta})$
\begin{eqnarray}\label{app:2cutrho}
\rho(\theta)
\!\!&=&\!\! \frac{zy(z+i\epsilon)}{2\pi} 
= \frac{\sqrt{\cos\theta - c_1}\cdot\sqrt{\cos\theta - c_2}}{\pi} \cdot
\sum_{n=0}^\infty Q_n \cos (n\theta) \nn\\
\!\!&=&\!\! \frac{\sqrt{\cos\theta - c_1}\cdot\sqrt{\cos\theta - c_2}}{\pi N \beta} \cdot
\int_\mathcal{T} \frac{\frac{1}{1-te^{i\theta}}+\frac{1}{1-te^{-i\theta}}-1}
{\sqrt{1-2c_1t+t^2} \cdot \sqrt{1-2c_2t+t^2}} dt \nn\\
\!\!&=&\!\! \frac{1}{\pi N\beta} \cdot \left[
\tanh^{-1} \frac{\sqrt{(\cos\theta-c_1)(\cos\theta-c_2)}}{\cos \theta - \sqrt{c_1c_2}}
- \tanh^{-1} \frac{\sqrt{(\cos\theta-c_1)(\cos\theta-c_2)}}{\cos \theta + \sqrt{c_1c_2}} \right]~.\qquad
\end{eqnarray}
Again, the last line contains branch cut ambiguities, which can in principle be fixed
unambiguously from the penultimate line.
In practice, however, we fix the ambiguities by choosing one that numerically yields a
sensible eigenvalue cut with continuous $\rho(\theta)$ and that connects the expected endpoints.
The last expression of \eqref{app:2cutrho} is already written in the form whose
standard branch will be the one appropriate for our purpose.

We have mentioned that the standard branches in the last expressions of
\eqref{app:2cutnorm} and of \eqref{app:2cutrho} are appropriate branch choices for our purpose.
Then it seems as if ${\rm Im}\,N\beta$ is only allowed between $\pm 2\pi i$.
However, as we shall find later, ${\rm Im}\,N\beta$ outside of this range can
actually be allowed by taking different sheets for the logarithm.
For example, consider modifying the integration contour $\mathcal{T}$ by attaching to it
an infinite-radius circle with an arbitrary wrapping number $k_\infty \in \mathbb{Z}$,
or a small circle around $t=0$ also with an arbitrary wrapping number $k_0 \in \mathbb{Z}$.
It is always possible for the contour to connect to/from the infinite circle
still without intersecting with the cut, as long as the cut is gapped.
Connection to the $t \sim 0$ circle is trivially possible because $\mathcal{T}$
is designed to pass through $t=0$.
Let the modified contour be $\mathcal{T}_{k_0,k_\infty}$.
Each wrap around the $t\sim 0$ circle adds $-2 \pi i$ to the integral in the first line of
\eqref{app:2cutnorm}, while each wrap around the infinite circle adds $2\pi i$.
So using the modified contour $\mathcal{T}_{k_0,k_\infty}$ instead of $\mathcal{T}$,
we obtain a new value of $N\beta$,
\begin{equation}\label{app:2cutNbetak}
(N\beta)_{k_0,k_\infty} = N\beta + 2\pi i (k_\infty-k_0)~,
\end{equation}
for same $c_1$ and $c_2$.\footnote{\label{footnoteonk}
Curiously, such shifts of $N\beta$ are not possible for 1-cut saddles.
In the first line of \eqref{app:1cutnorm2}, the integrand is $O(t^0)$ around $t=0$,
not yielding a residue.
For $|t| \gg 1$, the correct branch of $\sqrt{1-2c_1t+t^2}$ is $-t$
if $\mathcal{T}$ is continued from $t=0$ without intersecting with the cut,
so again there is no residue at $t=\infty$.}

We can similarly re-evaluate $\rho(\theta)$ with the modified contour;
we revisit the second line of \eqref{app:2cutrho} because the geometric series
in the third line adds an issue with analytic continuations.
Under the addition of the $t \sim 0$ circle to the contour, 
the integral in \eqref{app:2cutQgenintegral} does not change,
while the addition of the infinite circle with wrapping number $k_\infty$ changes it by
\begin{equation}
\int_{\mathcal{T}_{0,k_\infty} - \mathcal{T}_{0,0}}
\frac{t^n dt}{\sqrt{1-2tc_1 + t^2} \cdot \sqrt{1-2tc_2 + t^2}}
= 2\pi i k_\infty \cdot
\left[ \frac{1}{\sqrt{1-2tc_1+t^2} \sqrt{1-2tc_2+t^2}}\right]_{t^{n-1}}~,
\end{equation}
where $[ \cdots ]_{t^{n-1}}$ refers to the coefficient of $t^{n-1}$ in $[ \cdots ]$
when series expanded around $t=0$.
This combines with the summation over $n$ in the second line of \eqref{app:2cutrho},
such that (the integral for $n=0$ is not changed, so the summation starts from $n=1$)
\begin{eqnarray}
&& \sum_{n=1}^\infty \int_{\mathcal{T}_{0,k_\infty} - \mathcal{T}_{0,0}}
\frac{t^n \cos(n\theta) dt}{\sqrt{1-2tc_1 + t^2} \cdot \sqrt{1-2tc_2 + t^2}} \\
&&= \pi ik_\infty \cdot \left[
\frac{e^{i\theta}}{\sqrt{1-2e^{i\theta}c_1+e^{2i\theta}} \sqrt{1-2e^{i\theta}c_2+e^{2i\theta}}}
+ (\theta \to -\theta) \right] 
= \frac{\pi ik_\infty}{\sqrt{\cos\theta - c_1}\cdot\sqrt{\cos\theta - c_2}}~.\nonumber
\end{eqnarray}
As a result, the new density function is written in terms of the original $\rho_{0,0}(\theta)$ as
\begin{equation}\label{app:2cutrhok}
\rho_{k_0,k_\infty}(\theta) = \frac{N\beta \rho_{0,0} (\theta) + 2ik_\infty}{N\beta + 2\pi i (k_\infty - k_0)} ~.
\end{equation}
To conclude, once we have a 2-cut saddle for some input value of $N\beta$
with endpoints parametrized by $c_1$ and $c_2$ and the density function $\rho_{0,0}(\theta)$,
we also obtain candidate saddles for different input values $N\beta + 2\pi i(k_\infty - k_0)$
that have identical $c_1$ and $c_2$, and the density function $\rho_{k_0,k_\infty} (\theta)$
given by \eqref{app:2cutrhok}.
We emphasize that these are only candidates;
in practice, it remains to check which of these give sensible saddles.
We will demonstrate such a process later in this subsection.

Compatibility of \eqref{app:2cutQgen} and \eqref{app:2cutQ1}, thus \eqref{app:2cutnorm},
gave one constraint on two variables $c_1$ and $c_2$ that represent the endpoints of the cuts.
Thus, one combination of them still remains as a free parameter.
In the real matrix model with all $g_n \in \mathbb{R}$ and all eigenvalues $e^{i\theta}$
on the unit circle without contour deformation \cite{Jurkiewicz:1982iz},
this free parameter is precisely the filling fraction between the two cuts.
That is, the force-free equation \eqref{app:FFE} governs extremization of the action
with respect to local displacement of an eigenvalue,
but it does not guarantee extremization under moving an eigenvalue from one disjoint cut to another,
thereby changing the filling fractions of each cut.
(In some context, this effect is known as eigenvalue instantons.)
In some sense, one obtains an $O(N)$ number of local saddles,
and $\log Z$ would be a sum over some of them
through which the steepest descent contour is made to pass, see \eqref{filling-fraction-sum}.
The resolution taken in \cite{Jurkiewicz:1982iz} is to once more extremize $\log Z$
over the filling fraction, which amounts to identifying the chemical potential $\mu$
on each cut. (Force-free equation guarantees that $\mu$ be constant along each cut.)
This extra equation, together with the normalization condition $\int_{\calC} \rho(\theta)d\theta = 1$,
fixes the two endpoint variables $c_1$ and $c_2$.
More generally, the filling fraction extremization yields $m-1$ equations for $m$-cut solutions,
which is the correct number of equations needed for fixing all $m$ endpoint variables
together with the overall normalization condition.

However, in the model with complex coefficients and therefore
generically complex eigenvalue saddles, this argument faces a conceptual puzzle.
Namely, the extremization over filling fraction requires that only the
real part of the chemical potential on each cut is equal
(equivalently, only ${\rm Re}(\log Z)$ is to be maximized),
and one real component out of two complex variables $c_1$ and $c_2$ still remains free.
On the other hand, there is an additional constraint that is not present for the real model,
imposed by the condition that the filling fraction is real, namely
$\nu \equiv \int_{\calC_1} \rho(\theta)d\theta = 1-\int_{\calC} \rho(\theta)d\theta \in \mathbb{R}$.
Unless this condition is met, $\theta_{1,2}$ cannot be true endpoints of the respective cuts,
along which $\rho(\theta) d\theta$ must be real.
This seems to give one much needed real constraint to finally fix $c_1$ and $c_2$ completely.
The problem is that, since now one combination of $c_1$ and $c_2$ are fixed by
two completely different real conditions, the formula for $\rho$,
and more importantly for $\log Z$,
seem unlikely to be holomorphic in the input variable.

In Section \ref{sec:lowTsaddle}, this puzzle was discussed in detail.
As in the main text the focus is on the thermodynamics,
we took a microcanonical viewpoint where the `charge' $j$ is fixed.
Then, the physical (inverse) temperature ${\rm Re}\, \beta$ is dual to $j$
and ${\rm Im}\, \beta$ must be tuned so as to extremize $\log Z$ for given $\nu$,
which corresponds to minimizing the cancellations in the index to represent
the true partition function of the thermodynamic system.
Then one should maximize $\log Z$ over $\nu$.
However, it was also argued around \eqref{maximal-order} that one has freedom to change the
order between maximizing over ${\rm Im}\, \beta$ and over $\nu$,
so that one may equally well maximize ${\rm Re}(\log Z)$ over $\nu$ first.
This is a useful viewpoint in treating the (grand-)canonical ensemble where
$\beta$ is the fixed parameter.

The goal of this appendix is to study the matrix model with little regard to thermodynamics.
Thus, we shall take the approach just mentioned,
where we extremize ${\rm Re}(\log Z)$ or equivalently equate ${\rm Re}\,\mu$ between
disjoint pieces of the cut, over real filling fractions $\nu \in \mathbb{R}$
to obtain the most dominant saddle for a matrix model with fixed (complex) $\beta$.

Unfortunately, the bulk density function \eqref{app:2cutrho} is already too involved
for further progress to be made analytically.
Instead, in the rest of this appendix we take a limit that should connect to the
extreme low temperature limit, namely when ${\rm Re}\, \gamma = {\rm Re}\, N\beta \gg 1$.
This limit is sufficient for our purpose for studying the 2-cut saddles in Section \ref{sec:lowTsaddle}, 
which is to bridge between the 1-cut saddles discussed in
Section \ref{sec:highTsaddle} and Appendix \ref{sec:app1cut}
that connects to the extreme high temperature limit,
and the extreme low temperature limit of the uniform confined saddle.

For this limit, it is convenient to reparametrize $(c_1, c_2)$ by $(c_0, \eps)$:
\begin{equation}\label{app:c1andc2}
c_1 = c_0(1-i\eps)~, \qquad c_2 = c_0(1+i\eps)~,
\end{equation}
and assume that $|\eps|$ is small.
The latter assumption means that both endpoints $c_1$ and $c_2$ come very close to $c_0$,
and thus to each other.
This is expected for the 2-cut saddle to continuously connect to the uniform gapless confined saddle,
because $c_1 \to c_2$ signals vanishing of the gap.
Then \eqref{app:2cutnorm} gives
\begin{equation}\label{app:2cutlowTnorm}
N\beta = - 2 \log \left( i\cdot \frac{-i\eps + \frac{i}{8} \eps^3 + \cdots}{2 + \frac14 \eps^2 + \cdots} \right)
= 2 \log \frac{2}{\eps} + \frac{\eps^2}{2} + O(\eps^4)~.
\end{equation}
From now on, we consistently suppress any subleading powers of $\eps \sim e^{-\frac{N\beta}{2}}$,
but retain (sometimes up to certain powers of) the logarithmic divergence $\log\eps^{-1} \sim N\beta$.

We also evaluate the density function \eqref{app:2cutrho} under this approximation.
The first $\tanh^{-1}$ term leads to a logarithmic divergence,
\begin{eqnarray}\label{app:2cutarctanhterm}
\tanh^{-1} \frac{\sqrt{(\cos\theta-c_1)(\cos\theta-c_2)}}{\cos \theta - \sqrt{c_1c_2}}
&=& \tanh^{-1} \left( 1+\frac{\cos\theta \cdot c_0 \eps^2}{2(\cos\theta - c_0)^2} + O(\eps^4) \right) \nn\\
&=& \frac12 \log 2 - \frac12 \log
\left( - \frac{\cos\theta \cdot c_0 \eps^2}{2 (\cos\theta - c_0)^2} + O(\eps^4) \right) \nn\\
&=& \log \frac{2}{\eps} + \frac12 \log
\left( -\frac{(\cos\theta - c_0)^2}{\cos\theta \cdot c_0} \right) + O(\eps^2) ~.
\end{eqnarray}
Combined with the finite second $\tanh^{-1}$, we obtain\footnote{At
some point during this evaluation, one expands in powers of $\frac{c_0 \eps}{\cos \theta - c_0}$.
For $\theta$ very close to either endpoints $\theta_{1,2}$,
this factor is enhanced and is not as suppressing as $\eps$.
However, this enhancement happens for only a small ($\sim O(\eps)$) range of $\theta_0$,
and thus \eqref{app:2cutlowTrho} is valid insofar as the error in $\int d\theta \rho(\theta) f(\theta)$
is suppressed as $O(\eps)$.\label{footnote:c0expansion}}
\begin{eqnarray}\label{app:2cutlowTrho}
\rho(\theta) &=& \frac{1}{2\pi N\beta} \cdot \left[
2 \log\frac{2}{\eps} + \log \left( -\frac{(\cos\theta - c_0)^2}{\cos\theta \cdot c_0} \right)
- \log\frac{1+\frac{\cos\theta - c_0}{\cos\theta + c_0}}{1-\frac{\cos\theta - c_0}{\cos\theta + c_0}}
\right] + O(\eps^2) \nn\\
&=& \frac{1}{2\pi} + \frac{1}{2\pi N\beta} \cdot \log \left( -\frac{(\cos\theta - c_0)^2}{\cos^2\theta} \right) + O(\eps^2)~.
\end{eqnarray}

At this point, we make a remark on large imaginary parts of $N\beta$.
The `first sheets' in the last expressions of \eqref{app:2cutlowTnorm}
and of \eqref{app:2cutlowTrho} yield legitimate 2-cut saddles.
Then it seems as if ${\rm Im}\,N\beta$ is only allowed between $\pm 2\pi i$.
However, as we have suggested in the paragraph containing \eqref{app:2cutNbetak},
saddles for $N\beta$ differing by multiples of $2\pi i$ might be obtained with slight modification.
Specifically, we look for saddles for ${\rm Im}\,N\beta \notin (-2 \pi i, 2\pi i]$
that are continuously connected to those obtained from the first sheets.
Thus, consider fixing small $|\eps|$ and continuously rotating the phase of $\eps$
so that ${\rm Im}\,N\beta$ changes continuously with fixed ${\rm Re}\,N\beta$.
This allows us to go to the next sheet of \eqref{app:2cutlowTnorm}
where ${\rm Im}\,N\beta$ can be outside of the range $(-2 \pi i, 2\pi i]$.
For $\rho(\theta)$ to also vary continuously as the phase of $\eps$ is rotated,
one must take always the first sheet for the $\log$ but the new value of $N\beta$
in the second line of \eqref{app:2cutlowTrho}.
As we shall show explicitly (e.g. in Figure \ref{fig:2cut2}),
this indeed gives 2-cut saddles for ${\rm Im}\,N\beta \notin (-2 \pi i, 2\pi i]$
justifiable within perturbative orders of $(N\beta)^{-1}$.

This way of obtaining saddles for ${\rm Im}\,N\beta \notin (-2 \pi i, 2\pi i]$ is in fact
of the type of modification discussed between \eqref{app:2cutNbetak} and \eqref{app:2cutrhok}.
For $-k_0 = k_\infty = k$, \eqref{app:2cutNbetak} becomes
\begin{equation}\label{app:2cutlowTNbetak}
(N\beta)_{-k,k} = N\beta + 4 \pi ik~,
\end{equation}
and \eqref{app:2cutrhok} with the original $\rho_{0,0}$ given in \eqref{app:2cutlowTrho} becomes
\begin{eqnarray}\label{app:2cutlowTrhok}
\rho_{-k,k}(\theta) &=& \frac{N\beta}{N\beta + 4\pi ik} \cdot
\left( \rho_{0,0} (\theta) + \frac{2ik}{N\beta} \right) \nn\\
&=& \left( 1 - \frac{4\pi i k}{N\beta+4\pi ik} \right) \cdot \frac{1}{2\pi}
+ \frac{\log \left(-\frac{(\cos\theta - c_0)^2}{\cos^2\theta}\right) + 4\pi ik}
{2\pi(N\beta+4\pi ik)} + O(\eps^2) \nn\\
&=& \frac{1}{2\pi} + \frac{1}{2\pi (N\beta+4\pi ik)} \cdot
\log \left( -\frac{(\cos\theta - c_0)^2}{\cos^2\theta} \right) + O(\eps^2)~,
\end{eqnarray}
justifying the treatment of \eqref{app:2cutlowTrho} that only $N\beta$ is
replaced with $N\beta + 4\pi ik$.
With this discussion in mind, we now simply interpret \eqref{app:2cutlowTnorm} as
allowing arbitrary sheets for the logarithm,
thereby removing the restriction on ${\rm Im}\,N\beta$,
and take \eqref{app:2cutlowTrho} with its principal branch.

\begin{figure}[t]
\centering
\hspace{-0.03\textwidth}
\includegraphics[width=0.32\textwidth]{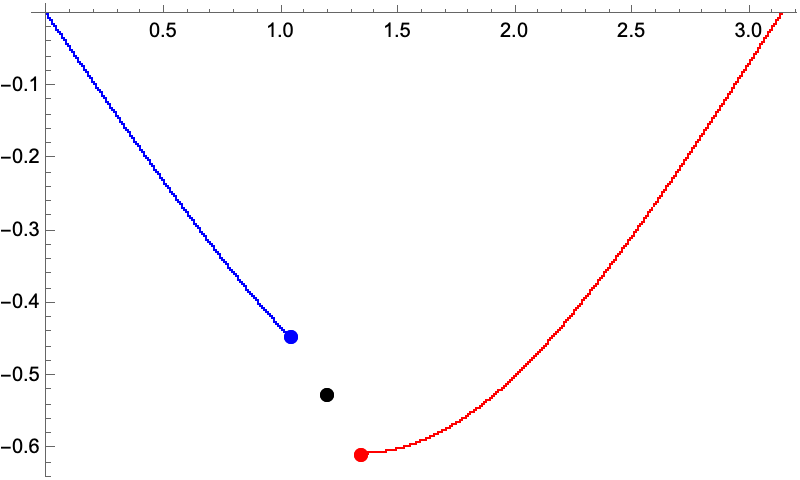}
\hspace{0.02\textwidth}
\includegraphics[width=0.32\textwidth]{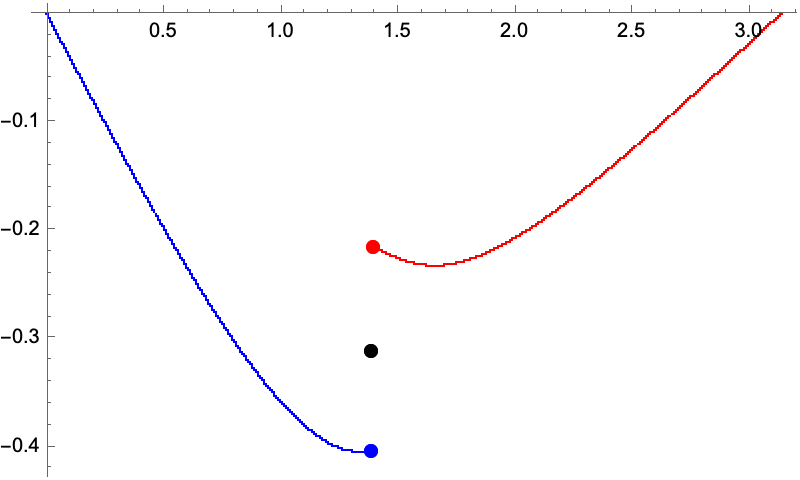}
\hspace{0.02\textwidth}
\includegraphics[width=0.32\textwidth]{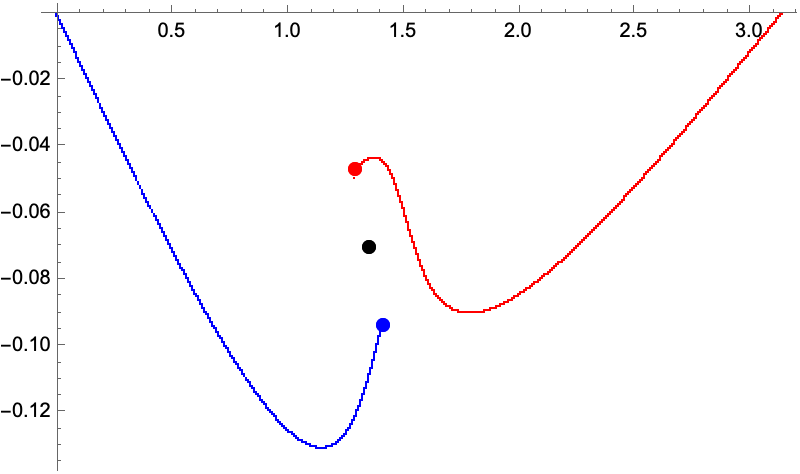}
\hspace{-0.03\textwidth}
\caption{\label{fig:spiral}
Two-cut saddles for $N\beta = 4$, $4-4i$ and $4-8i$ respectively,
with fixed filling fractions $(\nu_1,\nu_2) = (0.288,0.712)$,
numerically obtained with $N = 10^5$.
Left to right, the endpoints are rotating counter-clockwise,
but soon after $N\beta = 4-8i$ we do not find a continued two-cut saddle
as both cuts come close to each other rather than spiraling.
}\end{figure}

We admit that this treatment for arbitrary ${\rm Im}\,N\beta$ is justified only
within perturbative orders of $(N\beta)^{-1}$,
as opposed to the non-perturbative corrections $\eps \sim e^{-N\beta}$.
Note from \eqref{app:c1andc2} that the endpoints for both cuts, namely $\theta_{1,2}$,
will be roughly opposite to each other centered at $\cos^{-1} c_0$.
As ${\rm Im}\,N\beta$ is varied and $\eps$ rotates, the endpoints also rotate around $\cos^{-1} c_0$.
For $\rho(\theta)$ and thus the eigenvalue cuts to change continuously under this rotation,
the eigenvalue cuts would have to eventually spiral around $\cos^{-1} c_0$
in order to not intersect with each other, which sounds unrealistic.
In fact, for ${\rm Re}\,N\beta = 4$ for which $|e^{-N\beta/2}| = 0.135$ is small but not negligible,
we are able to find a two-cut saddle for some ${\rm Im}\,N\beta$ outside of the range
$(-2 \pi i, 2\pi i]$, but for more extreme values of ${\rm Im}\,N\beta$ we are not able to
find a two-cut saddle, see Figure \ref{fig:spiral}.
Such spiral effect is only visible at the non-perturbative level $\eps \sim e^{-N\beta}$,
because the distances between the two endpoints or from the center scale as $\eps$.
Thus, within perturbative orders of $(N\beta)^{-1}$ to which we will restrict in this subsection,
this issue is hidden and we are in fact able to obtain the two-cut saddles reliably for
fairly large $|{\rm Im}\,N\beta|$, as we shall show later in Figure \ref{fig:2cut2}.
We conjecture that in the latter case, the eigenvalue cuts do not change continuously
as $\eps$ rotates, in the non-perturbative order that we neglect.
That is, at some point as the endpoints rotate with negligible radius,
the exact cuts will jump from one cut being on top of the other cut
to it being at the bottom of the other cut.

Of two complex parameters $c_0$ and $\eps$ that describe the endpoints,
the latter is fixed via \eqref{app:2cutlowTnorm}
given the input $\beta$ for the matrix model.
It remains to determine complex $c_0$.
As we have explained on general ground, we use two separate real conditions to do so.
The first condition is that the filling fraction is real, namely
$\nu \equiv \int_{\calC_1} \rho(\theta)d\theta \in \mathbb{R}$.
Unless this condition is satisfied, $c_1$ cannot parametrize a true endpoint of an eigenvalue cut.
Because $\int_{\calC} \rho(\theta)d\theta = 1$ over both cuts is guaranteed,
the filling fraction over the second cut is simply $1-\nu$
so we do not impose any extra condition on it.
Using \eqref{app:2cutlowTrho}, $\nu$ is computed to be
\begin{equation}\label{app:nucond}
\nu = \frac{\theta_0}{\pi}
+ \frac{i}{\pi N\beta} \left( \pi \theta_0 + {\rm Li}_2(e^{2i\theta_0}) - {\rm Li}_2(-e^{2i\theta_0})
+ 2{\rm Li}_2(-e^{i\theta_0}) - 2{\rm Li}_2(e^{-i\theta_0}) + \frac{\pi^2}{4}\right) + O(\eps^2)~.
\end{equation}
The second condition is that the real parts of chemical potentials on both cuts are equal.
The definition of chemical potential, i.e. the amount of free energy that costs
to remove a particle from an ensemble, is straightforward from the action \eqref{app:ZandS}:
\begin{equation}\label{app:2cutmudef}
\mu(\theta) = \int_{\calC} d\theta' \rho(\theta') \log\left(4\sin^2\frac{\theta-\theta'}{2}\right)
+ \frac{1}{N} \sum_n \frac{g_n}{n} (e^{in\theta} + e^{-in\theta})~.
\end{equation}
The force-free equation is precisely about constancy of this chemical potential along a
continuous cut. Using this property, we only consider the difference of its values
at representative points of both cuts, namely $\theta = 0$ and $\theta = \pi$.
So we would like to compute $\Delta \mu \equiv \mu(0) - \mu(\pi)$:
\begin{equation}\label{app:mucond}
\Delta\mu = \frac{1}{2\pi N\beta} \cdot \int_{\calC} d\theta
\log \left(-\frac{(\cos\theta-c_0)^2}{\cos^2\theta} \right) \cdot \log\frac{1-\cos\theta}{1+\cos\theta}
+ \frac{8iG}{N\beta} + O(\eps^2)~.
\end{equation}
$G \approx 0.91597$ in the last term is the Catalan's constant,
but the first term remains unevaluated.

The simultaneous solution of ${\rm Im}\,\nu = 0$ and ${\rm Re}\,\Delta\mu = 0$
with respect to complex $c_0 = \cos \theta_0$,
cannot be directly and analytically obtained from \eqref{app:nucond} and \eqref{app:mucond}.
However, we find that the numerical solution exists close to
$c_0 = 0 ~\leftrightarrow~ \theta_0 = \frac{\pi}{2}$.
Under this behavior, the first cut ends at $\frac{\pi}{2}$ and the second starts at $\frac{\pi}{2}$,
making it look like a confined saddle with uniform distribution $\rho(\theta) = \frac{1}{2\pi}$
on the entire unit circle $\theta \in (-\pi, \pi]$.
Deviation from this limit is parametrized by $(N\beta)^{-1}$.
In fact, it will turn out that $c_0 = O(N\beta)^{-1}$,
so we expand the two equations in powers of $(N\beta)^{-1}$ assuming $c_0 \sim (N\beta)^{-1}$,
and try to solve perturbatively in $(N\beta)^{-1}$.
Note that we have been neglecting powers of $\eps \sim e^{-N\beta}$,
but we can still consistently expand in any desired powers of $N\beta \sim \log \eps^{-1}$.

First, let us expand \eqref{app:nucond} by substituting
$\theta_0 = \frac{\pi}{2} - c_0 - \frac16 c_0^3 + O(N\beta)^{-5}$. It gives
\begin{eqnarray}\label{app:nupert}
\nu &=& \frac12 + \frac{i\pi}{2N\beta}
- \frac{c_0}{\pi} - \frac{ic_0}{N\beta} - \frac{2c_0}{\pi N\beta} \left( 1 + \log\frac{2}{c_0} \right)
- \frac{c_0^3}{6\pi} + O(c_0^4)~.
\end{eqnarray}

Expanding in small $c_0 \sim (N\beta)^{-1}$ makes evaluation of \eqref{app:mucond} possible,
because we can then expand $\log \left(-\frac{(\cos\theta-c_0)^2}{\cos^2\theta} \right)$
in the integrand into polynomials (a similar comment to footnote \ref{footnote:c0expansion} applies).
For example, at the leading order of this expansion, the log (treated with the first sheet)
is $i\pi$ in the first cut and $-i\pi$ in the second, so one needs to evaluate
\begin{eqnarray}
\left[ \int_0^{\theta_0} - \int_{\theta_0}^\pi \right] \log\frac{1-\cos\theta}{1+\cos\theta} d\theta
&=& 2\theta_0 \log \tan^2\frac{\theta_0}{2} + 4i {\rm Li}_2 \left(i \tan \frac{\theta_0}{2} \right)
- 4i {\rm Li}_2 \left(-i \tan \frac{\theta_0}{2} \right) \nn\\
&=& -8G + 2c_0^2 + O(c_0)^4~.
\end{eqnarray}
The next orders involve (here, the two integration ranges $\int_0^{\theta_0}$
and $\int_{\theta_0}^{\pi}$ are merged)
\begin{eqnarray}
\int_0^\pi \left( - \frac{c_0}{\cos\theta} - \frac{c_0^2}{2\cos^2\theta}
- \frac{c_0^3}{3\cos^3\theta} + \cdots \right)
\log\frac{1-\cos\theta}{1+\cos\theta} d\theta
&=& c_\mu \left[c_0 + \frac{c_0^3}{6} + O(c_0)^4 \right]~,
\end{eqnarray}
where
\begin{eqnarray}\label{app:cmu}
c_\mu \!&=&\! -(\log 2)^2 + 3 \log 2 \cdot \log (2-\sqrt{2}) - 2 \log(-2+\sqrt{2}) \log(2-\sqrt{2})
+ 2 {\rm Li}_2(1) - {\rm Li}_2(2)\nn\\
&&\! - 4 {\rm Li}_2(-1-\sqrt{2}) + 2{\rm Li}_2(2-\sqrt{2})  + 2{\rm Li}_2(1-\sqrt{2}) - 2 {\rm Li}_2(-1+\sqrt{2}) + 2{\rm Li}_2(1+\sqrt{2}) \nn\\
\!&\approx&\! 9.8696\ .
\end{eqnarray}
Combining these results, we get
\begin{eqnarray}\label{app:mupert}
\Delta\mu &=& \frac{2 c_0}{N\beta} \cdot \left[
\frac{c_\mu}{\pi} + ic_0 + \frac{c_\mu}{6\pi} c_0^2 \right] + O(c_0^5)~.
\end{eqnarray}
Now with \eqref{app:nupert} and \eqref{app:mupert}, the two conditions
${\rm Im}\,\nu = 0$ and ${\rm Re}\,\Delta\mu = 0$ are solved by
\begin{equation}\label{app:2cutc0}
c_0 = \frac{i\pi^2}{(N\beta)^*} \cdot \left[ \frac12 -
\frac{1}{{\rm Re}\,(N\beta)} \cdot \left( 1 + \log \frac{4|N\beta|}{\pi^2} \right) \right]
- \frac{\pi^5}{4c_\mu |N\beta|^2} \cdot \frac{{\rm Im}\,(N\beta)}{{\rm Re}\,(N\beta)}
+ O(N\beta)^{-3}~.
\end{equation}
This equation is highly non-holomorphic in $N\beta$, highlighting the holomorphic anomaly
discussed in Section \ref{sec:lowTsaddle} as well as earlier in this subsection.

The two complex endpoint variables $c_0$ and $\eps$ are finally fixed
(up to truncations in $(N\beta)^{-1}$ that we have made)
by \eqref{app:2cutlowTnorm} and two more real conditions culminating in \eqref{app:2cutc0}.
Together with $\rho$ given by \eqref{app:2cutlowTrho},
we now have a complete description of the 2-cut saddle.
Before turning to the free energy, or the on-shell action evaluation,
we show examples in part to visualize the solutions as well as to
ascertain correctness of the branch choices made.

\begin{figure}[t]
\centering
\hspace{-0.03\textwidth}
\begin{subfigure}[t]{0.32\textwidth}
\includegraphics[width=\textwidth]{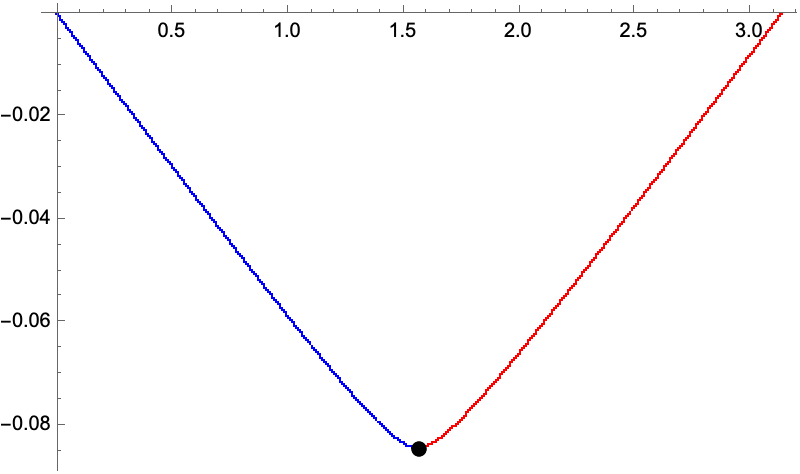}
\caption{}
\end{subfigure}
\hspace{0.02\textwidth}
\begin{subfigure}[t]{0.32\textwidth}
\includegraphics[width=\textwidth]{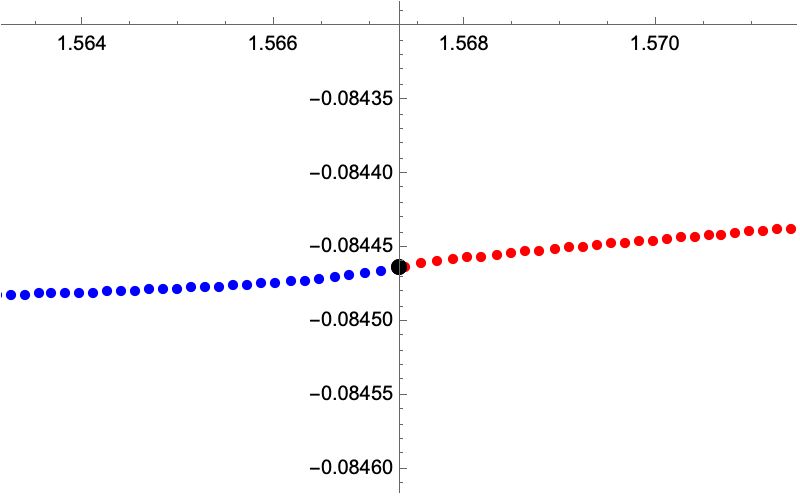}
\caption{}
\end{subfigure}
\hspace{0.02\textwidth}
\begin{subfigure}[t]{0.32\textwidth}
\includegraphics[width=\textwidth]{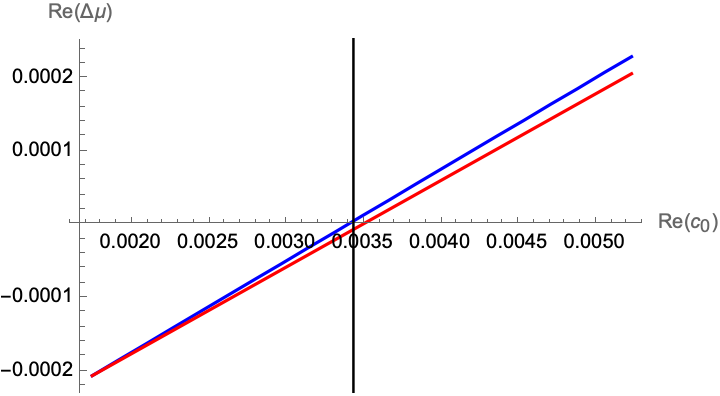}
\caption{}
\end{subfigure}
\hspace{-0.03\textwidth}
\caption{\label{fig:2cut1}
The 2-cut saddle obtained for $N\beta = 50-2\,i$ numerically ($N=50002$),
by solving ${\rm Im}\,\nu = 0$ and ${\rm Re}\,\Delta\mu = 0$ given
\eqref{app:nucond} and \eqref{app:mupert} which gives $c_0 = 0.00348787 + 0.08456350\,i$.
(a,b) The eigenvalue distribution drawn on the complex plane for $\theta$ with different scales,
blue and red representing the cuts centered around $0$ and $\pi$ respectively
and the black dot representing the endpoint $\theta_0 = \cos^{-1} c_0$.
(c) ${\rm Re}\,\Delta\mu$ computed for candidate saddles with different values of ${\rm Re}\,c_0$,
by evaluating numerically using discrete summation (blue) or by \eqref{app:mupert} (red).
Black vertical line marks the value of $c_0$ used for (a,b).}\end{figure}

Take for example $N\beta = 50 - 2\,i$.
We shall take negative imaginary parts for $N\beta$;
solutions for $N\beta$'s with positive imaginary parts are related by
\eqref{eff-action-pm} so they give the same physical results,
although some formulae need to be slightly modified due to branch issues.
$\eps$ is determined by \eqref{app:2cutlowTnorm} to be of order $10^{-11}$,
so it is very well justified to neglect powers of $\eps$.
Solving numerically two real equations ${\rm Im}\,\nu = 0$ with $\nu$ given in \eqref{app:nucond}
and ${\rm Re}\,\Delta\mu = 0$ with $\Delta\mu$ given in \eqref{app:mupert},
we obtain the complex value $c_0 = 0.00348787 + 0.08456350\,i$.
Note that this is slightly different from what \eqref{app:2cutc0} gives,
which is $c_{0,\eqref{app:2cutc0}} = 0.00343319 + 0.08273418\,i$,
with the difference $\Delta c_0 = 0.00005467 + 0.00182932\,i$.
This difference can be understood as the $O(N\beta)^{-3}$ correction in \eqref{app:2cutc0},
because we treat $N\beta$ as being of same order of magnitude as $|c_0| \sim 10^{-1}$.

We take the former value of $c_0 = 0.00348787 + 0.08456350\,i$,
because it should be more accurate given that \eqref{app:nucond} is exact to all orders of
$(N\beta)^{-1}$ and \eqref{app:mupert} is expanded up to higher order than \eqref{app:2cutc0}.
Then we find the eigenvalue cuts numerically with $N = 50002$,
see Figure \ref{fig:2cut1}(a) and (b).
That is, we assume an eigenvalue at $\theta = 0$ and at $\theta = \pi$ (thus $50000+2$)
and determine subsequent complex eigenvalues by requiring
$\rho(\theta)\cdot \Delta\theta = \frac{1}{N}$ between adjacent eigenvalues.
Once each sequence of eigenvalues coincides (within numerical tolerance set to $10^{-4}$ here)
with the expected endpoint $\cos^{-1}(c_0)$, the sequence is terminated.
Number of eigenvalues in each sequence it took to reach the endpoint determines the
filling fraction of the respective cut. They turn out to be
\begin{equation}
\nu_1 = \frac{24885}{50002} = 0.49768~, \qquad
\nu_2 = \frac{25117}{50002} = 0.50232~,
\end{equation}
The fact that both cuts indeed end up at the expected endpoint,
and that the two filling fractions add up to 1
(indeed, the two discrete numbers of eigenvalues add up to $N=50002$ exactly),
consist a highly non-trivial test for correctness of the solution and the branch choices made;
the first fact confirms that ${\rm Im}\,\nu = 0$ was solved correctly
and the second fact confirms the overall normalization that led to \eqref{app:2cutlowTnorm}.
One thing that remains worth checking is the equation ${\rm Re}\,\Delta\mu = 0$.
For this, we find similar saddles with different values of ${\rm Re}\,c_0$ around its correct value,
by determining ${\rm Im}\,c_0$ only using the condition ${\rm Im}\,\nu = 0$
which ensure that they are at least valid candidate saddles.
Then for each of these saddles, we compute $\Delta\mu$ in two ways;
first from the primitive definition \eqref{app:2cutmudef} by replacing the integral
as a discrete summation over $N$ eigenvalues,
second using the perturbative formula \eqref{app:mupert} for the respective value of $c_0$,
see Figure \ref{fig:2cut1}(c).
The two computations give similar results, adding the final touch of confidence.
Note that the scale of the imaginary axis is significantly smaller;
the eigenvalue distribution is close to the uniform confined saddle
which lies entirely on the real axis.

\begin{figure}[t]
\centering
\hspace{-0.03\textwidth}
\begin{subfigure}[t]{0.32\textwidth}
\includegraphics[width=\textwidth]{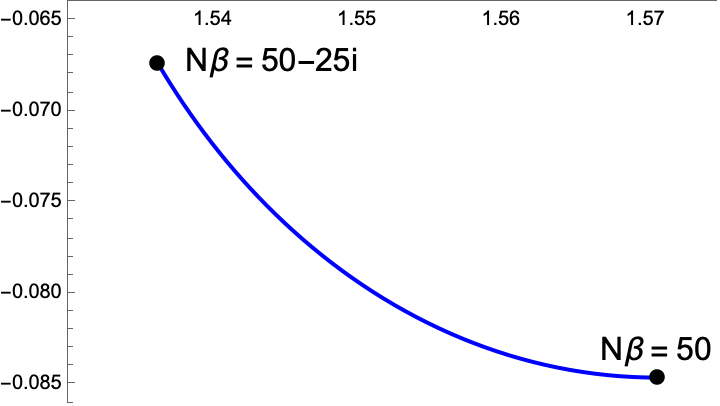}
\caption{}
\end{subfigure}
\hspace{0.02\textwidth}
\begin{subfigure}[t]{0.32\textwidth}
\includegraphics[width=\textwidth]{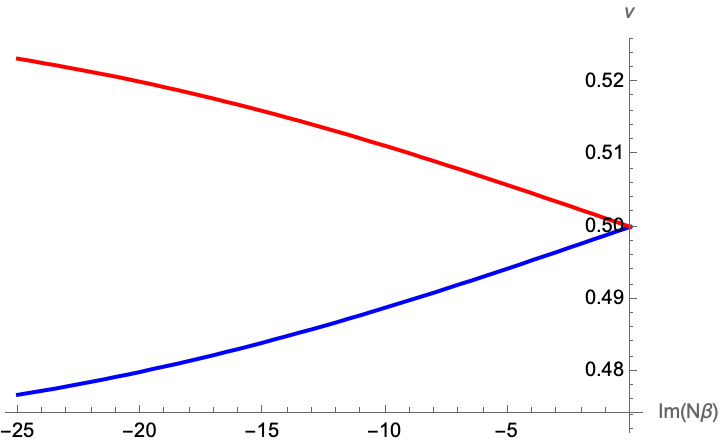}
\caption{}
\end{subfigure}
\hspace{0.02\textwidth}
\begin{subfigure}[t]{0.32\textwidth}
\includegraphics[width=\textwidth]{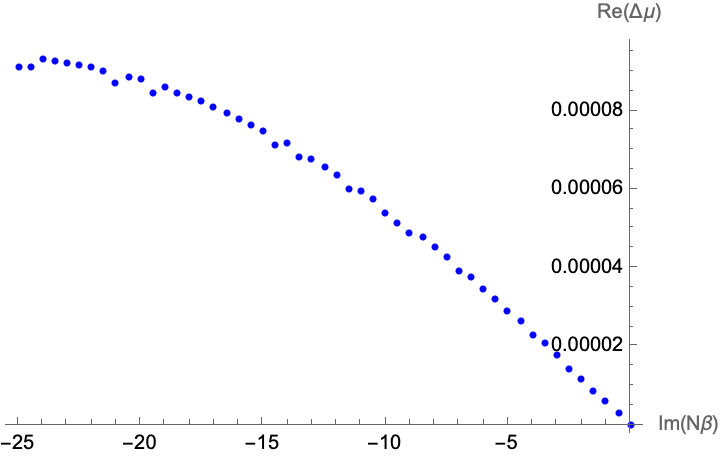}
\caption{}
\end{subfigure}
\hspace{-0.03\textwidth}
\caption{\label{fig:2cut2}
The 2-cut saddles for fixed ${\rm Re}\,N\beta = 50$ and varying ${\rm Im}\,N\beta \in [-25,0]$.
(a) The endpoint $\theta_0 = \cos^{-1} c_0$ determined by solving ${\rm Im}\,\nu = 0$
and ${\rm Re}\,\Delta\mu = 0$ given \eqref{app:nucond} and \eqref{app:mupert}.
(b) The filling fractions of both cuts and
(c) ${\rm Re}\,\Delta\mu$ evaluated numerically for each ${\rm Im}\,N\beta$.}
\end{figure}

To further verify our claim around \eqref{app:2cutlowTNbetak} regarding
addition of $4 \pi i$'s to $N\beta$, we repeat the exercise just described for
$N\beta = 50 + ({\rm Im}\,N\beta)\,i$ with various values of ${\rm Im}\,N\beta \in [-25,0]$.
With the treatment of large $|{\rm Im}\,N\beta|$ mentioned below \eqref{app:2cutlowTrhok},
we find a smooth series of 2-cut saddles with respective values of ${\rm Im}\,N\beta$.
As we depict in Figure \ref{fig:2cut2},
the value of $c_0$ moves continuously as ${\rm Im}\,N\beta$ changes,
but both cuts continue to keep end at $\cos^{-1} c_0$ safely
and the two filling fractions always add up to 1.
For each 2-cut saddle, ${\rm Re}\,\Delta\mu$ is evaluated numerically from \eqref{app:2cutmudef}
as a discrete summation over $N$ eigenvalues.
The values of ${\rm Re}\,\Delta\mu$ are continuous in ${\rm Im}\,N\beta$
(up to small fluctuations that can be accounted for by numerical errors),
and they stay very small $\lesssim 10^{-5}$.
We believe that this is sufficiently small to be argued as arising from the
$O(c_0^5)$ correction in \eqref{app:mupert} which was used to determine $c_0$,
as $|c_0| \sim 10^{-1}$ for all cases.

We end this subsection by computing the free energy or equivalently the on-shell action,
namely $S \sim - \log Z$ (see \eqref{app:ZandS}), for the 2-cut saddles.
We shall perform this computation up to certain orders of $(N\beta)^{-1}$,
so we truncate \eqref{app:2cutlowTrho} accordingly as
\begin{eqnarray}
\rho(\theta) &=&
\frac{1}{2\pi} \pm \frac{i}{2N\beta} - \frac{c_0}{\pi N\beta} \cdot \sec\theta + O(N\beta)^{-3}~,
\end{eqnarray}
where the $\pm$ sign applies to the first cut ($\calC_1$) and the second cut ($\calC_2$) respectively.

Similarly to what we did for the 1-cut saddles towards the end of the last subsection,
let us separate the effective action into $S_2$ involving the two-body interaction
and $S_1$ involving the potential. That is, rewriting \eqref{app:ZandS},
\begin{equation}
\frac{\log Z}{N^2} = -\frac{S[\rho(\theta)]}{N^2} =
-\frac{S_2[\rho(\theta)]}{N^2} -\frac{S_1[\rho(\theta)]}{N^2}~,
\end{equation}
where
\begin{eqnarray}
- \frac{S_2[\rho(\theta)]}{N^2} &=&
\frac12 \int d\theta_a d\theta_b \rho(\theta_a) \rho(\theta_b) \log \left( 4\sin^2 \frac{\theta_a-\theta_b}{2} \right)~, \nn\\
- \frac{S_1[\rho(\theta)]}{N^2} &=& \frac{1}{N} \sum \frac{g_n}{n} \int d\theta \rho(\theta) \left( e^{in\theta} + e^{-in\theta} \right)~.
\end{eqnarray}
First, we evaluate the two-body interaction term $S_2$.
Omitting terms of $O(N\beta)^{-3}$, we can write
\begin{eqnarray}
- \frac{S_2[\rho(\theta)]}{N^2}
&=& \frac12 \int_{-\theta_0}^{\theta_0} \!\!\!\! d\theta_a \int_{-\theta_0}^{\theta_0} \!\!\!\! d\theta_b
\left[ \left( \frac{1}{2\pi} + \frac{i}{2N\beta} \right)^2 - \frac{c_0(\sec \theta_a + \sec \theta_b)}{2\pi^2N\beta} \right]
\log 2(1-\cos (\theta_a-\theta_b)) \nn\\
&& + \frac12 \int_{\theta_0-\pi}^{\pi-\theta_0} \!\!\!\!\!\! d\theta_a \int_{\theta_0-\pi}^{\pi-\theta_0} \!\!\!\!\!\! d\theta_b
\left[ \left( \frac{1}{2\pi} - \frac{i}{2N\beta} \right)^2 + \frac{c_0(\sec \theta_a + \sec \theta_b)}{2\pi^2N\beta} \right]
\log 2(1-\cos (\theta_a-\theta_b)) \nn\\
&& + \int_{-\theta_0}^{\theta_0} \!\!\!\! d\theta_a\int_{\theta_0-\pi}^{\pi-\theta_0} \!\!\!\!\!\!\! d\theta_b
\left[ \frac{1}{(2\pi)^2} + \frac{1}{(2N\beta)^2} - \frac{c_0(\sec \theta_a - \sec \theta_b)}{2\pi^2N\beta} \right]
\log 2(1+\cos (\theta_a-\theta_b))~. \nn\\
\end{eqnarray}
One can show that all contributions from the terms involving $\sec$ cancel each other,
because these terms can be evaluated at their leading order in $c_0$,
for which $\theta_0 = \pi - \theta_0 = \frac{\pi}{2}$ for the integration range.
We are left with constants in the square brackets.
Then we need the following integrals:
(note that for the first integral, the integration range is the $\theta_a > \theta_b$
half of the square in the $(\theta_a,\theta_b)$-plane.)
\begin{eqnarray}
\int_{-\theta_0}^{\theta_0} d\theta_a \int_{-\theta_0}^{\theta_a} \!\!\!\! d\theta_b \log 2(1-\cos (\theta_a-\theta_b))
&=& \frac{4i\pi^3}{3} \cdot B_3 \left(\frac{\theta_0}{\pi} \right) + 2 {\rm Li}_3 (e^{-2i\theta_0}) - 2 \zeta(3) \nn\\
&=& {\rm Li}_3 (e^{2i\theta_0}) + {\rm Li}_3 (e^{-2i\theta_0}) - 2 \zeta(3)~, \nn\\
\int_{-\theta_0}^{\theta_0} \!\!\!\! d\theta_a \int_{\theta_0-\pi}^{\pi-\theta_0} \!\!\!\! d\theta_b \log 2(1+\cos (\theta_a-\theta_b))
&=& - 2 {\rm Li}_3 (e^{2i\theta_0}) - 2 {\rm Li}_3 (e^{-2i\theta_0}) + 4 \zeta(3)~,
\end{eqnarray}
where $B_3$ is the Bernoulli polynomial and we used
$ {\rm Li}_3 (e^{2\pi i x}) - {\rm Li}_3 (e^{-2\pi i x}) = - \frac{(2\pi i)^3}{6} \cdot B_3(x)$.
Then after simple algebra, we have
\begin{eqnarray}
- \frac{S_2[\rho(\theta)]}{N^2}
=\frac72 \zeta(3) \cdot \frac{1}{(N\beta)^2} + O(N\beta)^{-3}~.
\end{eqnarray}
Evaluation of the potential term $S_1$ is much easier.
Because there is an external factor of $(N\beta)^{-1}$ already, we only need the constant part of $\rho$:
($O(N\beta)^{-3}$ is omitted in intermediate expressions for brevity)
\begin{eqnarray}
- \frac{S_1[\rho(\theta)]}{N^2}
&=& \frac{4}{N\beta} \sum_{n:\text{ odd}}\frac{i^n}{n^2} \bigg[
\int_{-\theta_1}^{\theta_1} d\theta \left( \frac{1}{2\pi} + \frac{i}{2N\beta} + O(N\beta)^{-2} \right) \cos(n\theta) \nn\\
&& \hspace{3cm}
+ \int_{\theta_2}^{2\pi - \theta_2} d\theta \left( \frac{1}{2\pi} - \frac{i}{2N\beta} + O(N\beta)^{-2} \right) \cos(n\theta) \bigg] \nn\\
&=& \frac{2i}{(N\beta)^2} \sum_{n:\text{ odd}}\frac{i^n}{n^2}
\left[ \int_{-\pi/2}^{\pi/2} d\theta \cos(n\theta) - \int_{\pi/2}^{3\pi/2} d\theta \cos(n\theta) \right] +\mathcal{O}(N\beta)^{-3}\nn\\
&=& -\frac{8}{(N\beta)^2} \sum_{n:\text{ odd}}\frac{1}{n^3} +\mathcal{O}(N\beta)^{-3}
= -7\zeta(3) \cdot \frac{1}{(N\beta)^2} + O(N\beta)^{-3}~.
\end{eqnarray}
Combining the two terms, we obtain
\begin{equation}\label{app:2cutlogZ}
\frac{\log Z}{N^2}
= - \frac{S_2[\rho(\theta)] + S_1[\rho(\theta)]}{N^2}
= - \frac72 \zeta(3) \cdot \frac{1}{(N\beta)^2} + O(N\beta)^{-3}~.
\end{equation}
%

\section{Free partition function}\label{sec:appC}

In this appendix, we construct the gapped saddle point solutions of the free $U(N)$ vector model partition function  in the large $N$ high temperature scaling limit. (The gapless 
solutions of this model are studied in \cite{Shenker:2011zf}.)
A purpose of this section is to illustrate that the methods used in this paper for the index 
extend to the partition functions. We believe that the same techniques 
will be applicable, to certain extent, to the interacting vector model partition function.

The partition function is given by
\begin{equation} \label{Shenker:Z}
    Z(N,\beta ) = \frac{1}{N!} \int \prod_i d \alpha_i \exp \left[ \sum_{i<j} 2 \ln \bigg|  2 \sin \frac{\alpha_i - \alpha_j}{2} \bigg|    + 2 N_f \sum_{m=1}^\infty \frac{1}{m}z_S (x^m) \sum_i \cos (m\alpha_i )      \right]~,
\end{equation}
where $z_S(x) = x^{\frac{1}{2} } \frac{1+x}{(1-x)^{2}}$ is the letter partition function, 
and $N_f$ is the number of fundamental scalar fields.
In the $N\to \infty$ limit with $\beta\sim N^{-\frac{1}{2}}\rightarrow 0$ (where $x=e^{-\beta}$), 
the partition function (\ref{Shenker:Z}) and the chemical potential 
can be written in terms of the eigenvalue density $\rho(\theta)$ as
\begin{align}\label{action-chemical}
    \log Z &=  N^2 \int d \theta_1 d \theta_2 \rho(\theta_1) \rho (\theta_2) \ln \bigg|  2 \sin  \frac{\theta_1- \theta_2 }{2}  \bigg| + \frac{2N_f N}{\beta^2} \int d\theta \rho(\theta)   \left( {\rm Li}_3 (e^{i\theta}) +  {\rm Li}_3 ( e^{-i\theta} )  \right)  \ , \nn \\
    \mu& \equiv \mu(\alpha) =  2 \int d\theta \rho (\theta ) \ln \bigg| 2 \sin \frac{\alpha- \theta}{2}   \bigg| + \frac{2N_f}{N \beta^2} \left(  {\rm Li}_3 (e^{i\alpha}) + {\rm Li}_3 (e^{-i\alpha})   \right)  \ .
\end{align}
The saddle point equation is given by
\begin{equation} \label{Shenker:eom}
    0 = \int d\theta \rho (\theta) \cot \left( \frac{\alpha- \theta}{2}   \right) + \frac{2 N_f }{N\beta^2} \left( i\, {\rm Li}_2( e^{i\alpha} ) -i\, {\rm Li}_2(e^{-i\alpha})    \right)  \ .
\end{equation}
If $\rho(\theta)$ satisfies this equation, the chemical potential $\mu$ 
does not depend on $\alpha\in [-\theta_0,\theta_0]$.

Again employing the general results of \cite{Aharony:2003sx},
the gapped solution for $\rho(\theta)$ is given by
\begin{align} \label{Shenker:rhoformula}
    \rho (\theta ) &=  \frac{1}{\pi} \sqrt{\sin^2 \frac{\theta_0}{2} - \sin^2 \frac{\theta}{2} } \sum_{n=1}^\infty  Q_n \cos \left[ \left( n-\frac{1}{2} \right) \theta   \right]  \ , \nn \\
    Q_n &=  \frac{2N_f}{N\beta^2} \sum_{l=0}^\infty \frac{2}{(n+l)^2} P_l (c_0)  \ \ (n \neq 0) \ ,  \nn \\
    Q_0 &= \frac{2N_f}{N\beta^2} \sum_{l=1}^\infty \frac{2}{l^2} P_l(c_0) \ , \ \ \ \  \ Q_1-Q_0=2  \ , 
\end{align}
where $c_0= \cos \theta_0$, and $\pm\theta_0$ are endpoints of the eigenvalue cut. 
For further calculations, we define $Q_n(z)$ with an auxiliary variable $z$ as
\begin{align}
    Q_n(z) &= \frac{2 N_f}{N \beta^2} \sum_{l=0}^\infty \frac{2 z^{n+l}}{(n+l)^2} P_l(c_0)  \ , 
    \qquad (n\geq 1) \nn\\
    Q_0(z) &= \frac{2N_f}{N \beta^2} \sum_{l=1}^\infty \frac{2z^l}{l^2}P_l(c_0)  \ .
\end{align}
Note that $Q_n(1)=Q_n$. One finds the following closed-form 
expressions for the second logarithmic derivatives of $Q_n(z)$:
\begin{align} \label{Shenker:twoderiQ}
    \left( z \frac{d}{dz} \right)^2 Q_n(z) &=  \frac{2N_f}{N\beta^2} \sum_{l=0}^\infty 2 z^{n+l} P_l(c_0)  =  \frac{4N_f}{N\beta^2} \frac{z^n}{\sqrt{1-2c_0z +z^2}}  \ ,    \nn\\
    \left( z \frac{d}{dz} \right)^2 Q_0(z) &= \frac{2N_f}{N \beta^2} \sum_{l=1}^\infty 2 z^l P_l (c_0) =  \frac{4N_f}{N\beta^2} \left( \frac{1}{\sqrt{1-2c_0z +z^2}} - 1   \right) \ .
\end{align}

We first calculate the relation between $\gamma= \frac{N\beta^2}{N_f}$ 
and $c_0$, from the condition $Q_1-Q_0=2$. From (\ref{Shenker:twoderiQ}), one obtains
\begin{align}\label{Q1-Q0-der}
    \left( z \frac{d}{dz} \right) \left( Q_1(z)-Q_0(z) \right) &=  \frac{8N_f}{N\beta^2} \log \left[  \frac{c_0+1}{2} \cdot \frac{1-z+\sqrt{1-2c_0z +z^2}}{c_0 - z + \sqrt{1- 2c_0 z +z^2}}   \right]~,
\end{align}
by integrating $(z\frac{d}{dz})^2(Q_1(z)-Q_0(z))$ once. 
Further integrating both sides of (\ref{Q1-Q0-der}) with $\int_0^1\frac{dz}{z}$ and recalling 
that $Q_1-Q_0=2$, $Q_1(0)-Q_0(0)=0$, one obtains 
\begin{align}\label{gamma-integral}
    \frac{N\beta^2 }{N_f} &= 4 \int_0^1 \frac{dz}{z}   \log \left[  \frac{c_0+1}{2} \cdot \frac{1-z+\sqrt{1-2c_0z +z^2}}{c_0 - z + \sqrt{1-
    2c_0 z +z^2}}   \right]   \ . 
\end{align}
This gives an expression for $\gamma(c_0)=\frac{N\beta^2}{N_f}$ by an integral. 
To evaluate it, one first differentiates (\ref{gamma-integral}) with $c_0$ and then 
integrates in $z$ to obtain 
\begin{eqnarray}
    \frac{d \gamma (c_0)}{dc_0} &=& 
    4 \int_0^1 \frac{dz}{z}  \frac{d}{dc_0} \log \left[ 
    \frac{ (c_0+1)(1-z+\sqrt{1-2c_0z+z^2})}{2(c_0 -z +\sqrt{1-2c_0z +z^2}) } \right]
    \\
    &=& 4 \left[ \frac{\log(1-z+\sqrt{1-2c_0 z+z^2})}{1+c_0}  \right]_{z=0}^{z=1}
    =\frac{2 \log \left( \frac{1-c_0}{2} \right)}{1+c_0} \ .  
    \nonumber
\end{eqnarray}
After integrating this with respect to $c_0$ and demanding $\gamma(1)=0$ (i.e. the 
cut shrinks, $\theta_0\rightarrow 0$, in the high temperature limit $\gamma\rightarrow 0$), 
one obtains
\begin{align} \label{Shenker:gamma}
    \gamma(c_0) &=  -2\, {\rm Li}_2 \left( \cos^2 \frac{\theta_0}{2} \right) + \frac{\pi^2}{3} \ .
\end{align}
This expression relates the `inverse temperature' $\gamma$ and the endpoint $\theta_0$ of the cut.

To compute $\rho(\theta)$, we define
\begin{equation}
    f(\theta,z) \equiv \sum_{n=1}^\infty  Q_n (z) \cos \left[ \left( n-\frac{1}{2} \right) \theta   \right] \ ,
\end{equation}
which from (\ref{Shenker:rhoformula}) is related to $\rho(\theta)$ by
\begin{align} \label{Shenker:rhoandf}
    \rho(\theta) = \frac{1}{\pi} \sqrt{\sin^2 \frac{\theta_0}{2}  - \sin^2 \frac{\theta}{2} }  \ f(\theta,z=1)  \ .
\end{align}
We first explicitly evaluate its second derivative using (\ref{Shenker:twoderiQ}): 
\begin{align}
    \left( z \frac{d}{dz} \right)^2 f( \theta, z ) &= \sum_{n=1}^\infty \frac{4N_f}{N\beta^2} \frac{z^n}{\sqrt{1-2c_0 z + z^2}}  \ \cos \left[ \left( n - \frac12 \right) \theta \right] \nn  \\
    &= \frac{4N_f}{N \beta^2} \cos \left( \frac{\theta}{2}   \right)  \cdot \frac{z(1-z)}{\sqrt{1-2c_0 z + z^2} \  (1-2\cos \theta z + z^2)}  \ .   
\end{align}
Computing its logarithmic integral and using 
$\frac{zdf}{dz}(\theta,0)=0$, one obtains 
\begin{align} \label{Shenker:1devf}
    \left( z \frac{d}{dz} \right) f(\theta, z)  &= \frac{4 N_f}{N \beta^2}  \cos \left( \frac{\theta}{2} \right) 
    \int_0^z dz^\prime \  \frac{1-z^\prime}{ \sqrt{1-2c_0z^\prime +z^{\prime 2}} 
    (1-2\cos \theta z^\prime +z^{\prime 2})}  \nn\\ 
    &= \frac{2}{\gamma} \cdot \frac{-2 e^{i\theta/2}}{\sqrt{-1 + 2c_0 e^{i\theta} -e^{2i\theta}}} \bigg(  \tan^{-1} \left[ \frac{1+ e^{i\theta} (-z+\sqrt{1-2c_0  z+ z^2})}{ \sqrt{-1+ 2c_0 e^{i\theta} -e^{2i\theta}}} \right] \\
    &\ \ \ \ \  +  \tan^{-1} \left[ \frac{e ^{i\theta} -z+\sqrt{1-2c_0  z+ z^2}}{ \sqrt{-1+ 2c_0 e^{i\theta} -e^{2i\theta}}} \right]   -2 \tan^{-1}  \left[ \frac{e^{i\theta}+1}{\sqrt{-1+2 c_0 e^{i\theta } -e^{2i\theta }}} \right] \bigg) \ .\nn 
\end{align}

Before considering its logarithmic integration once more, note that we know the explicit form 
of $f(\theta,z=1)$ when $\theta_0=\pi$, because this is the phase transition point 
at which the gap closes. $\rho(\theta)$ at this point is known as a limit of the 
gapless solution of  \cite{Shenker:2011zf}. In fact by inserting $c_0=-1$ to 
the first line of (\ref{Shenker:1devf}) (at $\gamma=\frac{\pi^2}{3}$), 
the $z^\prime$ integration can be performed explicitly to obtain $\frac{zdf}{dz}(\theta,z)$. 
Integrating it once more, one obtains the following  expression at $\theta_0=\pi$: 
\begin{align}\label{f-theta0-pi}
    f(\theta, z=1) &= -\frac{3}{ \pi^2  }\sec \left( \frac{\theta}{2} \right) \int_0^1 \frac{dz}{z} \left[ -2 \log (1+z)  + \log (1-2z\cos \theta +z^2)   \right ] \nn\\
    &= \frac{3}{\pi^2} \sec \left( \frac{\theta}{2} \right)  \left(  \frac{\pi^2}{6}   + {\rm Li}_2 \left( e^{i\theta} \right) + {\rm Li}_2 \left( e^{-i\theta} \right)  \right) \ .
\end{align}
Using (\ref{Shenker:rhoandf}), $\rho(\theta)$ at the phase transition point is given by
\begin{align} \label{Shenker:transrho}
     \rho(\theta)  &= \frac{1}{2\pi} + \frac{3}{\pi^3} \left( {\rm Li}_2 \left( e^{i\theta} \right) + {\rm Li}_2 \left( e^{-i\theta} \right)  \right)  \ , 
\end{align}
which agrees with \cite{Shenker:2011zf}.

Now we calculate $\rho(\theta)$ at general $\theta_0$. 
From (\ref{Shenker:rhoandf}) with $f(\theta,1)$ given by the logarithmic integral $\int_0^1 \frac{dz}{z}$
of (\ref{Shenker:1devf}), one obtains
\begin{align}\label{rho-partition-integral}
    \rho(\theta) &= \frac{2i}{\pi \gamma}   \int_0^1 \frac{dz}{z} \bigg(  \tan^{-1} \left[ \frac{1+ e^{i\theta} (-z+\sqrt{1-2c_0  z+ z^2})}{ \sqrt{-1+ 2c_0 e^{i\theta} -e^{2i\theta}}} \right] \nn \\
    &\ \ \ \ \ \  +  \tan^{-1} \left[ \frac{e ^{i\theta} -z+\sqrt{1-2c_0  z+ z^2}}{ \sqrt{-1+ 2c_0 e^{i\theta} -e^{2i\theta}}} \right]  -2 \tan^{-1}  \left[ \frac{e^{i\theta}+1}{\sqrt{-1+2 c_0 e^{i\theta } -e^{2i\theta }}} \right] \bigg) \ .
\end{align}
To compute the last integral easily, we define $g(c_0, \theta)$ by
\begin{align} \label{Shenker:defofg}
    \rho(\theta) = \frac{g(c_0,\theta)}{\gamma} \ .
\end{align}
$g(c_0,\theta)$ is given by the integral expression, (\ref{rho-partition-integral}) time $\gamma$.
After taking a $c_0$ derivative of $g(c_0,\theta)$, this integral over $z$ can be explicitly done 
and one obtains
\begin{align}\label{dg/dc0}
    \frac{dg(c_0,\theta)}{dc_0}  &= \frac{2i}{\pi}   \int_0^1 \frac{dz}{z} \frac{d}{dc_0} \bigg(  \tan^{-1} \left[ \frac{1+ e^{i\theta} (-z+\sqrt{1-2c_0  z+ z^2})}{ \sqrt{-1+ 2c_0 e^{i\theta} -e^{2i\theta}}} \right] \nn \\
    & \qquad  +  \tan^{-1} \left[ \frac{e ^{i\theta} -z+\sqrt{1-2c_0  z+ z^2}}{ \sqrt{-1+ 2c_0 e^{i\theta} -e^{2i\theta}}} \right]  -2 \tan^{-1}  \left[ \frac{e^{i\theta}+1}{\sqrt{-1+2 c_0 e^{i\theta } -e^{2i\theta }}} \right] \bigg)  \nn \\
    &= \frac{2 \log \left[ \frac{1-c_0}{2} \right]}{1+c_0} \cdot \frac{\cos \frac{\theta}{2}}{\pi \sqrt{2 \cos \theta- 2c_0}} \ .
\end{align}
Integrating this in $c_0$, one obtains 
\begin{align}\label{g-C}
    g(c_0 , \theta) &= \frac{1}{\pi} \bigg( {\rm Li}_2 \left[ - \frac{\sqrt{1+ \cos \theta} + \sqrt{-c_0 + \cos \theta}}{\sqrt{-1+ \cos \theta} - \sqrt{1+ \cos \theta }}  \right]   + {\rm Li}_2 \left[  \frac{\sqrt{1+\cos \theta} + \sqrt{-c_0 + \cos \theta}}{\sqrt{-1+ \cos \theta} + \sqrt{1+\cos \theta }}  \right]  \\
    & \quad - {\rm Li}_2 \left[  \frac{\sqrt{1+ \cos \theta} - \sqrt{-c_0 + \cos \theta}}{\sqrt{-1+\cos \theta} + \sqrt{1+\cos \theta} }  \right]   - {\rm Li}_2 \left[  \frac{-\sqrt{1+\cos \theta} + \sqrt{-c_0  + \cos \theta} }{\sqrt{-1+\cos \theta} - \sqrt{1+\cos \theta } }  \right]  \bigg)  + C(\theta) \nn  
\end{align}
where $C(\theta)$ is an integral constant. One finds $C(\theta)=0$ by comparing with (\ref{Shenker:transrho}) at $\theta_0=\pi$. The final expression for $\rho(\theta)$ is
\begin{align}\label{rho-final}
    \rho(\theta) &= \frac{1}{\pi \gamma} \bigg( {\rm Li}_2 \left[  \frac{\sqrt{1+ \cos \theta} + \sqrt{-c_0 + \cos \theta}}{   \sqrt{1+ \cos \theta } - \sqrt{-1+ \cos \theta} }  \right]   + {\rm Li}_2 \left[  \frac{\sqrt{1+\cos \theta} + \sqrt{-c_0 + \cos \theta}}{ \sqrt{1+ \cos \theta } + \sqrt{-1+ \cos \theta}}  \right] \nn\\
    & \quad - {\rm Li}_2 \left[  \frac{\sqrt{1+ \cos \theta} - \sqrt{-c_0 + \cos \theta}}{ \sqrt{1+ \cos \theta } + \sqrt{-1+ \cos \theta} }  \right]   - {\rm Li}_2 \left[  \frac{\sqrt{1+\cos \theta} - \sqrt{-c_0  + \cos \theta} }{\sqrt{1+ \cos \theta } - \sqrt{-1+ \cos \theta}}  \right]  \bigg) \ .
\end{align} 
Individual terms on the right hand side are complex, due to 
$\sqrt{-1+\cos\theta}=\pm i\sqrt{1-\cos\theta}$ in the argument of ${\rm Li}_2$, 
but they combine to yield real $\rho(\theta)$ on the real cut $\theta \in [-\theta_0, \theta_0 ]$. 
Although we did not care much about the reality of functions at all intermediate steps, 
it is clear how to ensure the reality from the the complex conjugate pairs
appearing in (\ref{rho-final}). The first and second terms in the parenthesis $(\ )$ are 
conjugate to each other by taking $\mp\sqrt{-1+\cos\theta}\rightarrow \mp i\sqrt{1-\cos\theta}$.
Similarly, the third and fourth terms are conjugate. This leads to 
\begin{align}
    \rho(\theta) &= \frac{2}{\pi \gamma} {\rm Re} \left[  {\rm Li}_2 \left(  \frac{\sqrt{1+ \cos \theta} + \sqrt{ -c_0 + \cos \theta } }{   \sqrt{1+ \cos \theta } + i \sqrt{1 - \cos \theta} }  \right) - {\rm Li}_2 \left(  \frac{\sqrt{1+ \cos \theta} - \sqrt{-c_0 + \cos \theta}}{ \sqrt{1+ \cos \theta } + i \sqrt{1-  \cos \theta} }  \right)  \right]  \ .
\end{align}
Note that the argument of the second ${\rm Li}_2$ function is always smaller than $1$, so 
it is given by the Taylor expansion ${\rm Li}_2(x)=\sum_{n=1}^\infty\frac{x^n}{n^2}$ within 
its radius of convergence. The argument of the first ${\rm Li}_2$ is smaller than $1$ 
at $\theta=\pm \theta_0$, admitting the Taylor expansion, but continuously changes and 
becomes larger than $1$ near $\theta=0$. However, the argument never hits the branch point $x=1$ 
so that the first term can be analytically continued without any ambiguity.

\begin{figure}[t]
	\centering
	\includegraphics[width=0.6\textwidth]{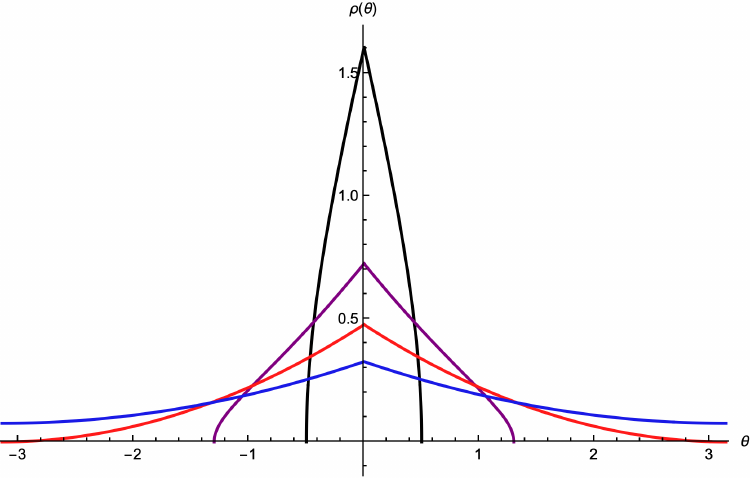}
	\caption{ Plots of eigenvalue density $\rho(\theta)$ at various temperatures. 
    The red curve is at the phase transition temperature, $\gamma=\frac{\pi^2}{3}=3.28987$. 
    The blue $\left( \gamma=6.25 \right)$ curve for a lower temperature, which exhibits no gap. 
    The purple $\left(\gamma=1.7297 \right)$ and black $\left( \gamma=0.477226  \right)$ curves 
    are gapped solutions at higher temperatures. }
	\label{Shenker:rhoplot}
\end{figure}
Fig. \ref{Shenker:rhoplot} shows the gapped distribution of this $\rho(\theta)$ at various 
temperatures. To compare, we also showed a gapless solution below the critical temperature 
(blue curve), given by \cite{Shenker:2011zf}
\begin{align}
    \rho(\theta) &= \frac{1}{2\pi} + \frac{1}{\pi \gamma} \left(  {\rm Li}_2 (e^{i\theta}) + {\rm Li}_2 (e^{-i\theta})  \right) = \frac{1}{2\pi}+\frac{1}{\pi\gamma}
    \left[-\frac{\pi^2}{6}+\frac{1}{2}(|\theta|-\pi)^2\right]  \ . 
\end{align}

Now we compute the free energy $\log Z$ for the gapless solutions. We first define
\begin{equation}
    f(\gamma) = \frac{\log Z}{N^2} \gamma^2 
    = \frac12 \int d\theta_1 d\theta_2 g(\theta_1)g(\theta_2) \log \left[4  \sin^2 \frac{\theta_1- \theta_2}{2}  \right]  +2 \int d\theta   g(\theta) \left[ {\rm Li}_3(e^{i\theta}) + {\rm Li}_3(e^{-i\theta})  \right]  
\end{equation}
where $g(\theta)$ denotes $g(c_0,\theta)$ of (\ref{Shenker:defofg}). 
First note that, from (\ref{g-C}) with $C(\theta=0)$, one finds 
\begin{equation}\label{g-1}
  g(\pm\theta_0)=0
\end{equation}
at the endpoints. With this and $\int_{-\theta_0}^{\theta_0} d\theta \rho(\theta)=1$, one also obtains
\begin{align}\label{g-2}
    \int_{-\theta_0}^{\theta_0} d\theta g(\theta) = \gamma \ \to \ \int_{-\theta_0}^{\theta_0} d\theta \frac{dg(\theta)}{d\gamma} =1 \ .
\end{align}
Using these properties, one obtains 
\begin{equation} \label{Shenker:devoff}
    \frac{d f(\gamma)}{d \gamma}=
    \int_{-\theta_0}^{\theta_0} d\theta_1\frac{dg(\theta_1)}{d\gamma}\left\{
    \int_{-\theta_0}^{\theta_0} 
    d\theta_2 g(\theta_2) \log \left[ 4 \sin^2 \frac{\theta_1-\theta_2}{2} \right] 
    + 2 \left[ {\rm Li}_3(e^{i\theta_1}) + {\rm Li}_3(e^{-i\theta_1})  \right] \right\}
\end{equation}
where we used (\ref{g-1}). Now we note that the expression in the 
curly bracket is related to the chemical potential (\ref{action-chemical}) by 
\begin{equation}\label{integral-mu}
  \mu\gamma=\int_{-\theta_0}^{\theta_0} 
    d\theta_2 g(\theta_2) \log \left[ 4 \sin^2 \frac{\theta_1-\theta_2}{2} \right] 
    + 2 \left[ {\rm Li}_3(e^{i\theta_1}) + {\rm Li}_3(e^{-i\theta_1})  \right]\ .
\end{equation}
Further noting that $\mu$ is $\theta_1$-independent 
at the saddle point and also using (\ref{g-2}), one obtains 
\begin{equation}\label{f-mu}
  \frac{df(\gamma)}{d\gamma}=\mu\gamma\ .
\end{equation}

Differentiating both sides of (\ref{integral-mu}) with $t=\cos^2 \frac{\theta_0}{2}$ 
(at $\theta_1=0$ for the RHS) and using (\ref{dg/dc0}) for $\frac{dg}{dc_0}=2\frac{dg}{dt}$, 
one obtains
\begin{equation}
  \frac{d(\mu\gamma)}{dt}=\int_{-\theta_0}^{\theta_0}d\theta \frac{dg(\theta)}{dt}
  \log\left[4\sin^2\frac{\theta}{2}\right]=\frac{2}{t} \left( \log [1-t] \right) ^2 \ .
\end{equation}
By Taylor-expanding the RHS and integrating it in $t$, 
one obtains
\begin{align}
    \mu \gamma= 2 \sum_{n,m=1}^\infty \frac{t^{n+m}}{(n+m)nm} +0 \ . 
\end{align}
Here we fixed the integral constant to $0$ using its value known at $c_0=-1$ (i.e. $t=0$). 
Using this expression for $\mu \gamma$ and (\ref{Shenker:gamma}), 
(\ref{f-mu}) can be rewritten as
\begin{equation}
    \frac{df}{dt}=\frac{d\gamma}{dt} \frac{df}{d\gamma}
    = \frac{2}{t} \log \left[  1-t \right]\cdot(\mu \gamma) 
    =- 4 \sum_{k,n,m=1}^\infty \frac{t^{n+m+k-1}}{(n+m)nmk} \ .
\end{equation}
By integrating this again, one obtains
\begin{align}
    f &=  -4 \sum_{k,n,m=1}^\infty \frac{t^{n+m+k}}{(n+m+k)(n+m)nmk}  + 4 \zeta (5) \nn\\
    &=  -4 \sum_{k,n,m=1}^\infty \frac{t^{n+m+k}}{(n+m+k)^2 (n+m)^2} \left( \frac{1}{n} + \frac{1}{m}  \right)     -4 \sum_{k,n,m=1}^\infty \frac{t^{n+m+k}}{(n+m+k)^2 nmk}  + 4 \zeta (5)   \nn \\
    &=  -8 \sum_{k,n,m=1}^\infty \frac{t^{n+m+k}}{(n+m+k)^2 (n+m)^2 n}  -4 \sum_{k,n,m=1}^\infty \frac{t^{n+m+k}}{(n+m+k)^2 nmk} +4 \zeta(5) \nn \\
    &= -8 \sum_{k>n>m>0} \frac{t^{k}}{k^2 n^2 m} -4 \sum_{k,n,m=1}^\infty \frac{t^{n+m+k}}{(n+m+k)^2 nmk} + 4 \zeta (5)   \nn \\
    &= -8 \ {\rm HPL} (2,2,1 ;t) - 24 S_{2,3}(t)  + 4 \zeta (5)
\end{align}
where $4\zeta(5)$ on the first line is chosen by the known value at $c_0=-1$ ($t=0$), 
and the harmonic polylogarithm (HPL) and Nielsen generalized polylogarithm $S_{a,b}(x)$ are defined by
\begin{eqnarray}
    {\rm HPL}(n_1,n_2, \cdots , n_k;x)& =& \sum_{m_1>m_2>\cdots >m_k>0} \frac{x^{m_1}}{m_1^{n_1} m_2^{n_2} \cdots m_k^{n_k}} \nn\\
    S_{a,b} (x) & =  &\frac{1}{b!} \sum_{n_1,n_2, \cdots, n_b=1}^\infty \frac{x^{n_1+n_2+\cdots +n_b}}{(n_1+n_2+\cdots +n_b)^a \ n_1 n_2 \cdots n_b}  \ .
\end{eqnarray}
Putting all together, one obtains 
\begin{align}
    \frac{\log Z}{N^2} = \frac{f}{\gamma^2} = \frac{-8 \  {\rm HPL} [2,2,1;\cos^2 \frac{\theta_0}{2}]  - 24 S_{2,3} \left( \cos^2 \frac{\theta_0}{2} \right) +4 \zeta(5)  }{  \left(  -2 {\rm Li}_2 \left(  \cos^2 \frac{\theta_0}{2}  \right) + \frac{\pi^2}{3}  \right)^2  } \ .
\end{align}

We expand $\log Z$ of our gapped saddle and the gapless saddle at the transition point, 
$T_c=\sqrt{\frac{3N}{\pi^2 N_f}}$ (i.e. $\gamma_{c}=\frac{\pi^2}{3}$), and obtain
\begin{equation}
  \frac{\log Z_{\rm gapped}-\log Z_{\rm ungapped}}{N^2}=
  -\frac{4\pi^5}{81}\left(\frac{N_f^3}{3N^3}\right)^{\frac{1}{2}}(T-T_c)^3+\cdots\ .
\end{equation}
This shows that the phase transition is of third order, as expected.

\end{document}